\documentclass[11pt]{cernrep}
\usepackage{graphicx}
\usepackage{cite}
\usepackage{epsfig}
\usepackage{amsmath,amssymb}
\begin{document}

\title{Higgs Working Group Summary Report }

\author{
{\underline{Convenors}}: 
S.~ Dawson$^1$, 
M.~ Grazzini$^2$, 
A.~ Nikitenko$^{3,a}$, 
M.~ Schumacher$^4$\\
\vskip .3in
\underline{Contributers}: 
N. E.~ Adam$^{5}$,
T.~Aziz$^{6}$,  
J.R.~Andersen$^{7}$,
A.~Belyaev$^8$,
T~Binoth$^{9}$, 
S.~Catani$^{2}$, 
M.~Ciccolini$^{10}$, 
J.E.~ Cole$^{11}$,
S.~Dawson$^{1}$, 
A.~Denner$^{10}$, 
S.~Dittmaier$^{12}$,
A.~ Djouadi$^{8,13,14}$,
M.~ Drees$^{14}$,
U.~ Ellwanger$^{13},$
C.~Englert$^{15}$, 
T.~Figy$^{16}$,
E.~Gabrielli$^{17}$, 
D.~Giordano$^{18}$,
S.~ Gleyzer$^{19}$,
R.~Godbole$^{20}$,
M.~Grazzini$^{2}$, 
S.~Greder$^{3}$, 
V. Halyo$^{5}$, 
M.~Hashemi$^{21}$,
S.~Heinemeyer$^{22}$, 
G.~Heinrich$^{16}$,
M.~ Herquet$^{23}$,
S.~ Hesselbach$^{16}$,
C.~ Hugonie$^{24}$,  
C.B.~Jackson$^{1}$,
N.~Kauer$^{25}$,
R.~Kinnunen$^{17}$,
S.F.~ King$^{8}$,
S.~ Lehti$^{17}$,
F.~Maltoni$^{23}$,
B.~ Mele$^{26}$,
P.~Mertsch$^{27}$,
M.~ Moretti$^{28}$, 
S.~ Moretti$^{8,29}$,
M.~M\"uhlleitner$^{30}$, 
A.K.~Nayak$^{6}$,
A.~Nikitenko$^{3,a}$, 
C.~Oleari$^{31}$,
F.~Piccinini$^{32}$, 
R.~ Pittau$^{33,34}$,
J. Rathsman$^{35}$,
I.~ Rottl\"{a}nder$^{14}$,
C.H.~Shepherd-Themistocleous$^{11}$,
M.~ Schumacher$^{4}$,
J.M.~Smillie$^{16}$,
A.~ Sopczak$^{36}$,
M.~Spira$^{10}$, 
M.~Takahashi$^{3}$,
A. M. Teixeira$^{13}$,
I.R.~Tomalin$^{11}$, 
M.~V\'azquez~Acosta$^{7}$, 
G.~Weiglein$^{16}$, 
C.D.~White$^{37}$
D.~Zeppenfeld$^{15}$
 }

\institute{
$^1$ Brookhaven National Laboratory, Upton, NY 11973, USA\\
$^2$ INFN, Sezione di Firenze, Via Sansone 1, I-50019 Sesto Fiorentino, Florence, Italy\\
$^{3}$ Imperial College, London, UK\\
$^a$ On leave from ITEP, Moscow\\
$^{4}$ Fachbereich Physik, University of Siegen, Walter-Flex-Stra\ss{}e 3, 57068 Siegen, Germany\\
$^{5}$ Department of Physics, Princeton University, Princeton, NJ 08544, USA\\
$^{6}$Tata Institute of Fundamental Research, Mumbai,India\\ 
$^{7}$ CERN, CH 1211 Geneva 23, Switzerland\\
$^{8}$School of Physics \& Astronomy, Southampton University, Southampton SO17 1BJ, UK\\
$^{9}$ School of Physics, The University of Edinburgh, Edinburgh EH9 3JZ, UK\\
$^{10}$ Paul Scherrer Institut, W\"urenlingen und Villigen, 5232 Villigen, Switzerland\\ 
$^{11}$ Rutherford Appleton Laboratory, Didcot, OX11 0QX, UK\\
$^{12}$ Max-Planck-Institut f\"ur Physik (Werner-Heisenberg-Institut),
80805 M\"unchen, Germany\\ 
$^{13}$ Laboratoire de Physique Th\'eorique, Universit\'e Paris--Sud, 
F--91405 Orsay Cedex, France.\\
$^{14}$ Physikalisches Institut, University of Bonn, Nussallee 12, 53115 Bonn, Germany
\\
$^{15}$ Institut f\"ur Theoretische Physik, Universit\"at Karlsruhe, P.O.Box
6980, 76128 Karlsruhe, Germany \\
$^{16}$ IPPP, University of Durham, Durham DH1~3LE, UK\\
$^{17}$ Helsinki Institute of Physics, Helsinki, Finland\\
$^{18}$Universit\`a degli Studi di Bari, INFN Sezione di Bari, Italy\\
$^{19}$ Department of Physics, Florida State University, Tallahassee, Florida 32306, USA\\
$^{20}$ Center for High Energy Physics, Indian Institute of Science, Bangalore 560
012, India.\\
$^{21}$ Universiteit Antwerpen, G.U.238, 
Groenenborgerlaan 171,
2020 Antwerpen, Belgium\\
$^{22}$ Instituto de Fisica de Cantabria (CSIC-UC), Santander, Spain\\
$^{23}$ Centre for Particle Physics and Phenomenology (CP3),
Universit\'{e} Catholique de Louvain, Louvain-la-Neuve, Belgium\\ 
$^{24}$ LPTA, Universit\'e de Montpellier II, 34095 Montpellier, France. \\
$^{25}$ Institut f\"ur Theoretische Physik, Universit\"at W\"urzburg, D-97074 W\"urzburg, Germany\\
$^{26}$ University of Rome La Sapienza and INFN, Sezione di Roma,  Rome, Italy\\
$^{27}$ Rudolf Peierls Centre for Theoretical Physics, University of Oxford, Oxford OX1 3NP, UK,\\
$^{28}$ Dipartimento di Fisica Universit\`a di Ferrara, and INFN,  
Sezione di Ferrara, Ferrara, Italy \\
$^{29}$Laboratoire de Physique Th\'eorique, Paris XI, 91405 Orsay, France \\
$^{30}$ LAPTH, 9 Chemin de Bellevue, 
Annecy-le-Vieux 74951, France\\
$^{31}$ Universit\`a di Milano-Bicocca and INFN Sezione di Milano-Bicocca,
20126 Milano, Italy \\
$^{32}$ INFN, Sezione di Pavia, via A. Bassi 6, I 27100, Pavia, Italy\\
$^{33}$ Dipartimento di Fisica Teorica, Universit\` a di Torino,
and INFN sezione di Torino, via P. Giuria 1, Torino, Italy \\
$^{34}$ Departamento de F\'{i}sica Te\'orica y del Cosmos,
Centro Andaluz de F\'{i}sica de Part\'{i}culas Elementales (CAFPE), 
Universidad de Granada, E-18071, Granada, Spain\\
$^{35}$ High Energy Physics, Uppsala University, Box 535, S-75121 Uppsala, Sweden \\
$^{36}$ Department of Physics, Lancaster University, Lancaster LA1 4YW, UK\\
$^{37}$ NIKHEF, Kruislaan 409, 1098 SJ Amsterdam, The Netherlands
}

\maketitle

 \begin{center}
   \textit{Report of the Working Group on  Higgs Bosons for the Workshop 
   ``Physics at TeV Colliders'', Les Houches, France, 11--29 June, 2007.  }
\end{center}
\newpage

%\begin{abstract}
 
%\end{abstract}

\setcounter{tocdepth}{1}
\tableofcontents
\setcounter{footnote}{0}

\part[INTRODUCTION]{INTRODUCTION}
\section[Foreword]{FOREWORD~\protect\footnote{Contributed 
by: S.~Dawson and M.~Grazzini}}
\def\ptmin{p_{T{\rm min}}}
\def\ptmax{p_{T{\rm max}}}
\newcommand\as{\alpha_{\mathrm{S}}} 
\def\ltap{\raisebox{-.6ex}{\rlap{$\,\sim\,$}}\raisebox{.4ex}{$\,<\,$}}
\def\gtap{\raisebox{-.6ex}{\rlap{$\,\sim\,$}}\raisebox{.4ex}{$\,>\,$}}
%\begin{document}

The elucidation of the mechanism of electroweak symmetry breaking is
one of the main goals of the LHC physics program.  In the Standard
Model (SM), mass generation is triggered by the Higgs mechanism,
which predicts the existence of one scalar state,
the Higgs boson \cite{Higgs:1964pj,Djouadi:2005gi}. The Higgs boson
couplings to fermions and gauge bosons are a prediction of the model
and the only unknown parameter is the Higgs boson mass. 

The Minimal Supersymmetric extension of the Standard Model (MSSM) requires the introduction
of two Higgs doublets, in order to preserve supersymmetry and give mass
to the fermions,
 and after spontaneous symmetry breaking
 five Higgs particles remain in the spectrum: two CP-even ($h$,$H$), one CP-odd ($A$) and two charged ($H^\pm$) Higgs bosons.
 At lowest order the MSSM Higgs sector can be described by
two parameters, generally chosen to be $m_A$, the mass of the pseudoscalar Higgs boson, and
$\tan\beta=v_2/v_1$, the ratio of the two vacuum expectation values.
The lowest order predictions receive large radiative corrections which must be included when calculating Higgs couplings or masses.
At tree level, the lightest neutral Higgs boson has an upper bound of $M_Z$, which is increased to $m_h\ltap 
{\cal O}(130-140)~{\rm GeV}$ when
radiative corrections are included \cite{Carena:2002es}.

The search
for the Higgs at collider experiments has now being on-going for two
decades. The present direct lower limit of the Higgs mass in the SM is
114.4 GeV (at 95\% CL) \cite{Barate:2003sz}, while precision measurements point to a rather
light Higgs, $m_h \ltap 180$ GeV \cite{Alcaraz:2007ri,lepewwg}.
The Tevatron has a chance to find evidence for a Higgs boson 
if enough integrated luminosity
can be accumulated. 
At present, the Tevatron is performing well, and it is approaching
the sensitivity limit required to exclude the existence of the SM Higgs for
$m_h \sim 160$~GeV\cite{:2007gz}.

If it exists, 
the Higgs boson will be seen at the LHC, which can provide a measurement 
of the Higgs mass at the per-mille level and of the Higgs boson 
couplings at the 5-20 \% level \cite{Duhrssen:2004cv}.
These tasks, however, require accurate theoretical predictions for 
both 
signal and background
cross sections and distributions, and this is true in particular for an accurate determination
of the properties of the discovered particle, such as spin, CP, and couplings.

In the following we review the status of theoretical predictions for
both signal and background at the LHC, with emphasis on recent developments
for the Standard Model Higgs boson.

\subsection{Gluon-Gluon Fusion}

The gluon fusion mechanism, mediated by a (heavy)-quark loop, provides the dominant production
mechanism of Higgs bosons at the LHC in the full mass range.

QCD corrections to this process at next-to-leading order (NLO) have been known for some time
\cite{Dawson:1991zj,Djouadi:1991tka,Spira:1995rr}
and their effect increases the leading order (LO) cross
section by about 80--100\%.  This calculation is very well approximated by the
large-$m_{top}$ limit. When the exact Born cross section (with full
dependence on the masses of top and bottom quarks) is used
to normalize the result,
the difference between the exact and the approximated NLO
cross sections ranges from 1 to $4\%$ when $m_h< 200$ GeV.
In recent years, the next-to-next-to-leading order (NNLO) corrections have been computed in this limit \cite{Harlander:2000mg,Catani:2001ic,Harlander:2001is,Harlander:2002wh,Anastasiou:2002yz,Ravindran:2003um},
leading to an additional increase of the cross section of about $10-15\%$.
Soft-gluon resummation leads to a further increase of about $6\%$ \cite{Catani:2003zt}.
The latter result is nicely confirmed by the more recent evaluation of the leading soft contributions at N$^3$LO \cite{Moch:2005ky,Laenen:2005uz,Idilbi:2005ni}.
Two loop EW effects are also known
%and increase the total rate by $5-10\%$
\cite{Djouadi:1994ge,Aglietti:2004nj,Degrassi:2004mx}.

In the MSSM, for large $\tan\beta$, the contribution from bottom quark loops becomes important
and the large-$m_{top}$ limit is not applicable. 
The full SUSY-QCD corrections are known in the limit of heavy squark
and gluino masses at 
NLO \cite{Harlander:2004tp,Harlander:2003bb,Harlander:2003kf}.
Recently, the exact contribution of squark loops
has been evaluated at NLO \cite{Muhlleitner:2006wx}
and is discussed in Sect. \ref{sec:sqcd} The massive virtual corrections
to the squark loops are given in Ref. \cite{Bonciani:2007ex,Anastasiou:2006hc}.

The higher order calculations mentioned above are certainly important
but they refer to total cross sections, i.e., the experimental cuts
are largely ignored. The impact of higher order corrections 
on the rate and the shape of the corresponding distributions
may be
strongly dependent on the choice of cuts.
In the case in which one \cite{deFlorian:1999zd} or 
two \cite{Campbell:2006xx} jets are tagged at large $p_T$ 
the NLO corrections for Higgs production from gluon fusion  are known
and implemented in parton level Monte Carlo programs. 
These predictions are obtained in the large $m_{top}$ limit, which is
a good approximation for small transverse momentum of the accompanying jet.
For Higgs plus
one jet production, there is a rather flat dependence of the $K$ factor on $p_T$
and rapidity for moderate $p_T$ and $y$.  
In the MSSM, the Higgs plus $1$ jet rate is known at lowest 
order only \cite{Brein:2003df,Field:2003yy}. 
 The Higgs plus $2$ jet
process from gluon fusion is a background for vector boson fusion, as discussed 
below.  Interference effects in the Higgs plus $2$ jet channel are discussed in
Sect. \ref{sec:int}

The NNLO inclusive cross section when a jet 
veto is applied \cite{Catani:2001cr} has been known for some time.
The first NNLO calculation that fully takes into account experimental cuts
was reported in Ref.~\cite{Anastasiou:2005qj}, in the case of the 
decay mode $h\to\gamma\gamma$ which is implemented in the FEHIP
Monte Carlo program.
In Ref.~\cite{Anastasiou:2007mz} the calculation was
extended to the decay mode $h\to W^+W^-\to l^+ l^-\nu{\overline \nu}$.
Recently, an 
independent NNLO calculation has been 
performed \cite{Catani:2007vq,Grazzini:2008tf},
including all the relevant decay modes of the Higgs boson:
$h\to\gamma\gamma$, $h\to W^+W^-\to l^+ l^-\nu{\overline \nu}$ 
and $h\to ZZ\to 4$ leptons.
Such a calculation is implemented in a Monte Carlo program and is 
documented in this report in Sect. \ref{sec:hnn}

Among the possible differential distributions, an important role
is played by the transverse momentum ($p_T$) spectrum of
the Higgs boson\cite{Langenegger:2006wu}. When $p_T\sim m_h$ the standard
fixed order expansion is applicable.
When $p_T\ll m_h$, large logarithmic contributions
appear that may invalidate the customary fixed order expansion.
The resummation of such contributions has been performed at 
different levels of theoretical
accuracy \cite{Balazs:2000wv,Berger:2002ut,Kulesza:2003wn,Bozzi:2003jy,Bozzi:2005wk,Bozzi:2007pn}.
In Refs.~\cite{Bozzi:2003jy,Bozzi:2005wk,Bozzi:2007pn} the resummed result up to next-to-next-to-leading logarithmic accuracy is matched
to the fixed order NLO result \cite{deFlorian:1999zd,Ravindran:2002dc,Glosser:2002gm} valid at large transverse momenta.
It is important to note that transverse-momentum resummation is approximately performed by standard Monte Carlo event generators. A comparison of results obtained with different tools was presented in Ref.~\cite{Balazs:2004rd}.

For Higgs boson masses below about 140 GeV the dominant decay mode $h\to b{\bar b}$ is swamped by the huge QCD background and the Higgs boson
can be found by looking at the rare $h\to \gamma\gamma$ decay mode.
The $\gamma\gamma$ background can be measured precisely 
from the data using sideband interpolation, but accurate theoretical predictions are useful to
estimate the expected accuracies and to better understand
detector performances.
The $h\to\gamma\gamma$ decay width is known
including full two loop QCD and EW effects \cite{Djouadi:1990aj,Degrassi:2005mc,Passarino:2007fp}.
The NNLO QCD effects are known in the large-$m_{top}$ limit \cite{Steinhauser:1996wy}.
The $\gamma\gamma$ irreducible background has been computed
up to NLO including
the fragmentation effects \cite{Binoth:1999qq}. The $gg\to \gamma\gamma$ contribution, which is formally NNLO,
is enhanced by the large gluon luminosity and
is known up to N$^3$LO (i.e. ${\cal O}(\alpha_s^3)$)\cite{Bern:2002jx}.

For Higgs masses between 140 and 180 GeV the 
$W^+W^-\to l^+ l^- \nu {\overline \nu}$ decay 
mode is the most important. Despite the absence of a mass peak, there are
strong angular correlations between the charged leptons \cite{Dittmar:1996ss}.
To suppress the $t{\bar t}$ background, a jet veto has to be applied
to cut events with high-$p_T$ $b$-jets from the decay of the top quark.
The impact of higher-order corrections on the Higgs signal is strongly reduced by the selection cuts \cite{Anastasiou:2007mz,Anastasiou:2008ik,Grazzini:2008tf}.
This channel appears to be one of the most promising for an early discovery \cite{Davatz:2004zg}, but at the same time it is the most challenging as far as the background is concerned.
Because of the missing energy, the Higgs mass cannot be directly reconstructed, and
a straightforward background extrapolation from sidebands is not possible. The background
has to be extrapolated from regions where the signal is absent and this 
requires a precise knowledge of the background distributions.
The $W^+W^-$ irreducible background is known
up to NLO \cite{Dixon:1999di,Campbell:1999ah} including spin correlations,
and the effects of multiple soft-gluon emissions has been included
up to NLL \cite{Grazzini:2005vw}.
Spin correlations in the $W$ decay are crucial
for a correct prediction of angular distributions
and are now implemented in the MC@NLO event generator \cite{Frixione:2002ik,Frixione:2003ei}.
The potentially large $gg\to W^+W^-$ 
contribution, formally NNLO,
has also been computed \cite{Binoth:2005ua,Duhrssen:2005bz}.
The $t{\bar t}$ background, including the effect of
spin correlations \cite{Bernreuther:2001rq}, is known up to NLO and is also included in MC@NLO.
A complete calculation including finite width effects (and thus $W^+W^-bb$, 
$Wtb$) is available at LO only \cite{Kauer:2001sp}.

When the Higgs mass is larger than about $180$ GeV, the decay $h\to ZZ\to 4$ leptons becomes dominant. 
This channel is much easier to observe  than the  $W^+W^-$ channel
because the invariant mass of the leptons can be reconstructed and
thus the background can be measured from the data.
Accurate predictions become important when the nature of the Higgs 
particles is studied.
The irreducible $ZZ$ background is known up to NLO including spin correlations \cite{Dixon:1999di,Campbell:1999ah}. The impact of soft-gluon effects on signal and background
has been studied recently \cite{Frederix:2008vb}.
The calculation of the $gg\to ZZ$ contribution is accounted for
in this report in Sects. \ref{sec:ggzz1} and \ref{sec:ggzz2}
We finally note that the full QCD+EW corrections to the decay modes
$h\to W^+W^-(ZZ)\to 4$ leptons
have been recently computed \cite{Bredenstein:2006rh,Bredenstein:2006ha}.

\subsection{Vector-Boson Fusion}

The vector boson fusion (VBF) process plays an important role
for a wide range of Higgs masses.
The VBF cross section is typically one
order of magnitude smaller than the one from gluon fusion, and
it becomes competitive with the latter for very large Higgs masses.

VBF occurs through the scattering of two valence quarks that exchange
a $W$ or a $Z$ boson. Since valence quark distributions in the proton
are peaked at relatively large Bjorken $x$ ($x\sim 0.1-0.2$), the scattered quarks emerge with very large longitudinal momentum and transverse
momentum of the order of a fraction of the boson mass. As a consequence, the typical signature of a VBF event is given by two hard jets with
a large rapidity interval between them, and since the exchanged
boson is colourless, there is no hadronic activity between them.
Although this channel has a smaller cross section with respect to gluon
fusion, it is very attractive both for discovery and for the measurement of the Higgs couplings.

The NLO QCD corrections to the total rate were computed some time ago
and found to be of the order of $5-10\%$ \cite{Han:1992hr}.
In recent years, these corrections have been implemented
for distributions \cite{Figy:2003nv,Figy:2004pt,Berger:2004pca}.
Recently, the full EW+QCD corrections to this process have been computed \cite{Ciccolini:2007jr,Ciccolini:2007ec}. A comparison of the different calculations is presented in this report in Sect. \ref{sec:vbf}

The $h$+2 jets final state can be produced also by gluon-gluon fusion.
This signature, although part of the inclusive Higgs boson signal,
represents a background when trying to isolate the $hWW$ and $hZZ$ couplings through VBF.
The gluon fusion contribution is known at LO with full top mass dependence \cite{DelDuca:2001fn}.
The kinematical distributions of the tagging jets show remarkable differences in the two production mechanisms. The $\Delta\phi$ distribution
of the tagging jets is rather flat for the VBF signal. By contrast,
the loop induced $hgg$ coupling leads to a
pronounced dip at $\Delta\phi=90^o$.
Another significant difference is found in the rapidity distribution
of the third hardest jet with respect to the rapidity average of
the other two. The VBF signal has a dip in the central region, where
the gluon fusion background is peaked.
As such, a cut on jets with $p_T>p_T^{\rm veto}$
in the central rapidity region (central jet veto)
enhances the relevance of the VBF signal.
Recently, NLO QCD corrections to the $h+2$ jets process
in the large $m_{top}$-limit have been computed \cite{Campbell:2006xx},
and also parton shower effects on the relevant distributions
have been evaluated \cite{DelDuca:2006hk}. 
These studies show that the discriminating power of previous
LO results is not significantly changed.
We note, however, that when the $p_T^{\rm veto}$ is much smaller
than the Higgs boson mass the coefficients of the perturbative series are enhanced by large logarithmic contributions that may
invalidate the fixed order expansion.
The latter point deserves more detailed investigation.
An experimental study of central jet veto efficiencies is presented
in this report in Sect. \ref{sec:crg}

The most important decay channels of the Higgs boson in VBF
are $h\to \tau^+\tau^-$ and $h\to W^+W^-\to l^+l^-\nu {\overline\nu}$.
The $h\to \tau^+\tau^-$ decay mode
provides an important discovery channel in the MSSM.
The $\tau^+\tau^-$ invariant mass can be reconstructed at the LHC
with an accuracy of a few GeV. This is possible because VBF 
typically produces Higgs bosons with large transverse momentum.
As a consequence, a sideband analysis can in principle be used to measure the background from the data.
The most important backgrounds are QCD $Zjj$ and EW $Zjj$ from VBF.
Both are known up to NLO \cite{Campbell:2002tg,Oleari:2003tc}. 

The $h\to W^+W^-\to l^+ l^- \nu {\overline\nu}$ decay mode
is the most challenging, because, as for
gluon fusion, it does not allow a direct Higgs mass reconstruction.
The irreducible $W^+W^-$ 
background is known up to NLO \cite{Jager:2006zc}.
The other important background is $t{\bar t}+{\rm ~jets}$, and has
the largest uncertainty. Recently, the NLO corrections
to $t{\bar t}+$ jet
have been computed \cite{Dittmaier:2007wz}. It will be essential to include the decay of the top quark with full spin correlations. 
In addition, finite width effects could be relevant.

\subsection{Associated Production With a $b{\bar b}$ Pair}

In the Standard Model, Higgs production in association with $b$ quarks is never important, since
this rate is suppressed by $m_b/v$.
This channel is important in MSSM scenarios at large $\tan\beta$,
 since the Higgs coupling to bottom quarks is enhanced in this regime.  
For $\tan\beta
\gtap 7$, Higgs production in association with a $b$ quark is the
dominant production mechanism at the LHC.
The cross section for $b$- Higgs production
 can be computed using two different formalisms, which represent different orderings
of perturbation theory.
In the four-flavour scheme the 
cross section starts with $gg\to b{\bar b}h$ at LO. 
The cross section for the associated production of the Higgs boson with zero, one or two high-transverse momentum $b$-jets is known up NLO \cite{Dittmaier:2003ej,Dawson:2003kb,Dawson:2004sh,Dawson:2005vi}.
In the five flavour scheme, the LO process is $b{\bar b}\to h$ and
bottom  quark parton distributions are introduced to sum the potentially large
logarithms, $\log(m_h/m_b)$.
The inclusive cross section has been computed up to NNLO \cite{Harlander:2003ai}, and the
cross section for the associated production with one high-$p_T$ $b$ jet
is known at NLO \cite{Campbell:2002zm}. In recent years, a detailed comparison between the results of the two approaches has been performed
with the conclusion that the two approaches lead to similar results. 
For a discussion see Ref. \cite{Campbell:2004pu,Buttar:2006zd}.  In addition, the electroweak
corrections to $b {\overline b}\rightarrow h$ \cite{Dittmaier:2006cz}, the dominant top quark contributions 
to the NNLO rate for the exclusive $b{\overline b} h$ process \cite{Boudjema:2007uh} 
and the SUSY QCD corrections to $b{\overline b}
\rightarrow h$, $b g\rightarrow bh$ are known \cite{Dawson:2007ur}.  
The effects of SUSY-QCD on $b$-Higgs production is discussed in
Sect. \ref{sec:bbh}

\subsection{Associated Production With a $t{\bar t}$ Pair}

The $h t {\bar t}$ channel offers the possibility of a clean
measurement of the top quark Yukawa coupling.
The NLO corrections to the $h t{\bar t}$ signal were 
independently
computed by two groups \cite{Beenakker:2001rj,Beenakker:2002nc,Dawson:2002tg,Dawson:2003zu},
and found to increase the signal cross section by $20-40\%$.
The $h t {\bar t}$ channel was initially thought to be an important
discovery channel in the low Higgs mass region, looking at
the $h\to b{\bar b}$ decay mode and triggering on the leptonic decay
of one of the top.
The main backgrounds are $t{\bar t}b{\bar b}$ and $t{\bar t}b jj$.
Recently, more detailed investigations based on a
more careful background evaluation and
full detector simulation lead to a more pessimistic view
on the possibility of observing the Higgs signal in this 
channel \cite{CMStdr}. This channel could be important for measuring
the top quark Yukawa coupling\cite{Duhrssen:2004cv,Belyaev:2002ua}.

\begin{figure}[!t]
\vskip 2in
       \centerline{\psfig{angle=-90, figure=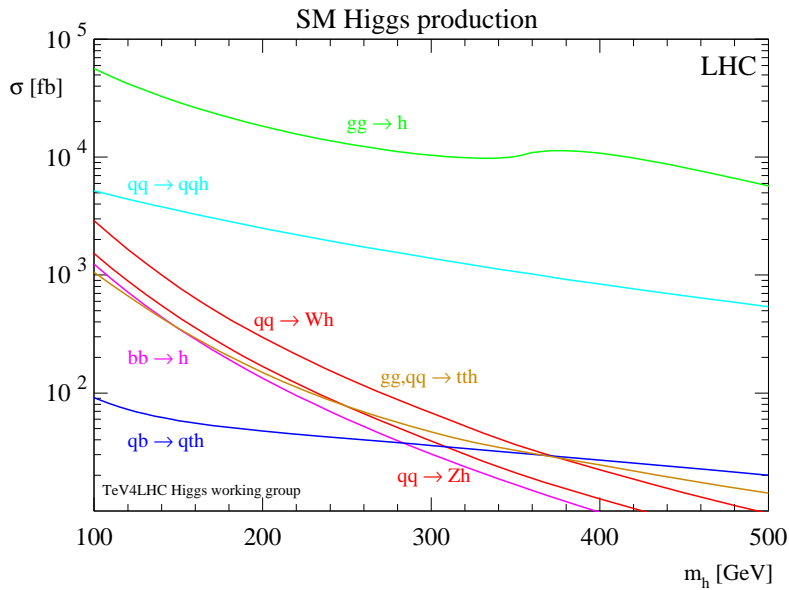,width=4in}}
       \caption{Total cross sections for Higgs production
at the LHC.  The gluon fusion result is NNLO QCD with soft gluon resummation effects
included at NNLL and uses MRST2002 PDFs with renormalization/factorization
scales equal to $m_h$.  The vector boson fusion curve is
shown  at NLO QCD with CTEQ6M PDFs and renormalization/factorization
scales equal to $m_h$.  The $Vh$ results ($V=W,Z$) include NNLO QCD corrections and
NLO EW corrections and use MRST2002 PDFs with the renormalization /factorization
scales equal to the $m_h-M_V$ invariant mass. The $b{\overline b}\rightarrow h$ result is NNLO
QCD, with MRST2002 PDFs, renormalization scale equal to $m_h$ and factorization
scale equal to $m_h/4$. The results for $t {\overline t} h$ production are NLO QCD, use
CTEQ6M PDFs and set the renormalization/factorization scale to $m_t+m_h/2$\cite{Aglietti:2006ne}.}
\label{fig:sigmas}
\end{figure}

\subsection{Associated production with a $W$ or a $Z$ boson}

This channel is essential for the Higgs search at the Tevatron for
Higgs masses below $130$ GeV. The leptonic decay of the vector boson
provides the necessary background rejection 
to allow for  looking at the $h\to b{\bar b}$ decay mode. 
The signal cross section is known up to NNLO in QCD, the
corrections being  about $+30\%$ \cite{Han:1991ia,Hamberg:1990np}.
These corrections are identical to those of Drell-Yan,
but in the case of $Zh$
 an additional contribution from the $gg$ initial state must be included \cite{Brein:2003wg}.
Full EW corrections are known and decrease the cross section by $5-10\%$ \cite{Ciccolini:2003jy}.

\subsection{Conclusions}

The important Higgs production channels are known  at NLO QCD and
in a few cases to NNLO and progress
is being made in implementing these results in Monte Carlo programs.
A summary of the total rates for the most important Higgs production channels is
shown in Fig. \ref{fig:sigmas} \cite{Aglietti:2006ne}.

%\bibliography{intro}

%\end{document}

\part[STANDARD MODEL HIGGS BOSONS]{STANDARD MODEL HIGGS BOSONS}
\section[HNNLO: A Monte Carlo Program for Higgs Boson Production at the LHC]
{HNNLO: A MONTE CARLO PROGRAM FOR HIGGS 
BOSON PRODUCTION AT THE LHC~\protect\footnote{Contributed by: S.~Catani and M.~Grazzini}}
\label{sec:hnn}
\def\ptmin{p_{T{\rm min}}}
\def\ptmax{p_{T{\rm max}}}
%\newcommand\as{\alpha_{\mathrm{S}}} 
%\begin{document}

%\title{HNNLO: a Monte Carlo program for Higgs boson production at the LHC}

%\author{S.~Catani, M.~Grazzini}
%\institute{INFN, Sezione di Firenze, Via Sansone 1, I-50019 Sesto Fiorentino, Florence, Italy}%

%\maketitle

%\begin{abstract}
%We consider Higgs boson production through gluon--gluon fusion in hadron collisions. We present a numerical program that computes the cross section up to NNLO in QCD perturbation theory. The program includes the decay modes $H\to\gamma\gamma$, $H\to WW\to l\nu l\nu$ and $H\to ZZ\to 4$ leptons,
%and allows the user to apply arbitrary cuts on the momenta of
%the partons and
%of the leptons (photons) produced in the final state.
%\end{abstract}

\subsection{Introduction}

Gluon-gluon fusion is the main production channel of
the Standard Model Higgs boson at the LHC.
At leading order (LO) in QCD perturbation theory, the cross section is proportional
to $\as^2$, $\as$ being the QCD coupling. The QCD radiative corrections to the total cross section are known at the
next-to-leading order (NLO) 
\cite{Dawson:1991zj,Djouadi:1991tka,Spira:1995rr}
and at the next-to-next-to-leading order (NNLO) 
\cite{Harlander:2000mg,Catani:2001ic,Harlander:2001is,Harlander:2002wh,Anastasiou:2002yz,Ravindran:2003um}.
The effects of a jet veto on the total cross section has been studied
up to NNLO \cite{Catani:2001cr}.
We recall that all the results at NNLO have been obtained by using 
the large-$M_t$ approximation, $M_t$ being the mass of the top quark.

These NNLO calculations are certainly important but they refer
to situations where the experimental cuts
are either ignored (as in the case of the total cross section) or taken into account
only in simplified cases (as in the case of the jet vetoed cross section).
The impact of higher-order corrections
may be strongly dependent on the details of the applied cuts and also the shape of the
distributions is typically affected by these details.

The first NNLO calculation that fully takes into account experimental cuts
was reported in Ref.~\cite{Anastasiou:2005qj}, in the case of the decay mode $H\to\gamma\gamma$.
In Ref.~\cite{Anastasiou:2007mz} the calculation is
extended to the decay mode $H\to WW\to l\nu l\nu$.

In Ref.~\cite{Catani:2007vq} we have presented an independent NNLO calculation
of the Higgs production cross section.
The method is completely different from
that used in Refs.~\cite{Anastasiou:2005qj,Anastasiou:2007mz}.
Our calculation is implemented
in a fully-exclusive parton level event generator.
This feature makes it particularly suitable for practical applications
to the computation of distributions in the form of bin histograms.
Our numerical program can be downloaded from \cite{hnnlo}.
The decay modes that are currently implemented
are $H\to \gamma\gamma$, $H\to WW\to l\nu l\nu$
and $H\to ZZ\to 4$ leptons \cite{Grazzini:2008tf}.

In the following
we present a brief selection of results that can be obtained
by our program.
We consider Higgs boson production at the LHC and
use the MRST2004 parton distributions \cite{Martin:2004ir},
with parton densities and $\as$ evaluated at each corresponding order
(i.e., we use $(n+1)$-loop $\as$ at N$^n$LO, with $n=0,1,2$). The 
renormalization and factorization scales are fixed to the value 
$\mu_R=\mu_F=M_H$, where $M_H$ is the mass of the Higgs boson.

\subsection{Results For the Decay Mode $H\to \gamma\gamma$}

%==========================================
\begin{figure}
\begin{center}
\includegraphics[width=0.5\textwidth]{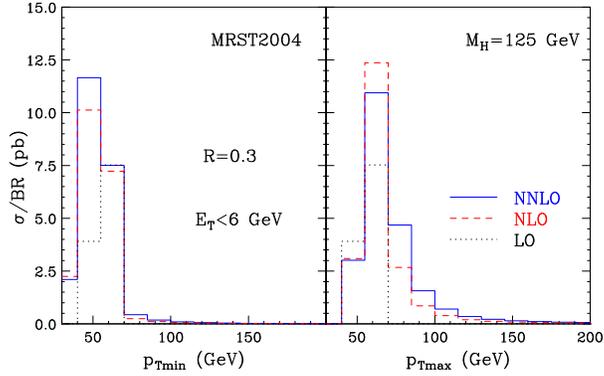}
 \caption{Distributions in $\ptmin$ and $\ptmax$ for the diphoton signal at the LHC. The cross section is divided by the branching ratio in two photons.}
\label{fig:isol}
\end{center}
\end{figure}
%==========================================

We consider the production of a Higgs boson of mass $M_H=125$ GeV in the
$H\to\gamma\gamma$ decay mode and follow
Ref.~\cite{CMStdr} to apply cuts on the photons.
For each event, we classify the photon transverse momenta according to their
minimum and maximum value,  
$\ptmin$ and $\ptmax$. The photons are required to be in the
central rapidity region, $|\eta|<2.5$, with  $\ptmin>35$~GeV
and $\ptmax>40$~GeV. We also require the photons to be isolated:
the hadronic (partonic) transverse energy in a cone of radius $R=0.3$ along the
photon direction 
has to be smaller than 6~GeV. By applying these cuts
the impact of the NNLO corrections on the NLO total cross section
is reduced from 19\% to 11\%.
 
In Fig.~\ref{fig:isol} we plot
the distributions in $\ptmin$ and $\ptmax$
of the signal process $gg\to H\to\gamma\gamma$.
We note that the shape of these distributions sizeably
differs when going from LO to NLO and to NNLO.
The origin of these perturbative instabilities is well known 
\cite{Catani:1997xc}.
Since the LO spectra
are kinematically bounded by $p_T\leq M_H/2$,
each higher-order perturbative contribution produces
(integrable) logarithmic singularities in the vicinity of
that boundary. More detailed studies are necessary to assess
the theoretical uncertainties of these fixed-order results
and the relevance of all-order resummed calculations.
 
\begin{figure}
\begin{center}
\includegraphics[width=0.5\textwidth]{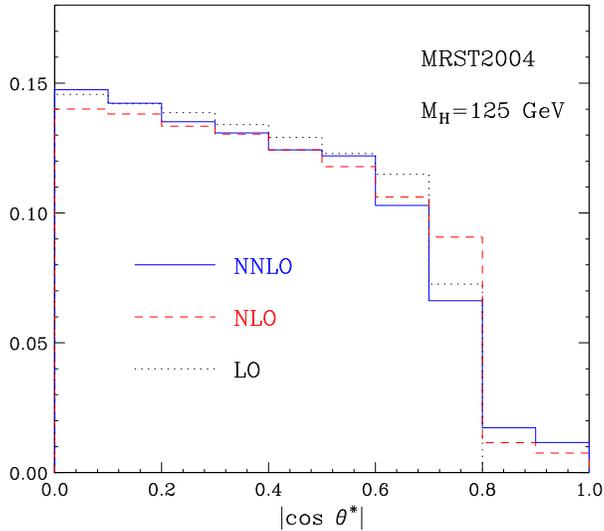}
 \caption{Normalized distribution in the variable $\cos\theta^*$.}
\label{fig:ctheta}
\end{center}
\end{figure}

In Fig.~\ref{fig:ctheta} we consider the (normalized) distribution in
the variable $\cos\theta^*$, where $\theta^*$ is the polar angle of one of the photons
in the rest frame of the Higgs boson
\footnote{We thank Suzanne Gascon and Markus Schumacher for suggesting
the use of this variable.}.
At small values of $\cos\theta^*$ the distribution
is quite stable with respect to higher order QCD corrections.
We also note that the LO distribution vanishes beyond the value $\cos\theta^*_{\rm max}<1$.
The upper bound $\cos\theta^*_{\rm max}$ is due to the fact that the photons
are required to have a minimum $p_T$ of $35$ GeV.
As in the case of Fig.~\ref{fig:isol}, in the vicinity of this LO kinematical boundary
there is an instability of the
perturbative results beyond LO.

%==========================================
%==========================================

\subsection{Results for the Decay Mode $H\to l\nu l\nu$}

We now consider the production of a Higgs boson with mass $M_H=165$ GeV in the decay mode $H\to l\nu l\nu$.
We apply a set of preselection cuts taken from the study of Ref.~\cite{Davatz:2004zg}. The charged leptons have $p_T$ larger than 20 GeV, and $|\eta|<2$. The missing $p_T$ is
larger than $20$ GeV and the invariant mass of the charged
leptons is smaller than $80$ GeV.
Finally, the azimuthal separation of the charged leptons in the
transverse plane ($\Delta\phi$) is smaller than $135^o$.
By applying these cuts the impact of the NNLO corrections
on the NLO result does not change and is of about $20\%$.

%==========================================
\begin{figure}
\begin{center}
\includegraphics[width=0.5\textwidth]{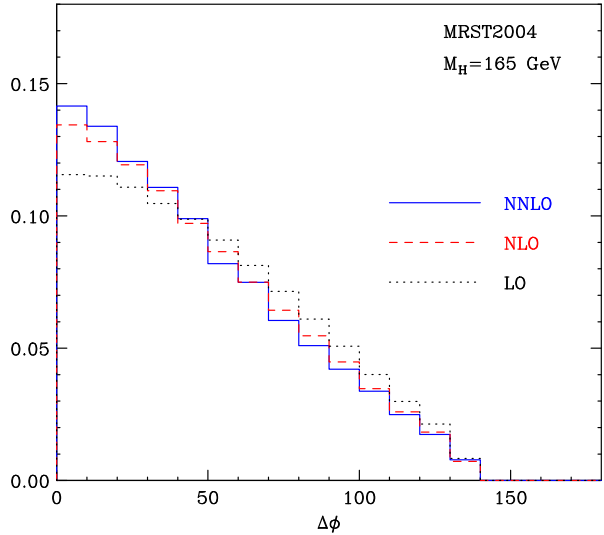}
 \caption{Normalized $\Delta\phi$ distribution at LO, NLO, NNLO.}
\label{fig:deltaphi}
\end{center}
\end{figure}
%==========================================
In Fig.\ref{fig:deltaphi} we
plot the $\Delta\phi$ distribution at LO, NLO and NNLO.
As is well known \cite{Dittmar:1996ss},
the charged leptons from the Higgs boson
signal tend to be close in angle,
and thus the distribution is peaked at small $\Delta\phi$.
We notice that the effect of the QCD corrections is to increase
the steepness of the distribution,
from LO to NLO and from NLO to NNLO.

\subsection{Conclusions}

We have illustrated a calculation of the Higgs boson production cross section at the LHC
up to NNLO in QCD perturbation theory. The calculation is implemented in the numerical program {\tt HNNLO},
which at present includes the decay modes $H\to\gamma\gamma$ and $H\to WW\to l\nu l\nu$ and $H\to ZZ\to 4$ leptons. The program allows the user to apply arbitrary cuts on the momenta of the partons and leptons (photons) produced in the final state, and to obtain the required distributions in the form of
bin histograms. We have presented a brief selection of numerical results that
can be obtained by our program. More detailed results for the decay modes $H\to WW$ and $H\to ZZ$ can be found in Ref.~\cite{Grazzini:2008tf}.
The fortran code {\tt HNNLO} can be downloaded from \cite{hnnlo}.

%\bibliography{hnnlolh}

%\end{document}

\section[Tuned Comparison of QCD Corrections to SM Higgs Boson Production 
via Vector Boson Fusion at the LHC]
{TUNED COMPARISON OF QCD CORRECTIONS TO
SM HIGGS-BOSON PRODUCTION VIA VECTOR BOSON FUSION AT THE 
LHC~\protect\footnote{Contributed by: M.~Ciccolini, A.~Denner, 
S.~Dittmaier, C.~Englert, T.~Figy, C.~Oleari, M.~Spira, and
  D.~Zeppenfeld}}
\label{sec:vbf}
%{\documentclass[11pt]{cernrep}
%\usepackage{graphicx,epsfig}
%\usepackage{cite}
%\bibliographystyle{lesHouches}%
%
%\begin{document}

%\title{ Tuned comparison of QCD corrections to SM Higgs-boson
%  production via weak-boson fusion at the LHC} 

%\author{M.~Ciccolini$^1$, A.~Denner$^1$, S.~Dittmaier$^{2,}$,
%  C.~Englert$^3$, T.~Figy$^4$, C.~Oleari$^5$, M.~Spira$^1$ and
%  D.~Zeppenfeld$^3$} 
%
%\institute{
%$^1$Paul Scherrer Institut, W\"urenlingen und Villigen, 5232 Villigen
%PSI, Switzerland\\ 
%$^2$Max-Planck-Institut f\"ur Physik (Werner-Heisenberg-Institut),
%80805 M\"unchen, Germany\\ 
%$^3$Institut f\"ur Theoretische Physik, Universit\"at Karlsruhe, P.O.Box
%6980, 76128 Karlsruhe, Germany \\
%$^4$IPPP, Durham University, Durham, DH1 3LE, UK \\
%$^5$Universit\`a di Milano-Bicocca and INFN Sezione di Milano-Bicocca,
%20126 Milano, Italy 
%}

%\maketitle

%\section{INTRODUCTION}
\subsection{Introduction}
The electroweak (EW) production of a Standard Model Higgs boson in
association with two hard jets in the forward and backward regions of
the detector---frequently quoted as the ``vector-boson fusion'' (VBF)
channel---is a cornerstone in the Higgs search both in the ATLAS
\cite{Asai:2004ws} and CMS \cite{Abdullin:2005yn} experiments at the
LHC.  Higgs production in the VBF channel also plays an important role
in the determination of Higgs couplings at the LHC (see e.g.\ Ref.
\cite{Duhrssen:2004cv}). Even bounds on non-standard couplings between
Higgs and EW gauge bosons can be imposed from precision
studies in this channel \cite{Hankele:2006ma}.

Higgs+2jets production in pp collisions proceeds through two
different channels.
The first channel corresponds to a pure EW process where the Higgs
boson couples to a weak boson. It comprises the scattering of
two (anti-)quarks mediated by $t$- and $u$-channel W- or Z-boson
exchange, with the Higgs boson radiated off the weak-boson propagator.
It also involves Higgs-boson radiation off a W- or Z-boson produced in
$s$-channel quark--antiquark annihilation (Higgs-strahlung process),
with the weak boson decaying hadronically.
The second channel proceeds mainly through strong interactions, the
Higgs boson being radiated off a heavy-quark loop that couples to any
parton of the incoming hadrons via gluons
\cite{DelDuca:2001fn,Campbell:2006xx}.

In the weak-boson-mediated processes, the two scattered quarks are
usually visible as two hard forward jets, in contrast to other jet
production mechanisms, offering a good background suppression
(transverse-momentum and rapidity cuts on jets, jet rapidity gap,
central-jet veto, etc.). Applying appropriate event selection criteria
(see e.g.~Refs. \cite{Barger:1994zq,Rainwater:1997dg,Rainwater:1998kj,
  Rainwater:1999sd,DelDuca:2006hk} and references in Refs.
\cite{Spira:1997dg,Djouadi:2005gi}) it is possible to sufficiently
suppress background and to enhance the VBF channel over the hadronic
Higgs+2jets production mechanism.

In order to match the required precision for theoretical predictions
at the LHC, QCD and EW corrections are needed. When VBF cuts are
imposed, the cross section can be approximated by the contribution of
squared $t$- and $u$-channel diagrams only, which reduces the QCD
corrections to vertex corrections to the weak-boson--quark coupling.
Explicit next-to-leading-order (NLO) QCD calculations in this
approximation exist since more than a decade
\cite{Spira:1997dg,Han:1992hr}, while corrections to distributions
have been calculated in the last few years
\cite{Figy:2003nv,Figy:2004pt,Berger:2004pca}.  Recently, the full NLO
EW and QCD corrections to this process have become available
\cite{Ciccolini:2007jr,Ciccolini:2007ec}. This calculation includes,
for the first time, the complete set of EW and QCD diagrams, namely
the $t$-, $u$-, and $s$-channel contributions, as well as all
interferences.

In this short note we compare the NLO QCD corrected cross-section
results obtained by three different calculations using a common set of
input parameters and a uniformly tuned setup. We also present, in
order to better understand the different approximations, the full NLO
QCD and EW corrected results as obtained in Refs.
\cite{Ciccolini:2007jr,Ciccolini:2007ec}.

In the next section, the different approaches that we compare are
briefly summarized. The precise setup is described in Section
\ref{hjjcomp_se:setup}, and Section \ref{hjjcomp_se:numerics} contains
the numerical results.

%\section{DIFFERENT APPROACHES AND CODES}
\subsection{Different Approaches and Codes}
\label{hjjcomp_se:approaches}

The following collaborations have contributed to the tuned comparison
of NLO QCD corrected results for Higgs-boson production via weak-boson
fusion at the LHC:
\begin{itemize}
\item {\tt CDD}: References \cite{Ciccolini:2007jr,Ciccolini:2007ec}
  present a detailed description of the calculation of the complete
  NLO EW and QCD corrections to Higgs-boson production in the VBF
  channel at the LHC. The NLO $\mathcal{O}(\alpha_s)$ corrections
  include the complete set of QCD diagrams, namely the $t$-, $u$-, and
  $s$-channel contributions, as well as all interferences. The
  integrated cross section (with and without dedicated VBF selection
  cuts) was calculated, as well as different Higgs-boson and
  tagging-jet observables. In the EW corrections, real corrections
  induced by photons in the initial state and QED corrections
  implicitly contained in the DGLAP evolution of PDFs were also taken
  into account. All EW contributions have been switched off for
  this comparison.
\item {\tt VBFNLO}\cite{Zeppenfeld:vbfnlo} is a NLO parton-level Monte
  Carlo program which implements one-loop QCD corrections for a
  collection of relevant VBF processes, of which Higgs-boson
  production, in the narrow resonance approximation, is the simplest
  example.  Higgs-boson production in weak-boson fusion is implemented
  following the results of Ref. \cite{Figy:2003nv}. {\tt VBFNLO}
  generates an isotropic Higgs-boson decay into two massless
  ``leptons'' (which represent $\tau^+\tau^-$ or $\gamma\gamma$ or
  $b\bar b$ final states), and imposes a cut on the invariant mass of
  the Higgs boson.  This feature has been disabled during this
  comparison, and only a non-decaying Higgs boson has been considered.
  We have employed {\tt VBFNLO-v.1.0}, and included only four flavours
  of the external quarks.
\item {\tt VV2H}\cite{Spira:vv2h} calculates the production cross
  section of Higgs bosons via $\mathrm{WW}/\mathrm{ZZ} \to
  \mathrm{h,H}$ at hadron colliders at NLO QCD according to the
  formulae presented in Refs. \cite{Spira:1997dg,Han:1992hr}.
  Interference effects between W- and Z-boson fusion are neglected.
  The program allows to calculate the total production cross section
  for the scalar Higgs bosons of the SM and MSSM.  For the present
  study we employed the {\tt VV2H} version dated July 23, 2007, which
  was modified in order to switch off the contributions from b quarks
  in the final and/or initial states.
\end{itemize}

%\section{COMMON SETUP FOR THE CALCULATION}
\subsection{Common Setup for the Calculation}
\label{hjjcomp_se:setup}

\subsubsection{Input parameters and scheme definitions}

We choose the following set of input parameters
\cite{Eidelman:2004wy}, which have also been used in
Refs.~\cite{Ciccolini:2007jr,Ciccolini:2007ec}:
\begin{eqnarray}
\begin{array}[b]{r@{\,}lr@{\,}lr@{\,}l}
G_{\mu} &= 1.16637\times 10^{-5}\,\mathrm{GeV}^{-2}, \quad 
& \alpha(0) &= 1/137.03599911, \quad 
%& \alpha_{\mathrm{s}}(M_\mathrm{Z}) & = 0.1187,
\\ 
M_\mathrm{W}^{\mathrm{LEP}} &= 80.425\,\mathrm{GeV}, 
& \Gamma_\mathrm{W}^{\mathrm{LEP}} &= 2.124\,\mathrm{GeV},  && \\
M_\mathrm{Z}^{\mathrm{LEP}}         &= 91.1876\,\mathrm{GeV},
& \Gamma_\mathrm{Z}^{\mathrm{LEP}} &= 2.4952\,\mathrm{GeV}, && \\
m_\mathrm{e}                &= 0.51099892\,\mathrm{MeV}, & m_\mu 
&= 105.658369\,\mathrm{MeV}, & m_\tau &= 1.77699\,\mathrm{GeV},\\
m_u &= 66\,\mathrm{MeV}, & m_c &=1.2\,\mathrm{GeV}, 
& m_t &= 174.3\,\mathrm{GeV}, \\
m_d &= 66\,\mathrm{MeV}, & m_s &=150\,\mathrm{MeV}, 
& m_b &= 4.3\,\mathrm{GeV}.
\end{array}
\end{eqnarray}

{\tt CDD} uses the complex-mass scheme \cite{Denner:2005fg}. This
requires a fixed width in the W- and Z-boson propagators in contrast
to the approach used at LEP to fit the W~and Z~resonances, where
running widths are taken. Following Ref. \cite{Bardin:1988xt} to
convert the ``on-shell'' values of $M_V^{\mathrm{LEP}}$ and
$\Gamma_V^{\mathrm{LEP}}$ ($V=\mathrm{W},\mathrm{Z}$) to the ``pole
values'' denoted by $M_V$ and $\Gamma_V$, leads to
\begin{eqnarray}
\begin{array}[b]{r@{\,}l@{\qquad}r@{\,}l}
M_\mathrm{W} &= 80.397\ldots\,\mathrm{GeV}, 
& \Gamma_\mathrm{W} &= 2.123\ldots\,\mathrm{GeV}, \\
M_\mathrm{Z} &= 91.1535\ldots\,\mathrm{GeV},
& \Gamma_\mathrm{Z} &= 2.4943\ldots\,\mathrm{GeV}.
\label{hjjcomp_eq:m_ga_pole_num}
\end{array}
\end{eqnarray}
In {\tt VV2H} and {\tt VBFNLO} the W- and Z-boson masses are fixed
according to Eq.\ (\ref{hjjcomp_eq:m_ga_pole_num}) and the
vector-bosons widths are set to zero. 

The masses of the light quarks are adjusted to reproduce the hadronic
contribution to the photonic vacuum polarization of Ref.
\cite{Jegerlehner:2001ca}. Since quark mixing effects are suppressed
we neglect quark mixing and use a unit CKM matrix.
All quark masses are set to zero in {\tt VBFNLO}.
We use the $G_\mu$ scheme, i.e.\ we derive the electromagnetic
coupling constant from the Fermi constant according to
\begin{equation}
\alpha_{G_\mu} = \sqrt{2}G_\mu
M_\mathrm{W}^2(1-M_\mathrm{W}^2/M_\mathrm{Z}^2)/\pi. 
\end{equation}

CTEQ6 parton distributions \cite{Pumplin:2002vw} are used. Processes
with external bottom quarks are not included in this comparison. As
discussed in Section 3.4 of Ref.  \cite{Ciccolini:2007ec} these
contribute at the level of a few per cent. We use $M_\mathrm{W}$ as
factorization scale both for QCD and QED collinear contributions. For
the calculation of the strong coupling constant we employ
$M_\mathrm{W}$ as the default renormalization scale, include 5
flavours in the two-loop running of $\alpha_{\mathrm{s}}$, and fix
$\alpha_{\mathrm{s}}(M_\mathrm{Z})= 0.118$, consistent with the CTEQ6M
distribution.

\subsubsection{Phase-space cuts and event selection}

We employ the same jet definition parameters, phase-space and event
selection cuts as described in Refs.
\cite{Figy:2004pt,Ciccolini:2007jr,Ciccolini:2007ec}. Jet
reconstruction from final-state partons is performed using the
$k_\mathrm{T}$-algorithm \cite{Catani:1992zp} as described in Ref.
\cite{Blazey:2000qt}. Jets are reconstructed from partons of
pseudorapidity $|\eta|<5$ using a jet resolution parameter $D=0.8$. In
the EW corrections, real photons are recombined with jets according to
the same algorithm. Thus, in real photon radiation events, final
states may consist of jets plus a real identifiable photon, or of jets
only.

We study total cross sections and cross sections for the set of
experimental ``VBF cuts''. These cuts significantly suppress
backgrounds to VBF processes, enhancing the signal-to-background
ratio. We require at least two hard jets with
\begin{equation}
\label{hjjcomp_eq:tagjetcust}
p_{\mathrm{Tj}}>20\,\mathrm{GeV},\qquad |y_{\mathrm{j}}|<4.5,
\end{equation}
where $p_{\mathrm{Tj}}$ is the transverse momentum of the jet and
$y_{\mathrm{j}}$ its rapidity. Two tagging jets $\mathrm{j}_1$ and
$\mathrm{j}_2$ are defined as the two jets passing the cuts
(\ref{hjjcomp_eq:tagjetcust}) with highest $p_{\mathrm{T}}$ such that
$p_{\mathrm{Tj}_1}>p_{\mathrm{Tj}_2}$. Furthermore, we require that
the tagging jets have a large rapidity separation and reside in
opposite detector hemispheres:
\begin{equation}
\Delta y_{\mathrm{jj}}\equiv | y_{\mathrm{j}_1}-y_{\mathrm{j}_2}| > 4, 
\qquad y_{\mathrm{j}_1}\cdot y_{\mathrm{j}_2}< 0.
\end{equation}

%\section{NUMERICAL RESULTS}
\subsection{Numerical Results}
\label{hjjcomp_se:numerics}

In this section we present, for a range of Higgs-boson masses, LO and
NLO QCD corrected results obtained by {\tt CDD},
$\sigma^{\mbox{\tiny{\tt CDD}}}_{\mathrm{LO/NLO}}$, with {\tt VV2H},
$\sigma^{\mbox{\tiny{\tt VV2H}}}_{\mathrm{LO/NLO}}$, and with {\tt
  VBFNLO}, $\sigma^{\mbox{\tiny{\tt VBFNLO}}}_{\mathrm{LO/NLO}}$.
These results were calculated approximating the cross section by the
contribution of squared $t$- and $u$-channel diagrams only, without
any interferences. We also present the QCD corrected results,
including all diagrams and interference contributions,
$\sigma^\mathrm{full\ QCD}_{\mathrm{LO/NLO}}$, together with the
results including both QCD and EW corrections,
$\sigma^\mathrm{QCD+EW}_{\mathrm{LO/NLO}}$, as obtained by {\tt CDD}.

Table \ref{hjjcomp_ta:nocuts} contains results for the total
integrated cross section without any cuts. The small difference
between the results obtained by {\tt VV2H} and {\tt VBFNLO} is due to
the different treatment of vector-boson widths. We observe that the
approximate LO cross sections agree within $5\times10^{-4}$, and the
NLO corrected results within $2\times10^{-3}$, a difference which is
of the order of the statistical error. The complete predictions
$\sigma^{\mathrm{QCD+EW}}$ differ from the results of {\tt VV2H} and
{\tt VBFNLO} by up to 30\% for low Higgs-boson masses and by a few per
cent for high Higgs-boson masses.  The bulk of this big difference for
small values of  $M_\mathrm{H}$ is due to the $s$-channel contributions,
which are only considered by {\tt CDD}.
\begin{table}
\centerline{
\begin{tabular}{|c|c|c|c|c|c|c|}
\hline
$M_\mathrm{H}\ [\mathrm{GeV}]$ &
120 &
150 &
170 &
200 &
400 &
700 \\ \hline
$\sigma^{\mbox{\tiny{\tt CDD}}}_\mathrm{LO}\ [\mathrm{fb}]$ &
4226.3(6)  &
3357.8(5)  &
2910.7(4)  &
2381.6(3)  &
 817.6(1)  &
 257.49(4) \\
$\sigma^{\mbox{\tiny{\tt VBFNLO}}}_{\mathrm{LO}}\ [\mathrm{fb}]$  &
4227.1(1) &
3358.0(1) &
2910.8(1) &
2380.79(8) &
817.48(3) &
257.444(9) \\
$\sigma^{\mbox{\tiny{\tt VV2H}}}_\mathrm{LO}\ [\mathrm{fb}]$ &
4226.2(4) &
3357.3(3) &
2910.2(3)  &
2380.4(2) &
 817.33(8) &
 257.40(3)  \\
$\sigma^\mathrm{QCD+EW}_\mathrm{LO}\ [\mathrm{fb}]$ &
5404.8(9) &
3933.7(6) &
3290.4(5) &
2597.9(4) &
 834.5(1) &
 259.26(4)\\\hline
$\sigma^{\mbox{\tiny{\tt CDD}}}_\mathrm{NLO}\ [\mathrm{fb}]$ &
4424(4)   &
3520(3)   &
3052(3)   &
2505(2)   &
 858.4(7) &
 268.2(2) \\
$\sigma^{\mbox{\tiny{\tt VBFNLO}}}_\mathrm{NLO}\ [\mathrm{fb}]$ &
4414.8(2) &
3519.8(2) &
3055.9(2) &
2503.3(1) &
858.73(4) &
268.02(1) \\
$\sigma^{\mbox{\tiny{\tt VV2H}}}_\mathrm{NLO}\ [\mathrm{fb}]$ &
4415(1)   &
3519.7(8) &
3055.8(7)  &
2503.4(6) &
858.8(2)  &
268.03(6)  \\
$\sigma^\mathrm{full\ QCD}_\mathrm{NLO}\ [\mathrm{fb}]$ &
6030(4)   &
4313(3)   &
3579(2)   &
2802(2)   &
 878.9(6) &
 269.9(2) \\
$\sigma^\mathrm{QCD+EW}_\mathrm{NLO}\ [\mathrm{fb}]$ &
5694(4)   &
4063(3)   &
3400(3)   &
2666(2)   &
 839.0(7) &
 285.9(3) \\\hline
\end{tabular}
}
\caption{Total integrated cross section for $\mathrm{pp} \to
  \mathrm{H} + 2\mathrm{jets} + X$ in LO and NLO without any cuts,
  calculated by {\tt CDD}, $\sigma^{\mbox{\tiny{\tt
        CDD}}}_{\mathrm{LO/NLO}}$, with {\tt VV2H},
  $\sigma^{\mbox{\tiny{\tt VV2H}}}_{\mathrm{LO/NLO}}$, and with {\tt
    VBFNLO}, $\sigma^{\mbox{\tiny{\tt VBFNLO}}}_{\mathrm{LO/NLO}}$,
  for the setup defined in the text.} 
\label{hjjcomp_ta:nocuts}
\end{table}

Table \ref{hjjcomp_ta:vbfcuts} shows results for the integrated
cross section after imposing VBF selection cuts. We observe that the
approximate LO cross sections agree within $8\times10^{-4}$, and the
NLO corrected results within $1\times10^{-3}$, a difference which is
of the order of the statistical error.  The difference between the
complete predictions $\sigma^{\mathrm{QCD+EW}}$ and the results of
{\tt VBFNLO} is half a per mille or less in LO, and 6--8\%, the size
of the EW corrections, in NLO. This shows that, in this configuration,
$s$-channel and interference contributions can be safely neglected,
but EW corrections are as large as QCD corrections.
\begin{table}
\centerline{
\begin{tabular}{|c|c|c|c|c|c|c|}
\hline
$M_\mathrm{H}\ [\mathrm{GeV}]$ &
120 &
150 &
170 &
200 &
400 &
700 \\ \hline
$\sigma^{\mbox{\tiny{\tt CDD}}}_\mathrm{LO}\ [\mathrm{fb}]$ &
1686.2(3) &
1433.4(2) &
1290.3(2) &
1106.8(1) &
 451.27(5)&
 153.68(2)\\
$\sigma^{\mbox{\tiny{\tt VBFNLO}}}_{\mathrm{LO}}\ [\mathrm{fb}]$  &
1686.90(5) &
1433.79(4) &
1290.42(4) &
1106.97(3) &
451.31(1) &
153.689(4) \\
$\sigma^\mathrm{QCD+EW}_\mathrm{LO}\ [\mathrm{fb}]$ &
1686.5(3) &
1432.7(2) &
1289.8(2) &
1106.4(1) &
 451.16(5)&
 153.66(2)\\\hline
$\sigma^{\mbox{\tiny{\tt CDD}}}_\mathrm{NLO}\ [\mathrm{fb}]$ &
1728(2) &
1463(1) &
1313(2) &
1121(1) &
 444.8(3) &
 147.2(1)\\
$\sigma^{\mbox{\tiny{\tt VBFNLO}}}_\mathrm{NLO}\ [\mathrm{fb}]$ &
1728.8(2) &
1461.7(2) &
1311.7(1)&
1119.8(1) &
444.71(3) &
147.14(1) \\
$\sigma^\mathrm{full\ QCD}_\mathrm{NLO}\ [\mathrm{fb}]$ &
1738(2) &
1468(2) &
1318(1) &
1122(1) &
 445.0(4) &
 147.23(9)\\
$\sigma^\mathrm{QCD+EW}_\mathrm{NLO}\ [\mathrm{fb}]$ &
1599(2) &
1354(2) &
1230(1) &
1048(1) &
 419.2(4) &
 155.8(1)\\\hline
\end{tabular}
}
\caption{Integrated cross section for $\mathrm{pp} \to \mathrm{H} +
  2\mathrm{jets} + X$ in LO and NLO, including VBF selection cuts,
  calculated by {\tt CDD}, $\sigma^{\mbox{\tiny{\tt
        CDD}}}_{\mathrm{LO/NLO}}$, and with {\tt VBFNLO},
  $\sigma^{\mbox{\tiny{\tt VBFNLO}}}_{\mathrm{LO/NLO}}$, for the setup
  defined in the text.} 
\label{hjjcomp_ta:vbfcuts}
\end{table}

%\section{CONCLUSIONS}
\subsection{Conclusions}

We have presented results for NLO cross sections of Standard Model
Higgs-boson production via weak-boson fusion at the LHC. A tuned
comparison of QCD corrected results obtained by three different
calculations has been performed. Taking into account only $t$- and
$u$-channel diagrams we found good agreement. We have also presented
full NLO EW and QCD corrected results to gain some insight into the
nature of this approximation. We found agreement between the
approximate and full $\mathcal{O}(\alpha_s)$ results when VBF cuts are
applied. On the other hand, for the total integrated cross section,
there is a sizeable difference between those results, which arises
almost exclusively from $s$-channel contributions. Furthermore, EW
corrections are in general as large as QCD corrections.

\section*{Acknowledgements}
This work is supported in part by the European Community's Marie-Curie
Research Training Network under contract MRTN-CT-2006-035505 ``Tools
and Precision Calculations for Physics Discoveries at Colliders''.

%\bibliography{hjjcomp}

%\end{document}

\section[Loop Induced Interference Effects in Higgs Plus Two Jet 
Production at the LHC]
{LOOP INDUCED INTERFERENCE EFFECTS IN HIGGS PLUS TWO 
JET PRODUCTION AT THE LHC
~\protect\footnote{Contributed by J.~R.~Andersen, T.~Binoth, G.~Heinrich, 
J.~M.~Smillie}
}
\label{sec:int}
%\documentclass{cernrep}
%\usepackage{graphicx,epsfig}
%\usepackage{amsmath,amssymb}
%\usepackage{cite}
%\bibliographystyle{lesHouches}

%\begin{document}

%\title{Loop induced interference effects in Higgs plus two jet production at the LHC}

%\author{J.~R.~Andersen$^a$, T.~Binoth$^b$, G.~Heinrich$^c$, J.~M.~Smillie$^c$}
%\institute{$^a$ Theory Division, Physics Department, CERN, CH-1211 Geneva 23,
%  Switzerland\\
%$^b$School of Physics, The University of Edinburgh, Edinburgh EH9 3JZ, UK \\
%$^c$Institute of Particle Physics Phenomenology, Department of Physics, University of
%Durham, \\ Durham, DH1 3LE, UK}
%\maketitle

%\begin{abstract}
%  We calculate the order $\mathcal{O}(\alpha^2\alpha_s^3)$
%  interference between the gluon fusion and weak boson fusion
%  processes allowed at the one-loop level in Higgs boson plus two jet
%  production at the LHC.  
%  We discuss the various mechanisms which conspire to make this
%  contribution numerically negligible for experimental studies at the
%  LHC.
%\end{abstract}

\subsection{Introduction}
\label{sec:OLI_introduction}

\begin{figure}[hbp]
  \hspace{3.7cm} \includegraphics[height=2.7cm]{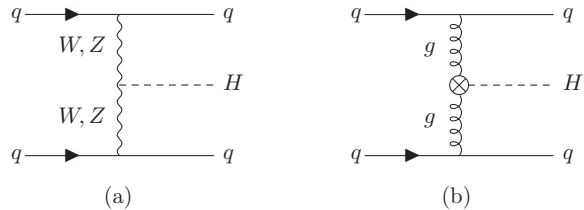}
  \caption{(a) The WBF process for Higgs production in the Standard Model and
    (b) the equivalent gluon-fusion diagram mediated through a top-loop.}
   \vspace{-0.5cm}
   \label{fig:OLI_VBF}
\end{figure}

Understanding the mechanism of electro-weak symmetry breaking is 
one of the primary goals at the CERN Large Hadron Collider.
Central to
this study is  the measurement of the couplings of any observed Higgs scalar to the
electro-weak bosons. A useful production process in this context is $pp\to
Hjj$~\cite{Cahn:1983ip,Dicus:1985zg,Altarelli:1987ue} through weak boson fusion
(WBF)~\cite{Plehn:2001nj}, as shown in Fig.~\ref{fig:OLI_VBF}(a),  with contributions from all
identifiable decay channels. The Higgs plus two jet signature also receives contributions
from Higgs boson production through gluon-fusion mediated through a top-loop, as
illustrated in Fig.~\ref{fig:OLI_VBF}\,(b).
However, the Higgs plus dijet-sample can be biased towards WBF by suppressing
the gluon-fusion channel through a combination of cuts.

The next-to-leading order corrections to Higgs plus two jet 
production are considered to be well under
control.  For WBF, both the radiative corrections within
QCD~\cite{Han:1992hr,Djouadi:1991tka,Figy:2003nv,Berger:2004pc} and the
electro-weak sector~\cite{Ciccolini:2007jr,Ciccolini:2007ec} have been calculated; for
the gluon fusion process, the first radiative corrections have been
calculated within QCD~\cite{Campbell:2006xx,DelDuca:2004wt} using the
heavy top mass effective
Lagrangian\cite{Wilczek:1977zn,Dawson:1991zj,Djouadi:1991tka}.  The
radiative corrections to the WBF channel are small, $3 \% - 6\% $, and
there is even partial numerical cancellation between the QCD and
electro-weak contributions.  It would therefore seem that the Higgs
coupling to electro-weak bosons can be very cleanly studied with a
$Hjj$-sample.

However, there is an irreducible (i.e.~unaffected by the WBF cuts)
contamination in the extraction of the $ZZH$-coupling from
interference between the gluon fusion and WBF processes, which  was ignored in
the literature until recently. At tree level, such interference is only allowed in
amplitudes where there is a $t\leftrightarrow u$-channel crossing
which leads to a high level of kinematic
suppression~\cite{Andersen:2006ag}. These and other
crossing-suppressed one-loop amplitudes were later calculated 
together with the electro-weak corrections~\cite{Ciccolini:2007jr,Ciccolini:2007ec}.

Here we will report on the calculation of the processes allowed at the
one-loop level which do not suffer from the kinematic suppression stemming from the
requirement of a $t\leftrightarrow u$-crossing~\cite{Andersen:2007mp}.  
At  order $\mathcal{O}(\alpha^2\alpha_s^3)$, one finds an interference
term between the gluon- and $Z$-induced amplitude which is not allowed at
$\mathcal{O}(\alpha^2\alpha_s^2)$ by colour conservation.  
The $W$-induced
amplitudes are crossing-suppressed and therefore not taken into
account.  The diagrams where the vector boson is in the $s$-channel
can also be safely neglected because they are strongly suppressed by
the WBF cuts.  As discussed in Ref.\cite{Andersen:2006ag}, 
%this is the lowest order contribution to the interference between $ZZH$ and
%$ggH$-processes for non-identical quark flavours and helicity
%configurations, and 
for identical quark flavours 
the loop amplitudes are the first order which does not require a
kinematically disfavoured crossing.
% Given that electroweak corrections to the WBF process (an order
% $\mathcal{O}(\alpha)$ correction to an $\mathcal{O}(\alpha^4)$
% process) have been shown to be relevant for this important process
% \cite{Ciccolini:2007jr}, simple power counting alone suggests that the
% size of the irreducible contamination due to the discussed
% interference effect should be checked.

In the following section we will briefly sketch the calculation before discussing our
results in section \ref{sec:OLI_results}, which are summarized in the conclusions.

\subsection{The Calculation}
\label{sec:OLI_calculation}
\begin{figure}[tbp]
  \hspace{2.2cm} \includegraphics[height=3cm]{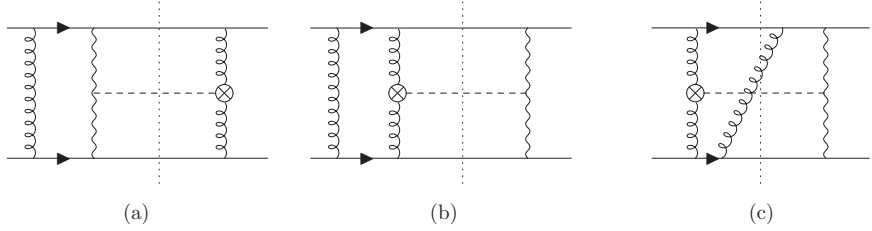}
  \caption{\label{fig:OLI_interfdiags} Example of contributing
    one-loop interference terms: (a) $\mathcal{M}_{gZ}\mathcal{M}_g^*$
    and (b) $\mathcal{M}_{gg}\mathcal{M}_Z^*$. There are four
    contributing topologies for each gluon-fusion and $Z$-fusion
    process. (c) shows a real emission processes at matrix
    element squared level.}
\end{figure}

Our calculation of the 
loop interference terms and the real emission contributions is based on helicity 
amplitudes. 
The leading order amplitudes, denoted by $\mathcal{M}_Z$ 
and $\mathcal{M}_g$ (Fig.~\ref{fig:OLI_VBF}(a) and (b)), 
are proportional to a colour singlet
and a colour octet term.  The colour singlet is formally of order
$\mathcal{O}({\alpha}^2)$ whereas the octet is of order
$\mathcal{O}(\alpha_s^2)$.  The one-loop amplitudes, which we call
$\mathcal{M}_{gZ}$ and $\mathcal{M}_{gg}$ respectively, are mixtures of octet and singlet
terms.  For the interference term we need to consider only the octet part of
$\mathcal{M}_{gZ}$ and the singlet part of $\mathcal{M}_{gg}$.  One finds that only four
one-loop five-point topologies for each amplitude survive this colour projection. 
Sample diagrams are shown in Fig.~\ref{fig:OLI_interfdiags}. 

The loop amplitudes require the evaluation of one-loop five-point tensor integrals with a
mixture of massless and massive configurations in both propagators and external legs. We
apply the reduction algorithm outlined in Ref.~\cite{Binoth:1999sp,Binoth:2005ff} to
express each Feynman diagram as a linear combination of 1-, 2-, and 3-point functions in
$D=4-2\epsilon$ dimensions and 4-point functions in $D$=6. The same algorithm has been
successfully applied to a number of one-loop
 computations~\cite{Binoth:2006mf,Binoth:2006ym,Binoth:2005ua,Binoth:2003xk}, 
 where further details can be found.
The algebraic
expressions were checked by independent implementations, both amongst the authors and with
another group~\cite{Dixon:priv}.

After the algebraic reduction, all helicity amplitudes are obtained
as linear combinations of  scalar integrals. 
% Schematically
% \begin{equation} 
%   \mathcal{M} =
%   \sum\limits_{j,\alpha} c_{j\alpha} I_j(\{s_\alpha,m_\alpha\}) \; , \quad I_j
%   \in \{I_2^D,I_3^D,I_4^{D+2} \} 
% \end{equation} 
% where the summation over $\alpha$
% indicates the summation over different argument lists $\{s_\alpha,m_\alpha\}$
% of a certain type of master integral (MI). 
No one-point functions appear in
the reduction, and also two-point functions are absent in the amplitudes of
$\mathcal{M}_{gZ}$. Furthermore, coefficients of some of the integrals which
arise in several topologies sum to zero. If the tree resulting from a certain
cut of a master integral corresponds to a helicity forbidden tree level
process, one can immediately infer the vanishing of the corresponding
coefficient. In our algebraic tensor reduction approach we  verify and use 
such cancellations before the numerical
evaluation of the cross section.  

As most of the required scalar integrals are not provided in the literature, we have evaluated
representations in terms of analytic functions valid in all kinematic regions.  These 
can be found in~\cite{Andersen:2007mp} for use in other calculations.
  
We used dimensional regularisation to extract the IR singularities from the 
divergent integrals.  The leading $1/\varepsilon^2$ poles cancel, but there remains a
$1/\varepsilon$ pole which is cancelled when the virtual corrections 
 are combined with the real emission part shown in Fig.~\ref{fig:OLI_interfdiags}(c).
As to be expected, 
the collinear IR divergences from the three-parton final states integrate to zero, leaving
only a soft divergence proportional to $1/\varepsilon$, 
which we isolated using the phase space slicing
method\cite{Giele:1991vf,Giele:1993dj}.
The phase space integration and the numerical evaluation
of integrals and coefficients is coded in a \texttt{C++} program allowing for a flexible
implementation of cuts and observables.

\subsection{Results}
\label{sec:OLI_results}

As the aim of our study was to investigate a possible pollution of the
clean extraction of the $ZZH$ vertex structure by the interference
terms, we apply the cuts summarised in Table \ref{tab:OLI_cuts}. These
are generally used to single out the WBF events from  the gluon
fusion ``background''~\cite{DelDuca:2001fn}.
\begin{table}[tbp]
  \centering
  \begin{tabular}{|rl||rl|}
    \hline
    $p_{a_T}$, $p_{b_T}$ & $> 20$ GeV & $\eta_a\cdot\eta_b$& $<0$ \\
    $\eta_j$ & $<$ 5 & $\vert \eta_a-\eta_b \vert$ & $> 4.2$ \\
    $s_{ab}$ & $>$ (600 GeV)$^2$ & & \\ \hline
  \end{tabular}
  \caption{The cuts used in the following analysis which bias the Higgs Boson
    plus dijet sample towards WBF.  The indices $a,b$ label the tagged jets.}
  \label{tab:OLI_cuts}
\end{table}
Our input parameters for the numerical studies are taken
from~\cite{Martin:2004ir} and~\cite{PDBook}.  In addition, we use a
Higgs boson mass of 115 GeV and the NLO parton distribution set from
Ref.\cite{Martin:2004ir}. We  use 2-loop running for $\alpha_s$,
in accordance with the chosen pdfs.

We observe that in all the flavour and helicity channels, the
contribution from the 3-parton final state is numerically negligible.
The only r\^ole of this real emission is
to cancel the divergences which arise from the one-loop diagrams.

As the interference effect is proportional to $2
\mathrm{Re}(\mathcal{M}_{gg}\mathcal{M}_Z^*+\mathcal{M}_{gZ}\mathcal{M}_g^*)$,
the result is not necessarily positive definite.  In fact, the sign of the
interference contribution depends on the azimuthal angle between the two
tagging jets, $\Delta\phi_{jj}$. As the event topology has two well
separated jets, it becomes possible to define an orientation of the azimuthal
angle which allows observability in the whole range of $[-\pi,\pi]$, as
pioneered in Ref.\cite{Plehn:2001nj,Hankele:2006ma}.  $\Delta\phi_{jj}$ is then defined
through
\begin{align}
  \begin{split}
    \label{eq:OLI_phijj}
  {|p_{+_T}||p_{-_T}|} \cos \Delta\phi_{jj}&= p_{+_T}\cdot
      p_{-_T},\\
    2 |p_{+_T}|| p_{-_T}| \sin \Delta\phi_{jj}&=\varepsilon_{\mu\nu\rho\sigma}
    b_+^\mu p_+^\nu
    b_-^\rho p_-^\sigma,
  \end{split}
\end{align}
where $b_+$ ($b_-$) are unit vectors in positive (negative) beam direction, and likewise
for the jet momenta $p_\pm$. The cuts ensure that the two tagging jets lie in opposite
hemispheres. Defined in this way, the observable $\Delta\phi_{jj}$ becomes a powerful
discriminator for different $C\!P$-structures of the Higgs Boson production
vertex~\cite{Hankele:2006ma}.

\begin{figure}[tbp]
  \centering
  \includegraphics[width=0.45\textwidth]{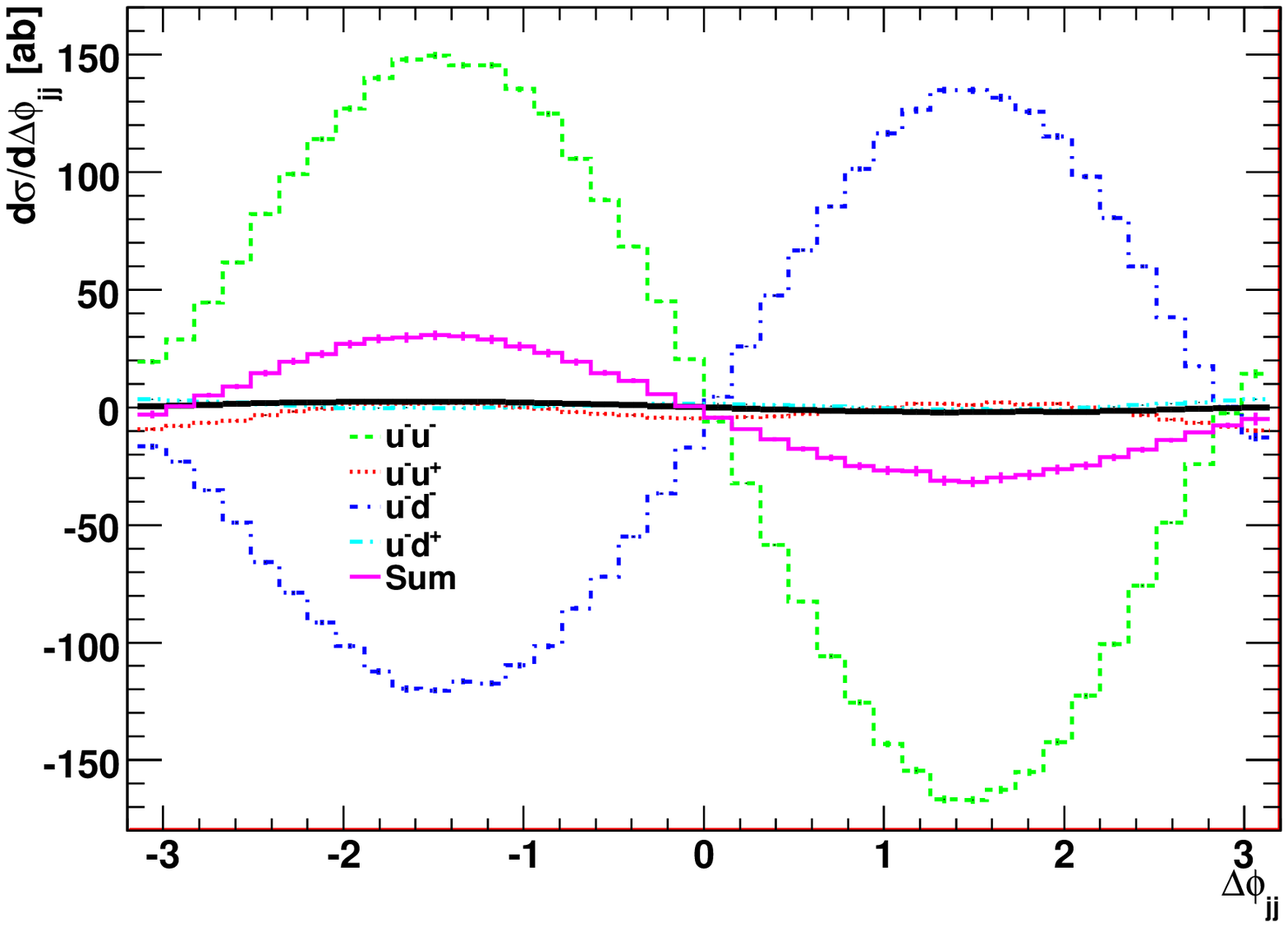} \hspace{0.3cm}
  \includegraphics[width=0.45\textwidth]{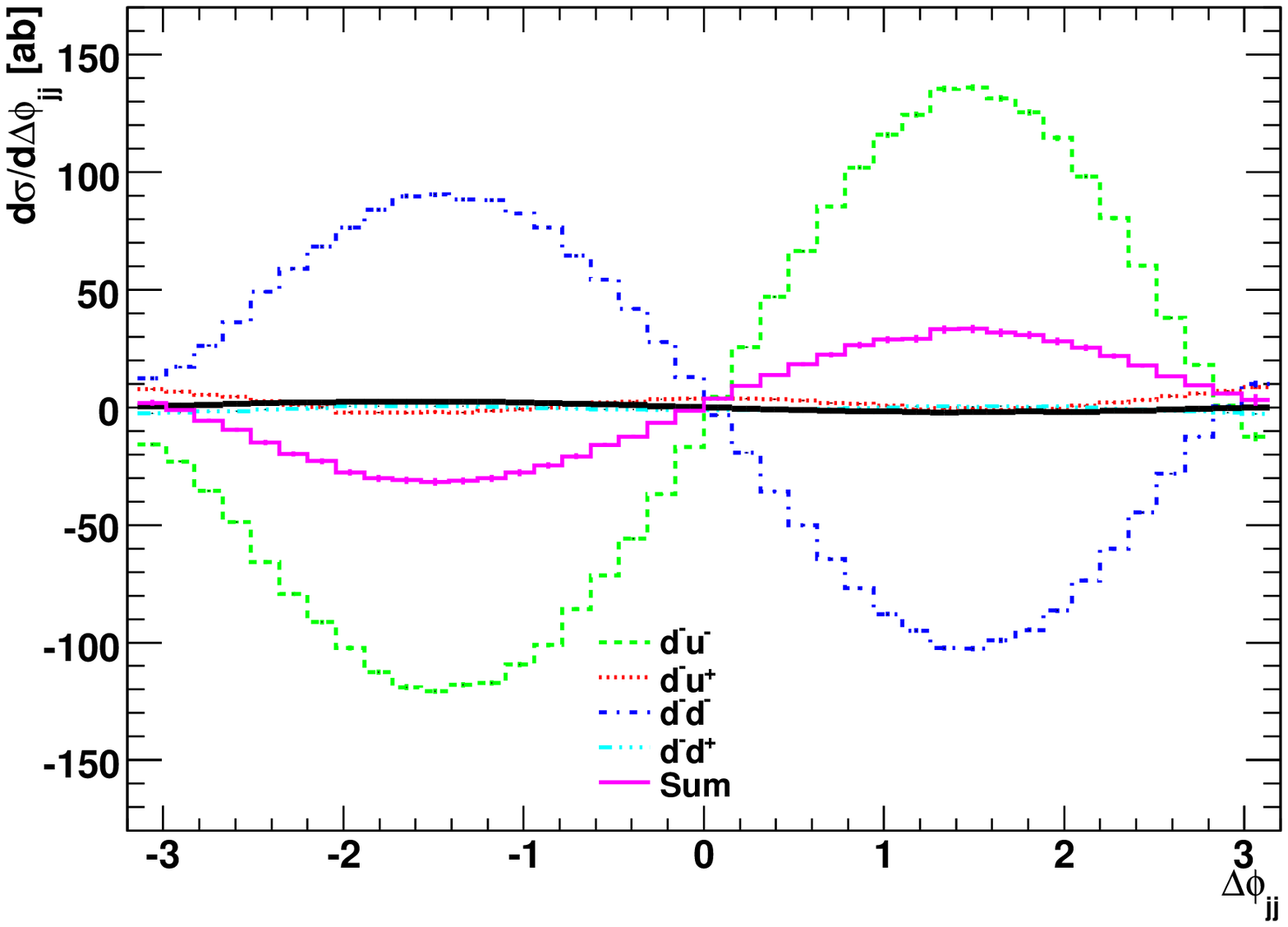}
  \caption{The $\Delta\phi_{jj}$-distribution for various flavour and
    helicity-configurations. The purple histogram labelled ``Sum'' indicates
    the sum over the four contributions shown. The sum over \emph{all}
    flavour and helicity assignments including all sea flavours is shown in the 
    black histogram.}
  \label{fig:OLI_flavhel_seaval}
\end{figure}

Figure~\ref{fig:OLI_flavhel_seaval} displays the contribution to the distribution in
$\Delta\phi_{jj}$ from the interference terms for various helicity and flavour
configurations.
There is an accidental cancellation of sea and valence quark
contributions which leads to the fact that the sum over all flavour
and helicity assignments peaks at around $2$~ab/rad only, with an
integrated effect of $1.19\pm0.07$~ab, where the error is due to the
numerical integration.

Due to the oscillatory behaviour, the total integrated cross section
does not represent the impact on the $\Delta\phi_{jj}$ distribution.
The integral of the absolute value of the $\Delta\phi_{jj}$ distribution, 
$\int_{-\pi}^\pi d\Delta\phi_{jj}\Big|\frac{d\sigma}{d\Delta\phi_{jj}}\Big|$, 
is a more useful measure of the impact of the interference effect on the
extraction of the $ZZH$-vertex.  This integral evaluates to $9.1\pm
0.1$~ab, an order of magnitude larger than the integral over the 
oscillating distribution.  The total integral over the
the absolute value of the fully differential cross section leads to
$29.59\pm 0.07$~ab.

Using the same cuts and value for the mass of the Higgs boson as in the
present study, we have checked that the total contribution to the
$\Delta\phi_{jj}$-distribution from the leading order WBF process (both $Z$
and $W^{+/-}$ included) is relatively flat at around 240~fb/rad. Therefore,
the result of the interference effect reported here is unlikely to be
measurable.

As can be readily seen in Fig.~\ref{fig:OLI_flavhel_seaval}, there is also 
a cancellation between the contribution from each flavour and helicity
assignment, as has also been pointed out 
in~\cite{Forshaw:2007vb}; this is because the sign of quark couplings to the
$Z$-boson becomes relevant as it is not squared for the
interference. The flavour- and helicity sum for each quark line
therefore leads to some cancellation, which amounts to roughly $20\%$
in the most relevant regions of the pdfs~\cite{Andersen:2007mp}. 
%There is also a change of sign in the
%relevant x-region which leads to a further reduction of the
%interference term after integration.

The complex phases arising from the full one-loop calculation of the
amplitudes also give rise to some suppression. We find that the
relevant products and sums for the interference effect project out
only about $20\%$ of the full complex loop amplitudes.

We chose the factorisation and renormalisation scales as in accordance with the
natural scales in the relevant high energy limit (as explained in
Ref.\cite{DelDuca:2006hk}), i.e. the factorisation scales are set equal to
the transverse momenta of the relevant jet, and the renormalisation scale for
the strong couplings are chosen correspondingly, i.e. one $\alpha_s$
evaluated at each value of the transverse momentum of the jets, and one at
the Higgs mass. 
However, we find that varying the choice of factorisation and
renormalisation scales, the exact numerical values of the cuts or the
parameters, or the choice of pdf set has no impact on the
conclusions: %
%%The effect on the interference and the WBF signal is
%%similar, so the relative
the numerical importance of the interference is basically unchanged.

\subsection{Conclusions}
\label{sec:OLI_conclusions}
We have outlined the calculation of the loop-induced  
$\mathcal{O}(\alpha^2\alpha_s^3)$ interference effect between the gluon 
fusion and weak boson fusion processes
in Higgs boson plus two jet production at the LHC.

We find by explicit calculation that this contribution is too
small to contaminate the extraction of the $ZZH$-coupling from WBF
processes.  Interestingly the effect which survives comes dominantly
from the virtual corrections.  We have analysed in detail why this
contribution is so small, and instead of a single effect we rather
find a conspiracy of several mechanisms:
\begin{itemize}
\item accidental cancellations between the sea quark 
      and valence quark contributions 
\item compensations between different weak isospin components of the valence quarks 
      due to their $SU(2)\times U(1)$ couplings,
      in combination with their weights from the (valence) quark content of the proton 
\item cancellations due to destructive interference of the phases from the 
      different contributions.
\end{itemize}
The exact impact of these partly accidental effects, in particular 
the latter, could not be assessed without  an
explicit calculation.

Finally we would like to point out that anomalous couplings which affect the
phases could change the interference pattern substantially. However, the first two
cancellation mechanisms still being present, we still expect the overall contribution to
be experimentally insignificant.
Please see Ref. \cite{Andersen:2008ue} for more details.

\section*{Acknowledgements}
We would like to thank Lance Dixon for enlightening and encouraging discussions and
important comments. The authors were all supported by the UK Science and Technology
Facilities Council. In addition, the work of TB was supported by the Deutsche
Forschungsgemeinschaft (DFG) under contract number BI 1050/2 and the Scottish Universities
Physics Alliance (SUPA).%
%
%\bibliography{bibli_OLI}

%\end{document}

%%% Local Variables: 
%%% mode: latex
%%% TeX-master: t
%%% End: 

\section[Higgs Boson Production in Association
With Multiple Hard Jets]
{HIGGS BOSON PRODUCTION IN ASSOCIATION WITH 
MULTIPLE HARD JETS~\protect
\footnote{Contributed by: J.R.~Andersen and C.D.~White}}
%\usepackage{graphicx}
%\usepackage{cite}
%\bibliographystyle{lesHouches}
%\begin{document}
%%
%
%\title{Higgs Boson Production in Association with Multiple Hard Jets}
%
%\author{J.R.~Andersen$^1$, C.D.~White$^2$}
%\institute{$^1$Theory Division, Physics Department, CERN, CH 1211 Geneva 23, Switzerland,
%\\$^2$NIKHEF, Kruislaan 409, 1098 SJ Amsterdam, The Netherlands}
%
%\maketitle
%
%\begin{abstract}
%  We present a new technique for estimating the emission of multiple (at
%  least two) hard jets in Higgs production via gluon-gluon fusion. It is
%  based upon high energy factorisation, with the region of applicability
%  extended by constraints on the behaviour of the scattering amplitudes
%  stemming from known all-order results. The resulting approximation is
%  matched to the known tree level matrix elements for 2 and 3 jet final
%  states, and implemented in a Monte Carlo generator. Example results are
%  presented. We find that multiple emission of hard jets leads to
%  significiant azimuthal decorrelation between the tagged final state
%  jets. The technique is readily extendable to other multi-jet processes.
%\end{abstract}

\subsection{Introduction}
% In the Standard Model, the Higgs boson has two production
% modes of significant cross-section and signal to background ratio. The
% highest cross-section arises in the gluon-gluon fusion (GGF) channel, where
% the Higgs couples to a $t$-channel gluon exchange via a top quark loop, where to a
% good approximation one may model this by a contact interaction. A significant
% background arises due to the strong nature of the interaction, and thus a
% second production mode - vector boson fusion (VBF) - is also considered
% widely in the literature as a discovery channel. Due to the electroweak
% nature of the $t$-channel exchange, the final state radiation properties,
% together with angular properties of the final state tagged jets, are
% significantly different.

% In order to the investigate the differences between these two processes, it
% is clearly desirable to understand them in as much detail as
% possible. Furthermore, the GGF process is itself a valuable discovery mode if
% one is able to reliably estimate its higher order corrections. Thus, it is
% necessary to consider the differential cross-section for Higgs production in
% GGF with an accompanying number of final state parton emissions i.e. the
% process $pp\rightarrow hj_1j_2\ldots j_n$, where $j_i$ represents a final
% state hadronic jet. 
It is widely hoped that the LHC will discover the source of electro-weak
symmetry breaking, mediated by the Higgs scalar within the context of the
Standard Model. In order to determine whether any observed fundamental scalar
is the Higgs Boson of the Standard Model, it is imperative to determine its
couplings, especially to the weak gauge bosons. This is possible both by
measuring the decay of the Higgs boson through the weak bosons, but also by
isolating the Higgs Boson production through weak boson fusion (WBF). This process
contributes to the signal for the production of a Higgs boson in association
with two jets. This channel also receives a significant contribution from
higher order corrections to Higgs boson production through gluon fusion. In
fact, it has recently been suggested\cite{Klamke:2007cu} that the increased
significance of the signal over the background obtained by requiring at least
two hard jets in association with a Higgs boson may decrease the necessary
integrated luminosity required for a discovery of the Higgs boson through
gluon fusion processes. However, in order to measure the Higgs boson
couplings to the weak bosons, it is necessary to suppress the gluon fusion
contribution to the production of a Higgs boson in association with two
jets. This is achieved\cite{DelDuca:2001fn} by applying the so-called
\emph{weak boson fusion}-cuts:
\begin{table}[tbp]
  \centering
  \begin{tabular}{|rl||rl|}
    \hline
    $p_{c_\perp}$, $p_{d_\perp}$, $p_{j_\perp}$ & $> 40$ GeV & $y_c\cdot y_d$& $<0$ \\
    $y_j$ & $<$ 5 & $\vert y_c-y_d \vert$ & $> 4.2$ \\
    $s_{cd}$ & $>$ (600 GeV)$^2$ & $y_c\le y_h\!\!\!\!\!$&$\le y_d$ \\ \hline
  \end{tabular}
  \caption{The cuts used in the following analysis which bias the Higgs boson
    plus dijet sample towards WBF.  The suffices $c,d$ label the tagged jets,
    $j$ any (possibly further) jet in the event.}
  \label{tab:cuts}
\end{table}
It is expected that the contribution from the gluon fusion process will be
further suppressed relative to WBF by vetoing further jet
activity\cite{Barger:1995zq}. The efficiency of such cuts can only be
assessed by calculating the higher order corrections to the gluon fusion
contribution to the $hjj$-channel. The first radiative corrections have
recently been calculated\cite{Campbell:2006xx}. While this fixed order
approach certainly is the best tested and understood approach for predicting
the first few perturbative corrections, the calculational complexity means
that currently the production of $hjj$ through gluon fusion has only been
calculated at next-to-leading order.

% Present methods for estimating higher jet rates in Higgs production include
% direct calculation of the associated matrix elements at up to NLO in QCD
% perturbation theory. This is the best approach, but calculational complexity
% increases rapidly with the number of external particles such that many
% desirable final states are beyond the present limit of exact perturbative
% calculation. It is also possible to estimate final state radiation by using
% parton shower algorithms, and a study of this kind has been presented in
% \cite{??}. Given that this amounts to a soft and collinear resummation of
% parton emission, however, one underestimates significantly the amount of hard
% radiation expected in the final state.

It is possible to estimate final state jet emission in this
process\cite{DelDuca:2006hk} using parton shower algorithms. In this
contribution we examine a different approach, and consider how to best
estimate hard jet emission in Higgs production via gluon fusion. We take as a
starting point a factorised form for the scattering amplitudes, which
formally applies in a certain kinematic limit (that of multi-Regge-kinematics
(MRK)). We extend the domain of applicability of the amplitudes from
Asymptotia into the region of relevance for the LHC by using known all-order
constraints of scattering amplitudes. We validate the approach by checking
the approximations in a comparison with fixed order results, where these are
available. Furthermore, the resulting estimate for the $n$-parton final state
(which includes some virtual corrections) is then matched to the known tree
level results for hjj and hjjj. Finally, we implement the description in a
Monte Carlo event generator for Higgs + multiparton production, and present a
sample of results.

\subsection{Estimating Multijet Rates}

% Start by calculating tree level hjj and hjjj to show the relevance of
% multi-parton states

\subsubsection{The FKL Amplitude}

Our starting point is the FKL factorised $(2\to n+2)$-gluon amplitudes\cite{Fadin:1975cb} adapted to
include also a Higgs boson
\begin{eqnarray}
i{\cal M}_{\mathrm{HE}}^{ab\rightarrow p_0\ldots p_jhp_{j+1}p_n}&=&2i\hat s
\left(i g_s f^{ad_0c_1} g_{\mu_a\mu_0}\right)\nonumber\\
&\cdot&
\prod_{i=1}^j \left(\frac{1}{q_i^2}\exp[\hat\alpha(q_i^2)(y_{i-1}-y_i)]\left(i g_s f^{c_id_ic_{i+1}}\right)C_{\mu_i}(q_i,q_{i+1})\right)\nonumber\\
&\cdot&\left(\frac 1 {q_h^2}\exp[\hat\alpha(q_i^2)(y_{j}-y_h)]C_{H}(q_{j+1},q_{h})\right)\label{FKL}
\\
&\cdot&\prod_{i=j+1}^n
\left(\frac{1}{q_i^2}\exp[\hat\alpha(q_i^2)(y'_{i-1}-y'_i)]\left(i g_s f^{c_id_ic_{i+1}}\right)C_{\mu_i}(q_i,q_{i+1})\right)\nonumber\\
&\cdot&\frac 1 {q_{n+1}^2}\exp[\hat\alpha(q_{n+1}^2)(y'_{n}-y_b)]\left(i g_s f^{bd_{n+1}c_{n+1}} g_{\mu_b\mu_{n+1}}\right)\nonumber
\end{eqnarray}
where $g_s$ is the strong coupling constant, and $q_i, q_h$ are the 4-momentum
of gluon propagators (e.g.~$q_i=p_a-\sum_{k=0}^{i-1}p_k$ for $i<j$), $C_{\mu_i}$ is the {\it Lipatov effective vertex}
for gluon emission, and $C_H$ is the effective vertex for the production of a
Higgs boson, as calculated in Ref.\cite{DelDuca:2003ba}. Furthermore, $\hat{\alpha}(q_i^2)$
occurs from the Reggeisation of the gluon propagator, and encodes virtual
corrections (see e.g. \cite{DelDuca:1995hf}). This approximation formally
applies in the MRK limit, which can be expressed in terms of the rapidities
$\{y_i\}$ of the outgoing partons and their transverse momenta
$\{p_{i\perp}\}$:
\begin{equation}
y_0\gg y_1\gg \ldots\gg y_{n+1};\quad p_{i\perp}\simeq p_{i+1\perp};\quad q_i^2 \simeq q_j^2.
\label{MRK}
\end{equation}
This limit is particularly well suited for studies within the WBF cuts of
Table~\ref{tab:cuts}, since a large rapidity span of the event is then guaranteed.

%% Relation to BFKL 
In the true limit of MRK, the squared 4-momenta $q_i^2\to-q_{\perp i}^2$, and
the square of the Lipatov vertices fulfil $-C_{\mu_i}
C^{\mu_i}\to4\frac{q_{\perp i}q_{\perp i+1}}{k_{\perp i}}$. Applying these
limits, the sum over $n$ to infinity of the amplitudes in Eq.~(\ref{FKL}),
integrated over the full phase space of emitted gluons can be obtained by
solving two coupled BFKL equations. This result would then apply to the
phase space of
\begin{equation}
y_0\gg y_1\gg \ldots\gg y_{n+1};\quad p_{i\perp}\simeq p_{i+1\perp};\quad q_{\perp i}^2 \simeq q_{\perp j}^2.
\label{BFKL-MRK}
\end{equation}

%% How we extend applicability away from strict MRK region
While both expressions are equally valid in the region of Asymptotia, we
extend the applicability of the results obtained in the High Energy Limit to
the region of interest for particle physics phenomenology by adhering to the
following guidelines:
\begin{enumerate}
\item \textsc{Do not introduce new divergences}: Using the expression
  in Eq.~(\ref{FKL}) corresponds to \emph{removing} some divergences from the
  full scattering amplitude (the collinear divergences), but not
  \emph{moving} any divergences. The expression in Eq.~(\ref{FKL}) is
  divergent only for momentum configurations for which the full scattering
  amplitude is also divergent. This is different to the case where the MRK
  limit of invariants has been substituted (resulting from the use of the
  BFKL equation), which displaces divergences within the phase space region
  of interest for the LHC.
\item \textsc{Do not apply the formalism where it fails}: We choose minimal
  interception by only removing the \emph{small} region on phase space where
  the expression of Eq.~(\ref{FKL}) results in unphysical (negative)
  differential cross sections. This happens when the effective Lipatov vertex
  is applied to momentum configurations very far from the MRK, where it is
  possible to obtain $-C_{\mu_i}C^{\mu_i}<0$. It is perhaps interesting to
  note that restricting the region of phase space where the formalism is
  applied is similar to the \emph{kinematic constraint} of
  Ref.\cite{Ciafaloni:1987ur,Catani:1989yc,Kwiecinski:1996td}, although in
  fact the latter fails to exclude all of the region where the formalism
  underpinning the BFKL equation fails.
\end{enumerate}

\begin{figure}
\begin{center}
\scalebox{0.5}{\includegraphics{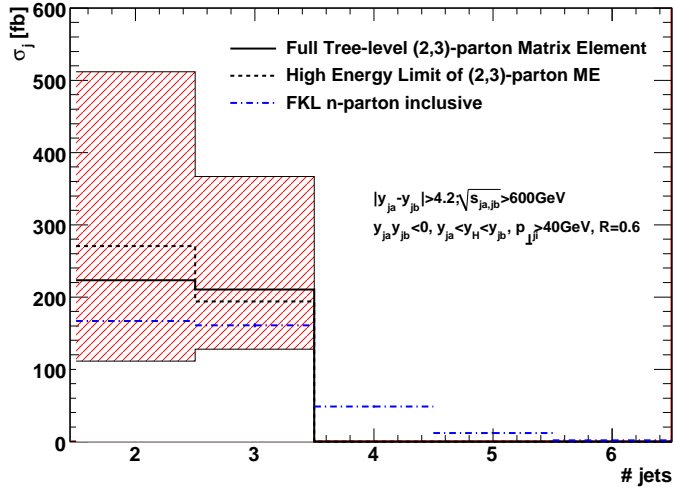}}
\caption{The 2 and 3 parton cross-sections calculated using the known LO
  matrix elements (solid), and the estimate gained from the modified high
  energy limit (dashed). One sees that the estimate is well within the scale
  variation of the LO result, obtained by varying the common renormalisation
  and factorisation scales in the range $0.5\leq\mu/\mu_0\leq2$, where
  $\mu_0$ is the default choice (indicated by the shaded regions). Also
  shown is the result obtained from the fully inclusive $n$-parton sample of
  Eq.~(\ref{FKL}).}
\label{LO}
\end{center}
\end{figure}

In figure \ref{LO} we compare the prediction for the production of a Higgs
boson in association with two and three partons (in a hard two-jet and
three-jet configuration respectively) within the WBF cuts of
table~\ref{tab:cuts}, obtained using both the full matrix element (extracted
from \texttt{MADEvent/MADGraph}\cite{Alwall:2007st}) and the relevant
expression of Eq.~(\ref{FKL}) for two and three parton production, with the
virtual corrections set to zero ($\hat\alpha=0$). We choose renormalisation
and factorisation scale in accordance with the study of
Ref.\cite{DelDuca:2006hk}.
One notes two things. Firstly, the approximation to the jet rates is well
within the scale uncertainty of the known tree level results. We have
therefore explicitly shown that the terms taken into account in this approach
indeed dominate . Secondly, the cross section for the
production of a Higgs boson in association with 3 jets is similar to the one
for the production of a Higgs boson in association with two jets. The large
size of the three-jet rate was already reported in Ref.\cite{DelDuca:2004wt},
and clearly demonstrates the necessity of considering hard multi-parton final
states in order to describe correctly the expected event topology and to
answer questions on e.g.~the effectiveness of a central jet veto in
suppressing the gluon fusion channel.

\subsection{All Order Results and Matching}

The divergence in Eq.~(\ref{FKL}) obtained when any $p_i\to 0$ is regulated
by the divergence of the virtual corrections encoded in $\hat\alpha$. By
implementing the regularisation through phase space slicing it becomes
possible to obtain the fully inclusive any-parton sample by summing
Eq.~(\ref{FKL}) over all $j,n$. This is very efficiently implemented by
following the method for phase space generation outlined in
Ref.\cite{Andersen:2006sp}. Furthermore, since we can trivially expand the
expressions to any order in $\alpha_s$, it is possible to check the
performance of the formalism against the available tree-level results, and to
implement matching to these. We choose to implement $\ln R$-matching at the
amplitude-level for channels which have a contribution in the high-energy
limit (e.g.~$ug\to hug$ and $gg\to hggg$), and $R$-matching for those which
do not (e.g.~$gg\to hu\bar u$ and $u\bar u\to hggg$).

% Having built an effective approximation to the multi-parton rates, one can
% improve the approximation by matching to the known tree level rates for lower
% orders. One must combine the LO results with the virtual and higher order
% emission estimates from the modified high energy approach, in such a way that
% one does not double-count terms occuring in both the high energy limit and
% the tree-level matrix elements. The matching procedure is not unique, but can
% be modified by terms that are subleading with respect to the high energy
% limit. Letting ${\cal M}_{ab}^{LO, HE}$ respectively denote the tree level
% and high energy approximations to a given multijet amplitude of partonic
% initial state ${ab}$, two possible methods are the so-called $R$- and
% $\log{R}$-matching schemes. In the first, one matches the matrix element
% itself:
% \begin{equation}
% {\cal M}_{ab}^{matched}={\cal M}_{ab}^{LO}+{\cal M}_{ab}^{HE}-{\cal M}_{ab}^{HE(0)},
% \label{R}
% \end{equation}
% where the final term denotes the high energy limit of the LO result and removes the double-counted piece. In $\log{R}$ matching, one applies equation (\ref{R}) to the logarithm of the amplitude. 

% Our prescription is to use logarithmic matching for all partonic subprocesses that have a non-vanishing high energy limit. For all other processes, one must use $R$ matching.

% \section{RESULTS}
It is now possible to cluster each event in the inclusive sample of a Higgs
boson plus $n$ partons into jets according to a given algorithm. As an
example, we choose KtJet\cite{Butterworth:2002xg}. We use the parton
distribution functions of Ref.\cite{Martin:2004ir}. The distribution of final
state jets subject to the cuts of table \ref{tab:cuts} is shown with the
dashed histogram in Figure~\ref{LO}. One sees a significant number of events
with more than 3 hard ($p_\perp>40$GeV) jets. More importantly though, the
method outlined in this paper allows for an estimate of the emissions of
partons not quite hard enough to be classified as jets, but still causing
sufficient decorrelation. The azimuthal angular correlation between the
tagging jets has been suggested previously as a good observable for
differentiating between the GGF and WBF production modes. Furthermore, the
nature of the distribution of the azimuthal angle $\phi$ between the two
tagging jets can potentially be used to determine the nature of the Higgs
coupling to fermions \cite{Klamke:2007cu}. However, the usefulness of this
observable is threatened by hard jet emission which acts to decorrelate the
tagging jets. As suggested in Ref.\cite{Hankele:2006ja} the structure of the
distribution $d\sigma/d\phi_{j_aj_b}$ can be distilled into a single number
$A_\phi$ given by:
\begin{equation}
  \label{eq:Aphi}
  A_\phi=\frac{\sigma(\phi_{j_aj_b}<\pi/4)-\sigma(\pi/4<\phi_{j_aj_b}<3\pi/4)+\sigma(\phi_{j_aj_b}>3\pi/4)}{\sigma(\phi_{j_aj_b}<\pi/4)+\sigma(\pi/4<\phi_{j_aj_b}<3\pi/4)+\sigma(\phi_{j_aj_b}>3\pi/4)}
\end{equation}
The results using our approach are collected in Table~\ref{aphitab}.
\begin{table}
\begin{center}
\begin{tabular}{c|c}
&$A_\phi$\\
\hline
LO 2-jet&$0.504\pm 0.0013$\\
$\sum_n n$-parton, $=2$-jet & $0.267\pm 0.0034$\\
LO 3-jet&$0.228\pm 0.0018$\\
$\sum_n n$-parton, $\ge 2$-jet&$0.161\pm 0.0087$
\end{tabular}
\caption{The angular decorrelation parameter given by equation (\ref{eq:Aphi}), subject to the cuts of table \ref{tab:cuts}. Note that the 2 and 3-jet values are obtained from matrix elements matched to the known tree level results.}
\label{aphitab}
\end{center}
\end{table}
Of particular interest is the difference between the first two numbers. The
first ($A_\phi=0.504\pm 0.0013$) describes the result obtained in the two-jet
tree-level calculation. The second ($A_\phi=0.267\pm 0.0034$) is the result
obtained for events classified as containing only two hard jets, but
completely inclusive in the number of final state partons. The difference is
mostly due to the decorrelation caused by the additional radiation not
classified as hard jets. As expected, the further hard emissions have a
stronger effect than estimated using a parton shower
approach~\cite{DelDuca:2006hk}.

\subsection{Conclusions}

We have outlined a new technique for estimating multiple hard parton
emission, and demonstrated its application to Higgs boson production (via
GGF) in association with two jets. Our starting point is the FKL factorised
form of Higgs+multijet amplitudes, which formally applies in multi-Regge
kinematics (MRK). We extend the region of applicability of the formalism by
adhering to two simple rules. We compare the results obtained order by order
to those obtained in a fixed order approach and find very good agreement. The
approximations are well within the uncertainty estimated by varying the
renormalisation and factorisation scale by a factor of two in the tree level
results.

We have presented example results for the distribution of final state jets,
and for the azimuthal decorrelation parameter $A_\phi$. We find significant
decorrelation arising from additional hard final state radiation not captured
by present NLO calculations; significantly more than previously estimated
using parton shower algorithms.

The technique outlined here can be extended to e.g.~$W$+jet emission, as well
as pure multijet final states. It would be very interesting to interface the
final states found here with parton shower algorithms, thus resumming in
principle both the number of jets (hard partons) and the structure of each
(soft collinear radiation). Furthermore, the results presented here are based
upon effective vertices correct to leading logarithmic order. Work is in
progress towards extending the accuracy to next-to-leading logarithmic order.

\section*{Acknowledgements}
CDW is funded by the Dutch Organisation for Fundamental Matter Research
(FOM). He thanks Eric Laenen and Jos Vermaseren for helpful discussions. We
are also grateful to Vittorio Del Duca and Gavin Salam for encouraging
conversations.

%\bibliographystyle{lesHouches}
%\bibliography{Hnjets}

%\end{document}

\section[Gluon-Induced $Z$-boson Pair Production at the LHC: Parton Level
Results]
{GLUON- INDUCED $Z$-BOSON PAIR PRODUCTION AT THE LHC: PARTON LEVEL RESULTS~\protect
\footnote{Contributed by: T.~Binoth, N.~Kauer, and P.~Mertsch}}
\label{sec:ggzz1}
%\documentclass[11pt]{cernrep}
%\usepackage{graphicx}
%\bibliographystyle{lesHouches}
%\begin{document}

%\title{{\boldmath Gluon-induced $Z$-boson pair production at the LHC}}

%\author{T.~Binoth$^1$, N.~Kauer$^2$, P.~Mertsch$^3$ and D.~Giordano$^4$}
%\institute{$^1$School of Physics, The University of Edinburgh, Edinburgh EH9 3JZ, United
%Kingdom\\
%$^2$Institut f\"ur Theoretische Physik, Universit\"at W\"urzburg, D-97074 W\"urzburg, Germany\\
%$^3$Rudolf Peierls Centre for Theoretical Physics, University of Oxford, Oxford OX1
%3NP, United~Kingdom\\
%$^4$Universit\`a degli Studi di Bari, INFN Sezione di Bari, Italy}

%\maketitle%

%\begin{abstract}
%A calculation of the loop-induced gluon-fusion process 
%$gg \to Z^\ast(\gamma^\ast)Z^\ast(\gamma^\ast) \to \ell\bar{\ell}\ell'\bar{\ell'}$ 
%is presented, 
%which provides an important background for Higgs boson searches in the $H \to ZZ$
%channel at the LHC. 
%We find that the $gg$-induced process yields a correction of about $15\%$ relative to
%the NLO QCD prediction for the $q\bar{q}$-induced process.  
%The photon contribution is 
%important for Higgs masses below the $Z$-pair threshold.
%The $gg$-induced process can be taken into account 
%using our public program \texttt{GG2ZZ}.
%\end{abstract}

% ============================================================================

%\section[Parton-level results]{Parton-level results\footnote{Contributed by: T.%~Binoth, N.~Kauer and P.~Mertsch}}

% ============================================================================

\subsection{Introduction\label{ggZZ_sec:intro}}

The hadronic production of $Z$ boson pairs provides an important background for Higgs 
boson searches in the $H \to ZZ$ channel at the LHC.  It has been studied 
extensively in the literature including higher order 
corrections \cite{Mele:1990bq,Ohnemus:1994ff,Dixon:1999di,Campbell:1999ah}. 
Production of $Z$ boson pairs through gluon fusion 
contributes at ${\cal O}(\alpha_s^2)$ relative to $q\bar{q}$ annihilation, 
but its importance is enhanced by the large gluon flux at the LHC.
It was analyzed in Refs.~\cite{Dicus:1987dj,Glover:1988rg}. 
Leptonic $Z$ decays were subsequently studied for on-shell \cite{Matsuura:1991pj} and 
off-shell \cite{Zecher:1994kb} vector bosons.
In this note we present the first complete calculation of 
the gluon-induced loop process 
$gg \to Z(\gamma)Z(\gamma) \to \ell\bar{\ell}\ell'\bar{\ell'}$,
allowing for arbitrary invariant masses of the $Z$ bosons and including the 
$\gamma$ contributions.
Our calculation employs the same methods as Refs.~\cite{Binoth:2005ua,Binoth:2006mf}.
The tensor reduction scheme of Refs.~\cite{Binoth:1999sp,Binoth:2005ff} has been 
applied to obtain the amplitude representation implemented in our program.  
We compared it numerically with an amplitude representation based on FeynArts/FormCalc 
\cite{Hahn:1998yk,Hahn:2000kx} and found agreement.
%Fixed-width Breit-Wigner propagators are used for unstable gauge bosons.
Note that single resonant diagrams (in the case of massless leptons) and 
the corresponding photon exchange diagrams give a vanishing contribution.
A combination of the multi-channel \cite{Berends:1994pv} and phase-space-decomposition \cite{Kauer:2001sp,Kauer:2002sn} Monte Carlo
integration techniques was used with appropriate mappings to
compensate peaks in the amplitude.
%The phase space was also thoroughly
%checked, primarily by comparison with the double-checked implementation 
%of Ref.~\cite{Binoth:2006mf}.
A more detailed description of our calculation can be
found in a forthcoming article.

% ============================================================================

\subsection{Results\label{ggZZ_sec:results}}

In this section we present numerical results for the process $pp \to
Z(\gamma)Z(\gamma) \to \ell\bar{\ell}\ell'\bar{\ell'}$ at the LHC,
i.e.~for the production of two charged lepton pairs with different flavor.
Note that no flavor summation is applied.
First, we give the cross section when standard LHC cuts for $Z$ boson 
production \cite{Campbell:1999ah} are applied.
More precisely, we require $75$ GeV $< M_{\ell\ell} < 105$ GeV for the invariant 
masses of $\ell\bar{\ell}$ and $\ell'\bar{\ell'}$, which suppresses the 
photon contribution to less than 1\%.
Motivated by the finite acceptance and resolution of the 
detectors we further require $p_{T\ell} > 20$ GeV and 
$|\eta_\ell| < 2.5$ for all produced leptons.
To obtain numerical results we use the following set of input
parameters:
$M_W = 80.419$ GeV,  
$M_Z = 91.188$ GeV, 
$G_\mu  = 1.16639 \times 10^{-5}$ GeV$^{-2}$,
$\Gamma_Z  = 2.44$ GeV.
The weak mixing angle is given by $c_{\rm w} = M_W/M_Z,\ s_{\rm w}^2 =
1 - c_{\rm w}^2$.  The electromagnetic coupling is defined in the
$G_\mu$ scheme as $\alpha_{G_\mu} = \sqrt{2}G_\mu M_W^2s_{\rm
  w}^2/\pi$.  The masses of external fermions are neglected. The
values of the heavy quark masses in the intermediate loop are set to
$M_t = 170.9$~GeV and $M_b = 4.7$~GeV.  The $pp$ cross
sections are calculated at $\sqrt{s} = 14$~TeV employing the CTEQ6L1
and CTEQ6M \cite{Pumplin:2002vw} parton distribution functions at
tree- and loop-level, corresponding to $\Lambda^{\rm LO}_5 = 165$ MeV
and $\Lambda^{\overline{{\rm MS}}}_5 = 226$ MeV with one- and two-loop
running for $\alpha_s(\mu)$, respectively.  The renormalization and
factorization scales are set to $M_Z$.

We compare results for $\ell\bar{\ell}\ell'\bar{\ell'}$ production in gluon 
scattering with LO and NLO results for the quark scattering processes. Since 
we are interested in $Z(\gamma)Z(\gamma)$ production as a background, 
the $gg\to H \to ZZ$
signal amplitude is not included. The LO and NLO quark scattering
processes are computed with MCFM \cite{Campbell:1999ah}, which
implements helicity amplitudes with full spin correlations
\cite{Dixon:1998py} and includes finite-width effects and
single-resonant corrections.  Table~\ref{ggZZ_tbl:xsec} shows gluon
and quark scattering cross sections for the LHC.
% ----------------------------------------------------------------------
\begin{table}
\centerline{
\def\arraystretch{1.5}
\begin{tabular}{|c|c|cc|c|c|}
 \cline{2-6}
\multicolumn{1}{c|}{} & \multicolumn{5}{c|}{$\sigma(pp \to Z^\ast(\gamma^\ast)Z^\ast(\gamma^\ast) \to \ell\bar{\ell}\ell'\bar{\ell'})$~[fb]} \\ \cline{2-6}
\multicolumn{1}{c|}{} & & 
\multicolumn{2}{c|}{\raisebox{1ex}[-1ex]{$q\bar{q}$}}
& \multicolumn{1}{c|}{} &\multicolumn{1}{c|}{} \\[-1.5ex]
\cline{3-4}
\multicolumn{1}{c|}{} & 
\multicolumn{1}{c|}{\raisebox{2.7ex}[-2ex]{$gg$}} & 
\raisebox{0.9ex}{LO} & \raisebox{0.9ex}{NLO} 
& \raisebox{2.25ex}[-2ex]{$\frac{\sigma_{\rm NLO}}{\sigma_{\rm LO}}$} & 
  \raisebox{2.25ex}[-2ex]{$\frac{
 \sigma_{{\rm NLO}+gg}}{\sigma_{\rm NLO}}$}
\\[-1.5ex]
  \hline
 $\sigma_{\rm std}$ & $1.492(2)$ & $7.343(1)$ & $10.953(2)$ & 1.49 & 1.14 \\
 \hline
\end{tabular}}
\vspace*{.5cm}
\caption{\label{ggZZ_tbl:xsec}
  Cross sections for the gluon and quark scattering contributions to
  $pp \to Z^\ast(\gamma^\ast)Z^\ast(\gamma^\ast) \to \ell\bar{\ell}\ell'\bar{\ell'}$ at the LHC
  ($\sqrt{s} = 14$ TeV) where standard LHC cuts ($75$ GeV $< M_{\ell\ell} < 105$ GeV, 
  $p_{T\ell} > 20$ GeV, $|\eta_\ell| < 2.5$) are applied. The integration error is 
  given in brackets. We also show the ratio of the NLO to LO cross sections and the 
  ratio of the combined NLO+$gg$ contribution to the NLO cross section.}
\end{table}
% ------------------------------------------------------------------------
We find a NLO $K$-factor for $q\bar{q}\to ZZ$ of 1.5.
Enhanced by the large gluon flux at the LHC, 
the $gg$ process yields a 14\% correction to the total $ZZ$ cross section
calculated from quark scattering at NLO QCD.  
This is substantially higher than the corresponding $6\%$ increase for $WW$ 
production \cite{Binoth:2006mf}, where no right-handed $Vff$ coupling contributes.
Relative to the LO $q\bar{q}\to ZZ$ prediction the 
$gg$ contribution is about $20\%$ in agreement with the finding in 
Ref.~\cite{Zecher:1994kb}.
Without top and bottom quark contribution the $gg$ cross section is
$0.885(1)$ fb.
The massive bottom and top loops increase the result based on intermediate
light quarks by about $70\%$.  This is much more than the corresponding 
$15\%$ for $gg \to WW$ \cite{Binoth:2006mf}, where all quark loops are suppressed 
by at least one top propagator.  In the $gg \to ZZ$ case
on the other hand a pure $b$ quark loop occurs and gives rise to a contribution 
that is similar to the massless first or second generation down quark loop.
We estimate the remaining
theoretical uncertainty introduced by the QCD scale by varying
the renormalization and factorization scales independently 
between $M_Z/2$ and $2M_Z$.
For the $gg \to ZZ$ process we find a renormalization and
factorization scale uncertainty of approximately $20\%$. The scale
uncertainty of the $q\bar{q}\to ZZ$ process at NLO is approximately
$4\%$.  % 4.3\%
%We find in both cases that the largest cross
%section prediction corresponds to choosing $\mu_{\rm ren} = M_Z/2$ and
%$\mu_{\rm fac} = 2M_Z$, while the reverse combination $\mu_{\rm ren} =
%2M_Z$ and $\mu_{\rm fac} = M_Z/2$ yields the smallest value.
The scale uncertainties are thus similar for $gg \to ZZ$ and $gg \to WW$.

\begin{figure}
\begin{center}
\begin{minipage}[c]{.49\linewidth}
\flushleft \includegraphics[height=7.8cm,angle=90,clip=true]{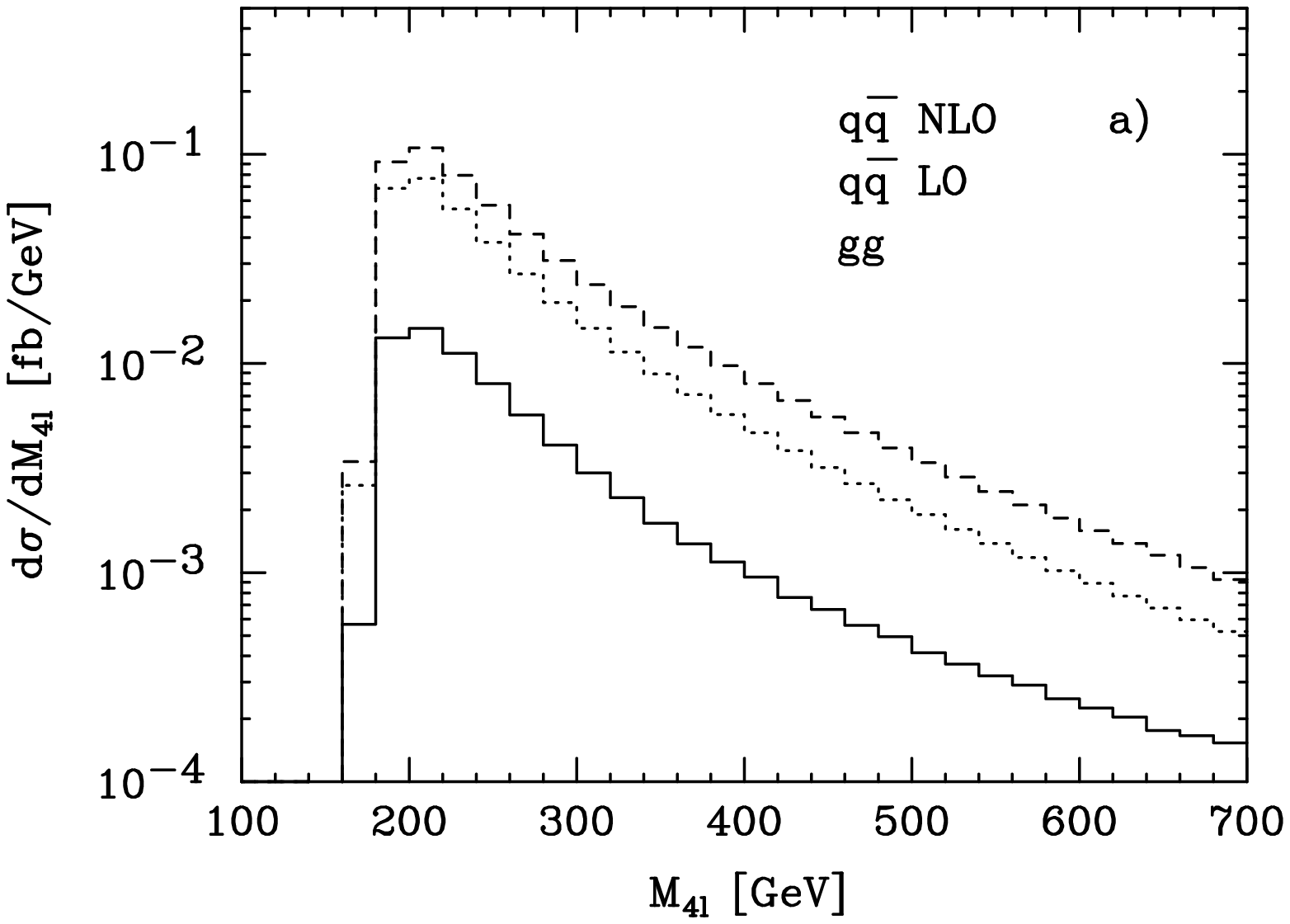}
\end{minipage} \hfill
\begin{minipage}[c]{.49\linewidth}
\flushright \includegraphics[height=7.cm,angle=90,clip=true]{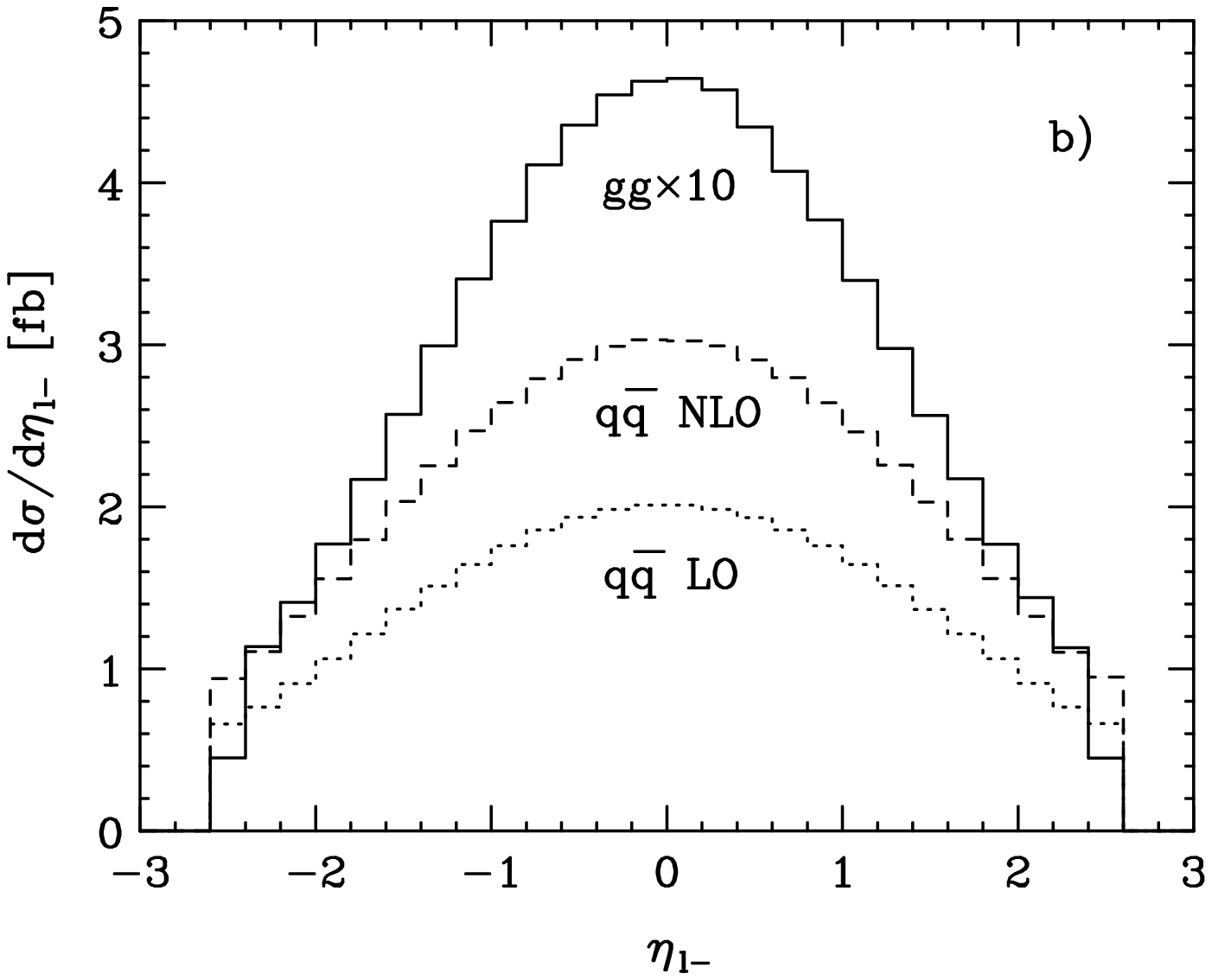}
\end{minipage}\\[0.2cm]
\caption{%
  Distributions in  the $\ell\bar{\ell}\ell'\bar{\ell'}$ invariant
  mass $M_{4l}$ (a) and the pseudorapidity $\eta_{\ell^-}$ of the
  negatively charged lepton (b) for the gluon scattering process (solid) 
  and the quark scattering processes in LO (dotted) and NLO QCD (dashed) of $pp
  \to Z^\ast(\gamma^\ast)Z^\ast(\gamma^\ast) \to \ell\bar{\ell}\ell'\bar{\ell'}$ at the LHC.
  Input parameters as defined in the main text. Standard LHC cuts are
  applied (see main text and Table \protect\ref{ggZZ_tbl:xsec}).
  The $gg$ distribution of $\eta_{\ell^-}$ is displayed after multiplication with a
  factor 10.
  \label{ggZZ_fig:stdcuts}}
\end{center}
\end{figure}

\begin{figure}
\begin{center}
\begin{minipage}[c]{.49\linewidth}
\flushleft \includegraphics[height=7.8cm,angle=90,clip=true]{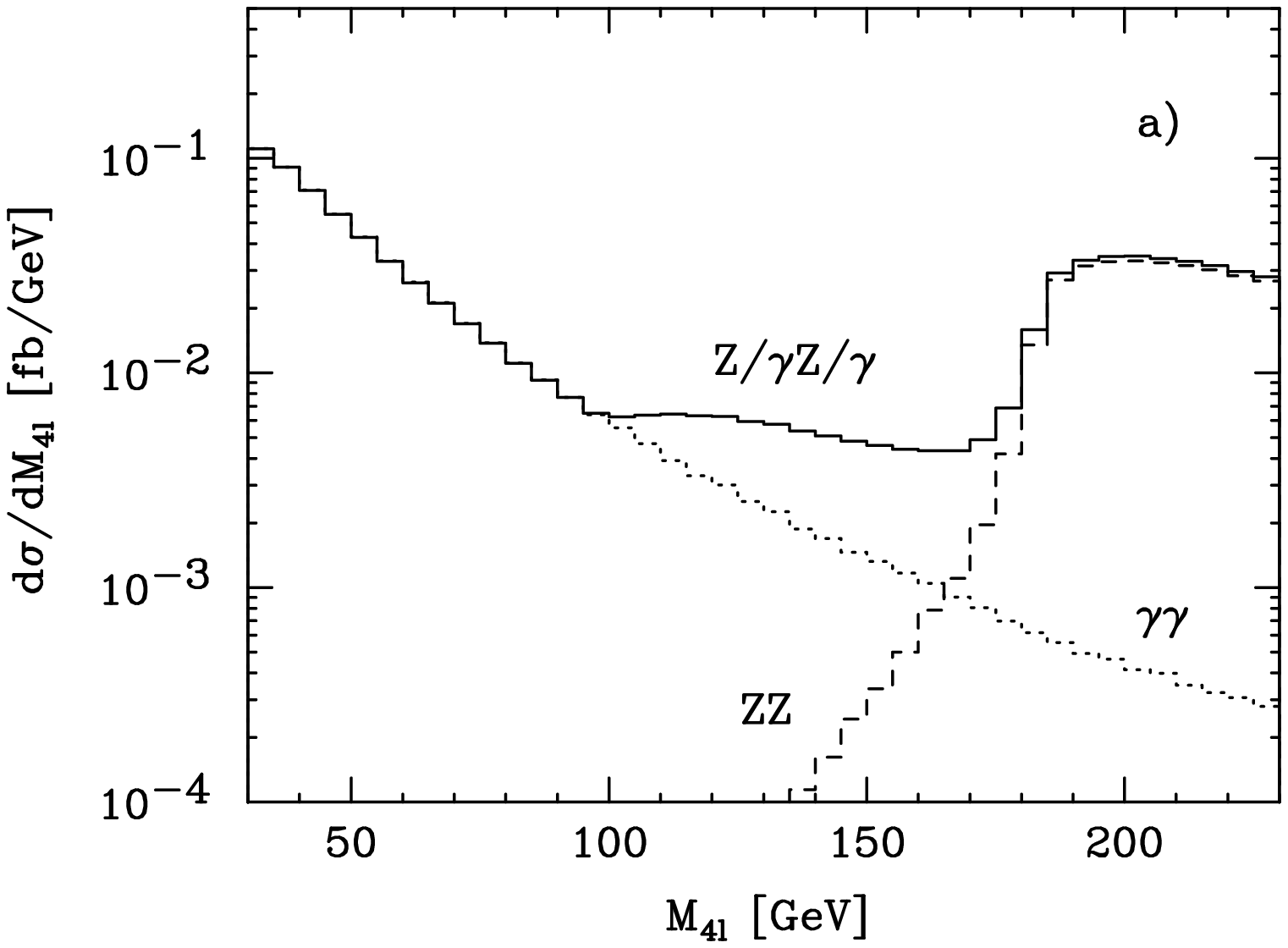}
\end{minipage} \hfill
\begin{minipage}[c]{.49\linewidth}
\flushright \includegraphics[height=7.45cm,angle=90,clip=true]{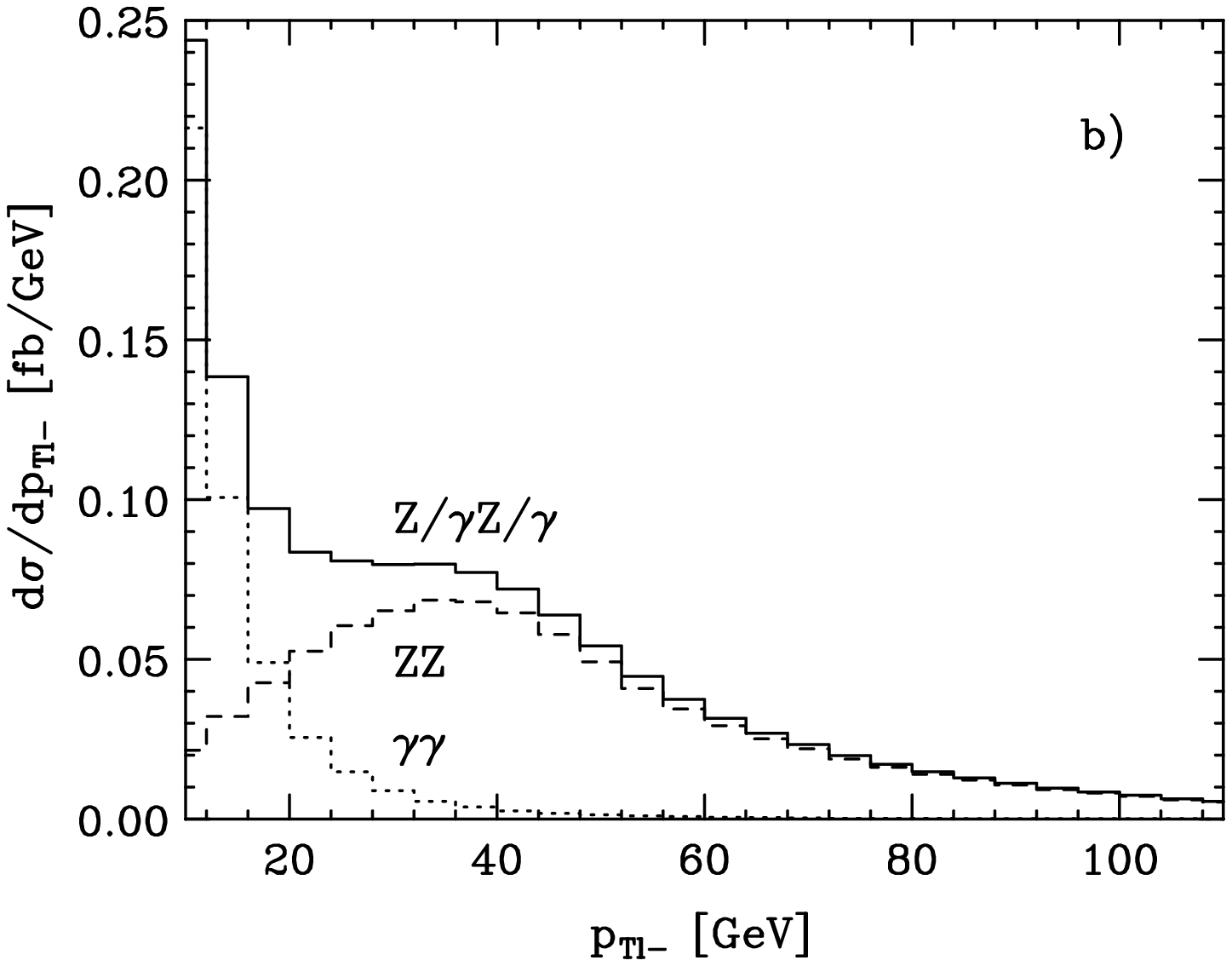}
\end{minipage}\\[0.2cm]
\caption{%
  Distributions in  the $\ell\bar{\ell}\ell'\bar{\ell'}$ invariant
  mass $M_{4l}$ (a) and the transverse momentum $p_{T\ell^-}$ of the
  negatively charged lepton (b) for the gluon scattering process 
  $gg \to Z^\ast(\gamma^\ast)Z^\ast(\gamma^\ast) \to \ell\bar{\ell}\ell'\bar{\ell'}$ at the LHC 
  with $ZZ$ contributions only (dashed), $\gamma\gamma$ contributions only 
  (dotted) and all contributions (solid).
  Input parameters and minimal set of cuts as defined in the main text.
\label{ggZZ_fig:basiccuts}}
\end{center}
\end{figure}

Selected differential distributions for 
$pp \to Z(\gamma)Z(\gamma) \to \ell\bar{\ell}\ell'\bar{\ell'}$ 
at the LHC are shown in Fig.~\ref{ggZZ_fig:stdcuts}, where the standard set of cuts defined above is applied.
%--------------------------------------------------------------------------
%--------------------------------------------------------------------------
Fig.~\ref{ggZZ_fig:stdcuts}a) shows the distribution in the invariant mass $M_{4l}$ 
of the four produced leptons. We compare the gluon-gluon induced
contribution with the quark scattering processes in LO and NLO. 
We observe that the invariant mass distribution of
the gluon-gluon induced process is similar in shape to the quark
scattering contributions and suppressed by about one order of
magnitude in normalization.
$Z$ boson pairs produced in quark-antiquark scattering at the LHC are
in general strongly boosted along the beam axis. Gluon-induced
processes on the other hand result in $ZZ$ events at more central
rapidities. This feature is born out by the distribution in the
pseudorapidity of the negatively charged lepton shown in
Fig.~\ref{ggZZ_fig:stdcuts}b). In order to distinguish the shapes of the various
contributions we have chosen a linear vertical scale and plot the
gluon-gluon contribution multiplied by a factor~10. Compared to LO
quark-antiquark scattering, the lepton distribution of the gluon-gluon
process shows a more pronounced peak at central rapidities. We also
observe an enhancement of the NLO corrections at central rapidities
which is due to the substantial contribution of gluon-quark processes
at NLO.
To demonstrate the impact of the photon contribution, 
we show in Fig.~\ref{ggZZ_fig:basiccuts} distributions for the gluon-gluon 
induced process that include only $ZZ$, only $\gamma\gamma$ and all contributions.
Here, a minimal set of cuts is applied, 
i.e.~only $M_{\ell\ell} > 5$ GeV in order to exclude the photon 
singularity.\footnote{As in Ref.~\protect\cite{Zecher:1994kb}, a technical cut $p_{TZ} >$ 2 GeV is employed with standard cuts to exclude critical configurations.  With 
minimal cuts, $p_{TZ} >$ 4 GeV is applied, except for $M_{\ell\ell} < 0.4M_Z$, 
where $p_{TZ} >$ 7 GeV is used.  An update of \texttt{GG2ZZ} that requires no technical cuts is 
in preparation.}
%--------------------------------------------------------------------------
%--------------------------------------------------------------------------
With this minimal set of cuts the 
LHC cross section for 
$gg \to Z(\gamma)Z(\gamma) \to \ell\bar{\ell}\ell'\bar{\ell'}$ 
increases to $7.874(5)$ fb. 
In Fig.~\ref{ggZZ_fig:basiccuts}a) the $Z$ and $\gamma$ contributions to 
the distribution in the invariant mass $M_{4l}$ are displayed.
We observe that for Higgs masses below the $Z$-pair threshold, where 
one $Z$ boson is produced off-shell, the photon contribution to the background 
is important.  In Fig.~\ref{ggZZ_fig:basiccuts}b) the contributions are 
shown for the distribution of the transverse momentum $p_{T\ell^-}$ of the negatively 
charged lepton.  For this observable, the photon contribution becomes non-negligible 
for values below 70 GeV.

% ============================================================================

\subsection{Conclusions}
We have calculated the loop-induced gluon-fusion process 
$gg \to Z^\ast(\gamma^\ast)Z^\ast(\gamma^\ast) \to \ell\bar{\ell}\ell'\bar{\ell'}$, 
which provides an important background for Higgs boson searches in the $H \to ZZ$
channel at the LHC. 
Our calculation demonstrates that the gluon-fusion contribution to 
the $ZZ$ background yields a correction of about $15\%$ to
the $q\bar{q}$ prediction at NLO QCD and that 
the photon contribution is important
for Higgs masses below the $Z$-pair threshold.
We conclude that the gluon-gluon induced background process
should be taken into account for an accurate determination of the 
discovery potential of Higgs boson searches in the $pp\to H\to ZZ \to$ 
leptons channel.
Our public program, named \texttt{GG2ZZ}, includes the $ZZ$, $Z\gamma$ and 
$\gamma\gamma$ contributions with full spin and polarization correlations, 
off-shell and interference effects, as well as finite top and bottom quark mass 
effects.  It is available on the Web \cite{GG2ZZ} and can be used as Monte Carlo 
integrator or to generate unweighted parton-level events in Les Houches standard 
format \cite{Boos:2001cv,Alwall:2006yp}.  ATLAS and CMS are currently using our
program to study the hadron-level impact of the $gg \to ZZ$ background on 
$H\to ZZ$ searches at the LHC.

% ============================================================================

\subsection*{Acknowledgements}
T.~Binoth and N.~Kauer would like to thank the Ecole de Physique des 
Houches and the
Galileo Galilei Institute for Theoretical
Physics for the hospitality and the INFN for partial support during the
completion of this work.  
This work was supported by the BMBF and DFG, Germany (contracts 05HT1WWA2
and BI 1050/2).

\section[Gluon-Induced $Z^\ast Z^\ast$ Background Simulation for Higgs Boson Search]
{GLUON- INDUCED $Z^\ast Z^\ast$ BACKGROUND SIMULATION FOR HIGGS
BOSON SEARCH~\protect\footnote{Contributed by: D.~Giordano}}
\label{sec:ggzz2}

\begin{figure}[h]
  \hfill
  \begin{minipage}[t]{.45\textwidth}
    \begin{center}  
      \includegraphics[width=\textwidth]{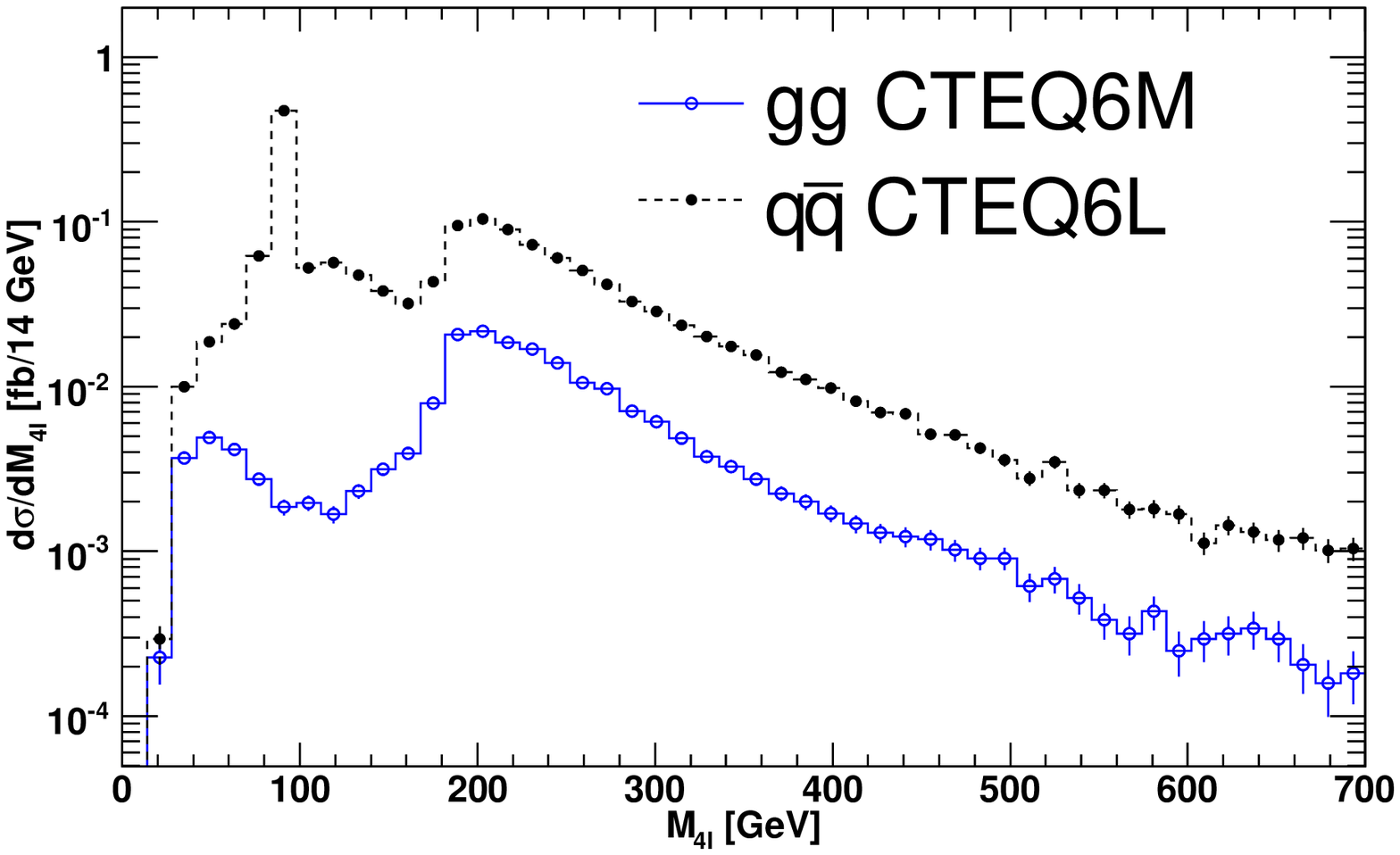}
      \caption{Distribution in the $\ell\bar{\ell}\ell'\bar{\ell'}$ invariant mass, $M_{4\ell}$, for the gluon scattering process (solid) and the quark scattering process (dashed) of $pp \to ZZ \to \ell\bar{\ell}\ell'\bar{\ell'}$ at the LHC, after applying pre-selection cuts.}
      \label{ggZZ_fig:M4l_compare}
    \end{center}
  \end{minipage}
  \hfill
  \begin{minipage}[t]{.4\textwidth}
    \begin{center}  
      \includegraphics[width=1\textwidth,height=.7\textwidth]{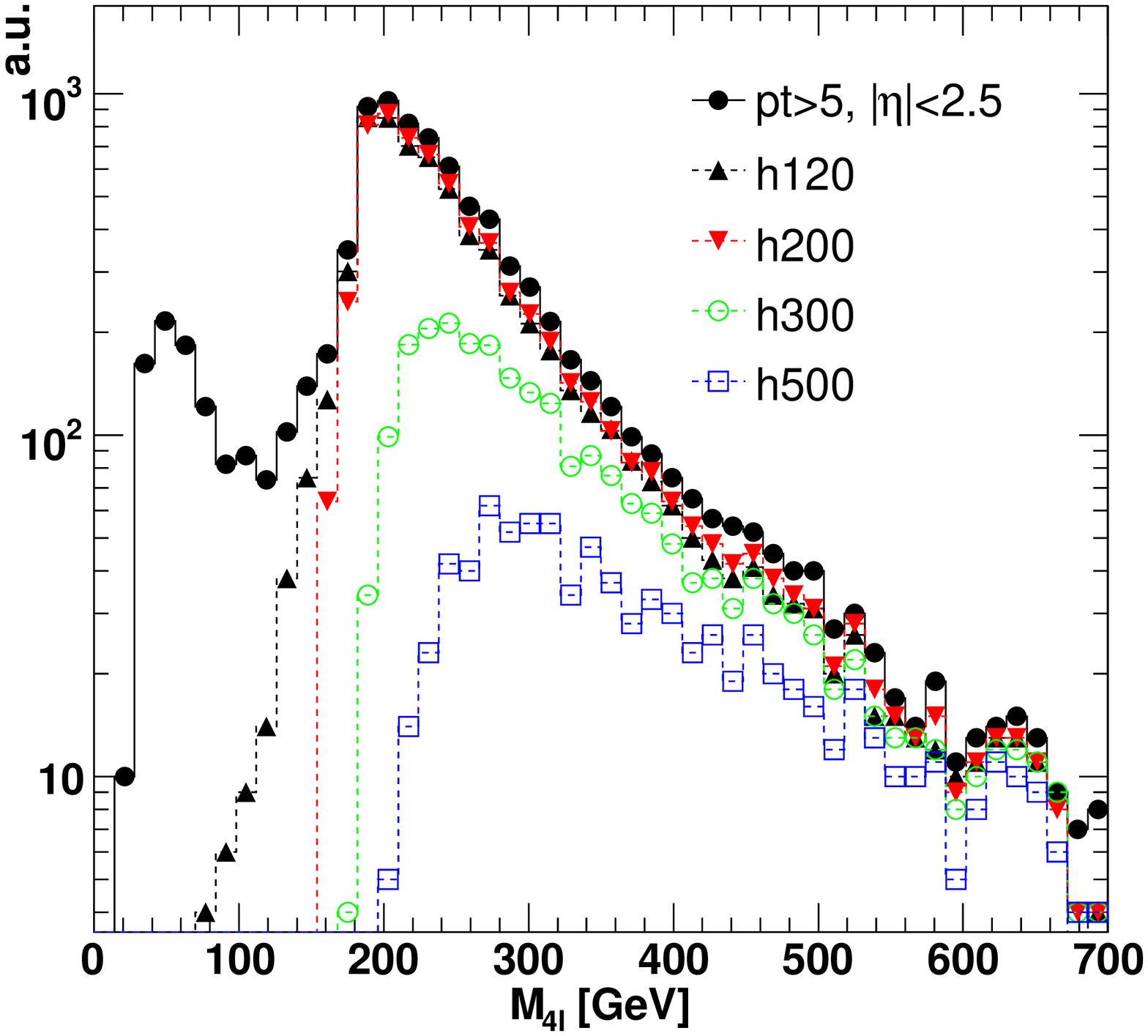}
      \caption{Selected distributions in the $\ell\bar{\ell}\ell'\bar{\ell'}$  invariant mass,  $M_{4\ell}$, for the gluon scattering process at the LHC, obteined applying the pre-selection cuts (solid) and the set of selection cuts optimized for the Higgs boson search in different mass scenarios from 120~GeV to 500~GeV.}
      \label{ggZZ_fig:M4l_cuts}
    \end{center}
  \end{minipage}
  \hfill
\end{figure}

The contribution of the  $gg \to  Z^\ast (\gamma ^\ast ) Z^\ast ( \gamma ^\ast ) \to \ell \bar{\ell} \ell' \bar{\ell'}$ process to the total $pp \to Z^\ast (\gamma ^\ast ) Z^\ast ( \gamma ^\ast ) \to \ell \bar{\ell} \ell' \bar{\ell'} $ production cross section has been evaluated after the application of the selection cuts adopted for the Higgs boson search through the $ H \to ZZ \to 2e 2\mu $ decay channel in the CMS experiment~\cite{Futyan:2007zz}.
The minimal set of kinematical cuts needed to maximize the discovery significance has been used: upper thresholds for the transverse momenta ($p_{T\ell}$) of the four produced leptons; upper threshold on the highest reconstructed $M_{\ell\bar{\ell}}$; lower threshold on the lowest reconstructed $M_{\ell\bar{\ell}}$; upper and lower thresholds on the $M_{4\ell}$. 
The values of the selection cuts are mass dependent, optimized for different Higgs boson mass scenarios, from 120 GeV to 500 GeV. 
The selection procedure and the cut values are described in details in Ref.~\cite{Futyan:2007zz}. 
A sample of 9k $gg \to  Z^\ast (\gamma ^\ast ) Z^\ast ( \gamma ^\ast ) \to \ell \bar{\ell} \ell' \bar{\ell'}$ events 
%(equivalent to an integrated luminosity of 1140~fb$^{-1})$ 
has been generated at parton level using the generator program \verb!GG2ZZ! here presented. 
For the simulation of the shower evolution we have interfaced the generated parton-level events to the PYTHIA Monte Carlo generator~\cite{sjostrand-2001-135}.
In order to increase the event statistics in the kinematical region of interest the following set of pre-selection cuts has been used: $p_{T\ell} > 5$~GeV, $|\eta_\ell|<2.5$, M$_{\ell\bar{\ell}} > 5$~GeV. 
The cross section for the selected events is 2.8~fb. %, the equivalent integrated luminosity of the produced sample is 1140~fb$^{-1}$ .
We compare the gluon induced contribution with 70k $q\bar{q} \to  Z^\ast (\gamma ^\ast ) Z^\ast ( \gamma ^\ast ) \to \ell \bar{\ell} \ell' \bar{\ell'}$  events generated with the MadGraph LO Monte Carlo generator~\cite{Maltoni:2002qb}. 
The LO cross section of this process is 27.67~fb, where the same set of pre-selection cuts has been applied. 
%The equivalent integrated luminosity of this sample is 2530~fb. 
In Fig~\ref{ggZZ_fig:M4l_compare} we compare the distribution of the invariant mass ($M_{4l}$) of the four leptons produced in the gluon-gluon and in the quark scattering processes respectively. 
The peak at $M_{4l} \sim M_Z$ in the $q\bar{q}$  distribution is due to the s-channel production process, that gives the main contribution to the cross section in the $M_{4l}$ mass region below and near $M_Z$. 
Since the Higgs mass region below 114.4~GeV has been excluded by the LEP data~\cite{Barate:2003sz}, 
almost all the events produced by the s-channel process are removed by the selection cuts adopted in the Higgs boson search analyses.
%The LHC cross section for the $gg \to  Z^\ast (\gamma ^\ast ) Z^\ast ( \gamma ^\ast ) \to \ell \bar{\ell} \ell' \bar{\ell'}$ and the $q\bar{q} \to  Z^\ast (\gamma ^\ast ) Z^\ast ( \gamma ^\ast ) \to \ell \bar{\ell} \ell' \bar{\ell'}$  processes is respectively of XXX~fb and YYY~fb. 
The effect of the mass dependent selection cuts on the $M_{4l}$ distribution is shown in Fig~\ref{ggZZ_fig:M4l_cuts}. 
The different curves correspond to the $gg \to  Z^\ast (\gamma ^\ast ) Z^\ast ( \gamma ^\ast ) \to \ell \bar{\ell} \ell' \bar{\ell'}$ events selected after the pre-selection cuts (solid curve) and for a few Higgs boson mass scenarios (dashed curves), when only the cuts on the four leptons transverse momenta and on the di-lepton invariant masses ($M_{\ell\bar{\ell}}$, $M_{\ell'\bar{\ell'}}$) have been applied. 
The photon contribution is strongly suppressed for the Higgs boson search above $2M_Z$. 

\begin{figure}[h]
\begin{center}
\includegraphics[width=0.5\textwidth]{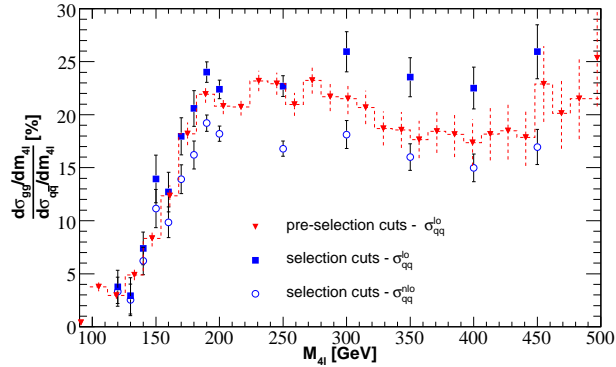} 
\caption{
Contribution of the gluon-induced background  $gg \to   Z^\ast (\gamma ^\ast ) Z^\ast ( \gamma ^\ast ) \to \ell \bar{\ell} \ell' \bar{\ell'} $ respect to the $q\bar{q} \to  Z^\ast (\gamma ^\ast ) Z^\ast ( \gamma ^\ast ) \to \ell \bar{\ell} \ell' \bar{\ell'}$ process, as a function of the four leptons invariant mass $M_{4\ell}$, after application of the pre-selection cuts only (dashed) and all selection cuts (other two curves).  The reference cross section of the $q\bar{q}$ scattering is evaluated at the LO (solid markers) and at the NLO (empty markers).
}
\label{ggZZ_fig:ratio}
\end{center}
\end{figure}

The contribution of the gluon scattering to the ZZ cross section is reported in Fig~\ref{ggZZ_fig:ratio}, in terms of the ratio of the selected $gg \to   Z^\ast (\gamma ^\ast ) Z^\ast ( \gamma ^\ast ) \to \ell \bar{\ell} \ell' \bar{\ell'} $ events respect to the LO $q\bar{q} \to  Z^\ast (\gamma ^\ast ) Z^\ast ( \gamma ^\ast ) \to \ell \bar{\ell} \ell' \bar{\ell'}$  selected events (solid square markers). 
The correction increases approximately linearly from 3\% to 24\% in the $M_{4l}$ region between 120~GeV and 200~GeV, and 
it is quite uniform, around $\sim 24\%$, in the $M_{4l}$ region above $200$~GeV.
Superimposing to the graphic the ratio of the distributions shown in Fig~\ref{ggZZ_fig:M4l_compare} (dashed curve), where only the pre-selection cuts were applied, we observe that the gluon induced contibution is enhanced by the selection cuts, especially in the mass region above 200~GeV. 
The empty round markers in Fig~\ref{ggZZ_fig:ratio} show the $gg$ contribution respect to the $q\bar{q}$ process calculated at the NLO. 
The mass dependent NLO k-factor evaluated in Ref~\cite{Bartalini:2006} 
has been used to rescale the quark scattering cross section at the NLO. 
The gluon-gluon contribution is reduced to 18\% in the $M_{4l}$ region above $200$~GeV, a value compatible with our previous evaluation.

% ============================================================================

%\bibliography{ggZZ,ggZZ_dg}

%\end{document}

\section[The Methods for the Central Rapidity Gap Selection in the Vector Boson Fusion Searches in CMS]
{THE METHODS FOR THE CENTRAL RAPIDITY GAP SELECTION IN THE
VECTOR BOSON FUSION
SEARCHES IN CMS~\protect\footnote{Contributed by: M.~V\'azquez~Acosta, 
S.~Greder, A.~Nikitenko, and M.~Takahashi}}
\label{sec:crg}
%\documentclass[11pt]{cernrep}
%\usepackage{graphicx,epsfig}
%\bibliographystyle{lesHouches}
%\begin{document}
%
%\title{The methods for the central rapidity gap selection in the VBF Higgs boson searches in CMS} 
%
%\author{M.~V\'azquez~Acosta$^1$, S.~Greder$^2$, A.~Nikitenko$^{2,~a}$, M.~Takahashi$^2$}
%\institute{$^1$CERN, European Organization for Nuclear Research, Geneva
%\\$^2$Imperial College, London
%\\$^a$on leave from ITEP, Moscow}

%\maketitle

%
\subsection{Introduction}

In the VBF Higgs boson searches at LHC a selection of the events with the central 
rapidity gap between the two tagging jets is aimed to reduce the QCD $Z$+jets and other 
backgrounds like W+jets and $t \bar{t}$ while keep a high efficiency for the Higgs 
boson signal from the VBF production, $VV \rightarrow H$. The central jet veto was 
proposed and used in the first VBF Higgs boson analyses ~\cite{Plehn:1999xi,Rainwater:1998kj} 
(see also references in it) and exploited in the recent, published ATLAS and 
CMS analyses ~\cite{Asai:2004ws,Takahashi:qqh}. The central calorimeter jet veto technique 
is suffering from the pile-up and the electronic noise in the calorimeters which could create 
the fake jets. The method of the reduction of the fake calorimeter jets using the information
from the event vertex and the tracks was proposed in ~\cite{Natasha:CJV} and successfully 
used in the CMS analyses ~\cite{Takahashi:qqh, ETH:HWW}. 
 
We consider three methods to perform the hadron activity veto in the 
central rapidity region: the (traditional) central calorimeter jet veto (CJV), 
the track counting veto (TCV) and the veto on jets made from the tracks only 
(TJV). The idea of the track counting veto is inspired by the paper 
~\cite{Dokshitzer:1991he} where it was proposed to distinguish between the gluon 
and vector boson fusion processes for the Higgs boson production. The performance 
of methods is compared in terms of the signal efficiency and the QCD $Z$+jets 
background rejection. 

\subsection{Studies at generator level}
\label{genlevel}

The QCD $Z$+jets events were generated using the ALPGEN~\cite{Mangano:2002ea} 
generator with the MLM prescription for jet-parton matching
\cite{Mangano:2006rw, Hoche:2006ph} in the PYTHIA6.4 shower generation 
\cite{Sjostrand:2006za}. The details on the ALPGEN generation and soft VBF 
preselections at the generator level can be found in~\cite{Takahashi:qqh}.

The final VBF selections similar to the ones used in the full simulation 
analysis~\cite{Takahashi:qqh} were applied to the PYTHIA particle level jets.
An event must have at least two leading $E_{T}$ jets reconstructed 
with a cone algorithm (cone size 0.5) that satisfy the following 
requirements:
\begin{enumerate}
\label{selection2}
\item $E_{T}^{j}>20$ GeV
\item $|\eta^{j}|~<~4.5$
\item $M_{j1j2}>1000$ GeV
\item $|\Delta\eta^{j1j2}|>4.2$
\item $\eta^{j1} \times \eta^{j2}<0$.
\end{enumerate}
where j1 and j2 are two leading $E_{T}$ jets ordered in $E_{T}$. 

The performance of the two methods, CJV and TCV was compared. The CJV requires to reject events with a 
third jet that satisfies
\begin{itemize}
\item $E_{T}^{j3}>20$ GeV
\item $\eta^{j~min} + 0.5 < \eta^{j3} < \eta^{j~max} - 0.5$,
\end{itemize}
where $\eta^{j~min}$ and $\eta^{j~max}$ are the minimum and maximum $\eta$ of the two leading jets (j1 and j2).
The TCV requires to reject events with a certain number of "tracks" (charged particles) within the tracker 
acceptance region, $|\eta|<$2.4 that satisfies                                   
\begin{itemize}
\item $p_{T}^{track}> p_{T}^{cut}$ GeV/$c$
\item $\eta^{j~min} + 0.5 < \eta^{track} < \eta^{j~max} - 0.5$,
\end{itemize}
The effect of multiple parton interactions generated with Tune DWT ~\cite{Acosta:2006bp} 
on the track counting veto was studied.

\begin{figure}[htb!]
\begin{center}
\includegraphics[width=.3\textwidth]{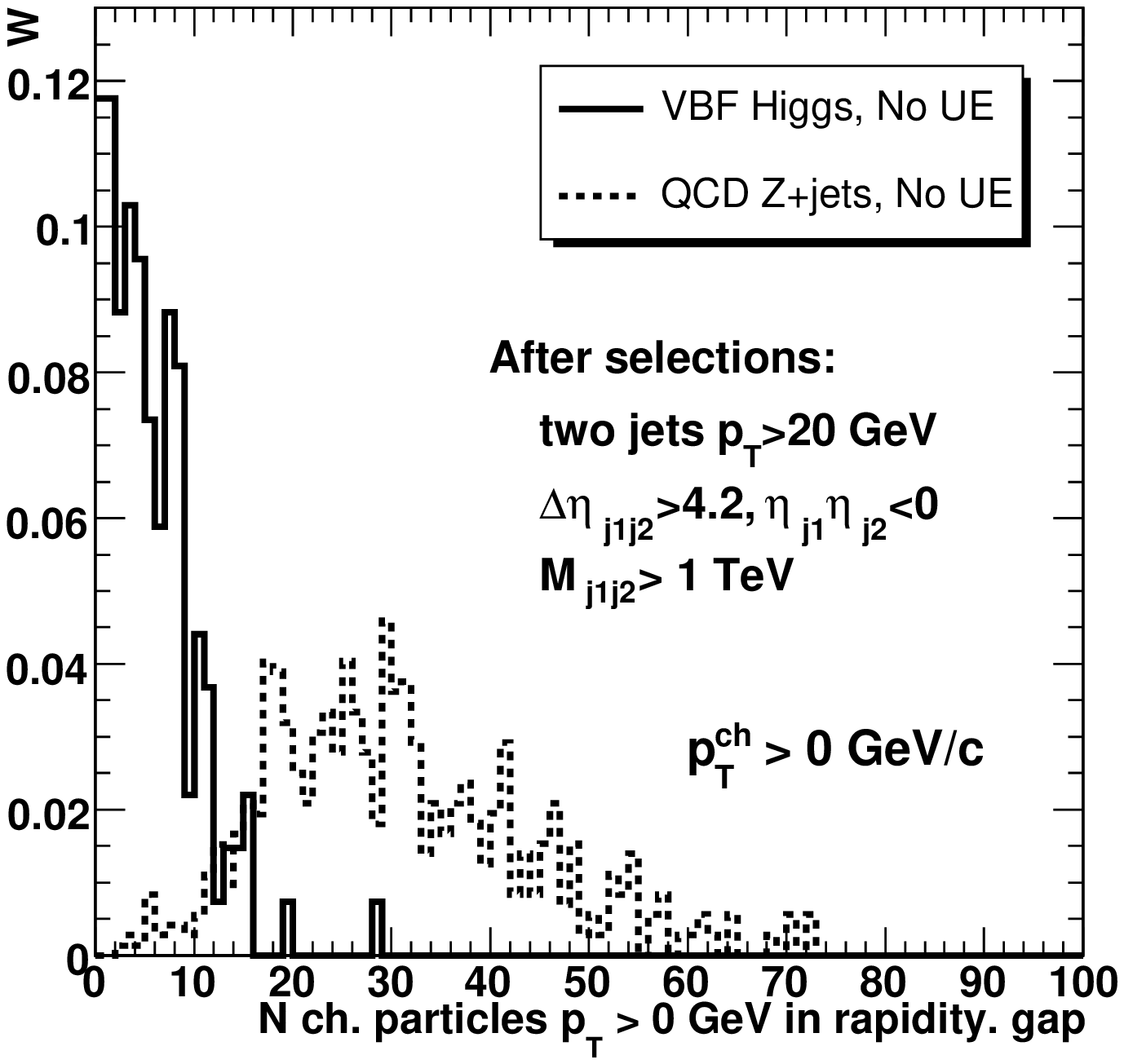}
\includegraphics[width=.3\textwidth]{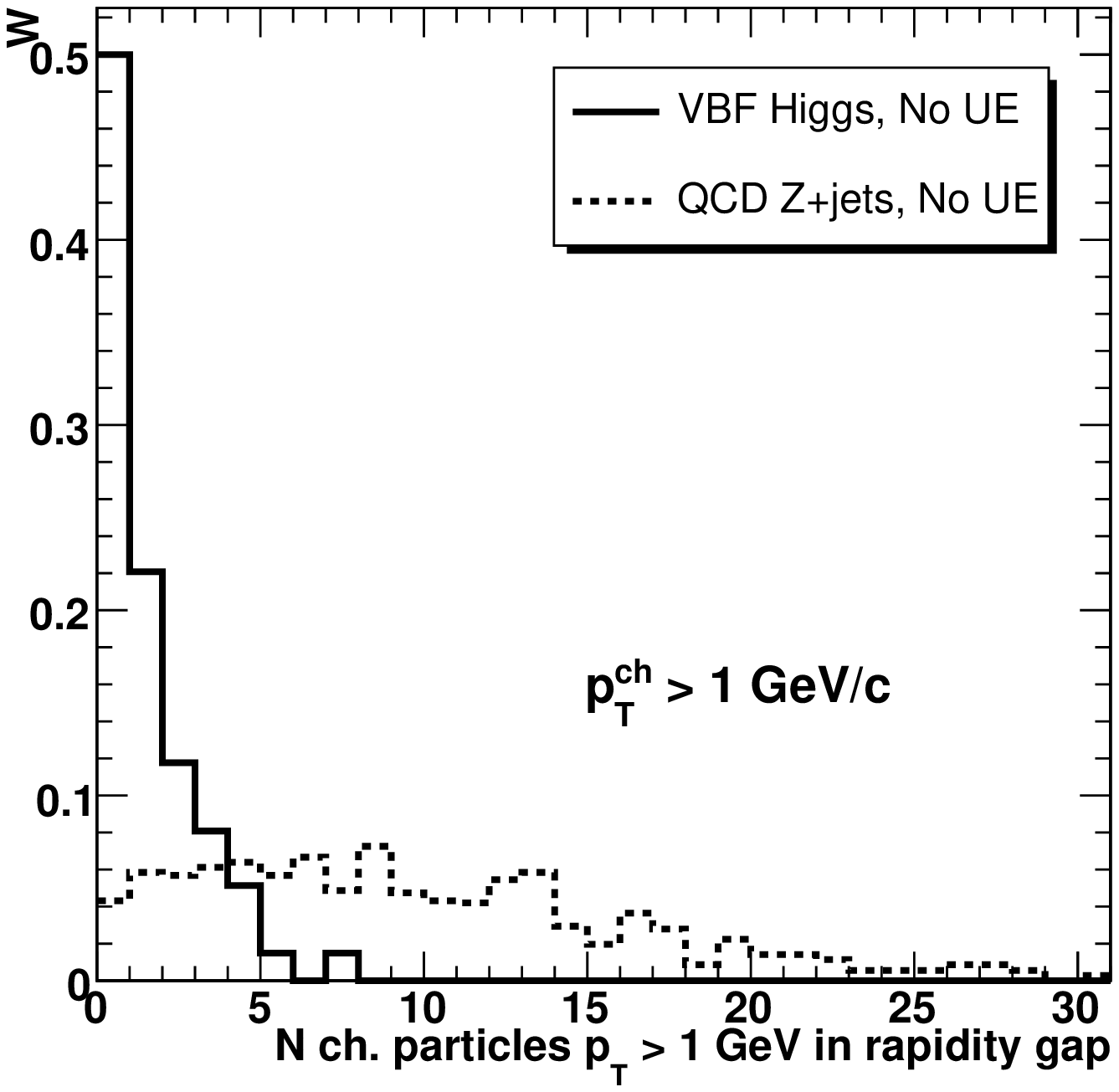}
\includegraphics[width=.3\textwidth]{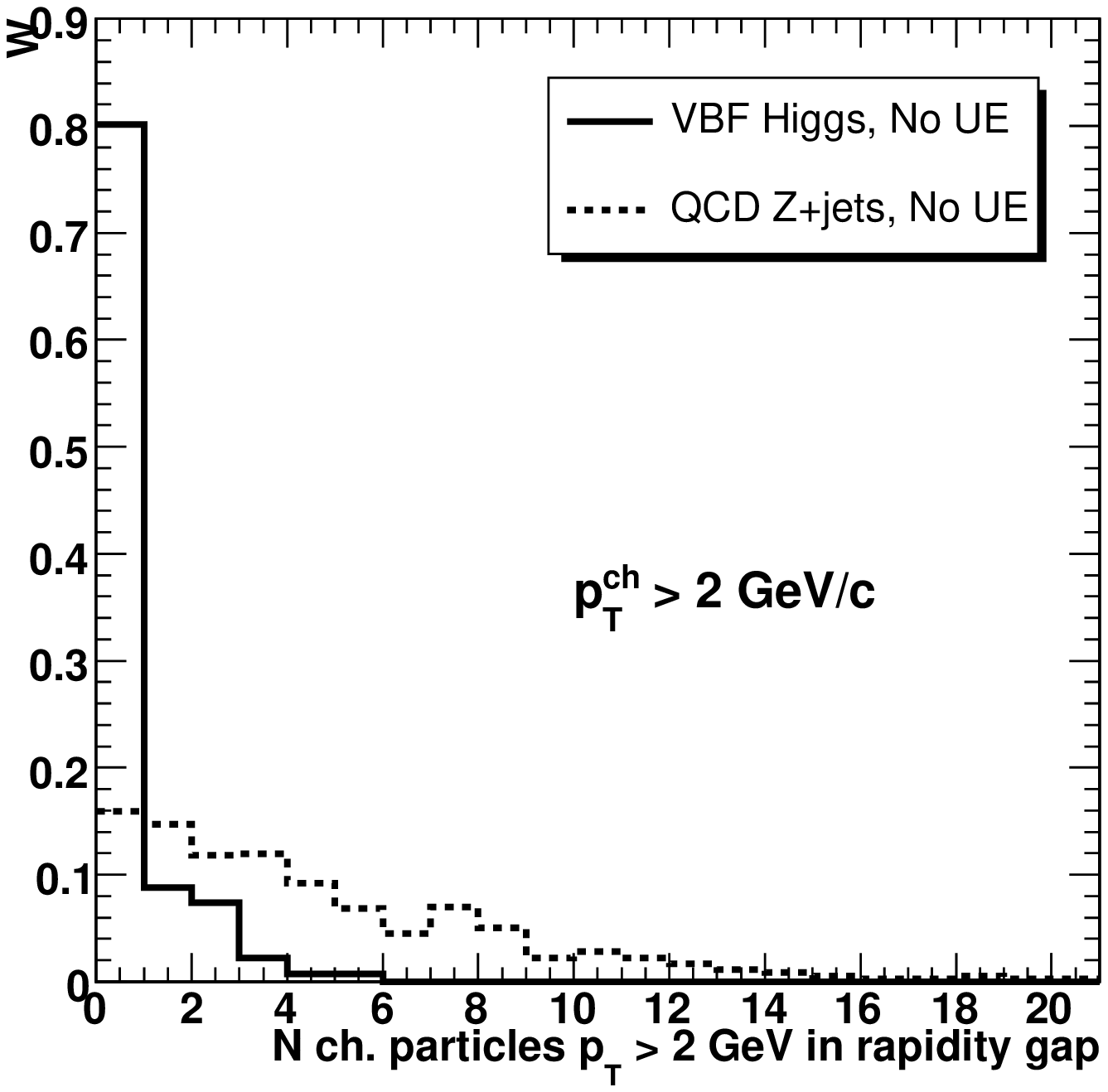}
\caption{The number of charged particles within the tracker acceptance and between 
two tagging jets ($\eta^{j~min} + 0.5 < \eta^{track} < \eta^{j~max} - 0.5$) with $p_{T}>$ 0 GeV/$c$ 
(left plot), $>$ 1 GeV/$c$ (middle plot) and $>2$ GeV/$c$ (right plot) for the signal (solid line) 
and the QCD $Z$+jets background (dashed line). The multiple parton interactions are switched off in PYTHIA.}
\label{fig:mctcv1}
\end{center}
\end{figure}

\begin{figure}[htb!]
\begin{center}
\includegraphics[width=.3\textwidth]{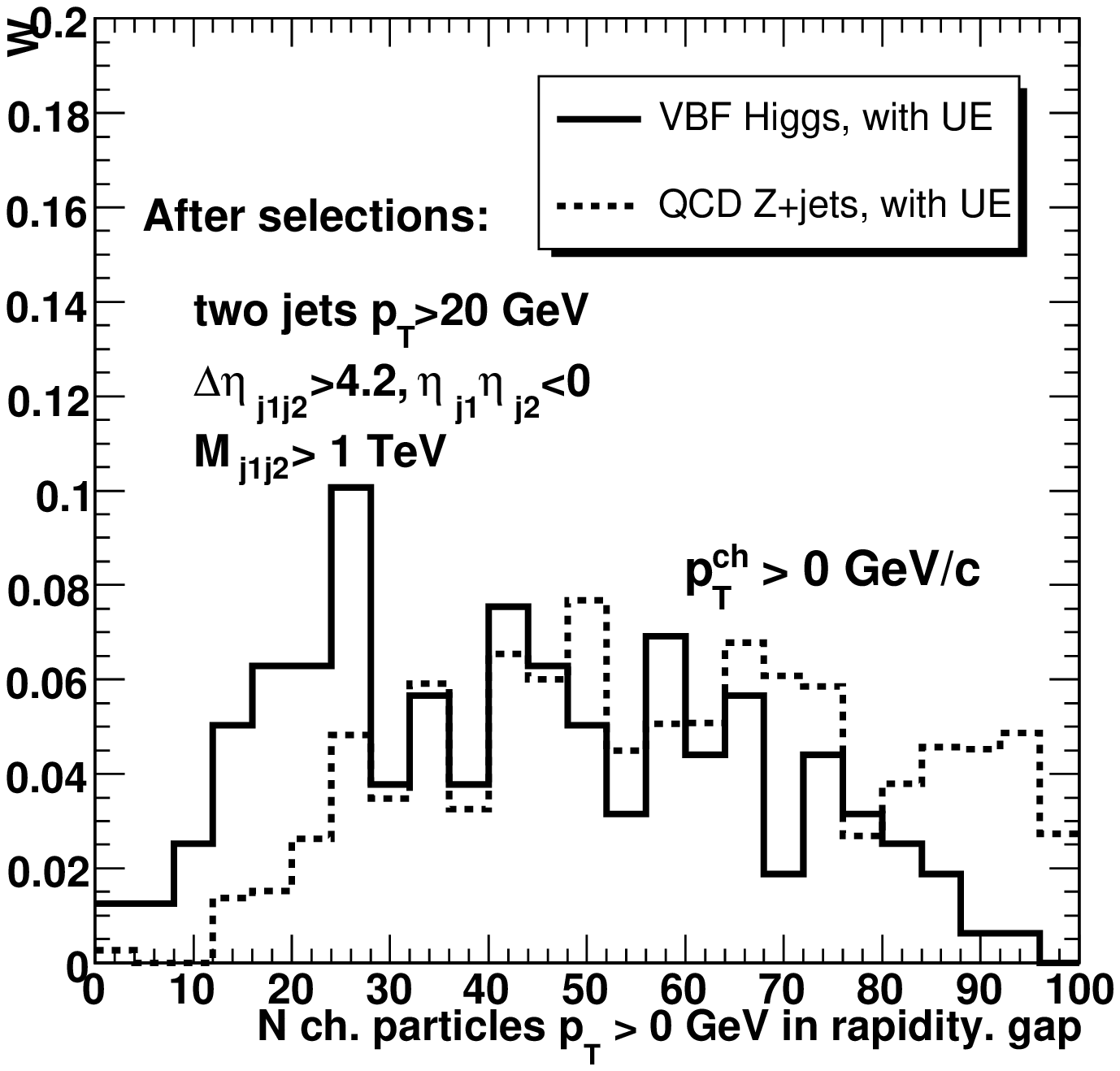}
\includegraphics[width=.3\textwidth]{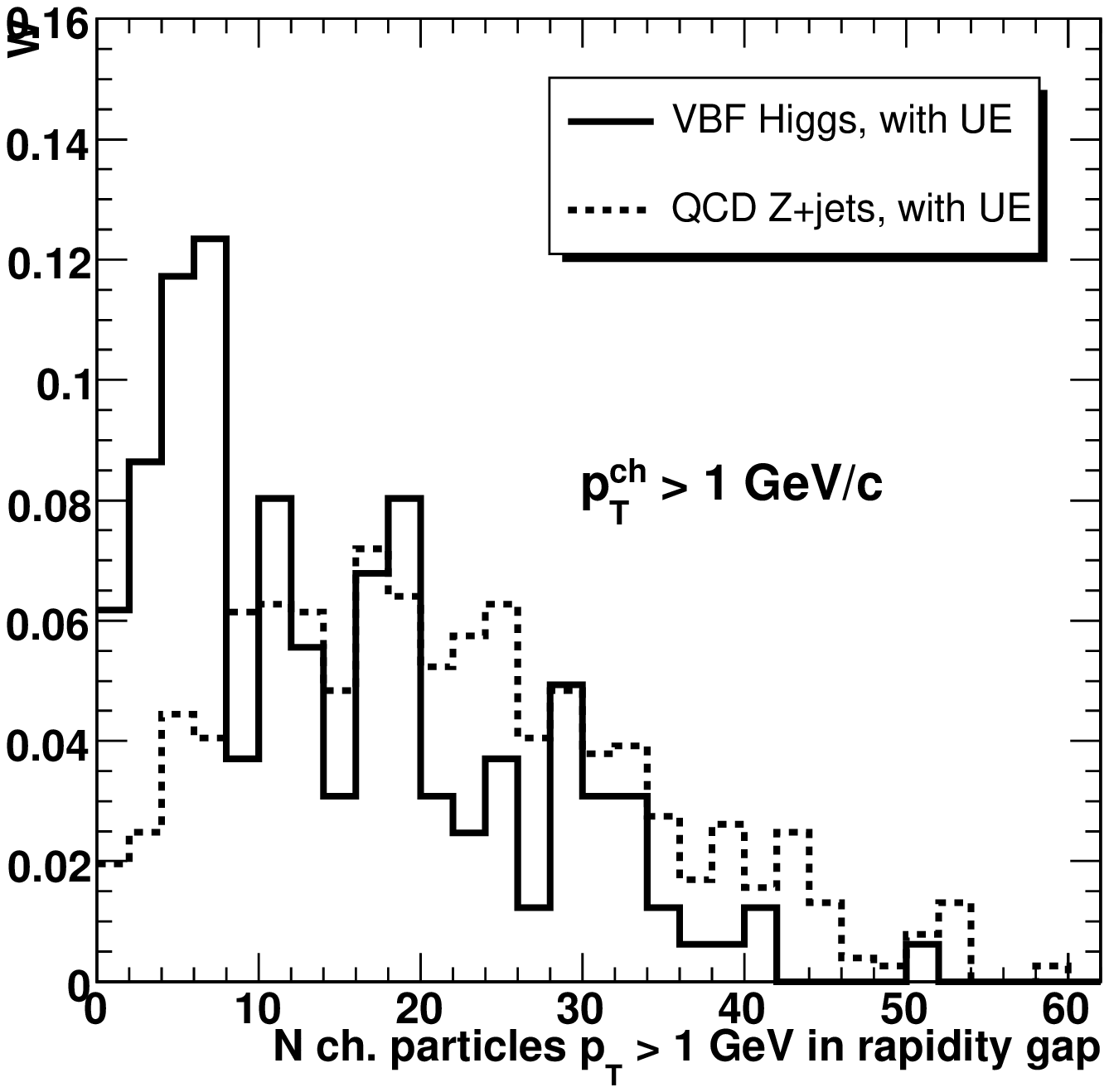}
\includegraphics[width=.3\textwidth]{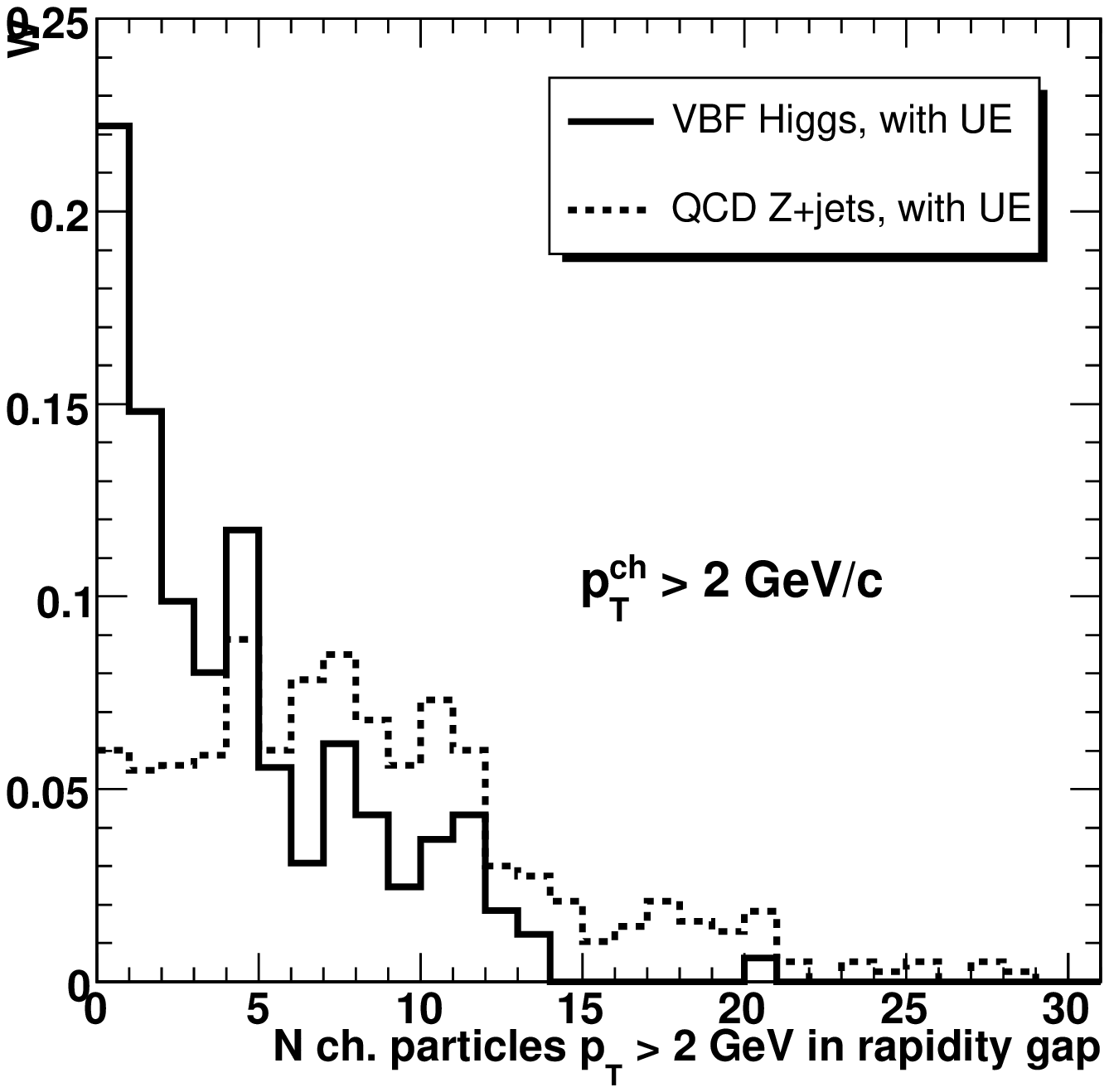}
\caption{The number of charged particles within the tracker acceptance and between two tagging jets 
($\eta^{j~min} + 0.5 < \eta^{track} < \eta^{j~max} - 0.5$) with $p_{T}>$ 0 GeV/$c$ (left plot), 
$>$ 1 GeV/$c$ (middle plot) and $>2$ GeV/$c$ (right plot) for the signal (solid line) 
and the QCD $Z$+jets background (dashed line). The multiple parton interactions generated with Tune DWT 
are switched on in PYTHIA.}
\label{fig:mctcv2}
\end{center}
\end{figure}

Fig.~\ref{fig:mctcv1} shows the number of charged particles within the tracker acceptance and between the
two tagging jets ($\eta^{j~min} + 0.5 < \eta^{track} < \eta^{j~max} - 0.5$) with $p_{T}>$ 0 GeV/$c$ 
(left plot), $>$ 1 GeV/$c$ (middle plot) and $>2$ GeV/$c$ (right plot) for the signal (solid line)
and the QCD $Z$+jets background (dashed line). The multiple parton interactions were switched off in PYTHIA.
%%%%%%%%%%%%%%%%%% F I G U R E %%%%%%%%%%%%%%%%%%%%%%%%%%%%%%%%%%%%%%%%%%%%%%
%%%%%%%%%%%%%%%%%% F I G U R E %%%%%%
One can see a clear difference between the signal and the QCD $Z$+jets background distributions. 
This difference, however is spoiled when the multiple parton interactions are switched 
on. Fig.~\ref{fig:mctcv2} shows the same distributions as in Fig.~\ref{fig:mctcv1} but
with the multiple parton interactions included in the generation. With no cut of the
charged particle $p_{T}$ applied, it is not possible to distinguish between the signal
and the background. The cut on the "track" $p_{T}$ removes charged particles from the underlying
event, thus giving the selection power for the TCV method. With the cut $p_{T}^{cut}$=2 GeV/$c$ 
the efficiency for the signal $VV \rightarrow H$ ($M_{H}$=120 GeV) is $\simeq$ 0.8 and 
for the QCD $Z$+jets background is 0.54.
%%%%%%%%%%%%%%%%%% F I G U R E %%%%%%%%%%%%%%%%%%%%%%%%%%%%%%%%%%%%%%%%%%%%%%
%%%%%%%%%%%%%%%%%% F I G U R E %%%%%%
For the same $\simeq$ 80 \% signal efficiency, the central jet veto efficiency for the background is smaller, 
0.44, thus leading to the conclusion that at the particle level simulation the central jet veto provides 
the better performance than the track counting veto. The final conclusion, however should be resulting from 
the full detector simulation including the detector and reconstruction effects, like fake jet contribution from 
the pile up and the electronic noise, the track and jet reconstruction inefficiency. 

\subsection{Studies with the full detector simulation}

The fully simulated datasets from the VBF Higgs boson analysis ($H\rightarrow \tau \tau \rightarrow \ell+\rm jet$)
~\cite{Takahashi:qqh} at an instantaneous luminosity $L = 2\times10^{33}$cm$^{-2}$s$^{-1}$ are used. The pile-up events 
(4.3 events per crossing) were included in the simulation. At the reconstruction level the same VBF selections 
2-4 on tagging jets as described in the section ~\ref{genlevel} were used and the tagging jets were required 
to have $E_{T}^{j}>$40 GeV. 
%The other event selections (not relevant for this study) were: 
%leptons ($e$ or $\mu$) with $p_{T}>12$ GeV/c,  $\tau$-jet with $E_{T}>20$ GeV, transverse invariant mass of 
%the lepton and the missing $<$40 GeV and $|\Delta\phi^{j_1j_2}|<$2.2. 
The CJV requires to reject events with a third jet that satisfies
\begin{itemize}
\item $E_{T}^{j3}>10$ GeV, where $E_{T}$ is a raw, non calibrated energy.
\item fake jet rejection parameter $\alpha^{j_3} = \sum p_T^{track}/E_T^{j_3} > 0.1$ 
      (see ~\cite{Takahashi:qqh} for details)
\end{itemize}
The TCV requires, on top of the selections mentioned in the previous section, the quality selections on the tracks: 
$\geq 8$ hits, $\Delta Z(\rm track,~ \rm vertex)<$2 mm, $\Delta R(\rm track,jet)>$0.5. The lepton and tracks from 
$\tau$ jet are not counted.
%%%%%%%%%%%%%%%%%% F I G U R E %%%%%%%%%%%%%%%%%%%%%%%%%%%%%%%%%%%%%%%%%%%%%%
\begin{figure}[htb!]
\begin{center}
\includegraphics[width=1.\textwidth]{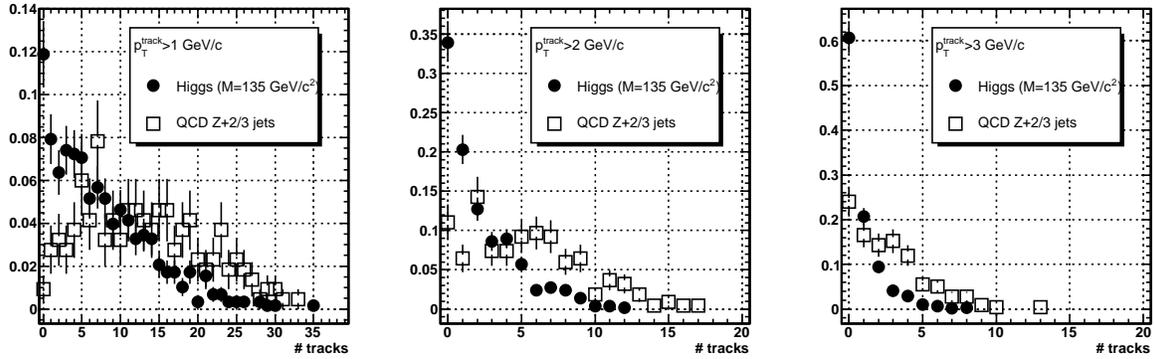}
\caption{Track multiplicity between the two forward tagging jets 
         ($\eta^{j~min} + 0.5 < \eta^{track} < \eta^{j~max} - 0.5$) 
         with $p_{T}>1$ GeV/$c$ (left plot), $>2$ GeV/$c$ (middle plot) and $>3$ GeV/$c$ 
         (right plot) for the signal (full circles) and the QCD Z+jets background (open squares).}
\label{fig:recotcv}
\end{center}
\end{figure}
%%%%%%%%%%%%%%%%%% F I G U R E %%%%%%

\begin{figure}[htb!]
\begin{center}
\includegraphics[width=0.7\textwidth]{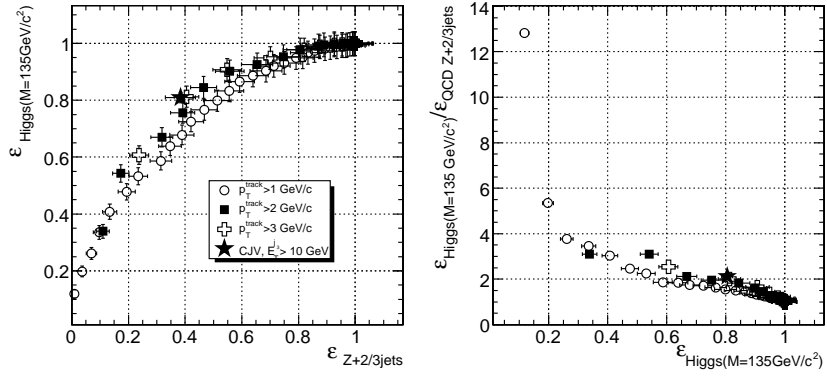}
\caption{The track counting veto (TCV) performances (see text) for the different $p_{T}^{track}$ and 
         track multiplicity thresholds.}
\label{fig:recotcvperf}
\end{center}
\end{figure}

Fig.~\ref{fig:recotcv} shows the number of reconstructed tracks between the two forward tagging jets
with $p_T>$ 1 GeV/$c$ (left plot), $>2$ GeV/$c$ (middle plot) and $>3$ GeV/$c$ (right plot). 
Both the Higgs boson signal (circles) and the QCD $Z$+jets background (squares) can be clearly separated when 
applying a cut on the track multiplicity and for different track $p_T$ thresholds. 
%One can see that if no track with $p_T>$ 3 GeV/c is allowed in the event, 
%already 60\% of the signal is selected whereas only 24\% of the background.
The left plot of Fig.~\ref{fig:recotcvperf} shows the performance of the algorithm, i.e the efficiency 
of selecting the signal versus the background. Starting from the bin 0 on the left bottom corner, 
the points correspond to an increasing cut on the track multiplicity and $p_{T}$ up to the right 
top corner where 100\% of events are selected. The black star indicates the performance of the 
central jet veto (CJV) based on calorimeter jets. One can notice that this latter achieves a good 
performance: 80.0$\pm$ 3.3\% efficiency for the signal and 39.7$\pm$ 5.\% efficiency for the backgrouind. 
The TCV algorithm can reach this discrimination power rejecting events with more that one track of
$p_{T}>3$ GeV/$c$. The right plot of Fig.~\ref{fig:recotcvperf} shows the ratio of the signal and the
background selection efficiencies as a function of the signal efficiency. It shows that the better 
ratio can be achieved with the TCV at the price of losing a bit of signal. This would obviously 
depend on the overall tuning of the analysis.
%%%%%%%%%%%%%%%%%% F I G U R E %%%%%%%%%%%%%%%%%%%%%%%%%%%%%%%%%%%%%%%%%%%%%%
%%%%%%%%%%%%%%%%%% F I G U R E %%%%%%
The third algorithm, the track-jet counting veto (TJV) is very similar to
the TCV. Tracks are first clustered along the beam axis ($Z$) starting from the track with the highest $p_T$ following the 
condition: $\Delta Z(\rm cluster,~\rm track)<$2 mm. Once a z-cluster is formed, the same procedure applies again 
with the remaining tracks. In a second step a traditional cone jet finding method is applied with 
$\Delta R = 0.5$ around seed tracks (with highest $p_{T}$). These jets are thus formed solely of tracks originating 
from the different z-clusters. They are finally associated with the signal vertex if their z-impact parameter 
is within 2 mm from the lepton z-impact parameter. This method allows to refine the description of the hadronization 
process usually producing several collimated particles with respect to the more exclusive approach of the TCV algorithm.
The discrimination variable are then the multiplicity and the minimum $p_T$ of the track-jets and its constituents
lying in between the two forward tagging jets. The performance of this algorithm has been found to be very
close to the TCV, reaching ~80\% for the signal efficiency and 40\% for the background when requiring no 
track-jet with $\Delta Z(\rm track-jet,~\rm lepton)<$ 2mm, $p_{T}^{jet}>3$ GeV/c and with at least one track of
$p_T>$0.9 GeV/c. 
%This is due to the fact
%that these track-jets are mainly isolated single-track jets. In fact the minimum cut $p_T>0.9$ GeV/c coming from
%track reconstruction ``cleans'' already the event from underlying activities and clusters of tracks are not
%likely to be reconstructed. Fig.~\ref{fig:tkjtkmult} shows the distribution of the track multiplicity of track-jets
%for signal and background. One can see that more than 88\% (85\%) are composed of one single track in signal
%(background) events hence reducing the possibility to exploit correlations between tracks originating from the 
%same interaction point.
%
%%%%%%%%%%%%%%%%%% F I G U R E %%%%%%%%%%%%%%%%%%%%%%%%%%%%%%%%%%%%%%%%%%%%%%
%\begin{figure}[htb!]
%\begin{center}
%\includegraphics[width=.4\textwidth]{tkj_trackmult.eps}
%\caption{Track multiplicty for track-jets within the two forward tagging jets for signal (full squares) and background (open circles) 
%         data.}
%\label{fig:tkjtkmult}
%\end{center}
%\end{figure}
%%%%%%%%%%%%%%%%%% F I G U R E %%%%%%

%\input{cjvmethods.tex}
%\input{cjvfromdata.tex}%
\subsection{The efficiency measurement of the central rapidity gap selection for Z $\rightarrow \tau \tau$ background.}

The efficiency of the central rapidity gap selection (CRGS) for the $Z$+jets, $Z \rightarrow \tau \tau$
background in the VBF $H\rightarrow \tau \tau$ search can be measured with the $Z$+jets, 
$Z \rightarrow \mu \mu$ events passed the "signal like" VBF jet selections. We 
estimated the expected number of such events and the statistical accuracy of the CRGS for 100 pb$^{-1}$
of integrated luminosity. Only QCD $Z$+jets events were used. The events from the EWK $Z$+2jets production 
still have to be added.  The fully simulated events with no pile-up were required to pass the di-muon trigger. 
In the off-line analysis the events with two muons $p_{T}>$ 10 GeV/$c$, $|\eta |<$2.4 isolated in the tracker were 
selected within the di-muon mass window 70$<M_{\mu \mu} <$110 GeV/$c^2$. The following VBF cuts relaxed for the 
early analysis with the first 100 pb$^{-1}$ of the data are used.  An event must have at least two leading 
$E_{T}$ jets that satisfy the following requirements: $E_{T}^{j}>$ 40 GeV, $|\eta^{j}|~<$4.5, $\eta^{j1} \times \eta^{j2}<0$ 
and:  
\begin{itemize}
\item soft VBF selections: $M_{j1j2}>400$ GeV/$c^2$, $|\Delta\eta^{j1j2}|>2.5$
\item hard VBF selections: $M_{j1j2}>800$ GeV/$c^2$, $|\Delta\eta^{j1j2}|>3.5$
\end{itemize}
The CJV used in this section requires to reject events with a third calorimeter jet that satisfies
\begin{itemize}
\item $E_{T}^{j3}>30$ GeV, where $E_{T}$ is the calibrated jet energy
\item $\eta^{j~min} + 0.5 < \eta^{j3} < \eta^{j~max} - 0.5$,
\end{itemize}

\begin{figure}[htb!]
\begin{center}
\includegraphics[width=.4\textwidth]{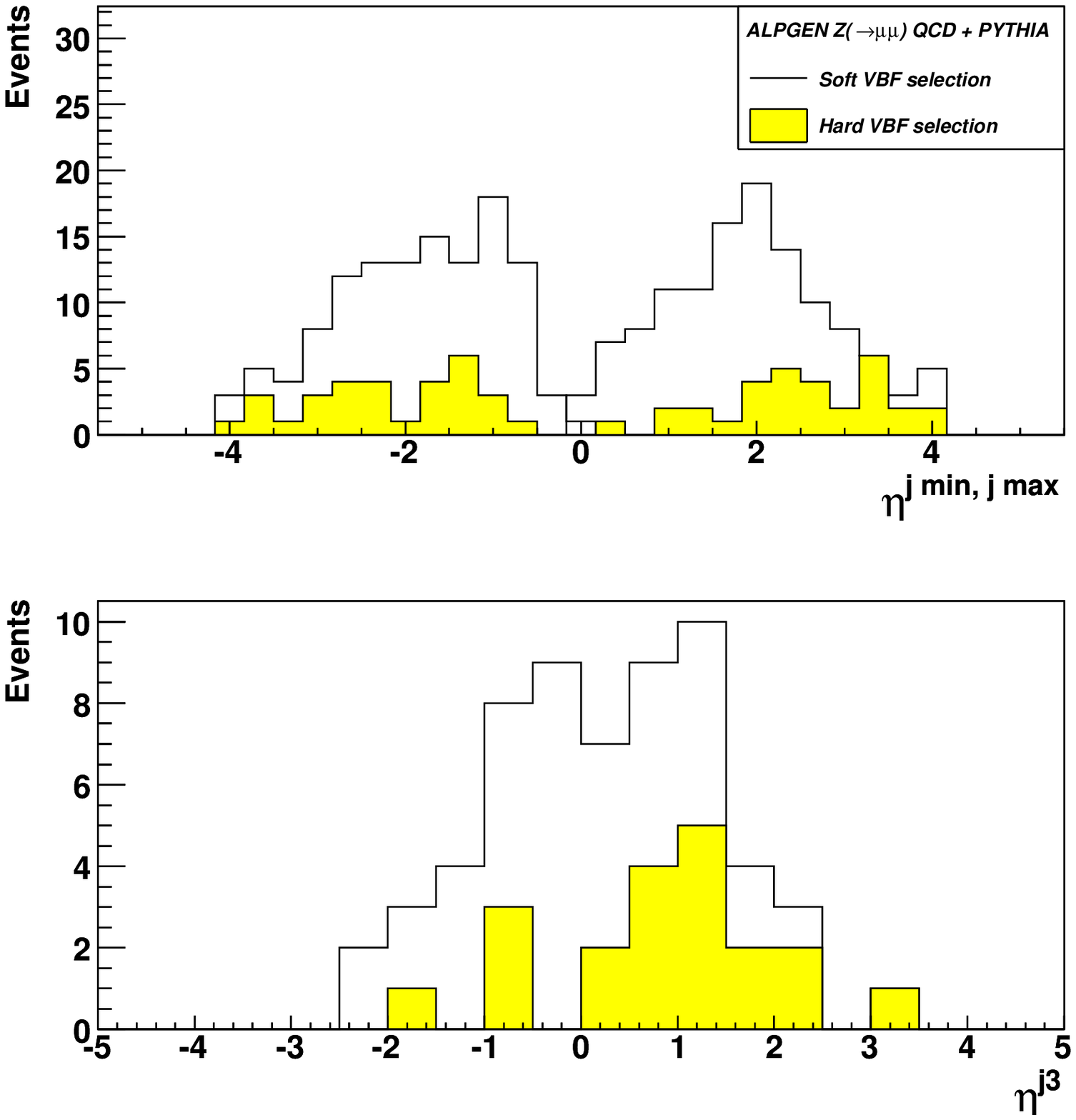}
\includegraphics[width=.4\textwidth]{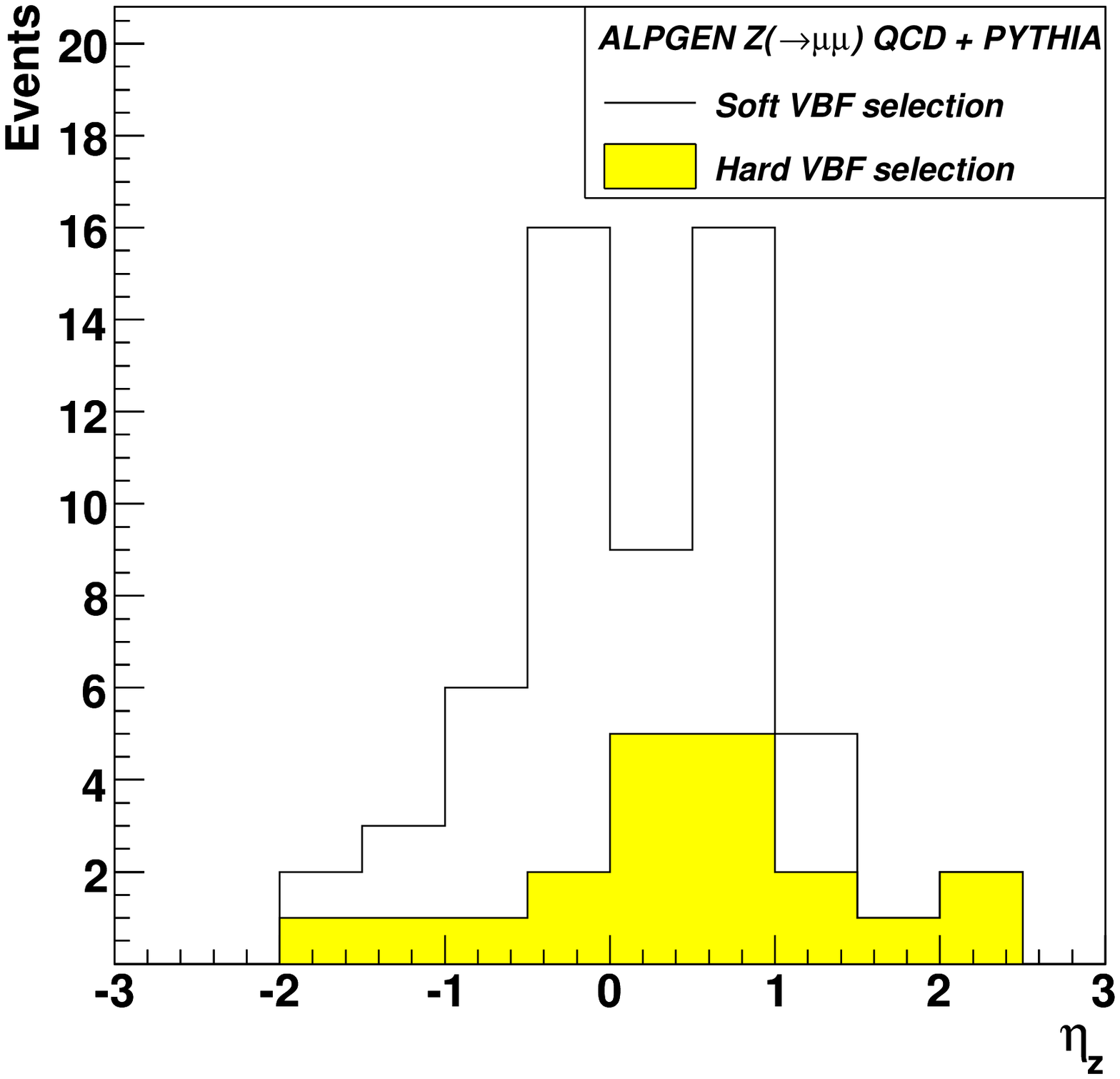}
\caption{The distribution of $\eta^{j~min}$ and $\eta^{j~max}$ (left-upper plot), the
         $\eta^{j3}$ (left-bottom plot) and the variable $\eta _{Z}$=$\eta ^{j3}$-0.5($\eta^{j~min}+\eta^{j~max})$ 
         (right plot).All selections except the CJV were applied.}
\label{fig:jets}
\end{center}
\end{figure}

Table ~\ref{table:CRGS} shows the expected number of events after selections for 100 pb$^{-1}$ and the efficiency and the 
statistical accuracy of the CJV.
\begin{table}[thb!]
\caption{The expected number of events after selections for 100 pb$^{-1}$ and the efficiency and the statistical accuracy of 
the CJV.}
\label{table:CRGS}
\begin{center}
\begin{tabular}[t]{|c|c|c|}
\hline
selections & number of events with 100 pb$^{-1}$ & CJV efficiency \\
\hline
"soft" VBF        & 121                          &                \\
\hline
"soft" VBF + CJV  &  61                          & 0.50 $\pm$ 0.06 \\
\hline
"hard" VBF        &  31                          &                 \\
\hline
"hard" VBF + CJV  &  11                          & 0.35 $\pm$ 0.11  \\ 
\hline
\end{tabular}
\end{center}
\end{table}
Fig.~\ref{fig:jets} shows the distribution of $\eta^{j~min}$ and $\eta^{j~max}$ (left-upper plot), the
$\eta^{j3}$ (left-bottom plot) and the variable $\eta _{Z}$=$\eta ^{j3}$-0.5($\eta^{j~min}+\eta^{j~max})$ 
(right plot) for 100 pb$^{-1}$ of the "data" for one random experiment. All selections except the CJV were 
applied.

\subsection{Conclusions}

With the full detector simulation is was shown that both the central jet veto and the track counting algorithms 
achieve very similar performance. The robustness and the stability of the methods under a variation of the run 
and detector conditions will be tested with the real data using $Z$+jets, $Z\rightarrow \mu\mu$
events. It is believed that the track counting algorithms relying on a single sub-detector, the tracking system, would 
perform with a higher reliability.

\section*{Acknowledgements}

We would like to thank T.~Sjostrand and V.~Khoze for very useful
discussions about the methods for the central rapidity gap selection.
A.N. and M.V.A. would like to thank the organizers of Les Houches Workshop 
2007 for the hospitality and the nice scientific atmosphere of the Workshop.

%\vspace*{-0.2cm}
%
%\bibliography{jetveto}

%\end{document}

\section[Production of a Higgs Boson and a Photon in 
Vector Boson Fusion at the LHC]
{PRODUCTION OF A HIGGS BOSON
AND A PHOTON IN VECTOR BOSON FUSION AT THE LHC
~\protect\footnote{Contributed by: E.~Gabrielli, 
F.~Maltoni, B.~ Mele, M.~ Moretti, 
F.~Piccinini, and R. Pittau}
}
%\documentclass[11pt]{cernrep}
%\usepackage{graphicx,epsfig}
%\bibliographystyle{lesHouches}
%\begin{document}
%
%
%
%
% commands
%
\def\beq{\begin{equation}}
\def\eeq{\end{equation}}
\def\bea{\begin{eqnarray}}
\def\eea{\end{eqnarray}}

\newcommand{\dedouble}{ \stackrel{ \leftrightarrow }{ \partial } }
\newcommand{\deR}{ \stackrel{ \rightarrow }{ \partial } }
\newcommand{\deL}{ \stackrel{ \leftarrow }{ \partial } }
\newcommand{\ci}{{\cal I}}
\newcommand{\ca}{{\cal A}}
\newcommand{\Wp}{W^{\prime}}
\newcommand{\vep}{\varepsilon}
\newcommand{\kk}{{\bf k}}
\newcommand{\pp}{{\bf p}}
\newcommand{\hs}{{\hat s}}
\newcommand{\proj}{\frac{1}{2}\;(\eta_{\mu\alpha}\eta_{\nu\beta}
+  \eta_{\mu\beta}\eta_{\nu\alpha} - \eta_{\mu\nu}\eta_{\alpha\beta})}
\newcommand{\projm}{\frac{1}{2}\;(\eta_{\mu\alpha}\eta_{\nu\beta}
+  \eta_{\mu\beta}\eta_{\nu\alpha}) 
- \frac{1}{3}\;\eta_{\mu\nu}\eta_{\alpha\beta}}

%%%%%%%%%%%%%%%%%%%%%%%%%%%%%%%%%%%%%%%%%%%%%%%%%%%%%%%%%%%%%%%%%%%%%%%
%\def\lsim{\raise0.3ex\hbox{$\;<$\kern-0.75em\raise-1.1ex\hbox{$\sim\;$}}} 

%\def\gsim{\raise0.3ex\hbox{$\;>$\kern-0.75em\raise-1.1ex\hbox{$\sim\;$}}}

%\def\Frac#1#2{\frac{\displaystyle{#1}}{\displaystyle{#2}}}

\newcommand{\gsim}
{\;\raisebox{-.3em}{$\stackrel{\displaystyle >}{\sim}$}\;}

\def\no{\nonumber\\}

\subsection{Introduction \label{sect1}}
Higgs boson search is one of the main tasks of
present and future collider experiments \cite{Assamagan:2004mu,Djouadi:2005gi}.
At the CERN Large Hadron Collider (LHC), 
the Higgs boson is expected to be produced with high rate 
via gluon or vector-boson fusion (VBF) mechanisms and associate 
$W(Z) H$ production. 
Apart from observing the Higgs signal, it would be crucial 
to make at the LHC also a measurement of the $Hb\bar{b}$ 
coupling~\cite{Duhrssen:2004cv}. 
To this aim, Higgs production via VBF, with the Higgs boson decaying into a 
$b\bar{b}$ pair, plays an important role \cite{Mangano:2002wn}. 
In this contribution, we consider a further process that could help in 
determining the $H b\bar{b}$ coupling, that is the
Higgs boson  production
in association with a large transverse-momentum  photon 
(with $p_{\rm T}\gsim 20$ GeV) and two {\it forward} jets 
\cite{Gabrielli:2007wf}
\begin{equation}
pp\to H\, \gamma\, jj\to b \bar b \,\gamma \,jj\, + X\, ,
\label{process}
\end{equation}
with $H$ decaying to $b\bar{b}$, where at the 
parton level the final QCD parton 
is identified with the corresponding jet. 
In our study,
we will not include diagrams where the photon is emitted from 
one of the two b-quarks, since the requirement of 
a large $p_{\rm T}$ photon would shift in that case the $b\bar b$ invariant 
mass outside the experimental $b\bar b$ mass resolution window around 
the Higgs mass. 

There are  a number of advantages in considering this QED higher-order 
variant of the VBF Higgs production process $pp\to H(\to b\bar b) \; jj$.
The fact that the production rate is penalized by the electromagnetic
coupling is compensated by a few peculiarities of the channel 
in Eq.~(\ref{process}).
First of all, the presence of an additional  high $p_{\rm T}$ photon can 
improve the triggering efficiencies for multi-jet final 
states, such as those needed to select $pp\to H(\to b\bar b) \; jj$  events. 
Second, there is a large gluonic component entering the partonic 
processes giving rise to the QCD backgrounds to the 
$b \bar b \,j j$ final state. As a consequence,
the QCD backgrounds are in general much less  {\em active}
in  radiating  a large $p_{\rm T}$ photon with respect to the VBF signal. 
In addition there are dynamical effects that 
dramatically suppress the radiation of a central photon in the 
irreducible QCD background to  $b \bar b \,\gamma \,jj$ with 
respect to the VBF channel. 
When the photon is forced to be emitted in the central 
rapidity region,
a destructive quantum interference arises between 
the photon emission 
off the initial quark exchanging a gluon 
(or any other neutral vector boson) in the $t$ channel, 
and the photon emission off the corresponding final quark. 
For the signal case of the $H\,\gamma\,jj$ production, 
the above mechanism of destructive interferences 
affects only the diagrams involving
the $ZZ$ fusion. On the other hand, in the diagrams involving 
$WW$ fusion (that are responsible for the dominant part of the 
basic VBF $H\,jj$ cross section)
the charged currents in the $qq^\prime W$ vertices
change the electric charges of the in-out
partons, and consequently 
the interference is now additive rather than destructive. 
Therefore, the cross section for 
$H\,\gamma\,jj$ is expected to follow the usual pattern 
of QED corrections as far as  its $WW$ fusion component 
is concerned. The relative 
contribution of the $ZZ$ fusion will be instead
remarkably smaller than in the case of the basic VBF $H\,jj$ 
process.

To summarize, a measurement
of the $b \bar b \,\gamma \,jj$ rate could lead to a combined 
determination of the Higgs boson couplings to  $b$ quarks and $W$
vector bosons, with less contamination from the $HZZ$ coupling 
uncertainties.

In Section~\ref{sect2}, we go through the main kinematical and 
dynamical characteristics of the process in Eq.~(\ref{process}). 
We also discuss the features of the main QCD irreducible background. 
In Section~\ref{sect3}, the signal rates are computed at parton level 
for a set of kinematical cuts 
that optimizes the signal/background ratio, restricting the 
analysis to the case of the irreducible background.
In Section~\ref{sect4}, the main reducible 
background channels are included in the analysis. 
Finally, in Section~\ref{sect5}, we draw our conclusions.
\subsection{Signal and Irreducible Background \label{sect2}}
%%%%%%%%%%%%%%%%%%%%%
\begin{table}
\begin{center}
\begin{tabular}{||l|l|l|l|l||}\hline 
 $m_H $~(GeV)  & 110 &  120 & 130 & 140
\\  \hline
$\sigma(H
\gamma jj) \; \; [fb]$  & 67.4  & 64.0   & 60.4  &  56.1
 \\ \hline
${\cal BR}(H\to b \bar b)$ & 0.770 & 0.678 & 0.525 & 0.341 \\ \hline
\end{tabular}            
\caption{\label{inclusive} Cross sections for the
$H\,\gamma\, jj$ signal at LHC, for $p_{\rm T}^\gamma \geq 20$~GeV, 
$\Delta R_{\gamma j}> 0.4$, and  a cut $m_{jj} > 100\, {\rm GeV}$ on 
the invariant mass of the final quark pair. Also shown are the 
Higgs boson branching ratios to $b\bar b$
(computed through HDECAY~\cite{Djouadi:1997yw}), that are not included in 
the cross sections shown.}
\end{center}
\end{table}
Cross sections for the $H\, \gamma\, jj$ production
at $\sqrt S =14$~TeV
are shown in Table~\ref{inclusive}.  In order to
present results as inclusive as possible only a minimal set of
kinematical cuts is applied
($\Delta R_{\gamma j}> 0.4$,
$p_{\rm T}^\gamma \geq 20$~GeV, and 
$m_{jj} > 100\, {\rm GeV}$). 
The Higgs boson branching ratios to $b\bar b$, which are
not included in the cross section results,
(computed through HDECAY~\cite{Djouadi:1997yw}), are also shown.
The full tree-level matrix elements for the electroweak process
$pp\to H\, \gamma\, jj\,$
%in Figure~1
have been computed independently with ALPGEN~\cite{Mangano:2002ea},
and MadEvent~\cite{Maltoni:2002qb}. Details on the values of input
parameters, such PDF's and scales are given in Section \ref{sect3}.

Before discussing the process in Eq.~(\ref{process}),
it is useful to  recall here the main kinematical properties 
of a typical VBF event, that is $pp\to H\, jj$, and 
the corresponding backgrounds.
For the Higgs boson decaying to a $b\bar{b}$ pair, the main 
background to the basic VBF process comes from the QCD
production of  the final state $b\bar{b} jj$,  whenever the 
$b\bar{b} jj$ kinematical characteristics approach the typical 
VBF configuration. 
By imposing a large invariant mass cut for the two-forward-jet 
system [i.e., $m_{jj}\gsim 
{\cal O}(1)$ TeV], a minimal $p_{\rm T}^{j}$ of a few tens GeV's, and 
requiring  the $b \bar b$ invariant mass to be  around $m_H$
within the $m_{b \bar b}$ experimental resolution, one can obtain
a signal significance ($S/\sqrt{B}$) 
of the order of $S/\sqrt{B}\sim 3-5$, assuming an integrated 
luminosity of 600 fb$^{-1}$~\cite{Mangano:2002wn}.

Let us now consider the  VBF Higgs production  when  a further 
central photon is emitted, namely $pp\to H\, \gamma\, jj$. 
According to the usual pattern of QED corrections,
one might expect the request of a further hard photon to keep
the  relative weight of signal and background quite stable. 
Were this the case, the rates for $pp\to H\, \gamma\, jj$ and
its background would be related to a ${\cal O}(\alpha)$ rescaling
of the rates for the $H\, jj$ signal  and its background, respectively,
where $\alpha$ is the fine electromagnetic structure constant.
On this basis,  one would conclude that 
there is no advantage in considering the $H\, \gamma \, jj$
variant of the  $H\, jj$ process, apart from the fact that the 
presence of a hard photon
in the final state can improve the triggering  efficiency of 
the detectors.
However, as we explained in the introduction, this pattern does not 
hold in general when restricted regions 
of phase space are considered. 

In the next section we will study 
this effect on a quantitative level, showing that
the requirement of a further central 
photon gives rise to a dramatic increase  (by more than 
one order of magnitude)
in the $S/B$ ratio, 
while the signal cross section roughly follows the naive 
QED rescaling.
%%%%%%%%%%%%%%%%%%%%%%%%%

\subsection{Cross Sections for the Signal versus the Irreducible Background
\label{sect3}}
%%%%%%%%%%%%%%%%%%%%

The numerical results presented in this section have been independently
obtained by the Monte Carlo event generators ALPGEN~\cite{Mangano:2002ea},
and MadEvent~\cite{Maltoni:2002qb}.
The signal is calculated in the narrow width approximation, 
{\sl i.e.} we computed  the exact lowest-order matrix element for the process 
$p p \to H \gamma \, jj $, and then let the Higgs boson 
decay into a $b \bar b$ pair according to its branching ratio 
and isotropic phase space. 
After the decay,  cuts on the $b-$quark jets are implemented. 
For the irreducible $p p \to b \bar b \gamma \, jj$ background,
we computed  all the matrix elements at ${\cal O}(\alpha_s^4 \alpha) 
$, neglecting 
${\cal O}(\alpha_s^2 \alpha^3)$,
${\cal O}(\alpha_s^3 \alpha^2)$,
${\cal O}(\alpha_s \alpha^4)$ and
${\cal O}(\alpha^5)$ contributions and their interference with the
${\cal O}(\alpha_s^4 \alpha)$ contribution. We checked that this 
has no numerical impact on the results.
The present study is limited at the parton level. A more complete 
simulation, that takes into account showering, 
hadronization and detector simulation, even if 
crucial for the assessment of the potential of this channel, 
is beyond the scope of the present contribution.  
As PDF's,  we use the parametric form of 
CTEQ5L~\cite{Lai:1999wy}, and the facto\-ri\-za\-tion/re\-norma\-li\-zation 
scales are fixed at 
$\mu_{\rm F}^2 = \mu_{\rm R}^2 = \sum E_t^2$ and 
$\mu_{\rm F}^2 = \mu_{\rm R}^2 = m_H^2 + \sum E_t^2$ for the 
backgrounds and signal, respectively ($E_t$ is the transverse 
energy of any QCD parton). The three
 Higgs-mass cases  120, 130 and 140~GeV are analysed. 
 
We start by the definition of two {\it basic} event selections 
(sets 1 and 2) that differ only by the
threshold on the photon transverse momentum $p_{\rm T}^\gamma$:
\begin{eqnarray}
&&p_{\rm T}^j \geq 30\, {\rm GeV}, \, \, \, \,\, 
p_{\rm T}^b \geq 30\, {\rm GeV}, \, \, \, \,\,
\Delta R_{ik} \geq 0.7,\, \nonumber \\
&&|\eta_\gamma|\leq 2.5, \, \, \,\,\,
|\eta_b|\leq 2.5, \, \, \,\,\, |\eta_j|\leq 5, \nonumber \\
&&m_{jj} > 400\, {\rm GeV}, \, \, \, \,\,\,\,  
m_H(1-10\%) \leq m_{b \bar b} \leq m_H(1+10\%), \nonumber \\
&& 1) \, \, \, p_{\rm T}^\gamma \geq 20\, {\rm GeV}, \nonumber \\
 &&  2) \, \, \, p_{\rm T}^\gamma \geq 30\, {\rm GeV},  
\label{eq:basic}
\end{eqnarray}
where $ik$ is any pair of partons in the final state, including 
the photon, and $\Delta R_{ik} =\sqrt{\Delta^2\eta_{ik}+\Delta^2\phi_{ik}}$,
with $\eta$ the pseudorapidity and $\phi$ the azimuthal angle.
The cross sections for the above {\it basic} event selections are 
reported 
in Table~\ref{basicxs}.
\begin{table}
\begin{center}
\begin{tabular}{||l|l|l|l|l||}\hline
& $p_{\rm T}^{\gamma, cut}$
 & $m_H = 120$~GeV  & $m_H = 130$~GeV  &  $m_H = 140$~GeV  
\\  \hline
$\sigma[H(\to b \bar b) 
\gamma jj] $ & 20~GeV& 9.3(1)~fb  & 7.4(1)~fb   & 4.74(7)~fb   \\ 
                      & 30~GeV &  6.54(7)~fb & 5.2(1)~fb     &   3.31(3)~fb
  \\  \hline
$\sigma[{b \bar b} \gamma jj] $ & 20~GeV & 406(2)~fb  
& 405(4)~fb & 389(1)~fb  \\ 
                             & 30~GeV   & 260.5(7)~fb  
& 257.9(6)~fb & 251.8(7)~fb  
\\ \hline \hline
$\sigma[H(\to b \bar b) jj] $ &  & 727(2)~fb  
& 566(2)~fb & 363(1)~fb 
\\ \hline
$\sigma[{b \bar b} jj] $ &  & 593.7(5)~pb  & 550.5(5)~pb &  505.6(4)~pb
\\ \hline
\end{tabular}            
\caption{\label{basicxs} Cross sections for  the signal 
and the irreducible
background for the {\it basic} event selections 
in Eq.~(\ref{eq:basic}). 
Higgs production cross sections include the Higgs branching 
ratios to $b\bar b$. The signal and irreducible background 
production rates for the VBF process without photon are also shown.}
\end{center}
\end{table}
Before comparing the signal and the background for the 
$H\, \gamma\, jj$ process, we tried to optimize our event 
selection in Eq.~(\ref{eq:basic}). Indeed,
the signal detectability can be further improved by 
imposing {\it optimized} 
cuts, that can be deduced by looking at the 
following kinematical distributions:
\bea
\frac{d\sigma}{d m_{jj}}, \, \, \, \, \, 
\frac{d\sigma}{d p_{\rm T}^{j1}}, \, \, \, \, \, 
\frac{d\sigma}{d p_{\rm T}^{b1}}, \, \, \, \, \, 
\frac{d\sigma}{d m_{\gamma H}}, \, \, \, \, \, 
\frac{d\sigma}{|\Delta \eta_{jj}|},   \nonumber
\eea
where $j1$ and $b1$ denote the leading ${p_{\rm T}}$ light jet 
and $b-$ jet, respectively, and $m_{\gamma H}$ is the invariant 
mass of the $\gamma b\bar b$ system.
By studying the variation of the significance $S/\sqrt{B}$ as a function 
of the cuts on the distributions 
(for more details see \cite{Gabrielli:2007wf}), 
we found an {\it optimized} event 
selection where, in addition to the {\it basic} cuts, we impose the 
following cuts
\begin{eqnarray}
&&m_{jj} \geq 800\, {\rm GeV}, \, \, \, \, \, 
p_{\rm T}^{j1} \geq 60\, {\rm GeV},  \, \, \, \, \, 
p_{\rm T}^{b1} \geq 60\, {\rm GeV}, \, \, \, \nonumber \\
&&|\Delta \eta_{jj}| > 4, \, \, \, \, \, 
m_{\gamma H} \geq 160\, {\rm GeV},  \, \, \, \, \, \Delta 
R_{\gamma b/\gamma j} \geq 1.2\, .
\label{eq:optimized}
\end{eqnarray}
With the above additional requirements, we find the cross 
sections reported in Table~\ref{optimizedxs}. 
\begin{table}
\begin{center}
\begin{tabular}{||l|l|l|l|l||}\hline
&  $p_{\rm T}^{\gamma, cut}$ & $m_H = 120$~GeV  & $m_H = 130$~GeV  
&  $m_H = 140$~GeV  
\\  \hline
$\sigma[H(\to b \bar b)  \gamma jj] $  & 20~GeV 
&3.59(7)~fb  &2.92(4)~fb &1.98(3)~fb  \\
                                       & 30~GeV
&2.62(3)~fb  &2.10(2)~fb &1.50(3)~fb  
\\ \hline
$\sigma[{b \bar b} \gamma jj] $ & 20~GeV    &33.5(1)~fb  
&37.8(2)~fb &40.2(1)~fb 
 \\ 
                          & 30~GeV   &25.7(1)~fb  &27.7(1)~fb &28.9(2)~fb
\\ \hline \hline
$\sigma[H(\to b \bar b)  jj] $ &  & 320(1)~fb  & 254.8(6)~fb & 167.7(3)~fb  
\\  \hline
$\sigma[{b \bar b} jj] $ &  & 103.4(2)~pb  & 102.0(2)~pb & 98.4(2)~pb 
\\ \hline
\end{tabular}            
\caption{\label{optimizedxs} Cross sections for  the signal and the irreducible
background for the {\it optimized} event selections 
of Eq.~(\ref{eq:optimized}), added to the {\it basic} selection
in Eq.~(\ref{eq:basic}).  Higgs production cross sections include 
the Higgs branching ratios to $b\bar b$. The
signal and irreducible background production rates for the basic 
VBF process are also shown.
}
\end{center}
\end{table}
One see that the requirement of the extra central 
photon with $p_{\rm T}^{\gamma}\gsim 20$~GeV in the final state 
involves a reduction factor 
of order 100 for the signal rate with respect to the final state 
without photon,  
according to the expectations of the 
 ${\cal O}(\alpha)$ QED naive scaling.
 On the other hand,
the radiative background is suppressed by a factor  of about 3000
with respect to the case of no photon radiation.
Finally, a summary of the statistical significances, including only 
the irreducible background, with an integrated luminosity of 
$100$~fb$^{-1}$ is given in Table~\ref{tab:sign1}. 
\begin{table}
\begin{center}
\begin{tabular}{||l|l|l|l|l||}\hline   &  $p_{\rm T}^{\gamma, cut}$
 & $m_H = 120$~GeV  & $m_H = 130$~GeV  &  $m_H = 140$~GeV  
\\  \hline
$S / \sqrt{B}|_{H\gamma\,jj}$ & 20~GeV & 2.6  & 2.0 & 1.3  
\\ \hline
$S / \sqrt{B}|_{H\gamma\,jj}$ & 30~GeV  & 2.2  & 1.7 & 1.2 
\\ \hline \hline
$S / \sqrt{B}|_{H\,jj}$ &  & 3.5  & 2.8 & 1.9 
\\ \hline
\end{tabular}            
\caption{\label{tab:sign1} Statistical significances 
with the event selection of Eq.~(\ref{eq:basic}) and 
(\ref{eq:optimized}), with an integrated  luminosity 
of $100$~fb$^{-1}$. The value $\epsilon_b = 60$\% for the 
$b-$tagging efficiency and a Higgs boson event reduction 
by $\epsilon_{b \bar b}\simeq$ 70\%, due to the finite ($\pm$10\%)
$b \bar b$ mass resolution,  have been assumed. Jet-tagging efficiency and 
photon-identification efficiency are set to 100\%. 
Only the irreducible background is included in this analysis.}
\end{center}
\end{table}
\subsection{Reducible Backgrounds \label{sect4}}
A complete analysis of the reducible backgrounds to the   
$H\,\gamma\,jj$    signal is beyond the scope of our study. 
However, in order to have a sensible estimate of the 
achievable $S/B$ ratio and statistical significance at parton level, 
we computed with ALPGEN the cross sections, assuming $m_H=120$~GeV and
with the optimized event selection of Eq.~(\ref{eq:basic}) 
and (\ref{eq:optimized}),  
for three main potentially dangerous processes
\begin{itemize}
\item $p p \to \gamma + 4 $~jets, where two among the light jets are 
fake tagged as $b-$jets;
\item $p p \to b {\bar b} + 3$~jets, where one of the light jets is 
misidentified as a photon;
\item $p p \to 5$~jets, where one of the light jets is misidentified as 
a photon, and two light jets are fake tagged as $b-$jets. 
\end{itemize}
By including 
also the reducible backgrounds, 
the statistical significance  decreases by 
about 14(12)\% for $p_{\rm T}^{\gamma, cut}=20(30)$~GeV)
with respect to Table~\ref{tab:sign1}, where only the 
irreducible background has been considered. The most dangerous 
contribution to reducible backgrounds comes from $p p \to b \bar b + 3 j$. 
\subsection{Conclusions \label{sect5}}
In this contribution, we studied the detectability of the Higgs boson 
production signal,
when the Higgs boson is accompanied by a high$-p_{\rm T}$ 
central photon and two forward jets at the LHC. The Higgs boson 
decay into a $b \bar b$ pair is considered. We analyzed the signal, 
the irreducible QCD background, 
and main reducible backgrounds at the parton level.
 The presence of a photon in the final state can improve 
the triggering efficiencies
 with respect to the basic VBF Higgs production without a photon. 
Moreover, we find that the requirement of a central 
photon in addition to the typical VBF final-state topology 
significantly suppresses the irreducible 
 QCD background. In particular, due to dynamical effects,
 the latter has rates that are lower than
 the expectations of the 
 ${\cal O}(\alpha)$ QED naive scaling by more than an order of magnitude.
 As a consequence, after optimizing kinematical cuts, we obtain a 
statistical significance $S/\sqrt{B}$ for the 
 $H(\to b\bar b)\gamma\,jj$   channel  that goes from around 3, 
if  $m_H\simeq 120$~ GeV, down  to
about  1.5,  if $m_H\simeq 140$~ GeV,
for an integrated  luminosity of $100$~fb$^{-1}$. 
These significances are not far from the corresponding values 
for  the basic $H(\to b\bar b)\,jj$ process without a photon. 
The latter estimates
are based on the irreducible QCD background. 
The impact of including a few main reducible backgrounds has 
found to be moderate.
The same dynamical effects that are responsible for the 
irreducible background suppression also remarkably curb 
the relative contribution of the 
$ZZ\to H$ boson fusion 
diagrams with respect to the $WW\to H$ ones  
in the process $pp\to H(\to b\bar b)\gamma\,jj$.
As a consequence, we think that the study of the 
$H(\to b\bar b)\gamma\,jj$  signal at the LHC could 
have a role in the determination of both the $Hbb$ and $HWW$ couplings.
%Further studies, including complete showering, hadronization, 
%and detector simulations, that are beyond the scopes of the 
%present contribution, will be needed to establish the actual 
%potential of the process $H(\to b\bar b)\gamma\,jj$ in this field.

\section[The $Z$ Plus Multi-Jet Background From
the Double Parton Interaction in the Vector Boson Fusion 
$H\rightarrow \tau^+\tau^-$ Search]
{THE $Z$ PLUS MULTI-JET BACKGROUND FROM THE DOUBLE PARTON INTERACTIONS
IN THE VECTOR BOSON FUSION $H\rightarrow \tau^+\tau^-$ SEARCH~\protect\footnote{Contributed by: 
A.~Nikitenko}}
%\documentclass[11pt]{cernrep}
%\usepackage{graphicx,epsfig}
%\bibliographystyle{lesHouches}
%\begin{document}

%\title{The $Z$ plus multi-jet background from the double parton interactions in the VBF $H\rightarrow \tau \tau$ searches.}

%\author{A.~Nikitenko$^{1,~a}$}
%\institute{$^1$Imperial College, London
%\\$^a$on leave from ITEP, Moscow}

%\maketitle

The $Z$+jets production is the dominant background obtained in the 
VBF $H\rightarrow \tau \tau$ searches at the LHC 
~\cite{Plehn:1999xi,Rainwater:1998kj,Asai:2004ws,Takahashi:qqh}. 
We estimated an additional $Z$+jets background originated from double parton interactions (DPI) 
in a proton-proton collision when the $Z$ boson is produced in one parton-parton interaction 
and the QCD di-jets are produced in the second parton-parton interaction. In that case 
the two choices of the tagging jets are possible: (a) one tagging jet is taken from the QCD di-jet 
production and the second one is taken from the Drell-Yan production and (b) two tagging jets are 
both selected from the QCD di-jet production.

The contribution from the double-parton interaction was estimated with PYTHIA6.4 ~\cite{Sjostrand:2006za} at 
the particle level ~\footnote{The recipe was kindly provided by T.~Sjostrand. The possibility to generate
two hard processes in the DPI is realized recently in PYTHIA8.}. At the first step the 
Drell-Yan and the QCD di-jet events were generated separately in PYTHIA. The Drell-Yan production
was generated with the full underlying event (UE) using Tune DWT~\cite{Acosta:2006bp}, while in the QCD 
di-jet production the UE was switched off (MSTP(81)=0). The Drell-Yan events were generated with
the di-lepton mass $m_{\ell \ell}>$ 70 GeV/$c^{2}$ and the QCD di-jet events were generated with 
$\hat{p_{T}}>$20 GeV/$c$.  The NLO cross section 2$\times$10$^{6}$ fb for the Drell-Yan 
production and the PYTHIA cross section 8.2$\times$10$^{11}$ fb for QCD di-jet production were 
used in the estimates presented. At the second step two events (Drell-Yan and QCD di-jets) were mixed together 
and analyzed as one event. Jets were found at the particle level by the simple cone algorithm 
(cone size 0.5) implemented in the PYTHIA PYCELL routine. 

The cross section for the double parton interactions was evaluated with the factorization formula
\begin{equation}
 \sigma _{A,B}~=~\frac{m}{2}~\frac{\sigma _{A} \times \sigma _{B}}{\sigma _{eff}},
\end{equation}
where $m$=1, for indistinguishable parton processes and $m$=2 for distinguishable parton processes
(in our case we use $m$=2). In the experimental study of double parton collisions CDF quotes
$\sigma _{eff}$=14.5 $mb$~\cite{Abe:1997xk}. For LHC energy we use currently the value $\sigma _{eff}$=20 $mb$ 
~\footnote{Private communication with T. Sjostrand.}. It gives the $\sigma _{A,D}$=8.2$\times$10$^{4}$ fb ($A$=Drell-Yan,
$B$=QCD di-jets). More pessimistic value of 12 $mb$ ~\footnote{Private communication with D.~Treleani.} 
will double our estimates of the $Z$+jets background from the double parton interactions. The longitudinal correlations in 
the double-parton structure functions neglected in the above formula can have a sizable effect at the LHC 
~\cite{Korotkikh:2004bz,Cattaruzza:2005nu}. 

We compare the $Z$+jets background from the double parton collisions with the "normal" QCD $Z$+jets background
from  one parton-parton collision. It was generated using the ALPGEN~\cite{Mangano:2002ea} generator 
with the MLM prescription for jet-parton matching \cite{Mangano:2006rw, Hoche:2006ph} at the PYTHIA6.4 shower 
simulation. We generated $\ell \ell$+2jets exclusive and $\ell \ell$+3 jets inclusive
samples with the ALPGEN settings: $m_{\ell \ell}>$ 70 GeV/$c^{2}$, $p_{T}^{j}>$20 GeV/$c$, 
$| \eta _{j}|<$ 5, $\Delta R_{jj}>$ 0.5. The user "soft" VBF pre-selections in ALPGEN generation
were: $\Delta \eta _{j1,j2}>$4, $\eta _{j1} \times \eta _{j2}<$0, $M_{j1j2}>$600 GeV/$c^{2}$,
where j1 and j2 are two leading $p_{T}$ partons ordered in $p_{T}$. The parameters for MLM jet-parton 
matching were: $E_{T}^{clus}$=20 GeV, $R^{clus}$=0.5 and $\eta ^{cl~max}=$5.0. 

The VBF selections similar to that were used in a full simulation analysis \cite{Takahashi:qqh} 
(except cut on $E_{T}$ of the tagging jets, which is lower here) were applied to 
the PYTHIA particle level jets. An event must have at least two leading $E_{T}$ jets 
reconstructed with a cone algorithm (cone size 0.5) that satisfy the following requirements:
\begin{itemize}
\label{selection1}
\item $E_{T}^{j}>20$ GeV
\item $\eta^{j}~<~5.0$
\item $M_{j1j2}>1000$ GeV
\item $|\Delta\eta^{j1j2}|>4.2$
\item $\eta^{j1} \times \eta^{j2}<0$.
\end{itemize}
where j1 and j2 are two leading $E_{T}$ jets ordered in $E_{T}$. The double parton scattering events
where the two leading jets were both originated from the Drell-Yan production (the fraction of
such events is $\simeq$ 20\%) were excluded from the consideration to avoid the double counting with 
"normal" QCD $Z$+jets background. 

\begin{figure}[htb!]
\begin{center}
\includegraphics[width=.4\textwidth]{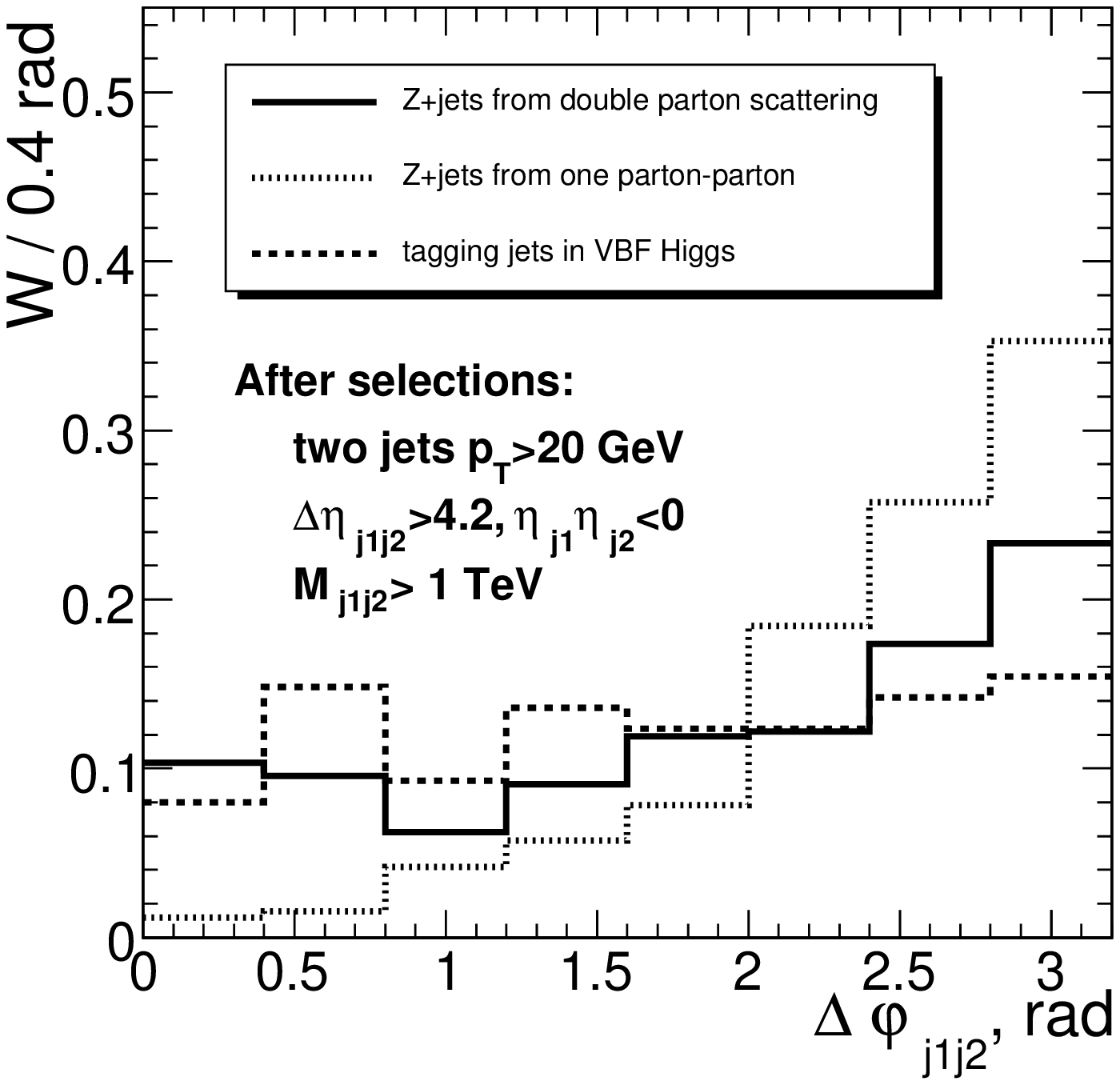}
\includegraphics[width=.4\textwidth]{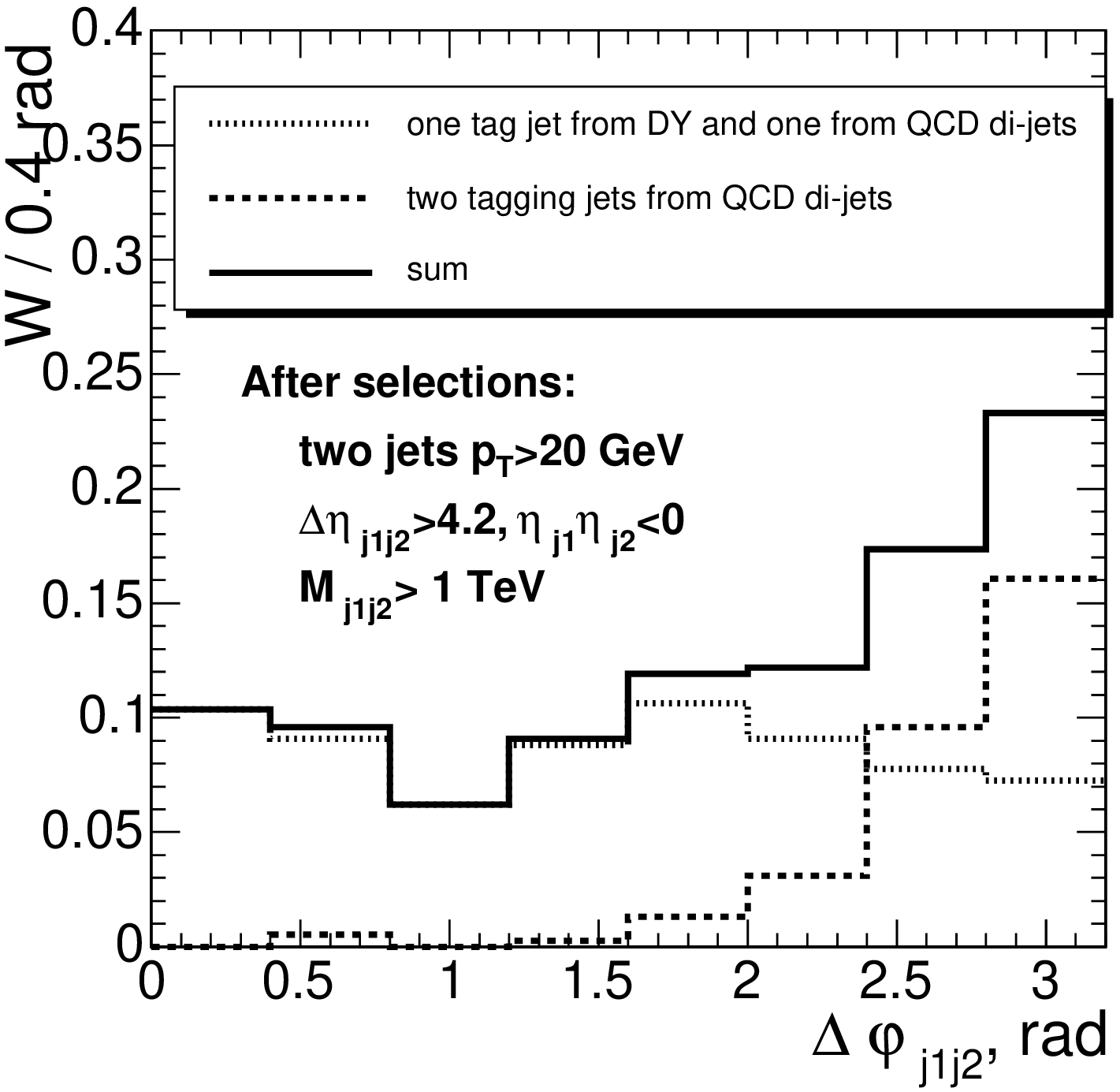}
\caption{Left plot: the angle in the transverse plane between two tagging jet for $Z$+jets 
         backgrounds and the signal $VV \rightarrow H$. 
         Right plot: the angle in the transverse plane between two tagging jet for $Z$+jets
         background from the DPI for the cases (a), (b) and total (see the text).
         All distributions are normalized to unit.}
\label{fig:dphiet}
\end{center}
\end{figure}

Table ~\ref{table:sigmas} shows the initial cross sections (in fb) for the $Z$+jets background
from one and two parton-parton interactions and cross sections after the VBF cuts.
\begin{table}[thb!]
\caption{The initial cross sections (in fb) for the $Z$+jets background from one and two parton-parton 
         interactions and cross sections after the VBF cuts.}
\label{table:sigmas}
\begin{center}
\begin{tabular}[t]{|c|c|c|c|c|}
\hline
interaction & \multicolumn{2}{|c|}{one parton-parton} & \multicolumn{2}{|c|}{two parton-parton}  \\
\hline
process & exclusive $\ell \ell$+2j & inclusive $\ell \ell$+3j & Drell-Yan & QCD di-jets \\
\hline
no cuts & 1.0$\times$10$^{3}$ & 2.0$\times$10$^{3}$ & \multicolumn{2}{|c|}{8.2$\times$10$^{4}$} \\
\hline
$\geq$ 2 jets, $E_{T}^{j}>20$ GeV  &               &                     & \multicolumn{2}{|c|}{4.0$\times$10$^{4}$} \\
$\Delta \eta _{j1,j2}>$4.2, $\eta _{j1} \times \eta _{j2}<$0 & 
          2.4$\times$10$^{2}$ & 5.3$\times$10$^{2}$ & \multicolumn{2}{|c|}{5.0$\times$10$^{3}$} \\
$M_{j1j2}>$1000 GeV/$c^{2}$ & &                     & \multicolumn{2}{|c|}{3.2$\times$10$^{2}$} \\
\hline
\hline
\end{tabular}
\end{center}
\end{table}
After selections the contribution from the double parton interactions is $\simeq$ 40\% (320 fb) 
of the "normal" $Z$+jets background from the one parton-parton interactions (770 fb). 
Fig.~\ref{fig:dphiet} (left plot) shows an angle in the transverse plane between
two tagging jets ($\Delta \varphi _{j1j2}$) for the $Z$+jets backgrounds and the signal 
$VV \rightarrow H$. All distributions are normalized to unit. 
We have obtained that the fraction of the DPI events when the one tagging jet is selected 
from the Drell-Yan and another from the QCD di-jet production (case (a)) is $\simeq$ 70\%; 
in the rest 30\% of the DPI events the both two tagging jets are selected from the QCD di-jet 
production (case (b)). Fig.~\ref{fig:dphiet} (right plot) shows the $\Delta \varphi _{j1j2}$ 
distributions for the cases (a) and (b) separately as well as their sum (the same curve
as in Fig.~\ref{fig:dphiet} (left plot)). One can see that in the case (a), as expected
there is no any correlations between two tagging jets, while in the case (b) they are
forming the back-to-back configuration.

\begin{figure}[htb!]
\begin{center}
\includegraphics[width=.4\textwidth]{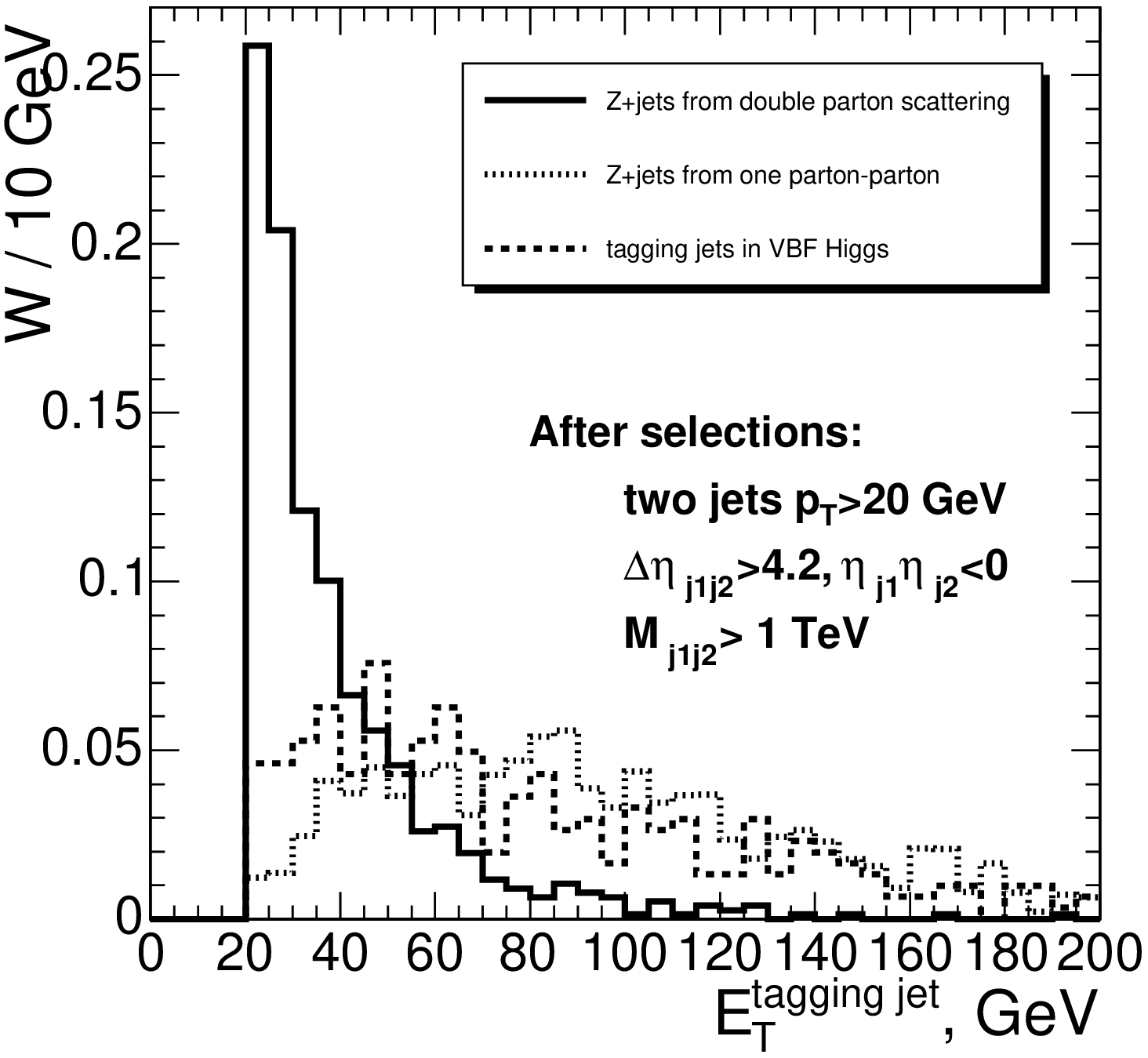}
\includegraphics[width=.4\textwidth]{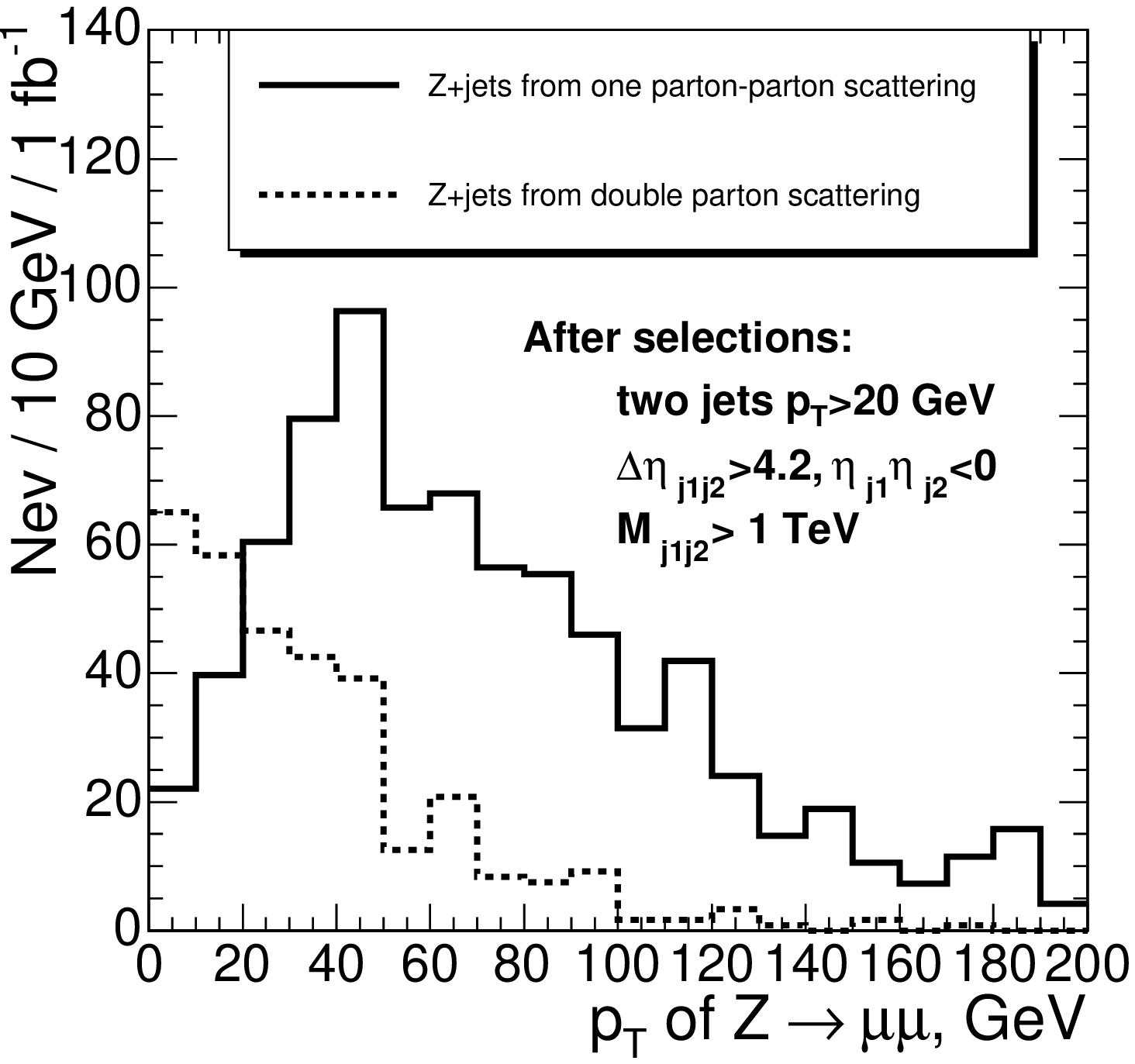}
\caption{Left plot: the transverse energy of the tagging jets from the $Z$+jets backgrounds and the 
         signal $VV \rightarrow H$; distributions are normalized to unit.
         Right plot: the distributions of the $Z$ boson transverse momentum $p_{T}^{Z}$ from
         the DPI and the "normal" $Z$+jets production are normalized on the expected 
         cross sections after the VBF cuts.}
\label{fig:ptz}
\end{center}
\end{figure}

Fig.~\ref{fig:ptz} (left plot) shows the transverse energy of the tagging jets 
from the $Z$+jets backgrounds and the signal $VV \rightarrow H$. One can see that 
the $Z$+jets background from the DPI can be largely suppressed with the cut on the tagging 
jet energy $E_{T}^{j}>$40 GeV. This cut was used in the full simulation analysis ~\cite{Takahashi:qqh}.
After applying this selection the cross section of the $Z$+jets from the DPI is $\simeq$ 100 fb
and the cross section of the "normal" $Z$+jets background is $\simeq$700 fb, thus the
relative contribution from the DPI is reduced to $\simeq$ 15 \%. The further reduction of the 
relative $Z$+jets contribution from the DPI is expected when the cuts on the momentum of the lepton 
(from $\tau \rightarrow \ell \nu \nu$ decay) and the hadronic $\tau$ jet 
(from $\tau \rightarrow \rm hadrons~\nu$ decay) will be applied. It is due to
the momentum of the $Z$ boson from the DPI is softer that the one from the "normal"
$Z$+jets production. It is shown in Fig.~\ref{fig:ptz} (right plot) where the distributions 
of $p_{T}^{Z}$ from DPI and the "normal" $Z$+jets production are normalized on the expected 
cross sections after the VBF cuts.

It is important to control and measure the $Z$+jets background from the double parton
interactions. The possibility of the usage of the $Z$+2jets, Z$\rightarrow \mu \mu$
events with the VBF jet selections and looking at the unbalance in $\overrightarrow{p}_{T}$ 
between the $Z$ boson and the jets is under investigation. 

\subsection{Conclusion}

The $Z$+jets background from the double parton interaction was estimated at the particle level 
to be less than 15\% of the "normal" QCD $Z$+jets background in the VBF $H \rightarrow \tau \tau$ 
searches at LHC after the experimental like event selections and assuming the $\sigma _{eff}$=20 $mb$
in the factorization formula. The fraction of the DPI events when the one tagging jet is selected 
from the Drell-Yan and another from the QCD di-jet production is $\simeq$ 70\% while in the 
rest 30\% of the DPI events the both two tagging jets are selected from the QCD di-jet 
production.

\section*{Acknowledgements}
I would like to thank T.~Sjostrand, P. Scands, D.~Zeppenfeld and A. Snigirev for 
very useful discussions and organizers of Les Houches Workshop 2007 for 
the warm hospitality and the friendly and stimulating atmosphere.

%\bibliography{dpszjets}
%
%\end{document}

\part[MSSM HIGGS BOSONS]{MSSM HIGGS BOSONS}
\section[SUSY--QCD Corrections to Squark Loops in Gluon Fusion to Higgs
Bosons]{SUSY--QCD CORRECTIONS TO SQUARK LOOPS IN GLUON FUSION TO
HIGGS BOSONS~\protect \footnote{Contributed by: M.~M\"uhlleitner and M.~Spira}}
\label{sec:sqcd}
%\documentclass[11pt]{cernrep}
%\usepackage{graphicx,epsfig}
%\bibliographystyle{lesHouches}
%\begin{document}

%\renewcommand{\textfraction}{0.01}

%\title{SUSY--QCD corrections to squark loops in gluon fusion to Higgs bosons}

%\author{M.~M\"uhlleitner and M.~Spira}
%\institute{
%$^1$CERN, Theory Division, CH-1211 Geneva 23, Switzerland\\
%$^2$LAPTH, 9 Chemin de Bellevue, B.P. 110,
%Annecy-le-Vieux 74951, France\\
%$^3$PSI, CH-5232 Villigen PSI, Switzerland}

%\maketitle

%\begin{abstract}
%In a large range of the minimal supersymmetric extension of the Standard 
%Model (MSSM), the Higgs bosons are dominantly produced 
%via the loop-induced gluon fusion processes, $gg \to h,H,A$ at 
%the Tevatron and LHC. Besides the top and bottom quark loops squark
%loops become important for squark masses below 
%$\sim 400$~GeV. We determine the QCD corrections to squark loops including 
%the full squark and Higgs mass dependences. The corrections are large and
%important for the Tevatron and LHC Higgs searches. The mass effects
%on the $K$ factor turn out to be $\sim 20$\%. We also derive the
%QCD corrections to the squark loops in the photonic Higgs couplings,
%which are important for the Higgs search at the LHC and a photon
%collider.
%\end{abstract}

\subsection{Introduction}
In the MSSM 2 Higgs doublets are introduced to generate masses of up and
down type quarks. After electroweak symmetry breaking this leads to five
physical Higgs particles, two light CP-even, $h,H$, one CP-odd, $A$, and
two charged $H^\pm$. At tree level the MSSM Higgs sector can be
described by 2 independent parameters, usually chosen as the
pseudoscalar mass $M_A$,
and the ratio of the 2 vacuum expectation values $\mathrm{tg}\beta =
v_2/v_1$. The MSSM Higgs couplings to quarks are modified such that the
couplings to down(up)-type quarks rise(decrease) with
$\mathrm{tg}\beta$.  The main neutral Higgs production at the Tevatron
and LHC proceeds via $gg$ fusion. The next-to-leading order (NLO) QCD
corrections to this process have been known for a long time
\cite{Spira:1993bb,Spira:1995rr} including the full quark mass
dependence. They turn out to be important, increasing the cross section
by up to 100\%. Next-to-next-to leading order (NNLO) corrections,
calculated in the large quark mass limit only
\cite{Harlander:2002wh,Harlander:2002vv,Anastasiou:2002yz,Anastasiou:2002wq,Ravindran:2003um},
add another 20-30\% and next-to-next-to-next-to leading order (NNNLO)
corrections have been estimated
\cite{Moch:2005ky,Ravindran:2005vv,Ravindran:2006cg}, indicating
improved perturbative convergence. NLO corrections to squark loops have
been known so far only in the heavy squark mass limit
\cite{Dawson:1996xz}, and the full SUSY-QCD corrections have been
obtained for heavy SUSY masses
\cite{Harlander:2003bb,Harlander:2003kf,Harlander:2004tp,Harlander:2005if}.
As a first step to a full SUSY-QCD result we present the QCD corrections
to squark loops including the full squark and Higgs mass dependences
\cite{Muhlleitner:2006wx}.

%\section{NLO QCD CORRECTIONS}
\subsection{NLO QCD Corrections}
For our calculation of the pure QCD corrections to squark loops we need a
modified MSSM Lagrangian which separates the gluon and gluino contributions in
a renormalizeable way. In this work we do not take into account the 
self-interaction among squarks, and the required Lagrangian is then given
by
\begin{eqnarray}
{\cal L} & = & -\frac{1}{4} G^{a\mu\nu} G^a_{\mu\nu} -\frac{1}{4}
F^{\mu\nu} F_{\mu\nu}
+ \frac{1}{2}\left[ (\partial_\mu {\cal H})^2 -
M_{\cal H}^2 {\cal H}^2 \right] \\ & + & \sum_Q \left[ \bar Q(i\!\! \not
\!\! D - m_Q) Q - g_Q^{\cal H} \frac{m_Q}{v} \bar QQ {\cal H} \right] +
\sum_{\tilde Q} \left[ |D_\mu \tilde Q|^2 - m_{\tilde Q}^2 |\tilde Q|^2
- g_{\tilde Q}^{\cal H} \frac{m_{\tilde Q}^2}{v} |\tilde Q|^2 {\cal H}
\right]
\nonumber
\end{eqnarray}
with the covariant derivative $D_\mu = \partial_\mu + ig_s G^a_\mu T^a +
ie A_\mu {\cal Q}$.  Here $G^a_{\mu\nu}$ denotes the gluon field
strength tensor and $G^a_\mu$ the gluon field accompanied by the color
$SU(3)$ generators $T^a$ $(a=1,\ldots,8)$, while $F_{\mu\nu}$ is the
photon field strength tensor and $A_\mu$ the photon field associated by
the electric charge operator ${\cal Q}$. The Higgs field ${\cal H}$
represents generically either the light scalar $h$ or the heavy scalar
$H$ Higgs boson of the MSSM\footnote{Since there are no squark loop
contributions to the pseudoscalar Higgs boson couplings to photons and
gluons at leading order (LO), in this paper we will only deal with the
scalar Higgs
bosons $h,H$.}. Since we do not take into
account gluino exchange contributions, the coefficients $g_Q^{\cal H}$ and 
$g^{\cal H}_{\tilde Q}$ are not renormalized, thus leading to a renormalizeable
model with strongly interacting scalars $\tilde Q$.
Gluino corrections are expected to be small 
\cite{Harlander:2003bb,Harlander:2003kf,Harlander:2004tp,Harlander:2005if}.

For our numerical results we choose the gluophobic Higgs scenario 
\cite{Carena:2002qg} which maximizes the destructive interference effects 
between top and stop loops in the light Higgs coupling to gluons. It is 
defined by the MSSM parameters [$m_t=174.3$~GeV] $M_{SUSY}= 350$~GeV,
$\mu=M_2=300$~GeV, $X_t=A_t - \mu/\mathrm{tg}\beta =-770$~GeV, $A_b=A_t$
and $m_{\tilde g} = 500$~GeV.
%\begin{eqnarray}
%\begin{array}{rclrcl}
%M_{SUSY} & = & 350~{\rm GeV} &
%\mu=M_2 & = & 300~{\rm GeV} \\
%X_t & = & -770~{\rm GeV} &
%m_{\tilde g} & = & 500~{\rm GeV} \, ,
%\end{array} 
%\end{eqnarray}
%where $X_t = A_t - \mu/\mathrm{tg}\beta$ and $A_b=A_t$,
The squark masses are given by
\begin{eqnarray}
\begin{array}{rrclrrcl}
\mathrm{tg}\beta=3: &       m_{\tilde t_1} & = & 156~{\rm GeV}  \qquad
\qquad &
\mathrm{tg}\beta=30: &      m_{\tilde t_1} & = & 155~{\rm GeV}  \\
        &       m_{\tilde t_2} & = & 517~{\rm GeV}  &
         &      m_{\tilde t_2} & = & 516~{\rm GeV}  \\
        &       m_{\tilde b_1} & = & 346~{\rm GeV}  &
         &      m_{\tilde b_1} & = & 314~{\rm GeV}  \\
        &       m_{\tilde b_2} & = & 358~{\rm GeV}  &
         &      m_{\tilde b_2} & = & 388~{\rm GeV} \, .
\end{array}
\end{eqnarray}
The results of this work look similar in other scenarios, whenever the
squark masses are of the order of the top mass, or the Higgs mass
reaches values beyond the corresponding squark-antisquark threshold.

\subsubsection{Scalar Higgs couplings to photons}
The leading order photonic Higgs couplings are mediated by top, bottom
and $W$ boson loops, with significant contributions from squark loops
for stop and sbottom masses below $\sim 400$~GeV
\cite{Spira:1993bb,Spira:1995rr,Spira:1997dg,Djouadi:2005gi,Djouadi:2005gj,Ellis:1975ap,Shifman:1979eb}.
The reverse processes $\gamma\gamma \to h,H$ play an important role for
the MSSM Higgs boson searches at a photon collider
\cite{Badelek:2001xb,Muhlleitner:2005pr,Muhlleitner:2001kw,Muhlleitner:2000jj,Asner:2001ia,Niezurawski:2005cr,Spira:2006aa}.
The two-loop diagrams of the QCD corrections to squark loops lead to
5-dimensional Feynman parameter integrals. We have reduced these
integrals in one calculation to 1-dimensional integrals which have been
integrated numerically. A second calculation has solved the integrals
purely numerically. The two calculations agree within integration
errors.  In order to improve the perturbative behaviour of the squark
loops they have been expressed in terms of the running squark masses
$m_{\tilde Q}(M_{\cal H}/2)$, which are related to the pole masses
$M_{\tilde Q}$ via $m_{\tilde Q}(\mu) = M_{\tilde Q} (
\alpha_s(\mu)/\alpha_s(M_{\tilde Q}) )^{6/\beta_0}$ where $\beta_0 =
33-2N_F$ with $N_F=5$ light flavors. Their scale is identified with
$\mu=M_{\cal H}/2$ within the photonic decay mode thus insuring a proper
definition of the $\tilde Q \bar{\tilde Q}$ thresholds $M_{\cal H} = 2
M_{\tilde Q}$. The LO scale dependence of the squark masses due to light
particle contributions has been taken into account.

\begin{figure}[hbt]
\begin{picture}(200,110)(0,0)
\put(-8,-85){\includegraphics{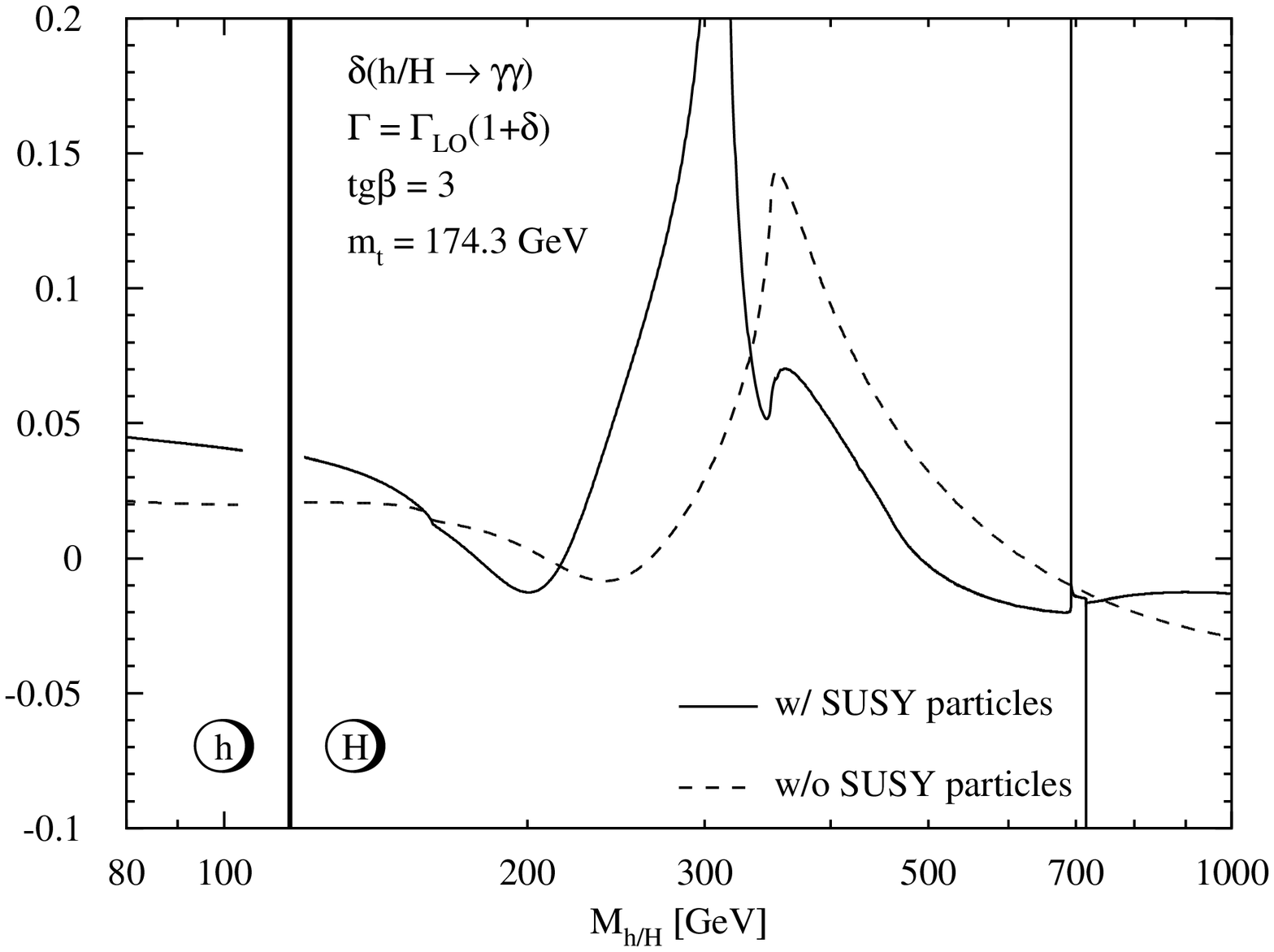}}
\put(227,-85.0){\includegraphics{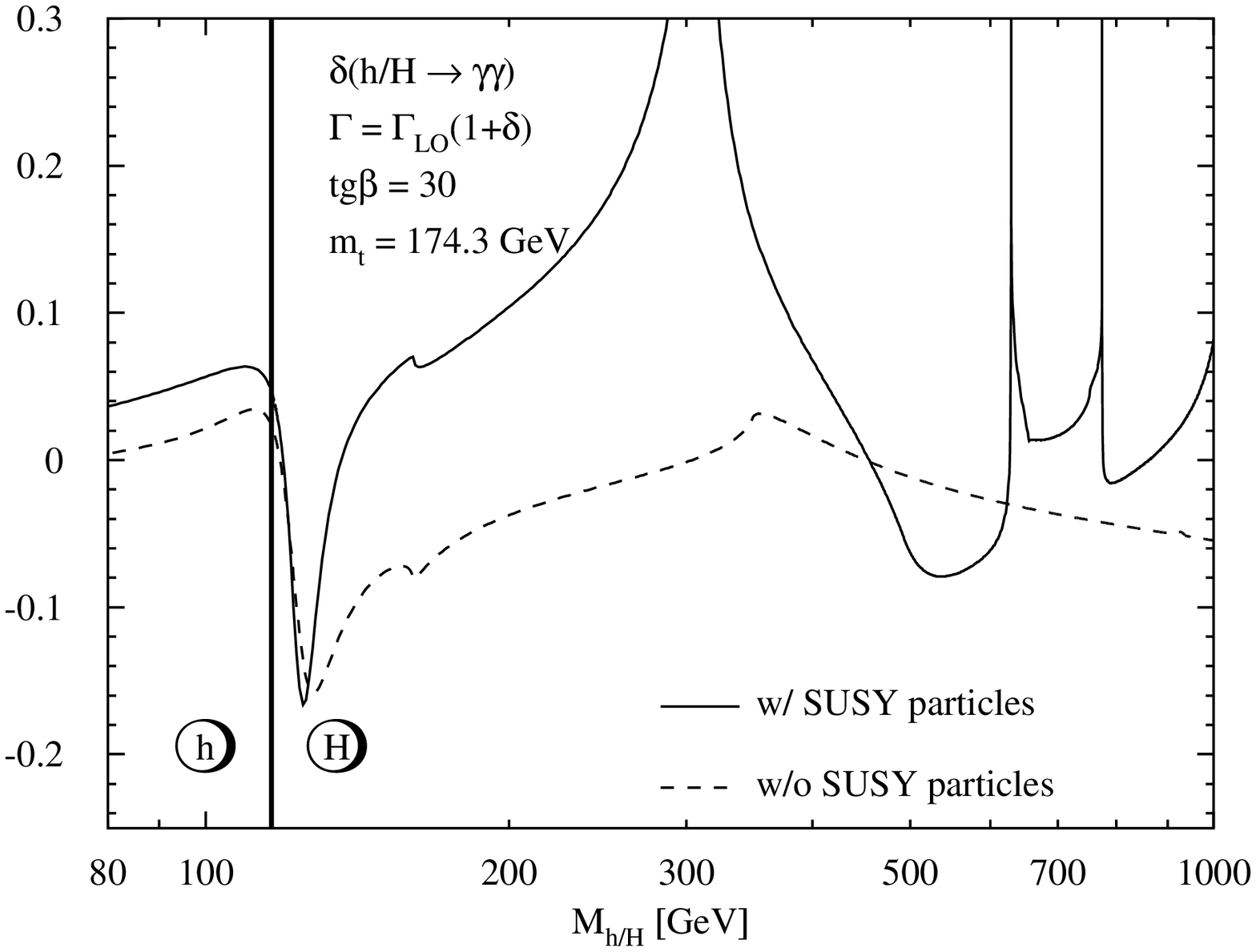}}
\end{picture}
\caption[]{\label{fg:dhgaga} \it Relative QCD corrections to the
scalar MSSM Higgs decay widths to two photons
for $\mathrm{tg}\beta=3$ and 30. The full curves
include all loop contributions while in the dashed lines the SUSY
contributions are omitted. The kinks and spikes correspond to the $WW,
\tilde t_1\bar{\tilde t}_1, t\bar t, \tilde b_1\bar{\tilde b}_1, \tilde
\tau_1 \bar{\tilde \tau}_1, \tilde \tau_2 \bar{\tilde\tau}_2$ and
$\tilde b_2\bar{\tilde b}_2$ thresholds in consecutive order with rising
Higgs mass. %The renormalization scale of the running quark and squark
%masses is chosen to be $M_{h/H}/2$, while the scale of $\alpha_s$ is
%taken to be the corresponding Higgs mass.
}
\end{figure}
Fig.\ref{fg:dhgaga} shows the relative QCD corrections to the photonic 
Higgs decay widths for the two cases, in which SUSY
particles have been taken into account or not. The spikes which appear at 
the $\tilde Q \bar{\tilde Q}$ thresholds are due to singularities originating 
from Coulomb singularities at the threshold since $\tilde Q \bar{\tilde Q}$ 
pairs can form $0^{++}$ states. This behaviour can be derived quantitatively
from the Sommerfeld rescattering corrections, and we checked explicitly 
that this agrees with our numerical results. As can be inferred from 
Fig.\ref{fg:dhgaga} the QCD corrections reach
a size of 10--20\% for moderate and large Higgs masses apart from the
threshold regions, where the perturbative results are unreliable due to
the Coulomb singularities. At a $\gamma\gamma$ collider the photon
fusion cross section can be measured with few per cent accuracy, and 
therefore these corrections have to be taken into account properly. The size 
of the QCD corrections with and without SUSY particle loops is of the same
order of magnitude, but they can be of opposite sign.

\begin{figure}[hbt]
\begin{picture}(200,110)(0,0)
\put(-8,-85){\includegraphics{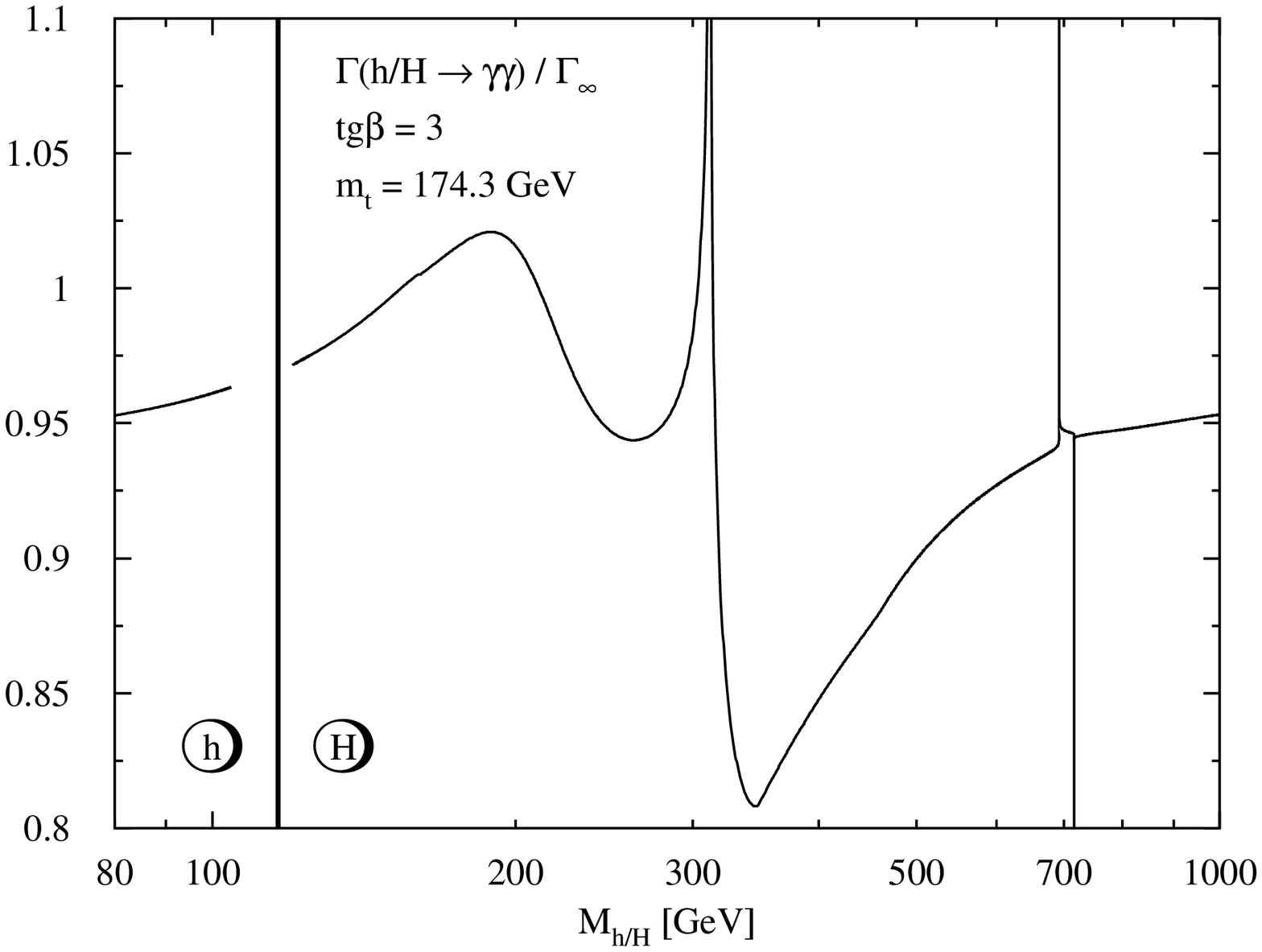}}
\put(227,-85){\includegraphics{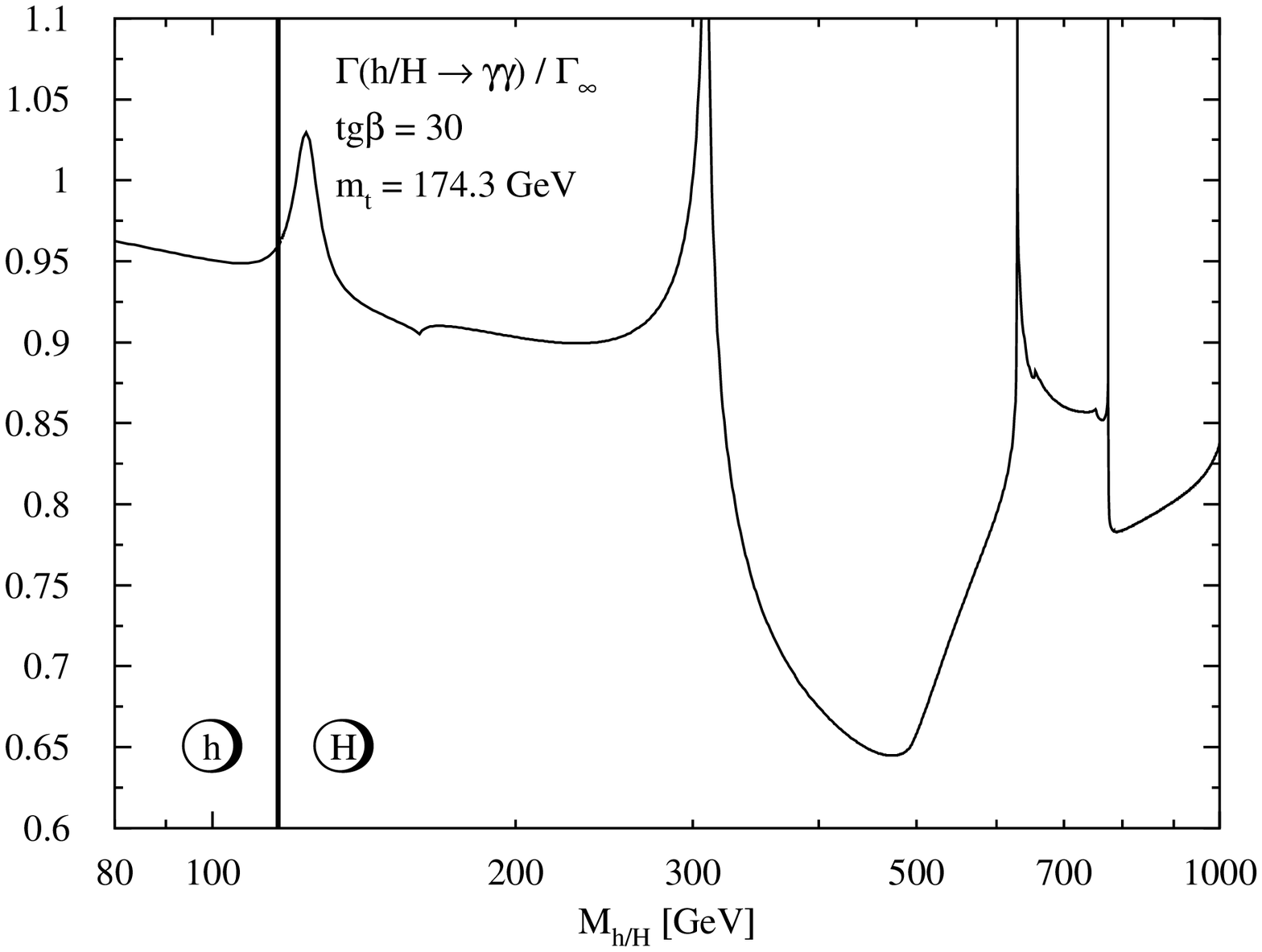}}
\end{picture}
\caption[]{\label{fg:mhgaga} \it Ratio of the QCD corrected partial
decay widths of the scalar MSSM Higgs bosons to two photons including
the full squark mass dependence and those obtained by taking the
relative QCD corrections to the squark loops in the heavy mass limit as
functions of the corresponding Higgs masses for $\mathrm{tg}\beta=3$ and 30.
%The scales have been chosen as in Fig.~\ref{fg:dhgaga}
}
\end{figure}
Fig.\ref{fg:mhgaga}, in which the ratio of the fully massive photonic decay 
width at NLO $\Gamma (h/H\to \gamma\gamma)$ and the NLO width with the
relative QCD corrections in the 
heavy squark mass limit $\Gamma_\infty$ is plotted, quantifies the size
of the squark mass effects beyond the heavy squark mass limit in the 
relative QCD corrections. (The full squark mass dependence in the LO width
has been kept in both expressions.) With a size of up to 
$\sim 30$\% the squark mass effects are larger than the expected
experimental uncertainty in the measurement of the Higgs production in 
$\gamma\gamma$ fusion and hence have to be taken into account in realistic
analyses.

\subsubsection{Gluon Fusion}
The gluon fusion processes $gg\to h,H$ are mediated by quark and squark 
triangle loops with the latter contributing significantly for squark 
masses below $\sim 400$~GeV. The NLO QCD corrections consist of virtual 
two-loop corrections and the real corrections from the radiation processes,
$gg\to g h/H$, $gq \to q h/H$ and $q\bar{q} \to g h/H$. 
The strong coupling constant $\alpha_s$ has been renormalized in the 
$\overline{\rm MS}$ scheme, 
with the top quark and squark contributions decoupled from the scale 
dependence, and the quark and squark masses in the on-shell scheme. 
The parton densities are defined in the $\overline{\rm MS}$
scheme with five active flavors, i.e. the top quark and the squarks are
not included in the factorization scale dependence. 

\begin{figure}[hbt]
\begin{picture}(200,110)(0,0)
\put(-8,-85){\includegraphics{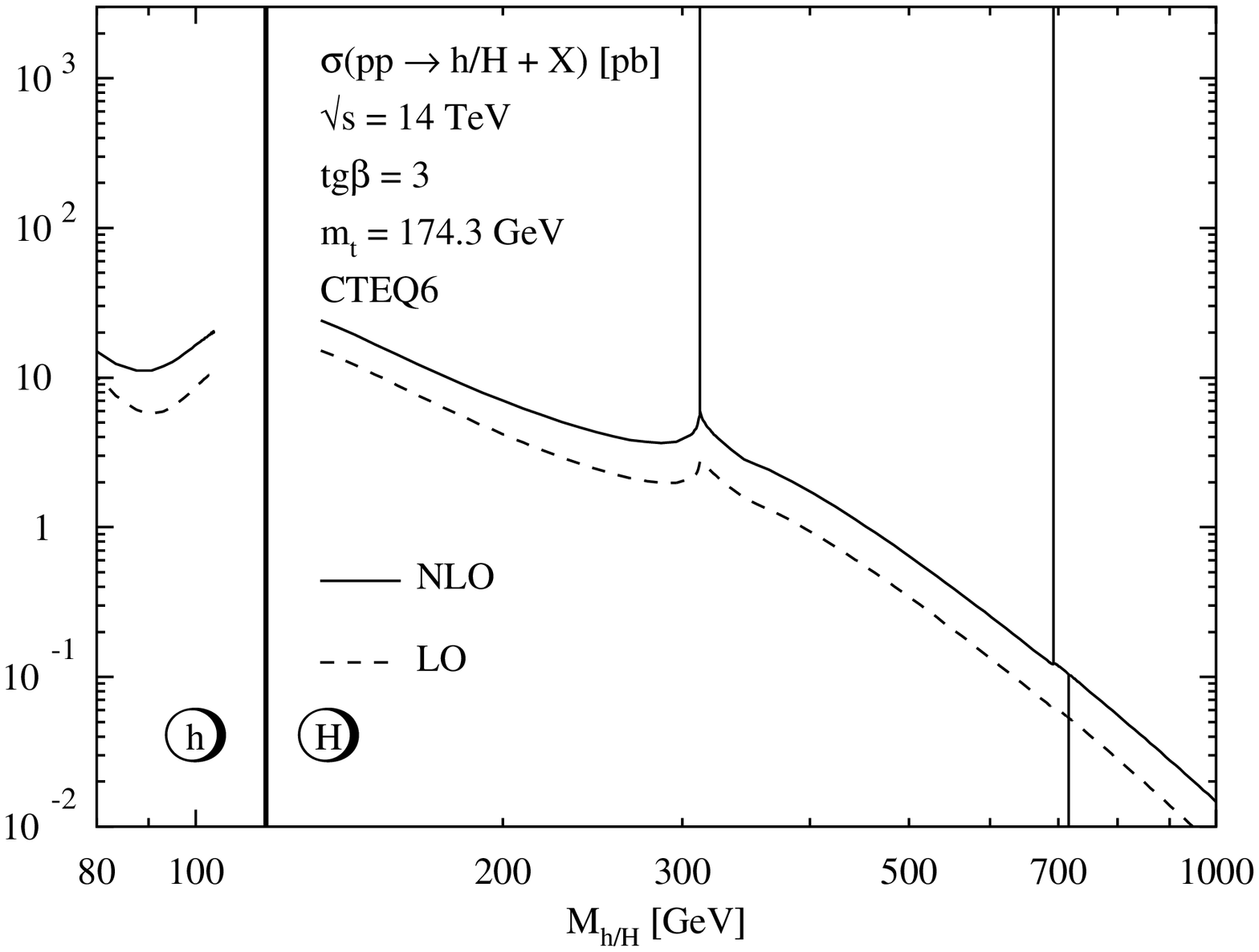}}
\put(227,-85){\includegraphics{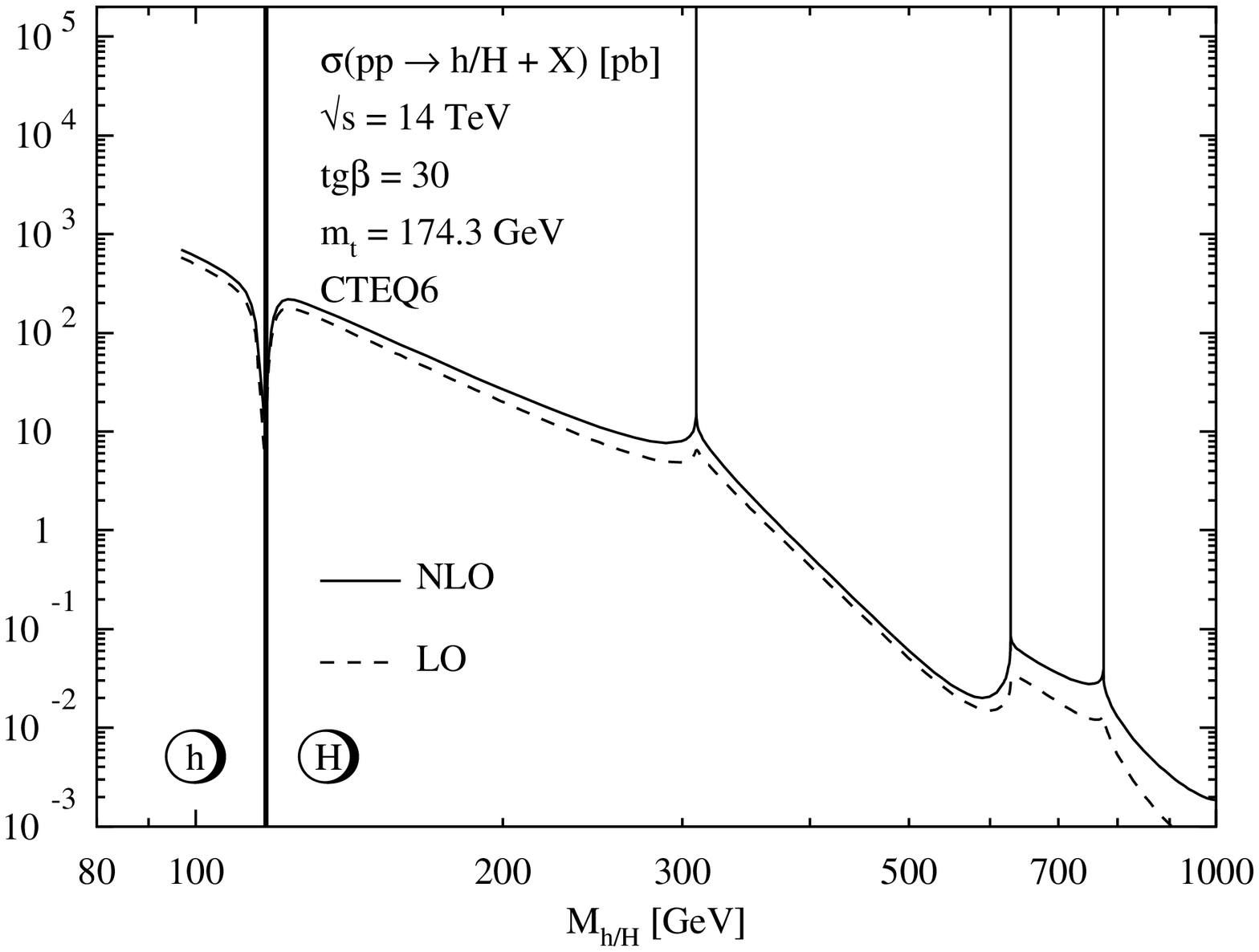}}
\end{picture}
\caption[]{\label{fg:gghnlo} \it Production cross sections of the scalar
MSSM Higgs bosons via gluon fusion as functions of the corresponding
Higgs masses for $\mathrm{tg}\beta=3$ and 30. The full curves include the QCD
corrections, while the dashed lines correspond to the LO
predictions.  The kinks and spikes correspond to the $\tilde
t_1\bar{\tilde t}_1, \tilde b_1\bar{\tilde b}_1$ and $\tilde
b_2\bar{\tilde b}_2$ thresholds in consecutive order with rising Higgs
mass. %The renormalization and factorization scales are chosen as the
%corresponding Higgs mass.
}
\end{figure}
Fig.~\ref{fg:gghnlo} shows the LO and NLO cross sections. The QCD
corrections increase the gluon fusion cross sections by 10--100\% and are
significantly larger in regions of large destructive
interferences between quark and squark loops. The corrections are of very 
similar size for the quark and squark loops individually. In spite of the 
large corrections the scale dependence is reduced from about 50\% 
at LO to $\sim 20\%$ at NLO thus indicating a significant stabilization of 
the theoretical predictions. Based on this and the approximate NNLO and
NNNLO results in the limit of heavy squarks and top quarks the residual theoretical
uncertainties of our NLO results can be estimated to less than about 20\%.
The spikes at the  $\tilde Q \bar{\tilde Q}$ thresholds are 
Coulomb singularities due to the formation of $0^{++}$ states.
\begin{figure}[hbt]
\begin{picture}(200,110)(0,0)
\put(-8,-85){\includegraphics{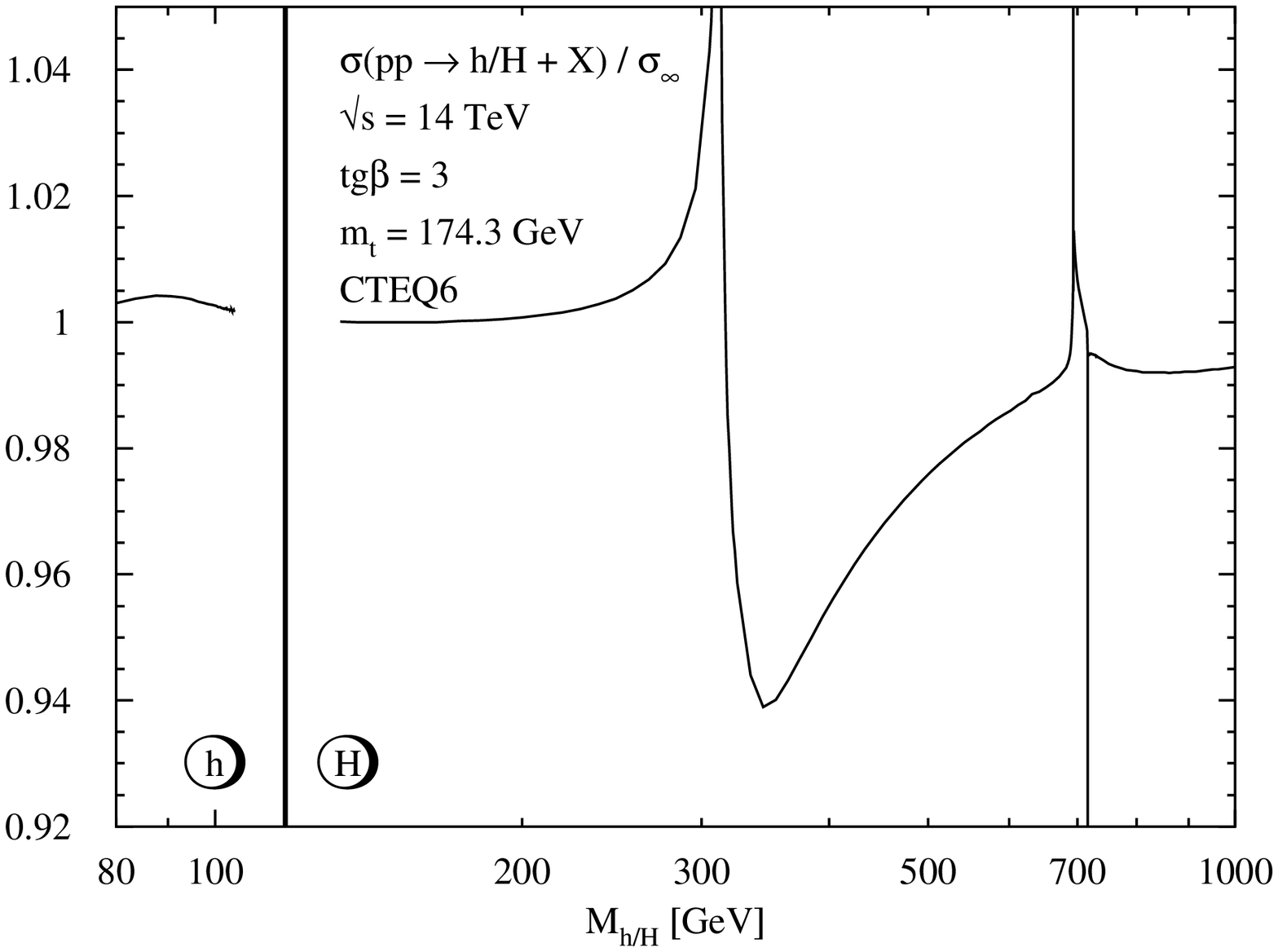}}
\put(227,-85){\includegraphics{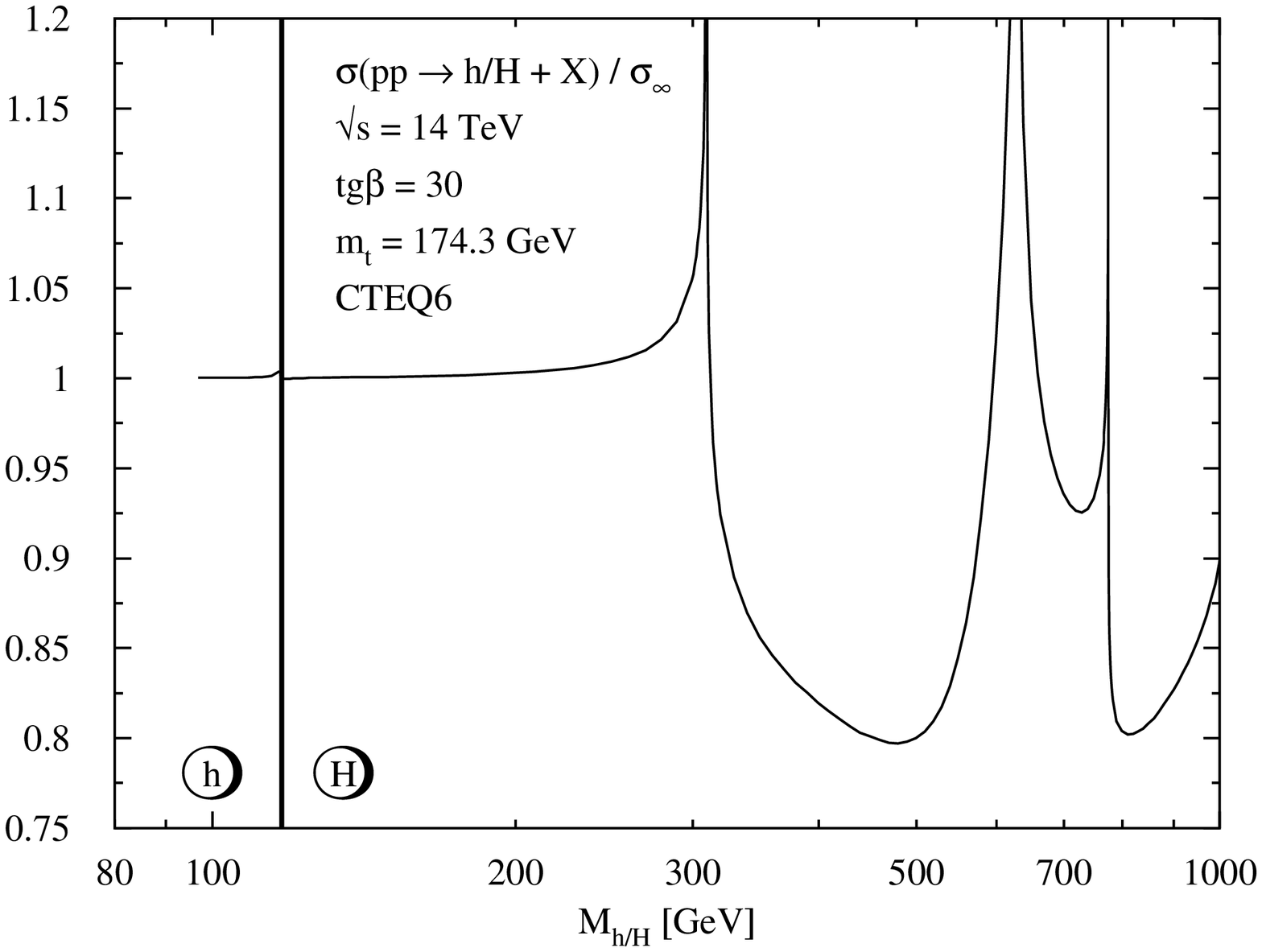}}
\end{picture}
\caption[]{\label{fg:gghmass} \it Ratio of the QCD corrected production
cross sections of the scalar MSSM Higgs bosons via gluon fusion
including the full squark mass dependence and those obtained by taking
the relative QCD corrections to the squark loops in the heavy mass limit
as functions of the corresponding Higgs masses for $\mathrm{tg}\beta=3$ and 
30. %Renormalization and factorization scales as in Fig.\ref{fg:gghnlo}.
}
\end{figure}

Fig.~\ref{fg:gghmass}, which shows the ratios of the NLO cross sections
including the full mass dependence and of the NLO cross sections in the
heavy squark limits, exemplifies the squark mass effects on the $K$
factors.  In addition to the LO squark mass dependence of the cross
section, the $K$ factors develop a squark mass dependence of up to about
20\% and hence support the relevance of our results compared to the
previous results of Ref.~\cite{Dawson:1996xz}. The squark mass effects
on the $K$ factors are larger than the corresponding quark mass effects
\cite{Kramer:1996iq}.  And they are larger than the residual theoretical
uncertainties so that they cannot be neglected in realistic analyses.
Since the gluino contributions are expected to be much smaller, the
squark mass dependence will be the dominant part of the differences
between the heavy mass limits and a full MSSM calculation at NLO.

\subsection{Conclusions}
We have discussed the NLO QCD corrections to the squark loop
contributions to neutral MSSM Higgs boson production in $gg$ fusion
at the LHC and their decay modes into photons, including the full
mass dependences. The corrections are sizeable and stabilize the theorectical
predictions compared to the LO results. Squark mass effects on the
relative QCD corrections are significant and larger than the
mass effects from quark loops. They are always relevant for Higgs masses
beyond the corresponding virtual squark-antisquark threshold or for
squark masses of the order of the top mass.  Since they are larger than
the experimental uncertainties and the approximative results beyond NLO
indicate sufficient perturbative convergence, the results of this work
have to be taken into account for realistic analyses.  
%
%
%\bibliography{sqcdggfus_bibli}
%
%\end{document}

\section[Higgs Boson Production in Association with $b$ Quarks: 
SUSY QCD Contributions]{HIGGS BOSON PRODUCTION IN ASSOCIATION WITH $b$ QUARKS:
SUSY QCD CONTRIBUTIONS~\protect\footnote{Contributed by: S.~Dawson and C.~B.~Jackson}}
\label{sec:bbh}

%\documentclass[11pt]{cernrep}
%\usepackage{graphicx}
%\usepackage{cite}
%\bibliographystyle{lesHouches}
%\begin{document}

%\title{Higgs Boson Production in Association with $b$ Quarks: SUSY QCD Contributions }

%\author{S.~Dawson and C.B~Jackson}
%\institute{Brookhaven National Laboratory, Upton, N.Y. 11973}

%\maketitle

%\begin{abstract}
% The associated production of a Higgs boson with a $b$ quark is a discovery mode for an MSSM Higgs boson at large values of 
%$\tan\beta$.  We investigate the contribution of squark and gluino loops to the $bg\rightarrow b\phi$ production rate.  
%We also consider the 
%associated production of a Higgs boson
%with a $b$ quark plus an additional jet.

%\end{abstract}

\subsection{Introduction}
  In the Standard Model, Higg
production in association with $b$ quarks is never important. 
 However, in the minimal supersymmetric model (MSSM),
 the couplings of the Higgs bosons to  $b$ quarks can be significantly
enhanced for large values of $\tan\beta$
and for a large range of parameter space, Higgs production in association
with $b$ quarks is the most likely discovery channel\cite{Dawson:2005vi,Carena:2007aq,Campbell:2004pu,Dittmaier:2003ej,Carena:1998gk}.  

The production of a Higgs boson in association with a $b$ quark 
has been extensively studied.  
In a $4$- flavor number scheme, the lowest order processes 
for producing a Higgs boson and a $b$ quark are $gg\rightarrow
 b {\overline b}\phi$ and 
$ q {\overline q}\rightarrow b {\overline b} 
\phi$\cite{Dittmaier:2003ej,Dawson:2004sh,Dicus:1998hs}
(The neutral
Higgs bosons of the MSSM
are generically $\phi=h^0,H^0,A^0$).
In a $5$- flavor number scheme, the $b$ quark appears as a parton
and potentially large logarithms of the form $\ln({M_\phi\over m_b})$
are absorbed into $b$ quark parton distribution functions. 
 Although
the $4$- and $5$- flavor number schemes represent different orderings of
perturbation theory, the two schemes have been shown to yield
equivalent numerical results.
In the $5$- flavor number scheme, the lowest order process for producing
a Higgs boson in association with $b$ quarks is $b{\overline b}
\rightarrow \phi$ when no $b$ quarks are tagged in the final state and 
 $b g\rightarrow b \phi$ when a single $b$ quark 
is tagged\cite{Dicus:1998hs,Dawson:2005vi,Campbell:2004pu,Dawson:2004sh,Dittmaier:2003ej}.  
 
In this note, we consider the production process, $bg\rightarrow b\phi$,
for which the NLO QCD corrections
are well understood,\cite{Dawson:2005vi,Campbell:2004pu,Maltoni:2003pn}. 
Here we present the ${\cal O}(\alpha_s^2)$
SUSY QCD (SQCD) corrections from gluino-squark
loops to the $b$- Higgs production cross section\cite{Dawson:2007ur,Brein:2007da}.
We compare the results from an  effective
Lagrangian approach  with those obtained from an exact one-loop
calculation.
Finally, we consider whether the process $bg\rightarrow b\phi $ +jet provides a
useful signature and compare this channel
 with the irreducible background from
$bg\rightarrow b Z$ +jet.

\subsection{Effective Lagrangian}

The MSSM contains two Higgs doublets, $H_u$ and $H_d$.
At tree level,
the $b$ quark couples to only one of the Higgs doublets ($H_d$) and
 there is no $ {\overline \psi}_L b_R H_u$
 coupling (where $\psi_L=(t_L,b_L)$).
The coupling of the $b$ quark to the ``wrong'' Higgs doublet at
one-loop leads to 
the  effective
interaction\cite{Carena:1999py,Hall:1993gn},
\begin{equation}
L_{eff}=-\lambda_b
{\overline \psi}_L\biggl(H_d+{\Delta m_b\over \tan\beta}
H_u\biggr)b_R+h.c. \, .
\label{effdef}
\end{equation}
This effective interaction
shifts the $b$ quark mass from its tree level value,
\begin{equation}
m_b={\lambda_b v_d\over \sqrt{2}} (1+\Delta m_b)\, ,
\end{equation}
and  also implies that the Yukawa couplings of the
Higgs bosons to the $b$ quark are shifted from the
tree level predictions.  The shift of the
Yukawa couplings\cite{Boudjema:2007uh}  can be accounted for using
an effective Lagrangian\cite{Carena:1999py,Guasch:2003cv},
\begin{eqnarray}
L_{eff}&=&-{m_b\over v_{SM}}\biggl({1\over 1+\Delta m_b}\biggr)
\biggl(-{\sin \alpha \over \cos\beta}\biggr)\biggl(1-{\Delta m_b\over \tan\beta
\tan \alpha}\biggr) {\overline b} b h^0\nonumber \\
&-&{m_b\over v_{SM}}\biggl({1\over 1+\Delta m_b}\biggr)
\biggl({\cos \alpha \over \cos\beta}\biggr)\biggl(1+{\Delta m_b \tan\alpha
\over \tan\beta}\biggr) {\overline b} b H^0\nonumber \\
&-&{m_b\over v_{SM}}\biggl({1\over 1+\Delta m_b}\biggr)
\biggl(-\tan\beta\biggr)\biggl(1-{\Delta m_b\over \tan^2\beta}\biggr) 
{\overline b}i\gamma_5 b A^0\nonumber \\
&\equiv & g_{b{\overline b}h} {\overline b} b h^0
+g_{b{\overline b}H} {\overline b} b H^0
+g_{b{\overline b}A} {\overline b}i\gamma_5 b A^0 \, ,
\label{mbdef}
\end{eqnarray}
where $v_{SM}=246$ GeV, $\tan\beta=v_u/v_d$, and $\alpha$ is the mixing
angle which diagonalizes the neutral Higgs boson mass matrix.  
The Lagrangian of Eq. \ref{mbdef} has been shown to
 sum all terms of ${\cal O}(\alpha_s^n\tan^n\beta)$ for large 
$\tan\beta$\cite{Carena:1999py}.

The expression for $\Delta m_b$ is found in the limit
 $m_b << M_\phi, M_Z <<m_{{\tilde b}_1}, m_{{\tilde b}_2}, m_{\tilde g}$,
(where $m_{{\tilde b}_1}, m_{{\tilde b}_2}, m_{\tilde g}$ are the
sbottom and gluino masses) .  The
contribution to $\Delta m_b$ from sbottom/gluino loops 
is\cite{Carena:1994bv,Hall:1993gn} 
\begin{equation}
\Delta m_b={2\alpha_s(\mu_R)\over 3 \pi} m_{\tilde g} \mu 
\tan\beta
I(m_{\tilde {b_1}},
m_{\tilde{ b_2}}, m_{\tilde g})\, ,
\label{db}
\end{equation}
where the function $I(a,b,c)$ is,
\begin{equation}
I(a,b,c)={1\over (a^2-b^2)(b^2-c^2)(a^2-c^2)}\biggl\{a^2b^2\log\biggl({a^2\over b^2}\biggr)
+b^2c^2\log\biggl({b^2\over c^2}\biggr)
+c^2a^2\log\biggl({c^2\over a^2}\biggr)\biggr\}\, .
\end{equation}
$\mu$ is the bilinear Higgs mixing parameter
and $\alpha_s(\mu_R)$ should be evaluated at a typical squark or gluino mass. 
Note that Eq. \ref{db} is valid for arbitrary values of $\tan\beta$.

Eq. \ref{db} is a non-decoupling effect: If  the 
 masses of the squarks and gluino, along with the mixing
parameter $\mu$, become
large for fixed $M_A$, $\Delta m_b$ does not vanish, 
\begin{equation}
\Delta m_b\rightarrow -sign(\mu){\alpha_s\over 3 \pi} \biggl(\tan\beta
+\cot \alpha\biggr)\, .
\label{decouple}
\end{equation}
In the large $M_A$ limit, 
\begin{equation}
\tan\beta +\cot\alpha \rightarrow -{2 M_Z^2\over M_A^2}\tan\beta \cos 2\beta
+{\cal O}\biggl({M_Z^4\over M_A^4}\biggr)\, ,
\label{decma}
\end{equation}
and the decoupling limit of the MSSM is recovered\cite{Haber:2000kq}. 

The effective Lagrangian can be used to approximate 
 the 
squark and gluino contributions to the rate for 
$bg\rightarrow b\phi$\cite{Dawson:2007ur}.
We define an Improved Born Approximation in which the Born amplitude is
normalized by the Yukawa couplings, $g_{b {\overline b}\phi}$, of 
Eq. \ref{mbdef},
\begin{equation} {d{\hat{\sigma}}_{IBA}\over dt}
\equiv {d{\hat{\sigma}}_{Born}\over dt}\biggl({g_{b {\overline b}\phi}
\over
g_{b {\overline b}\phi}^{LO}}\biggr)^2\, .
\label{sigiba}
\end{equation}
The Improved Born Approximation  incorporates the 
effective Lagrangian approximation to the  SQCD effects
on the $b {\overline b} \phi$ Yukawa couplings at low
energy, but does not include the full SQCD calculation.
In particular, the ``Improved Born Approximation'' does not
include contributions from box diagrams including internal squarks and gluinos or 
the full momentum dependence of the SQCD contributions.  

\subsection{Results}
\begin{figure}
\begin{center}
\scalebox{0.4}{\includegraphics{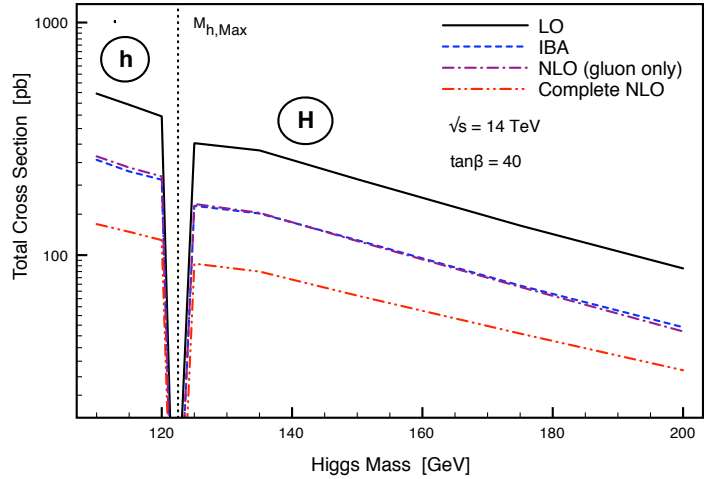}}
\caption{Total cross section for $pp\rightarrow b \phi$ ($\phi=h^0,H^0$)
at the LHC with $p_T(b)>20$ GeV and $\mid \eta_b \mid < 2.5$.
The curve labelled ``Complete NLO'' includes the full set of 
${\cal O}(\alpha_s^2)$ QCD and SQCD contributions, while the curve labelled ``NLO (gluon only)''
omits the SQCD contributions.  The MSSM parameters are
$m_{\tilde{g}}=250$ GeV, $m_{{\tilde b}_1}=250$ GeV, and $m_{{\tilde b}_2}=350$
 GeV.}
\label{final1}
\end{center}
\end{figure}

\begin{figure}
\begin{center}
\scalebox{0.4}{\includegraphics{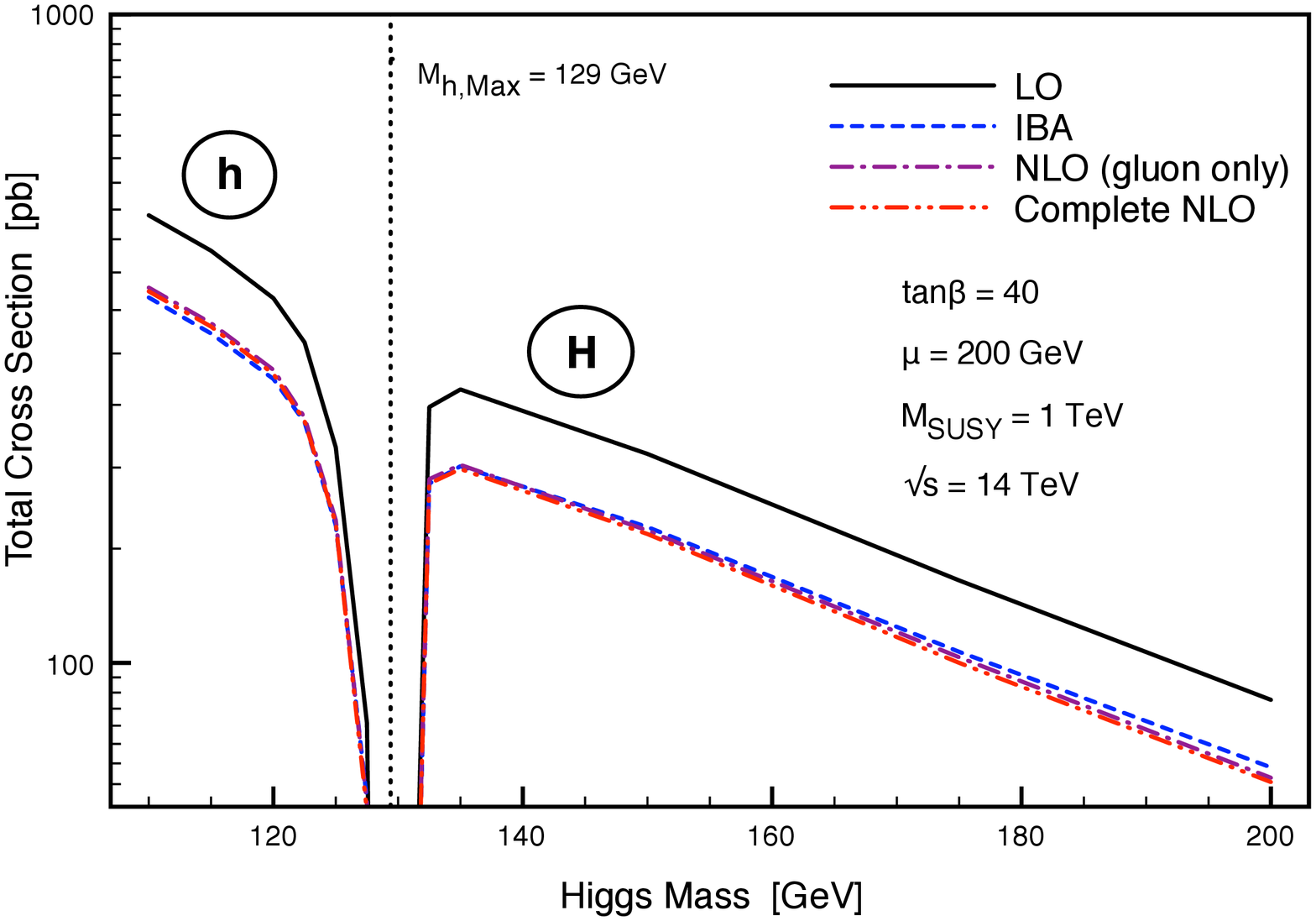}}
\caption{Total cross section for $pp\rightarrow b \phi$ ($\phi=h^0,H^0$)
at the LHC with $p_T(b)>20$ GeV and $\mid \eta_b \mid < 2.5$.
The curve labelled ``Complete NLO'' includes the full set of 
${\cal O}(\alpha_s^2)$ QCD and SQCD contributions, while the curve labelled ``NLO (gluon only)''
omits the SQCD contributions.  The MSSM parameters are
$m_{\tilde{g}}\sim m_{{\tilde b}_1}\sim m_{{\tilde b}_2}\sim 1$
 TeV.}
\label{final2}
\end{center}
\label{high}
\end{figure}

In Figs. \ref{final1} and \ref{final2}
 we compare the results for $bg\rightarrow
bh^0$ and $bg\rightarrow bH^0$ at the LHC\cite{Dawson:2007ur,Brein:2007ej}. 
The curves labelled ``LO''
use CTEQ6L PDFs, $\alpha_s(\mu_R)$ evaluated at $1$-loop, and use
the tree level Yukawa couplings.  The NLO results use CTEQ6M PDFs with
the $2$-loop evolution of $\alpha_s(\mu_R)$
and $\alpha_s^{NLO}(M_Z) = 0.118$.  The
Yukawa couplings of both the IBA
and NLO results are evaluated using the effective Lagrangian of 
Eq. \ref{mbdef}.
The outgoing $b$ quark is required to have $p_T (b)>20$ GeV and $\mid 
\eta_b \mid < 2.5$.  The renormalization and factorization scales are set to
be $M_\phi/4$.  The ``Improved Born Approximation'' (IBA) curves use 
NLO PDFS and the $2-$loop evolution of $\alpha_s(\mu_R)$. 
The $b$ quark mass in the Yukawa couplings is the running
$\overline{MS}$ mass evaluated at $2-$ loops for the NLO and IBA
results and at $1-$ loop for the LO results.
Finally, the MSSM parameters are evaluated using FeynHiggs to 
generate an effective Higgs mixing angle and radiatively corrected Higgs masses. 

From Fig. \ref{final1}, we see that for relatively light squark and
gluino masses, it is important to include the exact SQCD contributions and
that the Improved Born Approximation is a poor approximation to the complete
result. In this case, the SQCD contributions significantly reduce the rate.
On the other hand, for squark and gluino masses on the TeV scale, Fig.
\ref{final2} demonstrates that the effective Lagrangian approach to 
including the SQCD corrections is extremely accurate.  Both Figs.
\ref{final1} and \ref{final2} assume $\tan\beta=40$.  For small 
values of $\tan\beta$, the SQCD corrections are insignificant.

\begin{figure}
\begin{center}
\scalebox{0.4}{\includegraphics{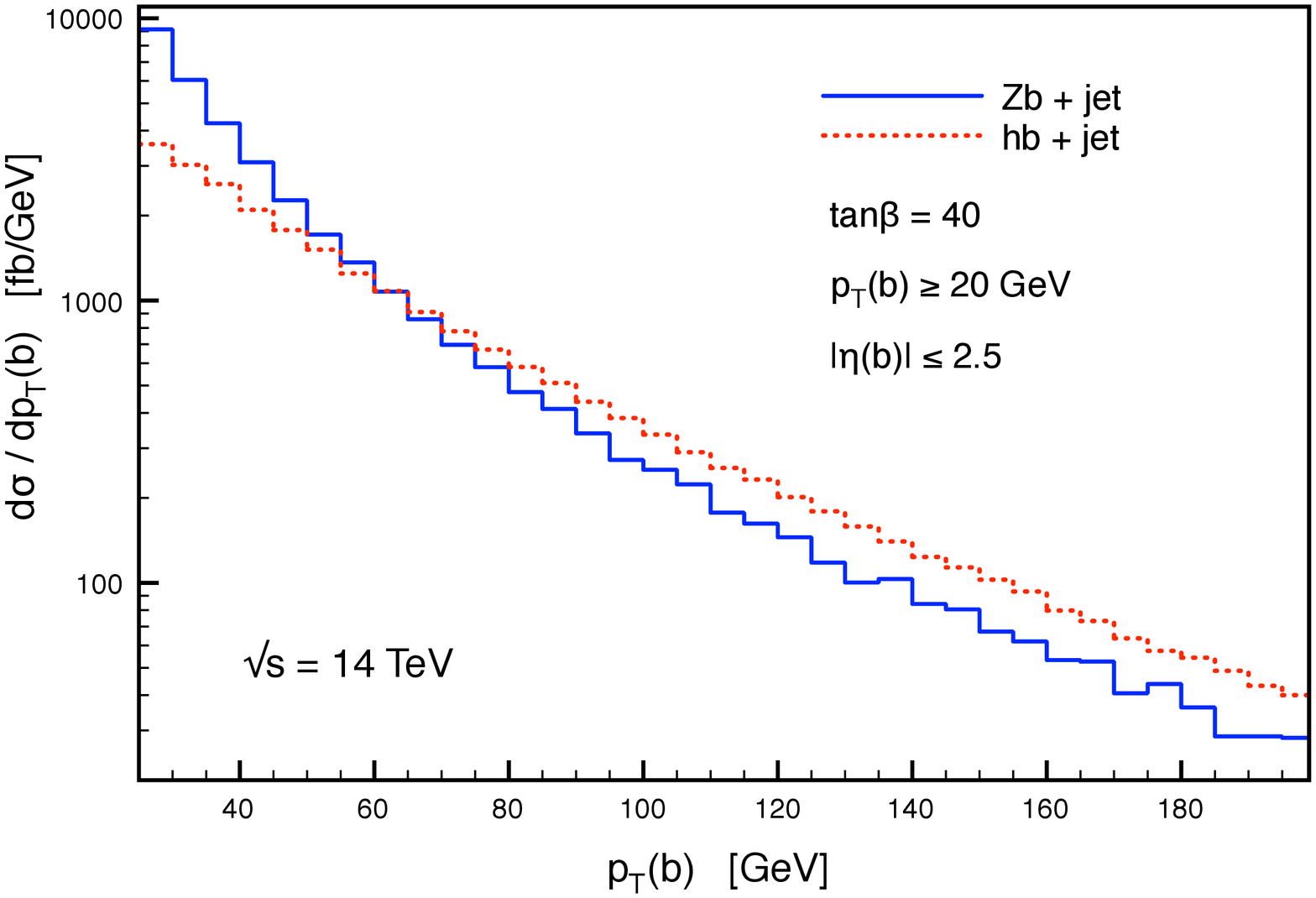}}
\caption{Comparison of the transverse momentum distributions for the 
bottom quark jet in the signal process $pp \to b \phi + j$ and background
process $pp \to b Z + j$ at the LHC.}
\label{bgj}
\end{center}
\end{figure}

In Fig. \ref{bgj}, we compare the tree level rate for $pp\rightarrow
b h^0 $ + jet, with the irreducible background from $pp
\rightarrow b Z$ + jet at the LHC for  $\tan\beta=40$.  
We require $p_T(b)$ and $p_T(jet)>20~GeV$
and $\mid \eta_b, \eta_{jet}\mid < 2.5$.  Unfortunately, the rate
is quite small.

%\bibliographystyle{lesHouches}
%\bibliography{bbh}
%\end{document}

\section[Charged Higgs Bosons in the MSSM
at CMS: Discovery Potential]
{CHARGED HIGGS BOSONS IN THE MSSM AT CMS: DISCOVERY
POTENTIAL~\protect\footnote{Contributed 
by M.~Hashemi, S.~Heinemeyer,R.~Kinnunen, 
A.~Nikitenko, and G.~Weiglein}}
%\documentclass[11pt]{cernrep}
%\usepackage{graphicx,epsfig}
%\bibliographystyle{lesHouches}
%\begin{document}

%{
\input paperdef

\graphicspath{{figs/}}

%\title{
%Charged Higgs Bosons in the MSSM at CMS:
%Discovery Potential}

%\author{M.~Hashemi$^1$, S.~Heinemeyer$^2$, R.~Kinnunen$^3$, 
%        A.~Nikitenko$^4$, G.~Weiglein$^5$}
%\institute{$^1$Universiteit Antwerpen,
%Middelheimcampus,
%G.U.238, 
%Groenenborgerlaan 171,
%2020 Antwerpen, Belgium\\
%$^2$Instituto de Fisica de Cantabria (CSIC-UC), Santander, Spain\\
%$^3$Helsinki Institute of Physics, Helsinki, Finland\\
%$^4$Imperial College, London, UK; on leave from ITEP, Moscow, Russia\\
%$^5$IPPP, University of Durham, Durham DH1~3LE, UK}

%\maketitle

%\begin{abstract}
%The search for MSSM Higgs bosons will be an important goal at the
%LHC. In order to analyze the search reach of the CMS experiment for the
%charged MSSM Higgs bosons, we combine the latest results for the
%CMS experimental sensitivities based on full simulation studies with 
%state-of-the-art theoretical predictions of MSSM Higgs-boson properties.
%The experimental analyses are done assuming an integrated luminosity of 
%30~\ifb\ for the two cases, $\MHp < \mt$ and $\MHp > \mt$.
%The results are interpreted as 5$\,\si$ discovery contours in MSSM 
%$\MA$--$\tb$ benchmark scenarios. Special emphasis is put on the variation 
%of the Higgs mixing parameter $\mu$.
%The variation of $\mu$ can shift the prospective discovery reach 
%by about $\De\tb = 40$.
%\end{abstract}

%%%%%%%%%%%%%%%%%%%%%%%%%%%%%%%%%%%%%%%%%%%%%%%%%%%%%%%%%%%%%%%%%%%%%%%%%%%%%%
%%%%%%%%%%%%%%%%%%%%%%%%%%%%%%%%%%%%%%%%%%%%%%%%%%%%%%%%%%%%%%%%%%%%%%%%%%%%%%

\subsection{Introduction}

Identifying the mechanism of electroweak symmetry
breaking will be one of the main goals of the LHC. 
The most popular models are the Higgs mechanism within the Standard
Model (SM) and within the Minimal Supersymmetric Standard Model
(MSSM)~\cite{Nilles:1983ge,Haber:1985rc,Barbieri:1987xf}. Contrary to
the case of the SM, in the MSSM  
two Higgs doublets are required.
This results in five physical Higgs bosons instead of the single Higgs
boson of the SM. These are the light and heavy $\cp$-even Higgs bosons, $h$
and $H$, the $\cp$-odd Higgs boson, $A$, and the charged Higgs boson,
$H^\pm$.
The Higgs sector of the MSSM can be specified at lowest
order in terms of the gauge couplings, the ratio of the two Higgs vacuum
expectation values, $\tb \equiv v_2/v_1$, and the mass of the $\cp$-odd
Higgs boson, $\MA$ (or $\MHp$, the mass of the charged Higgs boson).
Consequently, the masses of the $\cp$-even neutral Higgs bosons (and the
charged Higgs boson) are dependent quantities that can be
predicted in terms of the Higgs-sector parameters. The same applies to
the production and decay properties of the MSSM Higgs bosons%
\footnote{If the production or decay involves SUSY particles at
  tree-level, also other MSSM parameters enter the prediction.}%
.~Higgs-phenomenology
in the MSSM is strongly affected by higher-order corrections, in
particular from the sector of the third generation quarks and squarks,
so that the dependencies on various other MSSM parameters can be
important.

The charged Higgs bosons of the MSSM (or a more general Two Higgs
Doublet Model (THDM)) have been searched at
LEP~\cite{:2001xy}, yielding a bound of
$\MHp \gsim 80 \gev$~\cite{LEPchargedHiggsProc,LEPchargedHiggs}.
The Tevatron placed new bounds on the MSSM parameter space from charged
Higgs-boson searches at large $\tb$ and low $\MA$~\cite{Abulencia:2005jd}. At
the LHC the charged Higgs bosons will be accessible best at large $\tb$
up to $\MA \lsim 800 
\gev$\cite{ATLASTDR,CMStdr,Carena:2005ek}. At the ILC, for
$\MHp \lsim \sqrt{s}/2$ a high-precision determination of the charged
Higgs boson properties will be
possible~\cite{AguilarSaavedra:2001rg,tesla2,Abe:2001npb,Abe:2001gc,Heinemeyer:2005gs}.

The prospective sensitivities at
the LHC are usually displayed in terms of the parameters $\MA$ and $\tb$
(or $\MHp$ and $\tb$) that characterize the MSSM Higgs sector at lowest
order. The other MSSM 
parameters are conventionally fixed according to certain benchmark
scenarios~\cite{Carena:2002qg,Carena:2005ek}. We focus here~\cite{cmsHiggs2} on
the $5\,\si$ discovery contours for the charged MSSM Higgs boson
for the two cases $\MHp < \mt$ and $\MHp > \mt$, 
within the $\mhmax$~scenario and the no-mixing scenario.
For the interpretation of the exclusion bounds and prospective discovery
contours in the benchmark scenarios it is important to assess how
sensitively the results depend on those parameters that have been fixed
according to the benchmark prescriptions.
Consequently, we investigate how the 
5$\,\si$ discovery regions in the $\MHp$--$\tb$ plane 
for the charged MSSM Higgs boson obtainable with the CMS experiment at
the LHC depend on the other MSSM parameters, most prominently the Higgs
mixing parameter~$\mu$.

%%%%%%%%%%%%%%%%%%%%%%%%%%%%%%%%%%%%%%%%%%%%%%%%%%%%%%%%%%%%%%%%%%%%%%%%%%%%%%
%%%%%%%%%%%%%%%%%%%%%%%%%%%%%%%%%%%%%%%%%%%%%%%%%%%%%%%%%%%%%%%%%%%%%%%%%%%%%%

%\section{EXPERIMENTAL ANALYSIS}
\subsection{Experimental Analysis}
The main
production channels at the LHC are
\BE
pp \to t\bar t \to H^- \bar b \; t \mbox{~~or~~} \bar t \; H^+ b~
\label{pp2Hpm}
\EE
and
\BE
gb \to H^- t \mbox{~~or~~} g \bar b \to H^+ \bar t~.
\label{gb2Hpm}
\EE
The decay used in the analysis to detect the charged Higgs boson is
\BE
H^\pm \to \tau \nu_\tau~.
\label{Hbug}
\EE
The analysis described below correspond to 
CMS experimental sensitivities based on full simulation studies, 
assuming an integrated luminosity of 30~\ifb.

%%%%%%%%%%%%%%%%%%%%%%%%%%%%%%%%%%%%%%%%%%%%%%%%%%%%%%%%%%%%%%%%%%%%%%%%%%%%%%

\subsubsection{The light charged Higgs Boson}
\label{sec:lightHpm}

The ``light charged Higgs boson'' is characterized by $\MHp < \mt$. 
The main production channel is given in \refeq{pp2Hpm}. Close to
threshold also \refeq{gb2Hpm} contributes. The relevant (i.e.\
detectable) decay channel is given by \refeq{Hbug}.
The experimental analysis, based on 30~\ifb\ collected with CMS, is
presented in \citere{lightHexp}.

A total number of events leading to final states with the signal
characteristics is evaluated, including their respective experimental
efficiencies. The various channels and the corresponding efficiencies
can be found in \refta{tab:lightHp}.
A $5\,\si$ discovery can be achieved if a parameter point results in
more than 5260~events (with 30~\ifb). We furthermore used 
$\br(W^\pm \to l \nu_l) = 0.217$ ($l = \mu, e$), 
$\br(W^\pm \to \tau \nu_\tau) = 0.1085$,
$\br(W^\pm \to \mbox{jets}) = 0.67$,
$\br(\tau \to \mbox{hadrons}) = 0.65$. The next-to-leading order LHC
cross section for top quark pairs is taken to be 840~pb. For the $W$+3
jets background the leading order cross section for the process 
$pp \to W^{\pm} + \rm 3~jets$, 
$W^{\pm} \to \ell^{\pm} \nu$ ($\ell=e,~\mu$) as given by
MadGraph~\citere{Stelzer:1994ta,Maltoni:2002qb}   
generator (840 pb) was used.

%%%%%%%%%%%%%%%%%%%%%% T A B L E %%%%%%%%%%%%%%%%%%%%%%%%%%%%%%%%%%%%%%%%%
\begin{table}[htb!]
\renewcommand{\arraystretch}{1.2}
\BC
\begin{tabular}{|c|c|} \hline
channel & exp.\ efficiency \\ \hline\hline
$pp \to t \bar t \to H^+ b \; \bar t 
                 \to (\tau^+ \bar{\nu}_\tau) \; (W^+ b$);  
~$\tau \to \mbox{hadrons}$, $W \to l \nu_l$ & 0.0052 \\
\hline
$pp \to t \bar t \to W^+ \; W^-
                 \to (\tau \nu_\tau) \; (l \nu_l)$;
~$\tau \to \mbox{hadrons}$ & 0.00217 \\
\hline
$pp \to t \bar t \to W^+ \; W^-
                 \to (l \nu_l) \; (l \nu_l)$ & 0.000859 \\
\hline
$pp \to t \bar t \to W^+ \; W^-
                 \to (\mbox{jet jet}) \; (l \nu_l)$ & 0.000134 \\
\hline
$pp \to W + \rm 3~jets$, $W \to \ell \nu$ & 0.000013 \\
%\hline
%$g\bar b \to H^+ \bar t \to (\tau \nu_\tau) \; (W b)$;
%~$\tau \to \mbox{hadrons}$, $W \to l \nu_l$ & 0.0052 \\
\hline\hline
\end{tabular}
\EC
\vspace{-1em}
\caption{Relevant channels for the light charged Higgs bosons and their
  respective experimental efficiencies. The charge conjugated processes
  ought to be included. The efficiency for the charged Higgs production
  is given for $M_{H^{+}}$=160 GeV. $l$ denotes $e$ or $\mu$.
}
\label{tab:lightHp}
\renewcommand{\arraystretch}{1.0}
\end{table}
%%%%%%%%%%%%%%%%%%%%%% T A B L E %%%%%%%%%%%%%%%%%%%%%%%%%%%%%%%%%%%%%%%%%
%\begin{verbatim}
%  XSBR = sigmatt *
%       (  brthb * brtwb * brhtn * brwln * brtauhad * effs160 * 2
%        + brtwb^2 * brwtn * brwln * brtauhad * efflep * 2
%        + brtwb^2 * brwln^2 * effleplep
%        + brtwb^2 * brwjj * brwln * effin * 2)
%        + XSHp          * brhtn * brwln * brtauhad * effs160 * 2;
%  Nev = 30 * XSBR + 327;
%(Nev > 5260),
%brwln = 0.217;
%brwtn = 0.1085
%brwjj = 0.67;
%brtauhad = 0.65;
%sigmatt = 840 * 10^3;
%effs160 = 0.0052;
%efflep = 0.00217;
%effleplep = 0.000859;
%effin = 0.000134;
%\end{verbatim}

%%%%%%%%%%%%%%%%%%%%%%%%%%%%%%%%%%%%%%%%%%%%%%%%%%%%%%%%%%%%%%%%%%%%%%%%%%%%%%

\subsubsection{The heavy charged Higgs Boson}
\label{sec:heavyHpm}

The ``heavy charged Higgs boson'' is characterized by $\MHp > \mt$.
Here \refeq{gb2Hpm} gives the largest contribution, and very close to
threshold \refeq{pp2Hpm} can contribute somewhat. The relevant decay
channel is again given in \refeq{Hbug}.
The experimental analysis, based on 30~\ifb\ collected with CMS, is
presented in \citere{heavyHexp}.

The number of signal events is evaluated as
\BE
N_{\rm ev} = \cL \times \si(pp \to H^\pm + X) 
                 \times \br(H^\pm \to \tau \nu_\tau)
                 \times \br(\tau \to \mbox{hadrons})
                 \times \mbox{exp.\ eff.}~,
\EE
where $\cL$ denotes the luminosity and the experimental efficiency
is given in \refta{tab:heavyHp} as a function of $\MHp$.
A $5\,\si$ discovery corresponds to a number of signal events larger
than $14.1$.

%%%%%%%%%%%%%%%%%%%%%% T A B L E %%%%%%%%%%%%%%%%%%%%%%%%%%%%%%%%%%%%%%%%%
\begin{table}[htb!]
\renewcommand{\arraystretch}{1.2}
\BC
\begin{tabular}{|c||cccccc|} 
\hline\hline
$\MHp$ [GeV]            & 171.6 & 180.4 & 201.0 & 300.9 & 400.7 & 600.8 \\ 
\hline
exp.\ eff.\  [$10^{-4}$] & 3.5   & 4.0   & 5.0  & 23    & 32    & 42 \\
\hline\hline
\end{tabular}
\EC
\vspace{-1em}
\caption{Experimental efficiencies for the heavy charged Higgs boson
  detection. 
}
\label{tab:heavyHp}
\renewcommand{\arraystretch}{1.0}
\end{table}
%%%%%%%%%%%%%%%%%%%%%% T A B L E %%%%%%%%%%%%%%%%%%%%%%%%%%%%%%%%%%%%%%%%%
The charged Higgs boson production with the mass close to the top quark mass
(first column in \refta{tab:heavyHp}) was generated with the PYTHIA generator
processes 401 ($gg \to tbH^{\pm}$) and 402 ($qq \to tbH^{\pm}$) implemented as
described in ~\citere{Alwall:2004xw}. 

%\begin{verbatim}
%CMS = {{171.6, 3.5 10^-4},
%       {180.4, 4.0 10^-4}, 
%       {201.0, 5.0 10^-4},
%       {300.9, 2.3 10^-3}, 
%       {400.7, 3.2 10^-3},
%       {600.8, 4.2 10^-3}}; 
%  XSHp = (tHmLHC /. fhhiggsprod) * 2; (* factor 2 for H+ and H- *)
%  BRHp = (BR[HpFF] /. fhcouplings)[[1,3]]; 
%  MA = (MHiggs /. fhhiggscorr)[[3]];
%  Nev = 30 * XSHp * BRHp * 0.65 * eff; 
%(*      lumi               BR(tau -> jet) *)
% (Nev > 14.1) 
%
%\end{verbatim} 
%%%%%%%%%%%%%%%%%%%%%%%%%%%%%%%%%%%%%%%%%%%%%%%%%%%%%%%%%%%%%%%%%%%%%%%%%%%%%%
%%%%%%%%%%%%%%%%%%%%%%%%%%%%%%%%%%%%%%%%%%%%%%%%%%%%%%%%%%%%%%%%%%%%%%%%%%%%%%

\subsection{Calculation of Cross Section and Branching Ratios}

For the calculation of cross sections and branching ratios we use a
combination of up-to-date theory evaluations. The Lagrangian for the
interaction of the charged Higgs boson with the $t/b$~doublet is given
by~\cite{Carena:1999py}
\BE
\label{effL}
\cL = \frac{g}{2\MW} \frac{\mbms}{1 + \db} \Bigg[ 
    \wz \, V_{tb} \, \tb \; H^+ \bar{t}_L b_R \Bigg] + {\rm h.c.}
\EE
Here $\mbms$ denotes the running bottom quark mass including SM QCD
corrections. 
The prefactor $1/(1 + \db)$ in \refeq{effL} arises from the
resummation of the leading corrections to all orders. 
The explicit
form of $\db$ in the limit of heavy SUSY masses and $\tb \gg 1$
reads~\cite{Hempfling:1993kv,Hall:1993gn,Carena:1994bv}
\BE
\db = \frac{2\als}{3\,\pi} \, \mgl \, \mu \, \tb \,
                    \times \, I(\msbe, \msbz, \mgl) +
      \frac{\alt}{4\,\pi} \, \At \, \mu \, \tb \,
                    \times \, I(\mste, \mstz, \mu) ~.
\label{def:dmb}
\EE
Here $\mste$, $\mstz$, $\msbe$, $\msbz$ denote the $\Stop$ and
$\Sbot$~masses, $\mgl$ is the gluino mass.
Large negative $\mu$ can lead to a strong enhancement of the 
$H^\pm t b$ coupling, while a large positive value leads to a strong
suppression. 
Concerning the $\mhmax$ and the no-mixing benchmark scenarios, 
as discussed in \citeres{Gennai:2007ys,Carena:2005ek} the $\db$ effects are 
much more pronounced in the $\mhmax$ scenario, where the two terms in
\refeq{def:dmb} are of similar size. In the no-mixing scenario the first
term in \refeq{def:dmb} dominates, and the total effect is smaller.

For the production cross section in \refeq{pp2Hpm} we use the SM cross
section $\si(pp \to t \bar t) = 840~\rm{pb}$ times the 
$\br(H^\pm \to tb)$ including the $\db$ corrections described above.
The production cross section in \refeq{gb2Hpm} is evaluated as given in
\citeres{Plehn:2002vy,Berger:2003sm}. In addition also the $\db$ corrections of
\refeq{effL} are applied. Finally the $\br(H^\pm \to \tau \nu_\tau)$ is
evaluated taking into account all decay channels, among whom the most
relevant are $H^\pm \to tb, cs, W^{(*)}h$. For the decay to $tb$ again
the $\db$ corrections are included.
All the numerical evaluations are performed with the program 
{\tt FeynHiggs}~\cite{Heinemeyer:1998yj,Heinemeyer:1998np,Degrassi:2002fi,Frank:2006yh}.

%%%%%%%%%%%%%%%%%%%%%%%%%%%%%%%%%%%%%%%%%%%%%%%%%%%%%%%%%%%%%%%%%%%%%%%%%%%%%%
%%%%%%%%%%%%%%%%%%%%%%%%%%%%%%%%%%%%%%%%%%%%%%%%%%%%%%%%%%%%%%%%%%%%%%%%%%%%%%
\subsection{Numerical Analysis}
%\section{NUMERICAL ANALYSIS}
\label{sec:numanal}

The numerical analysis has been performed in the $\mhmax$~and the
no-mixing scenario~\cite{Carena:2002qg,Carena:2005ek} for 
$\mu = -1000, -200, +200, +1000 \gev$. In \reffi{fig:reach} we show the
results combined for the $5\,\si$ discovery contours for the light and
the heavy charged Higgs boson, corresponding to the experimental
analyses in \refses{sec:lightHpm} and \ref{sec:heavyHpm},
respectively. As described above, the analyses were performed for the
CMS detector and 30~\ifb. 
The top quark mass is set to $\mt = 175 \gev$. 

Within the $\mhmax$ scenario, shown in the left plot of
\reffi{fig:reach} the search for the light charged Higgs boson covers
the area of large $\tb$ and $\MHp \lsim 150 \gev$. The variation with
$\mu$ induces a strong shift in the $5\,\si$ discovery contours with 
$\De\tb = 15$ for $\MHp = 100 \gev$, rising up to $\De\tb = 40$ for
larger $\MHp$ values. The discovery region is largest (smallest) for 
$\mu = -(+)1000 \gev$, corresponding to the largest (smallest)
production cross section.

The $5\,\si$ discovery regions for the search for heavy charged Higgs
bosons show a similar behavior. For $\MHp = 170 \gev$ the accessible
parameter space starts at $\tb = 20 (58)$ for $\mu = -(+)1000 \gev$,
i.e.\ the variation of $\mu$ again induces a very strong shift in the
$5\,\si$ discovery contours. For $\MHp = 300 \gev$ the $5\,\si$ regions
vary from $\tb = 38$ to $\tb = 54$. For $\mu = -1000 \gev$ and larger
$\tb$ values the bottom Yukawa coupling becomes so large 
that a perturbative treatment would no longer be reliable in this
region.

%%%%%%%%%%%%%%%%%% F I G U R E %%%%%%%%%%%%%%%%%%%%%%%%%%%%%%%%%%%%%%%%%%%%%%
\begin{figure}[htb!]
\begin{center}
\includegraphics[width=0.45\textwidth]{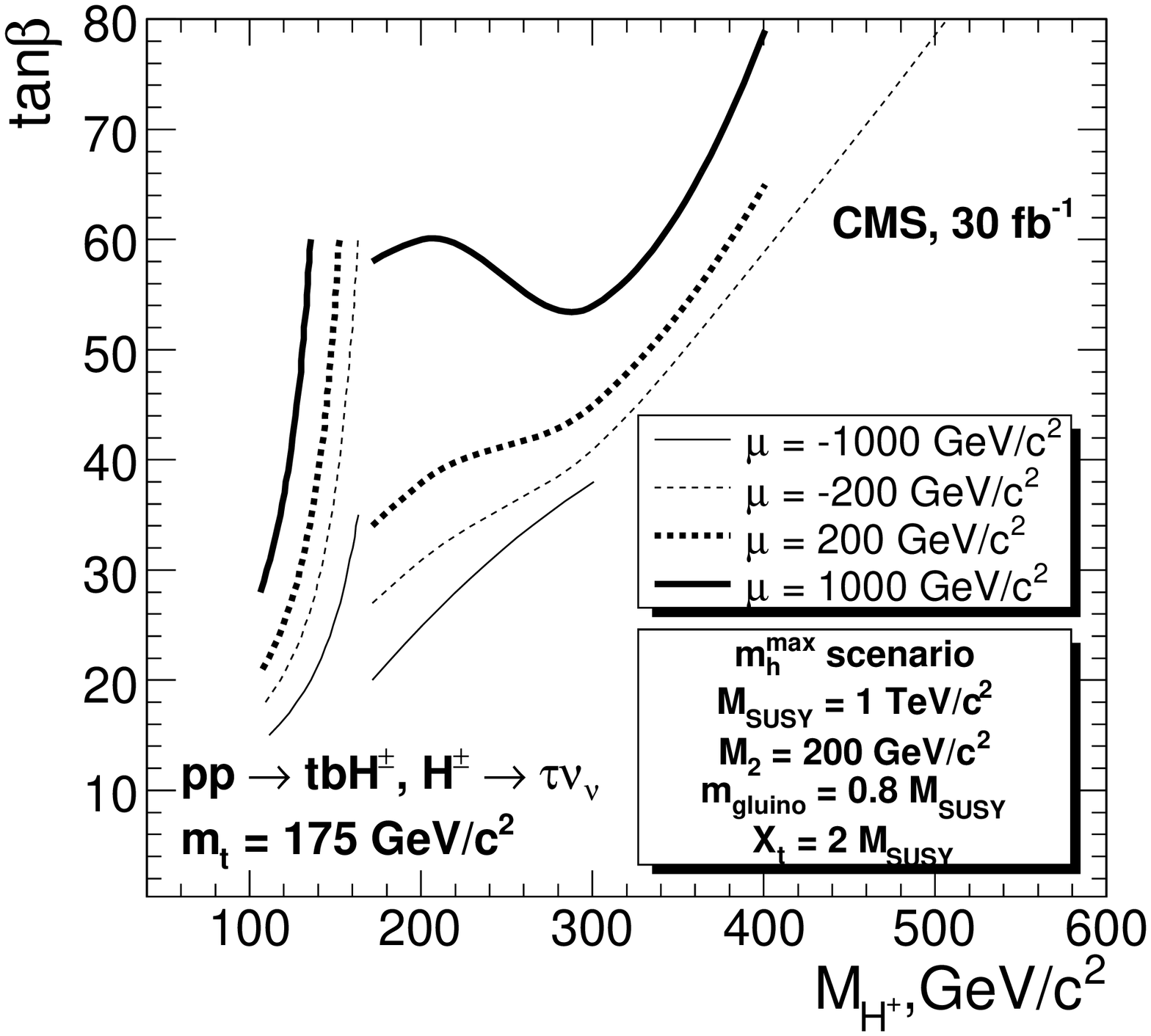}\hspace{1em}
\includegraphics[width=0.45\textwidth]{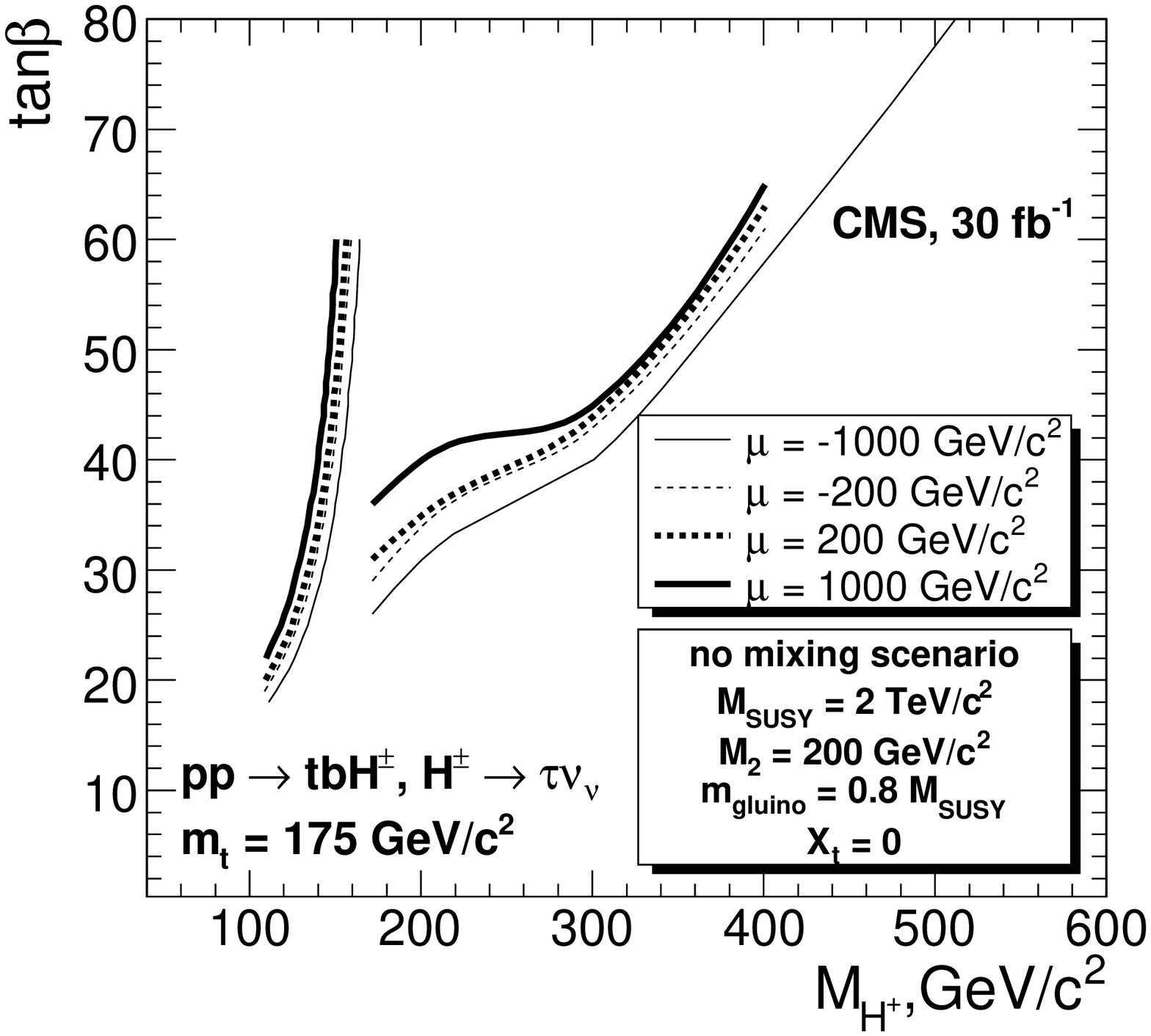}
 \caption{%
Discovery reach for the charged Higgs boson of CMS with 30~\ifb\ in the 
$\MHp$--$\tb$~plane for the $\mhmax$~scenario (left) and the no-mixing
scenario (right).
}
\label{fig:reach}
\end{center}
\end{figure}
%%%%%%%%%%%%%%%%%% F I G U R E %%%%%%%%%%%%%%%%%%%%%%%%%%%%%%%%%%%%%%%%%%%%%%

The no-mixing scenario is shown in the right plot of
\reffi{fig:reach}. The qualitative behavior is the same as for the
$\mhmax$~scenario. However, as discussed above, the effects from a
variation of $\mu$ are much less pronounced. The induced shifts stay
below $\De\tb = 20 (10)$ in the search for the light (heavy) charged
Higgs boson. The $5\,\si$ discovery areas are slightly larger than in
the $\mhmax$~scenario.

%%%%%%%%%%%%%%%%%%%%%%%%%%%%%%%%%%%%%%%%%%%%%%%%%%%%%%%%%%%%%%%%%%%%%%%%%%%%%%
%%%%%%%%%%%%%%%%%%%%%%%%%%%%%%%%%%%%%%%%%%%%%%%%%%%%%%%%%%%%%%%%%%%%%%%%%%%%%%

%\section*{CONCLUSIONS}
\subsection{Conclusions}
We have presented the $5\,\si$ discovery contours for the
search for the charged MSSM Higgs boson. 
We combine the latest results for the
CMS experimental sensitivities based on full simulation studies with 
state-of-the-art theoretical predictions of MSSM Higgs-boson properties.
The experimental analyses are done assuming an integrated luminosity of 
30~\ifb\ for the two cases, $\MHp < \mt$ and $\MHp > \mt$.

The numerical analysis has been performed in the $\mhmax$~and the
no-mixing scenario for $\mu = \pm 200, \pm 1000 \gev$.
The search for the light charged Higgs boson covers the are area of
large $\tb$ and $\MHp \lsim 160 \gev$. The search for the heavy charged
Higgs boson reaches up to $\MHp \lsim 400 \gev$ for large $tb$.
The variation of $\mu$ induces a very strong shift in the $5\,\si$
discovery contours of up to $\De\tb = 40$. The effect enters via the
variation of $\db$, affecting the charged Higgs production cross section
and branching ratios. Large negative $\mu$ values give the largest
reach, while large positive values yield the smallest $5\,\si$ discovery
areas.

\section*{Acknowledgements}
The work of S.H.\ was partially supported by CICYT (grant FPA~2007--66387).
Work supported in part by the European Community's Marie-Curie Research
Training Network under contract MRTN-CT-2006-035505
`Tools and Precision Calculations for Physics Discoveries at Colliders'.

%%%%%%%%%%%%%%%%%%%%%%%%%%%%%%%%%%%%%%%%%%%%%%%%%%%%%%%%%%%%%%%%%%%%%%%%%%%%%%
%%%%%%%%%%%%%%%%%%%%%%%%%%%%%%%%%%%%%%%%%%%%%%%%%%%%%%%%%%%%%%%%%%%%%%%%%%%%%%

%\bibliography{cmsHiggsLH}%

%}

%\end{document}

%%%%%%%%%%%%%%%%%%%%%%%%%%%%%%%%%%%%%%%%%%%%%%%%%%%%%%%%%%%%%%%%%%%%%%%%%%%%%%
%%%%%%%%%%%%%%%%%%%%%%%%%%%%%%%%%%%%%%%%%%%%%%%%%%%%%%%%%%%%%%%%%%%%%%%%%%%%%%

\section[Studies of Spin Effects in Charged Higgs Boson Production
 with an Iterative Discriminant Analysis at the Tevatron and LHC]
{STUDIES OF SPIN EFFECTS IN 
CHARGED HIGGS BOSON PRODUCTION WITH AN ITERATIVE DISCRIMINANT ANALYSIS
AT THE TEVATRON AND LHC~\protect\footnote{Contributed by: 
 S.~ Hesselbach, S.~ Moretti, J.~ Rathsman and A.~ Sopczak}}

\def\Ord{\lower .7ex\hbox{$\;\stackrel{\textstyle <}{\sim}\;$}}
\def\OOrd{\lower .7ex\hbox{$\;\stackrel{\textstyle >}{\sim}\;$}}

\subsection{Introduction}
The importance of charged Higgs boson searches 
has in the recent years been emphasized,
including in the `2005 Les Houches' proceedings~\cite{Buttar:2006zd}.
This work extends the charged Higgs boson `2005 Les Houches' studies.
It is the purpose of this note to outline the possible improvements 
that can be achieved at the Tevatron and LHC in the search for charged 
Higgs bosons focussing on the spin effects and the $H^\pm\to\tau\nu_\tau$ decay.
In order to quantify the spin effect an Iterative Discriminant Analysis
(IDA) method~\cite{ida} has been applied, which is a powerful tool to separate 
signal and background, even in cases such as the one presently under study when 
several selection variables with limited discriminant power are present.

\subsection{Tevatron Energy}

We start by studying charged Higgs production $q\bar q, gg \to  tbH^\pm$
with subsequent decays $t \to b W$, $H^\pm \to \tau \nu_\tau$
at the FNAL Tevatron with $\sqrt{s} = 1.96$~TeV.
In the following we analyze hadronic decays of
the $W^\pm$ boson and $\tau$ lepton
($W^\pm \to q\bar{q}'$, $\tau \to \mathrm{hadrons} + \nu_\tau$),
which results in the signature
$2 b + 2 j + \tau_\mathrm{jet} + p_t^{\rm miss}$
(2 $b$ jets, 2 light jets, 1 $\tau$ jet and missing transverse momentum).
The most important irreducible 
background process is $q\bar q, gg \to  t\bar{t}$
with the subsequent decays $t \to b W^+$ and $\bar{t} \to \bar{b} W^-$,
one $W^\pm$ boson decaying hadronically ($W^\pm \to q\bar{q}'$)
and one leptonically ($W^\mp \to \tau \nu_\tau$), which results in
the same final state particles as for the expected signal. 

\subsubsection{Simulation and Detector Response}
The signal process $q\bar q, gg \to  tbH^\pm$ is simulated with PYTHIA 
\cite{Sjostrand:2003wg}.
The subsequent decays $t \to b W^\pm$ (or its charge conjugate),
$W^\pm \to q\bar{q}'$ and
$H^\mp \to \tau \nu_\tau$ are also carried out within PYTHIA,
whereas the $\tau$ leptons are decayed externally with the program TAUOLA
\cite{Jadach:1990mz, Golonka:2003xt},
which includes the complete spin structure of the $\tau$ decay.
The background process $q\bar q, gg \to  t\bar{t}$ is also simulated
with PYTHIA with the built-in subroutines for $t\bar{t}$ production.
The decays of the top quarks and $W^\pm$ bosons are performed within PYTHIA
and that of the $\tau$ lepton within TAUOLA.

The momenta of the final $b$ and light quarks from the PYTHIA event
record are taken as the momenta of the corresponding jet, whereas for the
$\tau$ jet the sum of all non-leptonic final state particles as given by
TAUOLA is used.
The energy resolution of the detector and parton shower and hadronization effects
are emulated through a Gaussian
smearing $(\Delta(p_t)/p_t)^2 = (0.80/\sqrt{p_t})^2$
of the transverse momentum $p_t$
for all jets in the final state, 
including the $\tau$ jet~\cite{SHW}.
As typical for fast simulation studies, no effects of underlying events, are simulated.
Events are removed which contain jets with less than 20 GeV transverse 
momentum\footnote{In order to be largely independent of the specific 
detector performance, no requirement on the jet resolution is applied.}, 
corresponding to about $|\eta|>3$.
The transverse momentum of the leading charged pion in the $\tau$ jet
is assumed to be measured in the tracker independently of the
transverse momentum of the $\tau$ jet. The identification and momentum 
measurement of the pion is important to fully exploit the $\tau$ spin information.
In order to take into account the tracker performance
we apply  Gaussian smearing on $1/p_t^\pi$ with
$ \sigma(1/p_t^\pi)[\mathrm{TeV}^{-1}] =
 \sqrt{0.52^2 + 22^2/(p_t^\pi[\mathrm{GeV}])^2 \sin\theta_\pi}$,
where $\theta_\pi$ is the polar angle of the $\pi$.
The missing transverse momentum \mbox{$p_t^{\rm miss}$} is constructed from
the transverse momenta of all visible jets 
(including visible $\tau$ decay products)
after taking the modelling of the detector into account.
The generic detector description is a good approximation for both Tevatron experiments,
CDF and D0.

\subsubsection{Expected Rates}
For completeness we present a brief discussion of the expected cross section 
of the charged Higgs boson signature under investigation.
The signal cross section has been calculated for
$\tan\beta = 30$
and $m_{H^\pm} = 80, 100, 130$ and $150$~GeV with PYTHIA, version 6.325,
using the implementation described in~\cite{Alwall:2004xw},
in order to take the effects in the transition region into account. Furthermore,
it has been shown in~\cite{Alwall:2003tc} that the signal cross section
for $tbH^\pm$ agrees with the one from the top-decay approximation
$t\bar t \to tbH^\pm$ for charged Higgs boson masses up to about 160~GeV
if the same factorization and renormalization scales are used.
Thus, we have used everywhere in this study the factorization scale $(m_t + m_{H^\pm})/4$ and the renormalization scale $m_{H^\pm}$ for both signal and background (i.e., those recommended in~\cite{Alwall:2004xw} as most appropriate for the $tbH^\pm$ 
signal)\footnote{Clearly, for a proper
experimental study, factorization and renormalization scales for our background process 
$q\bar q$, $ gg \rightarrow t\bar t\to tbW^\pm$
ought to be chosen appropriately, i.e., unrelated to the charged Higgs boson mass.},
since the primary purpose of our study is to  single out variables that show a 
difference between our $W^\pm$ and $H^\pm$ data samples and that this
can unambiguously be ascribed to the different nature of the two kinds of bosons
(chiefly, their different mass and spin state).
In addition, the running $b$ quark mass entering in the Yukawa
coupling of the signal has been evaluated at $m_{H^\pm}$.
This procedure eventually results in a dependence of our background calculations on $\tan\beta$
and, especially, $m_{H^\pm}$ that is more marked than the one that would more naturally arise 
as only due to indirect effects through the top decay width.
Hence, the cross sections have been rescaled with a common factor
such that the total $t \bar t$ cross section is
$\sigma^{\rm prod}_{t\bar{t}} = 5.2$~pb \cite{ttbarxsec}.
To be more specific, we have first calculated the total cross section 
$\sigma^{\rm prod,PYTHIA}_{t\bar{t}}(m_{H^\pm})$ with the 
built-in routine for $t \bar t$ production
in PYTHIA for all $m_{H^\pm} = 80, 100, 130$ and $150$~GeV and then
calculated from this the respective rescaling factors
$c(m_{H^\pm}) = 
 5.2~\mathrm{pb}/\sigma^{\rm prod,PYTHIA}_{t\bar{t}}(m_{H^\pm})$
for each $m_{H^\pm}$.
Then we have calculated the background cross section 
for $m_{H^\pm}=80$~GeV into the final state with the signature
$2 b + 2 j + \tau_\mathrm{jet} + p_t^{\rm miss}$
by enforcing the respective decay channels in PYTHIA using the 
built-in routine for $t \bar t$ production and multiplied 
it with $c(80~\mathrm{GeV})$.
In the same manner we have calculated the signal cross sections with the
PYTHIA routines for $tbH^\pm$ production by enforcing the respective decay
channels in PYTHIA and multiplying with the rescaling factors
$c(m_{H^\pm})$ for $m_{H^\pm} = 80, 100, 130, 150$~GeV.
The resulting cross sections 
are given in Table~\ref{tab:hplus_crosssec} before 
($\sigma^{\rm th}$) and after ($\sigma$) applying the basic cuts 
$p_t^{\rm jets} > 20$~GeV and the hard cut $p_t^{\rm miss} > 100$~GeV.
For the four signal masses, the $tbH^\pm$ and $ t\bar t \to tbH^\pm$
cross section calculations agree numerically.

\begin{table}[htbp]
\caption{\label{tab:hplus_crosssec}
Tevatron cross sections of background $q\bar q, gg \to  t\bar{t}$
and signal $q\bar q, gg \to  tbH^\pm$
for $\tan\beta = 30$ and $m_{H^\pm} = 80, 100, 130$ and $150$~GeV
into the final state
$2 b + 2 j + \tau_\mathrm{jet} + p_t^{\rm miss}$
before ($\sigma^{\rm th}$) and after ($\sigma$)
the basic cuts ($p_t > 20$~GeV for all jets)
and the hard cut ($p_t^{\rm miss} > 100$ GeV).
}
\centering
\begin{tabular}{c|c|c|c|c|c}
 & $q\bar q, gg \to  t\bar{t}$ &
  \multicolumn{4}{c}{$q\bar q, gg \to  tbH^\pm$} \\
$m_{H^\pm}$ (GeV) & 80 & 80 & 100 & 130 & 150\\
  \hline
$\sigma^{\rm th}$ (fb) & 350 &  535    &  415    & 213 & 85  \\
$\sigma$ (fb) for $p_t^\mathrm{jets} > 20$ GeV 
         & 125   & 244     &   202   &  105 & 32 \\
$\sigma$ (fb) for $(p_t^\mathrm{jets},p_t^{\rm miss}) > (20,100)$ GeV 
         & 21   &   30   &  25    &  18 & 7 \\
\end{tabular}
\end{table}

\subsubsection{Event Preselection and Discussion of Discriminant Variables}
The expected cross sections of the 
$2 b + 2 j + \tau_\mathrm{jet} + p_t^{\rm miss}$ signature 
are of the same order of magnitude for the signal and background 
reactions, as shown in Table~\ref{tab:hplus_crosssec}.
Thus, the same number of signal and background events is assumed for the 
analysis of different kinematic selection variables.
For the signal $5\cdot 10^5$ events have been simulated with PYTHIA for each 
charged Higgs mass at the Tevatron energy of 1.96~TeV using the built-in 
$t\bar t$ routine in the $t\bar t \rightarrow tbH^\pm$ approximation, 
while for the $t\bar t$ background also $5\cdot 10^5$ events have been 
simulated using the built-in $t\bar t$ routine.
Then the basic cuts $p_t^{\rm jets} > 20$~GeV are applied.
An additional hard cut on the missing transverse momentum
$p_t^{\rm miss} > 100$ GeV is used to suppress the QCD background, as 
for example demonstrated in Ref.~\cite{Assamagan:2002in}.
After the additional anti-QCD cut about 28000 to 42000 signal events,
depending on the simulated charged Higgs bosons mass, and about 30000 
$t\bar t$ background events remain.
Other background reactions, for example W+jet production, are expected to be 
negligible because they have either a much lower production cross section
or are strongly suppressed compared to $t\bar t$ background, 
as quantified for example in Ref.~\cite{Assamagan:2002in}.
In addition to the previous study 
(based on $5000\times \mathrm{BR}(\tau \to \mathrm{hadrons})$ events
each)~\cite{Buttar:2006zd},
the present one applies an IDA method~\cite{ida}
to explore efficiencies and purities. As already mentioned,
particular attention is devoted to the study of 
spin sensitive variables in the exploitation of polarization
effects for the separation of signal and background events.

Examples of the signal and background distributions of some of the kinematic 
variables used in the IDA method and the respective difference between
signal and background distributions are given in Ref.~\cite{Hesselbach:2007jj}, namely:
\begin{itemize}
\item the transverse momentum of the $\tau$ jet, $p_t^{\tau_{\rm jet}}$,
\item the transverse momentum of the leading $\pi^\pm$ in the $\tau$ jet, 
$p_t^{\pi^\pm}$,
\item the ratio $p_t^{\pi^\pm}/p_t^{\tau_{\rm jet}}$,
\item the transverse momentum of the second (least energetic) $b$ quark jet, $p_t^{b_2}$,
\item the transverse mass\footnote{Strictly speaking this is not the transverse mass since there are two neutrinos in the decay chain of the charged Higgs boson we are considering, even so the characteristics of this mass are very similar to that of the true transverse mass.} in the $\tau_{\rm jet} + p_t^{\rm miss}$ system,
      $m_t = \sqrt{2 p_t^{\tau_\mathrm{jet}} p_t^{\rm miss}
             [1-\cos(\Delta\phi)]}$, where $\Delta\phi$ is the azimuthal angle
      between $p_t^{\tau_\mathrm{jet}}$ and $p_t^{\rm miss}$,
\item the invariant mass distribution of the two light quark jets and the
      second $b$ quark jet, $m_{jjb_2}$,
\item the spatial distance between the $\tau$ jet and the second $b$ quark jet, 
      $\Delta R(\tau,b_2) = \sqrt{(\Delta\phi)^2 + (\Delta\eta)^2}$,
      where $\Delta\phi$ is the azimuthal angle between the $\tau$ and $b$ jet,
      and 
\item the sum of the (scalar) transverse momenta of all the quark jets, 
      $H_{\rm jets} = p_t^{j_1} + p_t^{j_2} + p_t^{b_1} + p_t^{b_2}$.
\end{itemize}
The distributions of signal and background events are normalized 
to the same number of $10^4$ events, 
in order to make small differences better visible.

The signal and background distributions 
for the variables shown in Ref.~\cite{Hesselbach:2007jj}
are as expected rather similar for $m_{H^\pm}=m_{W^\pm}$
and are hence mostly important to discriminate between signal and
background in the IDA for $m_{H^\pm} > m_{W^\pm}$.
Especially the transverse mass, 
shows a large variation with the charged Higgs boson mass.
However, the different spin
of the charged Higgs boson and the $W^\pm$ boson has a large effect on
the $\tau$ jet variables $p_t^{\tau_{\rm jet}}$ and $p_t^{\pi^\pm}$
resulting in significantly different distributions of signal and background
even for $m_{H^\pm}=m_{W^\pm}$.
Moreover, the spin effects in the $p_t^{\tau_{\rm jet}}$ and $p_t^{\pi^\pm}$ 
distributions are correlated
where the distributions of the ratio 
$p_t^{\pi^\pm}/p_t^{\tau_{\rm jet}}$~\cite{Roy:1991sf,Raychaudhuri:1995cc,Roy:1999xw}
show even larger differences~\cite{Hesselbach:2007jj}.
This highlights the importance
of the additional variable $p_t^{\pi^\pm}$ (and hence
$p_t^{\pi^\pm}/p_t^{\tau_{\rm jet}}$),
compared to a previous study \cite{Buttar:2006zd}.
The large separation power of this variable
is indeed due to the different $\tau$ polarizations
in signal and background~\cite{Hesselbach:2007jj}.
There the signal and background distributions for $p_t^{\tau_{\rm jet}}$,
$p_t^{\pi^\pm}$ and $p_t^{\pi^\pm}/p_t^{\tau_{\rm jet}}$ are shown
for reference samples where the $\tau$ decay has been performed without the
inclusion of spin effects with the built-in routines of PYTHIA and
hence the differences between signal and background nearly vanish.

\subsubsection{Iterative Discriminant Analysis (IDA)}

The IDA method is a modified Fisher Discriminant Analysis~\cite{ida}
and is characterized by 
the use of a quadratic, instead of a linear, discriminant function and 
also involves iterations in order to 
enhance the separation between signal and background.

In order to analyze our events with the IDA method, signal
and background  have been split in two samples of equal size.
With the first set of samples the IDA training has been performed and then
the second set of samples has been analyzed.
We have used the following 20 variables in the IDA study:
the transverse momenta 
$p_t^{\tau_{\rm jet}}$, $p_t^{\pi^\pm}$,
$p_t^{\rm miss}$,
$p_t^{b_1}$, $p_t^{b_2}$, $p_t^{j_1}$, $p_t^{j_2}$, $p_t^{j j}$;
the transverse mass $m_t$;
the invariant masses
$m_{jj}$, $m_{jjb_1}$, $m_{jjb_2}$, $m_{bb}$ and
$\hat{s}=m_{jjbb\tau}$;
the spatial distances $\Delta R(\tau,b_1)$, $\Delta R(\tau,b_2)$,
$\Delta R(\tau,j_1)$, $\Delta R(\tau,j_2)$;
the total transverse momenta of all quark jets $H_{\rm jets}$ 
and of all jets $H_{\rm all} = H_{\rm jets} + p_t^{\tau_{\rm jet}}$.
In the analysis of real data, b-quark tagging probabilities and 
the reconstruction of $t$ and $W$ masses could be used to improve
the jet pairing, and replace the allocation of least and most energetic
$b$-jet by a probabilistic analysis.

The results of the IDA study 
are obtained for the event samples with spin effect in the
$\tau$ decays for $m_{H^\pm}=80, 100, 130, 150$~GeV and for the
reference samples without the spin effect for $m_{H^\pm}=80$~GeV
in order to illustrate the spin effect.
In all plots of the IDA output variable the number of 
background events has been normalized to the number of signal events.
Two IDA steps have been performed.
After the first step, 90\% of the signal is retained when a cut at zero 
is applied on the IDA output variable.
The signal and background events after this cut are then passed to the 
second IDA step. 
A cut on IDA output variable distributions after the second step
leads to the efficiency and purity (defined as ratio of the number of signal events
divided by the sum of 
signal and background events) combinations. 
These combinations define the working point (number of expected
background events for a given 
signal efficiency) and the latter can be optimized to maximize the
discovery potential.

In order to illustrate the effect of the hard cut on the missing
transverse momentum ($p_t^{\rm miss}>100$~GeV), which is imposed to
suppress the QCD background, 
the final efficiency-purity plot of the IDA analysis is shown 
in Fig.~\ref{fig:hplus_ida_hard_cut} for $m_{H^\pm}=80$~GeV for two
reference samples (red, long dashed: with spin effects in the $\tau$
decay; red, dotted: without spin effects) without imposing the hard
cut. 
As expected the achievable purity for a given efficiency decreases
with the hard cut,
therefore the spin effects become even more important to
separate signal and background.
In principle, by choosing the signal reduction rates in the previous IDA iterations, 
the signal and background rates in the final distributions can be varied appropriately.
However, we have checked that a different number of IDA iterations and/or different efficiencies 
for the first IDA iteration have only a minor effect on the final result.

\begin{figure}[htp]
\vspace*{-5mm}
\begin{minipage}{0.59\textwidth}
\begin{center}
\epsfig{file=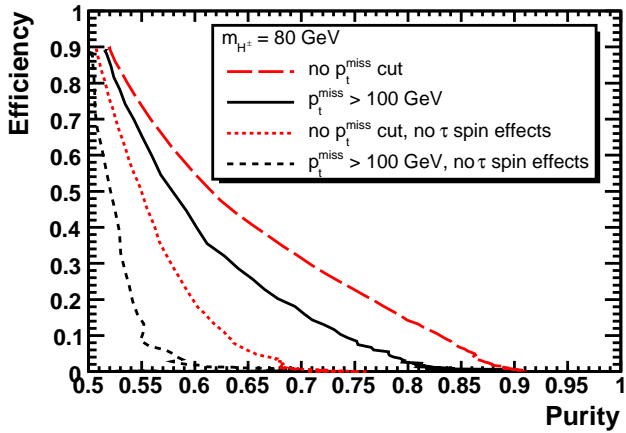, width=\textwidth}
\end{center}
\end{minipage}\hfill
\begin{minipage}{0.39\textwidth}
\caption{
Efficiency as a function of purity for $m_{H^\pm}=80$~GeV and
$\sqrt{s}=1.96$ TeV.
The black lines are the results after applying the hard cut
\mbox{$p_t^{\rm miss} > 100$~GeV}
when not taking the spin effects in the $\tau$ decay into account
(dashed) and with spin effects in the $\tau$ decay (solid).
The red lines are the results without applying the hard cut
on \mbox{$p_t^{\rm miss}$}
when not taking the spin effects in the $\tau$ decay into account
(dotted) and with spin effects in the $\tau$ decay (long dashed).
}
\label{fig:hplus_ida_hard_cut}
\end{minipage}
\vspace*{-10mm}
\end{figure}

\subsection{LHC Energy}

The simulation procedure and the emulation of the detector response are the same as 
those outlined in Sect.~2.1 for the Tevatron, as well as, for the preselection and IDA method, 
as described in Sects. 2.3 and 2.4, respectively. Hence, only the expected LHC rates are 
discussed, followed by the description of changes in the distributions of the variables and 
the final IDA results.

Unlike the case of the Tevatron, where only charged Higgs masses smaller
than the top quark mass can be explored, and 2HDM/MSSM signatures
practically rely on $\tau\nu_\tau$ pairs only, at the LHC the phenomenology 
is more varied. Here, the  search strategies depend strongly 
on the charged Higgs boson mass.
If $m_{H^\pm} < m_{t} - m_{b}$ (later referred to as a light Higgs boson), 
the charged Higgs boson can be produced in top \mbox{(anti-)}\-quark decay. The main source of 
top (anti-)quarks at the LHC is again $t \bar{t}$ pair production ($\sigma_{t\bar{t}}=850$ pb at 
NLO)~\cite{beneke00}. 
For the whole ($\tan\beta, m_{H^\pm}$) parameter space there is a competition between the $ bW^\pm$ 
and $ bH^\pm$ channels in top decay keeping the sum 
$\mathrm{BR}(t \to b W^+) + \mathrm{BR}(t \to b H^+)$
at almost unity.
The top quark decay to $ bW^\pm$ is however the dominant mode for most of the parameter space. 
Thus, the best way to search for a (light) charged Higgs boson is by requiring that the top
quark produced in the $tbH^\pm$ process decays to a $W^\pm$. 
While in the case of $H^\pm$ decays $\tau$'s will be tagged via their hadronic decay producing low-multiplicity narrow jets 
in the detector, there are two different $W^\pm$ decays that can be explored. The leptonic signature
$ b \bar{b} H^\pm W^\mp \to b \bar{b} \tau \nu l \nu $ provides a clean selection 
of the signal via the identification of the lepton $l=e,\mu$.
In this case the charged Higgs transverse mass cannot be reconstructed because
of the presence of two neutrinos with different origin. In this channel charged Higgs 
discovery will be determined 
by the observation of an excess of such events over SM expectations through a simple counting experiment. In the case of hadronic decays 
$ b \bar{b}H^\pm W^\mp  \to  b \bar{b}\tau \nu jj$ the transverse mass can instead be 
reconstructed since all neutrinos are arising from the charged Higgs boson decay. 
This allows for an efficient separation of the signal and the main 
$t\bar{t} \to b \bar{b}W^\pm W^\mp  \to  b \bar{b}\tau \nu jj$ background
(assuming $m_{H^\pm}\OOrd m_{W^\pm}$). 
The absence of a lepton ($e$ or $\mu$) provides a less 
clean environment but the use of the transverse mass makes it possible to reach the same mass discovery region as 
in the previous case and also to extract the charged Higgs boson mass. Both these channels show that after an 
integrated luminosity of 30 fb$^{-1}$ the discovery could be possible up to a mass of 150 GeV 
for all tan$\beta$ values in both ATLAS and CMS~\cite{Mohn:2007fd,biscarat,abdullin}. 

If the charged Higgs is heavier than the top quark, the dominant decay
channels are 
$H^\pm \to \tau \nu$ and $H^\pm \to tb$ depending on $\tan\beta$. They
have both been studied by  
ATLAS and CMS~\cite{assamagan,kinnunen,salmi,lowette}.
The charged Higgs bosons are produced in the $pp \to tbH^\pm$ channel. For the 
$H^\pm \to tb$ decay, a charged Higgs boson can be discovered up to 
high masses ($m_{H^\pm} \sim 400$~GeV) in the case of very large $\tan\beta$ values and this reach
cannot be much improved because of the large multi-jet environment. For the 
$H^\pm \to \tau \nu$ decay mode this reach is larger due to a cleaner signal despite a 
lower BR. In this case the 5$\sigma$ reach ranges from $\tan\beta=20$ for 
$m_{H^\pm}=200$ GeV to $\tan\beta=30$ for $m_{H^\pm}=400$ GeV.

For the LHC, signal and background events have been simulated in the
same way as for the
Tevatron as described before, however, without implying any rescaling factor to
match a measured $t\bar{t}$ cross section.
Table~\ref{tab:hplus_crosssecLHC} lists 
the resulting cross sections before 
($\sigma^{\rm th}$) and after ($\sigma$) applying the basic cuts 
$p_t^{\rm jets} > 20$~GeV and the hard cut $p_t^{\rm miss} > 100$~GeV.
The LHC rates allow for the discovery to be less challenging than at 
the Tevatron in the region $m_{H^\pm} \sim m_{W^\pm}$, 
yet the separation of  signal events from
background remains crucial for the measurement of the charged Higgs mass.

\begin{table}[htbp]
\vspace*{-0.4cm}
\caption{\label{tab:hplus_crosssecLHC}
LHC cross sections of background $q\bar q, gg \to  t\bar{t}$
and signal $q\bar q, gg \to  tbH^\pm$
for $\tan\beta = 30$ and $m_{H^\pm} = 80, 100, 130$ and $150$~GeV
into the final state
$2 b + 2 j + \tau_\mathrm{jet} + p_t^{\rm miss}$
before ($\sigma^{\rm th}$) and after ($\sigma$)
the basic cuts ($p_t > 20$~GeV for all jets)
and the hard cut ($p_t^{\rm miss} > 100$ GeV).
}
\centering
\begin{tabular}{c|c|c|c|c|c}
 & $q\bar q, gg \to  t\bar{t}$ &
  \multicolumn{4}{c}{$q\bar q, gg \to  tbH^\pm$} \\
$m_{H^\pm}$ (GeV) & 80 & 80 & 100 & 130 & 150 \\
  \hline
$\sigma^{\rm th}$ (pb) & 45.5 &  72.6    &  52.0    &  24.5 & 9.8  \\
$\sigma$ (pb) for $p_t^\mathrm{jets} > 20$ GeV 
   & 17.3 &  33.9   &  25.7    &  12.2  &  3.8 \\
$\sigma$ (pb) for $(p_t^\mathrm{jets},p_t^{\rm miss}) > (20,100)$ GeV
   & 4.6 &   6.0    &   4.8    &  2.9   &  1.2
\end{tabular}
\vspace*{-1mm}
\end{table}

The kinematic distributions 
for $\sqrt{s}=14$~TeV are shown in Ref.~\cite{Hesselbach:2007jj}.
The choice of variables is identical to the one for the Tevatron and
allows for a one-to-one comparison,
the differences being  due to a change in CM energy (and, to a
somewhat lesser extent, due to the leading partonic mode of the 
production process\footnote{As the latter
is dominated by $q\bar q$ annihilation at the Tevatron and $gg$ fusion
at the LHC.}). 
The main differences with respect to the Tevatron case
are that the various transverse momenta and
invariant masses have longer high energy tails. 
In particular, it should be noted
that the effect of the spin differences between $W^\pm$ and $H^\pm$
events can be explored very effectively also at LHC energies,
e.g. the ratio $p_t^{\pi^\pm}/p_t^{\tau_\mathrm{jet}}$ 
which is very sensitive to the spin effects.
These observations lead to the conclusion
that the same method using spin differences can be used to separate
signal from background at both the Tevatron and the LHC.

The distributions of the IDA output variables are shown in Ref.~\cite{Hesselbach:2007jj}
for the study at $\sqrt{s}=14$~TeV for two steps with 90\% efficiency in the first step.
These distributions are qualitatively similar to those for the Tevatron 
The final achievable purity for a given efficiency 
is shown in Fig.~\ref{fig:hplus_lhc_ida}. 
As for the Tevatron energy 
a good separation of signal and background events can be achieved with the spin 
sensitive variables and the IDA method even in case $m_{H^\pm} \sim m_{W^\pm}$.
For heavier $H^\pm$ masses the separation of signal and background events increases due to 
the kinematic differences of the event topology.

\begin{figure}[htbp]
\vspace*{-0.4cm}
\begin{minipage}{0.59\textwidth}
\epsfig{file=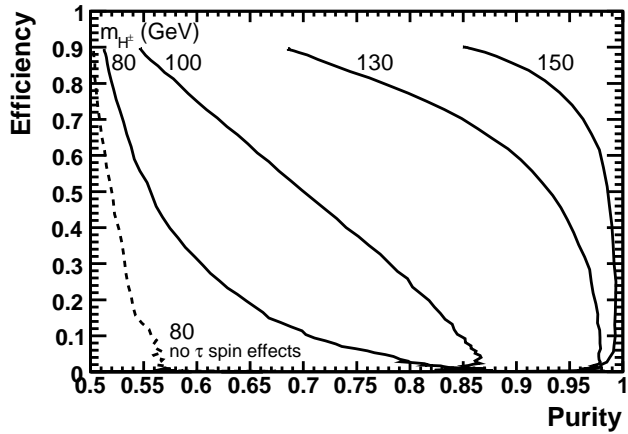,width=\textwidth}
\end{minipage}\hfill
\begin{minipage}{0.39\textwidth}
\caption{
Efficiency as a function of the purity
when not taking the spin effects in the $\tau$ decay into account for
$m_{H^\pm}=80$~GeV (dashed) and with spin effects in the $\tau$ decay for
$m_{H^\pm}=80,100,130,150$~GeV (solid, from left to right).
Results are for the LHC.
}
\label{fig:hplus_lhc_ida}
\end{minipage}\hfill
\vspace*{-0.7cm}
\end{figure}

\subsection{Conclusions}
The discovery of charged Higgs bosons 
would be a clear sign of physics beyond the SM.
In this case study we have investigated charged Higgs boson topologies 
produced at the current Tevatron and LHC energies and compared
them against the irreducible SM background due to top-antitop production
and decay. 
While sizable differences between signal and background
are expected whenever $m_{H^\pm}\ne m_{W^\pm}$, 
near the current mass limit of about $m_{H^\pm}\approx 80$ GeV the
kinematic spectra are very similar between SM decays and
those involving charged Higgs bosons. In this case,
spin information will significantly
distinguish between signal and irreducible SM background. In fact,
we have considered hadronic $\tau\nu_\tau$ decays of charged Higgs bosons, 
wherein the $\tau$ polarization induced by a decaying (pseudo)scalar object
is significantly different from those emerging in the vector ($W^\pm$) decays
onsetting in the top-antitop case. 
For a realistic analysis which is not specific for a particular detector, 
a dedicated Monte Carlo event generation and a simplified multipurpose 
detector response approximation have been applied.
The identification of a hadronic tau-lepton will be an experimental challenge 
in an environment with typically four jets being present. 
We have demonstrated how an IDA method can be an applied to separate signal and background
when the differences between the signal and background distributions are small. 
Our results show that the IDA method 
will be equally effective at both the Tevatron and LHC. 
While only the dominant irreducible $t\bar{t}$ background has been dealt 
with in detail, we have also specifically addressed the QCD background.
A suitably hard missing transverse momentum cut has been applied to
reject such jet activity 
and we have demonstrated that 
although the discriminative power is reduced by such a cut, the reduction is
small compared to the gain from including the $\tau$ polarization effects.
Using the differences in $\tau$
polarization between the signal and the dominant SM irreducible  $t\bar{t}$
background is crucial for disentangling the former
from the latter.

%\vspace*{-2mm}
\section*{Acknowledgements}
We would like to thank the organizers of the 2005 and 2007 editions of the
Les Houches workshops ``Physics at TeV Colliders' (where part of this work
was carried out) and Johan Alwall for fruitful discussions.%

%\bibliography{lh07_hplus}

%\end{document}

\part[CP VIOLATING HIGGS BOSONS]{CP VIOLATING HIGGS BOSONS}

\section[Jet Assignment Studies in the Search for the Decay 
$t \rightarrow b H^+$, 
$H^+ \rightarrow H_{1}^{0} W^+$, 
$H_{1}^{0} \rightarrow b \bar{b}$ 
in the CPX MSSM Scenario]
{JET ASSIGNMENT STUDIES IN THE SEARCH
FOR THE DECAY
$t \rightarrow b H^+$, 
$H^+ \rightarrow H_{1}^{0} W^+$, 
$H_{1}^{0} \rightarrow b \bar{b}$ 
IN THE CPX MSSM SCENARIO~\protect\footnote{
Contributed by: J.E.~ Cole, 
C.H.~Shepherd-Themistocleous,and I.R.~Tomalin}}
\label{sec:jets}
%\documentclass[11pt]{cernrep}
%\usepackage{graphicx,epsfig}
%\bibliographystyle{lesHouches}
%\begin{document}

%\title{Jet assignment studies in the search for the decay $t \rightarrow b H^+$, $H^+ \rightarrow H_{1}^{0} W^+$, $H_{1}^{0} \rightarrow b \bar{b}$ in the CPX MSSM scenario
%with the CMS detector}

%\author{J.E. Cole$^1$, C.H.~Shepherd-Themistocleous$^1$, I.R.~Tomalin$^1$}
%\institute{$^1$STFC Rutherford Appleton Laboratory, Harwell Science and Innovation Campus, Didcot, OX11 0QX, United Kingdom}

%\maketitle

\subsection{Introduction}

The Minimal Supersymmetric Standard Model (MSSM) can have loop-induced CP-violation (CPX) if the Higgsino mass parameter, the gaugino masses and the trilinear couplings are 
complex.  One of the key features of the CPX scenario is the suppression of the couplings of the neutral Higgs boson to both vector boson pairs and to $t \bar{t}$ pairs.
The suppression of the $H_{1}^{0} VV$ coupling effectively dilutes the limits set on the neutral Higgs using LEP data~\cite{Schael:2006cr}, allowing the existence of a light 
neutral Higgs boson ($40$ -- $50$ GeV) and a relatively light charged Higgs boson ($M(H^{\pm}) < M_{top}$) at low $\tan \beta$.  The suppression of the couplings also makes
the usual search methods at hadron colliders unviable.  However, the suppression of the $H_{1}^{0} VV$ leads to the enhancement of the $H_1 H^+ W^-$ coupling via a sum rule, 
making $t \bar{t}$ production events in which one of the top quarks decays via $t \rightarrow b H^+$, $H^+ \rightarrow H_{1}^{0} W^+$, $H_{1}^{0} \rightarrow b \bar{b}$ one
of the most promising search channels for the CPX scenario~\cite{Ghosh:2004cc}.

We present here a study of mass reconstruction and the impact of jet misassignment on this search using the CMS detector; A feasibility study for discovering the Higgs bosons 
in the CPX scenario also using the CMS detector is presented in Section \ref{sec:cpx}

\subsection{Event generation}

The signal event sample was generated using PYTHIA~\cite{Sjostrand:2006za} and assuming the following parameters: $M(H^{0}_{1} = 51$ GeV, $M(H^{\pm}) = 133$ GeV, 
$M_t = 175$ GeV, $\tan \beta = 5$ and $\Phi_{CP} = 90^{\circ}$.  In each event, one of the top quarks was forced to decay in the usual way, ie. $t \rightarrow b W$, while
the other was forced to decay via $t \rightarrow b H^+$, $H^+ \rightarrow H_{1}^{0} W^+$, $H_{1}^{0} \rightarrow b \bar{b}$.  All possible $W^{\pm}$ decays were allowed.
The relevant branching fractions were calculated using CPSuperH~\cite{Lee:2003nta} and were found to be:  $BR(t \rightarrow b H^+) = 0.01$, 
$BR(H^+ \rightarrow H_{1}^{0} W^+) = 0.99$ and $BR(H_{1}^{0} \rightarrow b \bar{b}) = 0.92$.  Taking the total $t \bar{t}$ production cross section to be 
$840$ pb~\cite{Beneke:2000hk}, this gives a cross section for this process of $8.68$ pb.

For the purposes of this study only the subset of signal events in which one $W^{\pm}$ decayed hadronically and the other decayed leptonically (electron or muon) were 
considered, as this is the experimental signature that will be used to identify events in this analysis.

\subsection{Event selection and mass reconstruction}

This study was performed using only generator-level information.  The iterative cone (IC) algorithm~\cite{DellaNegra:922757} with a radius of 0.5 was used for jet 
identification.  The jets are formed out of stable generator-level particles, although neutrinos and muons are explcitly excluded from the process.  Six or more jets must be 
found using the IC algorithm that satisfy the following requirements:  $p^{jet}_{T} > 20$ GeV and $|\eta_{jet}| < 2.4$.  Three of more must also satisfy 
$p^{jet}_{T} > 30$ GeV.  In addition, an electron or muon that satisfies $p^{l}_{T} > 20$ GeV and $|\eta_{l}| < 2.4$ must also be present and the missing $E_T$ reconstructed
from generator-level particles must be greater than $20$ GeV.

Events that pass these selection requirements then undergo the mass reconstruction procedure.  The events are searched for the two possible decay channels,
namely, $t \rightarrow b q \bar{q}^{\prime}$, $t \rightarrow bbbl \nu$ and $t \rightarrow bl \nu$, $t \rightarrow bbbq \bar{q}^{\prime}$ $+$ (c.c.).  As the study presented here
is performed using generator-level information, the true lepton and neutrino from the leptonically-decaying $W^{\pm}$ are used.  This means that during the mass reconstruction
procedure, the $W^{\pm}$ four-vector is calculated simply by summing the lepton and neutrino four-vectors.  When the $W^{\pm}$ decays hadronically, the mass is reconstructed
using jets and must lie within $25$ GeV of the nominal value.  The corresponding mass constraint is placed on all reconstructed top masses.  When reconstructing both of the 
top masses from a given jet combination, the jet associated with the b-quark ($t \rightarrow b W^{\pm}$ or $t \rightarrow bH^{\pm}$) must satisfy $p_T > 40$ GeV.  A number of 
jet combinations will pass these requirements in each event and therefore the best candidate for a given event is selected by the minimization of a $\chi^2$ based on the top 
masses and the mass of the hadronically-decaying $W^{\pm}$ candidate.

It should be noted that this mass reconstruction procedure results in three possible jet combinations associated with the best candidate $\chi^2$.  This is because the jets 
associated to the three b-quarks produced in the $t \rightarrow bbbW^{\pm}$ decay can be swapped around, but still give the same top mass value.  However, the stricter jet
$p_T$ requirement applied to the jet associated to the b-quark from the $t \rightarrow bH^{\pm}$ decay can cause one or possibly two of the three combinations to be rejected 
before the $\chi^2$ calculation is performed.  All the combinations corresponding to the best candidate $\chi^2$ that also satisfy the stricter jet $p_T$ requirement will be 
used when making the mass distributions.

\subsection{Mass reconstruction studies}
\label{cpvHiggs-mass-rec-studies}
Before attempting to reconstruct masses at the detector level, it is important to understand whether good mass reconstruction is possible.  This is done by identifying
the jets associated to the quarks produced in the decay channel (these quarks are hereafter referred to collectively as ``decay quarks'') and reconstructing the masses from 
these jets.

The association of jets with the decay quarks is done using two possible matching procedures:  Angular matching, in which the quantity 
$\Delta R = \sqrt{\Delta \eta^2 + \Delta \phi^2}$ is used to determine a unique set of jet-parton matches; or jet constituent matching, in which the
particles assigned to a given jet are classified according to the top quark decay from which they originated.  The fraction of the transverse momentum of a given jet, 
$p_{T}^{jet}$, carried by the constituents originating from each decay quark can then be determined and used to create a unique set of jet-parton matches.

\begin{figure}
\begin{center}
\mbox{
\includegraphics[width=0.35\textwidth]{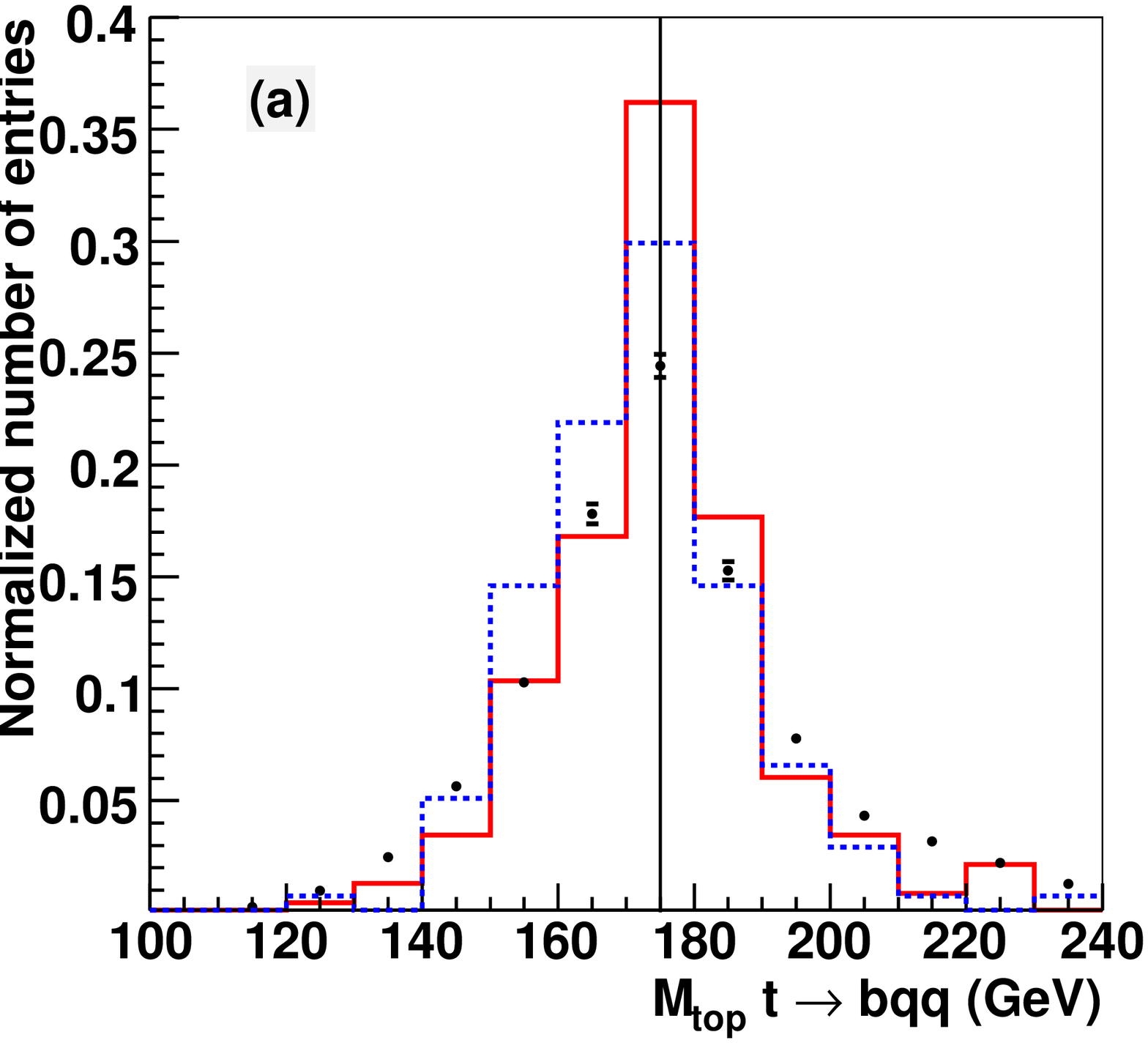}
\includegraphics[width=0.35\textwidth]{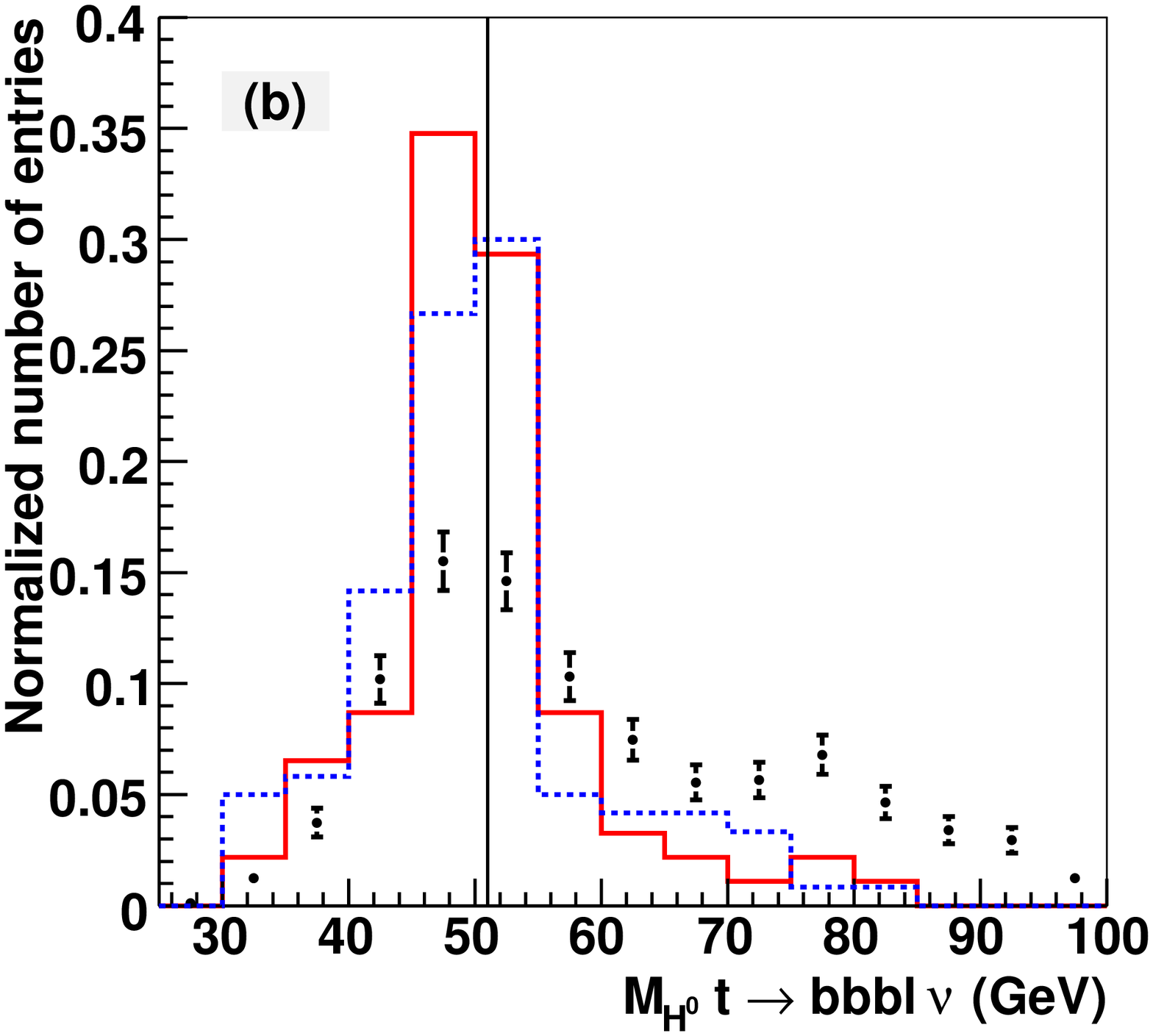}}
\caption{The top mass distribution from the decay $t \rightarrow bq \bar{q}^{\prime}$ and the $M(H^{0}_{1})$ distribution from the decay 
$t \rightarrow bbbl \nu$ reconstructed from angular-matched jets.  All distributions are made with angular-matched jets that satisfy $\Delta R < 0.5$.
The dashed histograms, in addition, have the top and $W^{\pm}$ mass constraints applied, while the solid lines have the $\Delta R$ requirements tightened
on for the decay products of the $W^{\pm} \rightarrow q \bar{q}^{\prime}$ and the $H^{0}_{1} \rightarrow b \bar{b} $.}
\label{cpvHiggs_fig1}
\end{center}
\end{figure}

Figure~\ref{cpvHiggs_fig1} shows the top mass distribution from the $t \rightarrow bq \bar{q}^{\prime}$ decay and the $H^{0}_{1}$ mass distribution from the
$t \rightarrow bbbl \nu$ decay reconstructed using angular-matched jets.  The points correspond to those made using only jets that satisfy
$\Delta R < 0.5$ and it can be seen that in both cases a clear peak is visible in the correct position, although the $H^{0}_{1}$ mass has a noticeable
high mass tail.  The dashed lines represent the distributions after some mass constraints have been applied:  in the case of the top mass from the 
$t \rightarrow bq \bar{q}^{\prime}$ decay, the light-quark jet pair must give $W^{\pm}$ mass within $25$ GeV of the nominal value, while the mass from the corresponding 
$t \rightarrow bbbl \nu$ decay must lie within $25$ GeV of the nominal value.  In the case of the $H_{1}^{0}$ mass distribution both top masses and the hadronically-decaying
$W^{\pm}$ must lie within $25$ GeV of their nominal values.  These mass constraints reduce slightly the high mass tail on the $H^{0}_{1}$ mass distribution.  The solid lines do
not have the mass constraints applied, but instead the $\Delta R$ requirement on the decay products of the $H_{1}^{0}$ and the hadronically-decaying $W^{\pm}$ boson have been 
tightened to $\Delta R < 0.1$.  This all but removes the high mass tail on the neutral Higgs mass distribution, suggesting that the tail is caused by problems in the 
jet-parton matching procedure.

Given the large number of jets in these events, the most likely reason for having problems with jet-parton matching (and potentially more generally with mass reconstruction)
is that the jets tend to overlap with each other.  This can be verified using the $p_{T}^{jet}$ fractions used for jet constituent matching.  These fractions are determined by
tracing all the particles associated to a given jet back to the top quark decay they came from.  The transverse momenta of the particles associated to a given decay quark are
then summed and the result divided by the jet transverse momentum, resulting in six $p_T$ fraction values per jet.

\begin{figure}
\begin{center}
\mbox{
\includegraphics[width=0.35\textwidth]{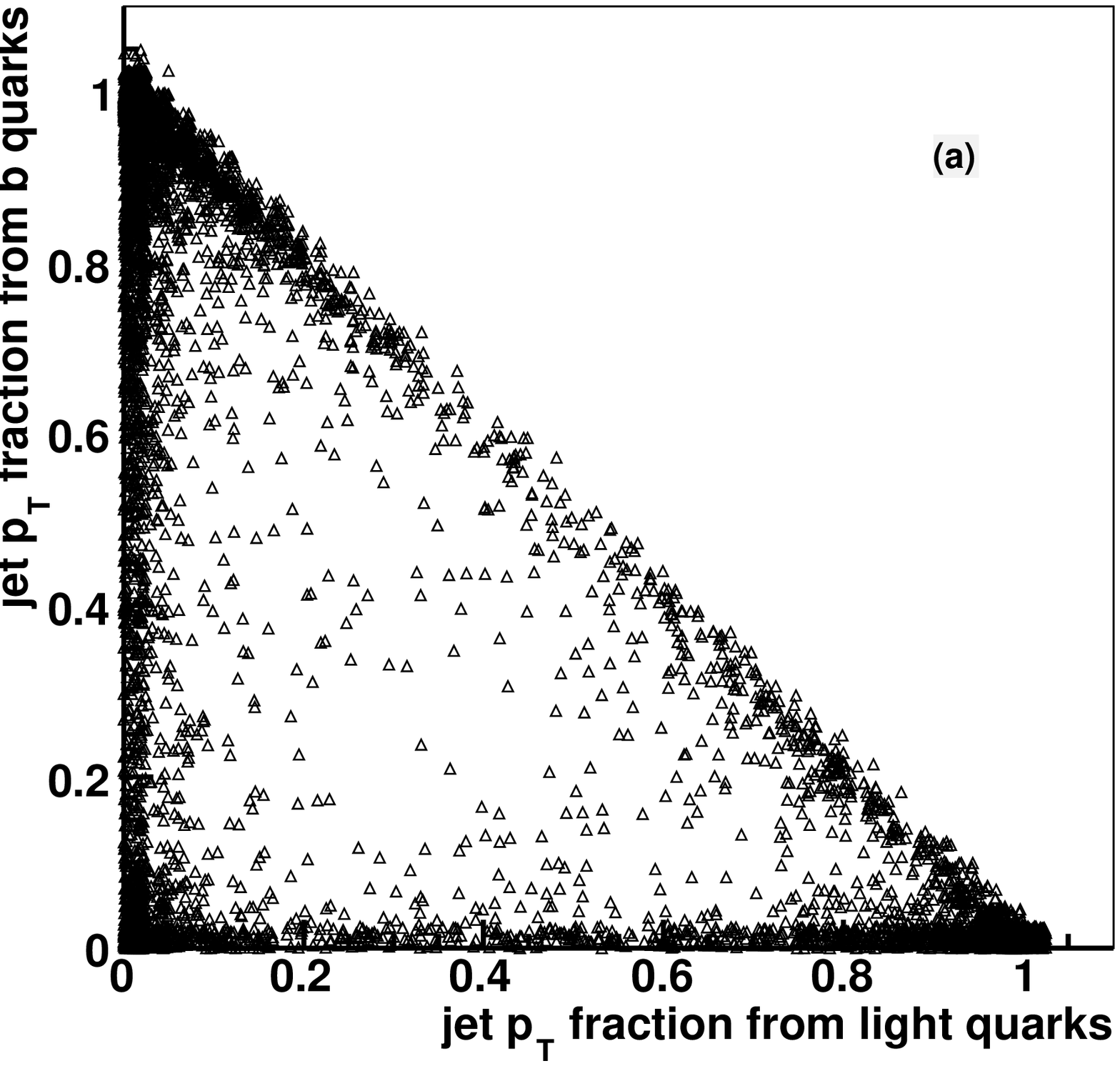}
\includegraphics[width=0.35\textwidth]{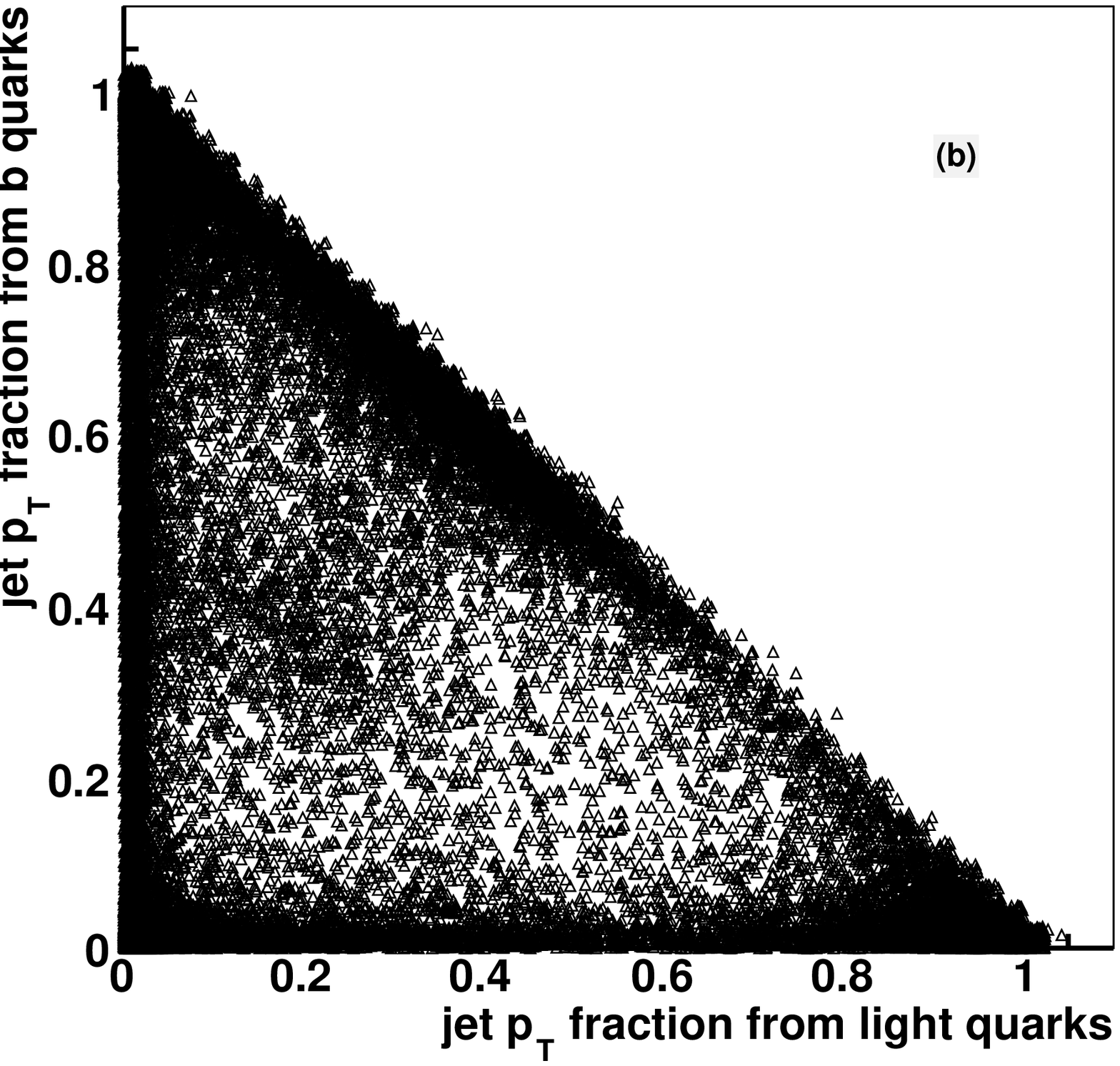}}
\caption{(a) A comparison of the jet transverse momentum fractions for constituents coming from the b-quark and the light quarks in the decay 
$t \rightarrow b q \bar{q}^{\prime}$ and (b) A comparison of the jet transverse momentum fractions for consituents coming from the decay products of the neutral
Higgs and the hadronically-decaying $W^{\pm}$ in the decay channel $t \rightarrow bbbq \bar{q}^{\prime}$.}
\label{cpvHiggs_fig3}
\end{center}
\end{figure}

Figure~\ref{cpvHiggs_fig3}(a) compares the $p^{jet}_{T}$ fractions for all jets with particles associated to the b-quark and to either of the
light quarks in the decay $t \rightarrow b q \bar{q}^{\prime}$.  Figure~\ref{cpvHiggs_fig3}(b) shows the equivalent distribution for the decay channel 
$t \rightarrow bbbq \bar{q}^{\prime}$, but compares the the fractions for all jets with particles associated to the $H^{0}_{1}$ decay products and the
light quarks coming from the $W^{\pm}$ decay.  No jet angular matching has been applied.  The two combinations are chosen because they represent 
the jets from decay quarks that are expected to be closest to each other.  
In the case of the SM top decay, the distribution shows that the jets are well separated, as the values are concentrated at very high or low values.  In the case of the 
$t \rightarrow bbbq \bar{q}^{\prime}$ decay, it is clear that the jets overlap significantly, as suspected.

\subsubsection{Jet assignment studies}

Although jet overlapping has been identified as a potential problem for mass reconstruction, the results in section~\ref{cpvHiggs-mass-rec-studies} show that it is
basically possible to reconstruct reasonable mass distributions.  However, the impact of jet misassignment on the mass distributions must also be understood and ways found
to minimize its effect.  Jet misassignment arises from two different sources:  the misassignment of jets associated to the 
decay quarks and the misassignment of jets associated to other hard partons in the event, for example, gluons from initial state radiation or produced 
during parton showering.  The contribution from these two sources can be studied by comparing the mass distributions from three different reconstruction 
procedures:  those produced using jets matched to the decay quarks (``fully-matched''), those produced using the subset of jets matched to the decay quarks, 
but without using the knowledge about which jet belongs to which quark, (``partially-matched'') and those produced using the standard mass reconstruction procedure 
(``unmatched'').  Comparisons of ``fully-matched'' and ``partially-matched'' distributions provide information about the misassignment of jets from decay quarks, while 
comparisons of ``partially-matched'' and ``unmatched distributions'' provide information about the misassignment of jets from other hard partons.

\begin{figure}
\vspace{-2.0cm}
\begin{center}
\includegraphics[width=0.7\textwidth]{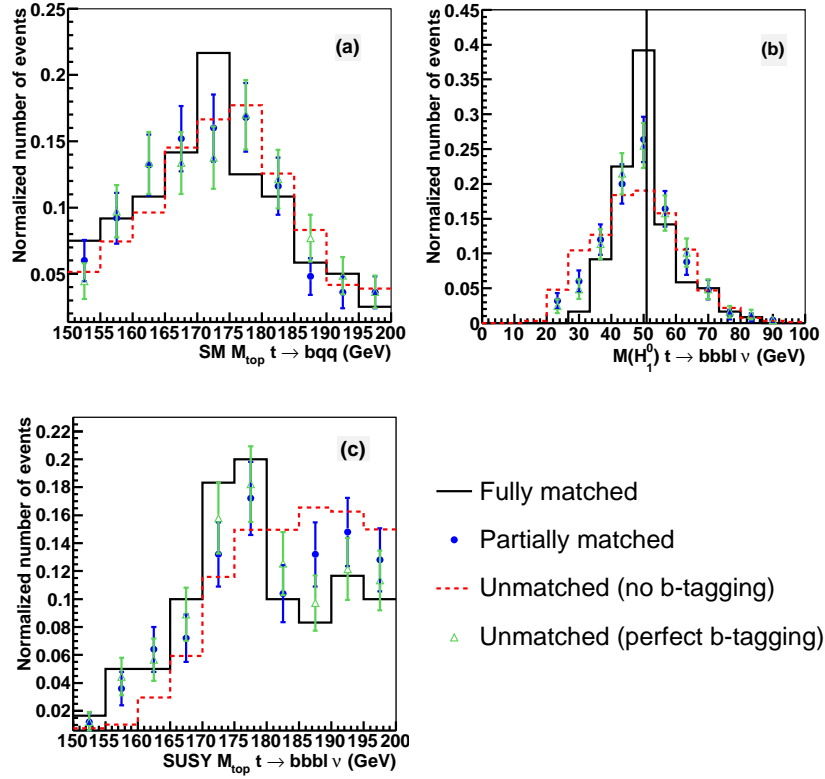}
\caption{A comparison of the fully-matched, partially-matched and unmatched (with and without b-tagging) mass distributions: (a) The top mass distribution from the decay 
$t \rightarrow b q \bar{q}^{\prime}$, (b) the $H_{1}^{0}$ mass distribution from the decay $t \rightarrow bbbl \nu$ and (c) the top mass distribution from the 
$t \rightarrow bbbl \nu$ decay.}
\label{cpvHiggs_fig5}
\end{center}
\end{figure}

Figure~\ref{cpvHiggs_fig5} shows the comparison of these three reconstruction methods for the top mass distribution from the $t \rightarrow b q \bar{q}^{\prime}$ decay and
the $H^{0}_{1}$ mass distribution and the corresponding top mass distribution from the $t \rightarrow bbbl \nu$ decay.  The three methods for the top mass from the SM top 
decay are broadly in agreement, indicating that the reconstruction procedure is working well.  The $H^{0}_{1}$ mass distribution shows differences between all three methods, 
indicating that there are contributions from both sources of misassigned jets.  However, in the case of the top mass distribution from the same decay the only difference 
is between the partially-matched and unmatched versions, ie. the contribution from the misassignment of jets associated to the decay quarks has disappeared.  This indicates
that the misassigned decay-quark jets observed in the $H^{0}_{1}$ mass distribution come from within the $t \rightarrow bbbl \nu$ decay chain.  The high mass tail observed on 
the unmatched top mass distribution is therefore partially caused by the misassignment of jets associated to other hard partons.  The remainder of the high mass tail, ie. the 
contribution that is also observed in the fully-matched distribution, is caused by overlapping jets, as discussed in section~\ref{cpvHiggs-mass-rec-studies}.

One possible method of improving the jet assignment during the mass reconstruction procedure is to use b-tagging.  To study what impact it may have 
at generator level, ``perfect'' b-tagging can be used.  Perfect b-tagging means using only jets that have been matched to one of the b-quarks if a 
b-tagged jet is required, while only jets not matched to a b-quark are used when a light-quark jet is required.  Perfect b-tagging has been applied 
to the unmatched distributions, as shown in Fig.\ref{cpvHiggs_fig5}, and it can be seen that the differences between the partially-matched and unmatched 
distributions are eliminated for all the mass distributions.  This is consistent with the conclusion that this difference is a result of the 
misassignment of jets associated to other hard partons in the event, as the other hard partons are more likely to be gluon or light-quark jets than b-quark jets.

\subsection{Conclusions}

A study of jet reconstruction and assignment has been performed at generator level for the analysis of CP-violating Higgs production at LHC via the decay
channel $pp \rightarrow t \bar{t} X$, $t \rightarrow b W^{\pm}$, $t \rightarrow bH^{\pm}$, $H^{\pm}\rightarrow H_{1}^{0} W^{\pm}$, 
$H_{1}^{0} \rightarrow b \bar{b}$.  It has been established that it is possible to reconstruct reasonable mass distributions for this decay channel, but studies
of jet-parton matching show that overlapping jets are a significant problem for the supersymmetric top decay.  This results in a high mass tail on the top mass 
distributions reconstructed from the $t \rightarrow bbbW^{\pm}$ decay.

Jet assignment has also been studied for this decay channel and it has been found to be good for the mass distributions reconstructed using the 
Standard Model top decay channels.  However, in the case of the supersymmetric top decay, the Higgs mass distributions show that there are contributions from both
the misassignment of jets associated to other decay quarks and from jets associated to other hard partons in the event.  However, only the latter 
contribution is observed in the corresponding top mass distributions, indicating that it is jets associated to the supersymmetric top decay that are being
misassigned, not those from the SM top decay.  The misassignment of jets from other hard partons also results in a high mass tail on the top mass 
distributions.  The use of perfect b-tagging (based on jet-parton matching) suppresses this effect.  This is consistent with the assumption that the other
hard partons come from initial state gluon radiation or parton showering, as in this case the misassigned jets are much more likely to be gluon- or light
quark-initiated jets.

It may be possible to reduce the impact of overlapping jets on the mass distributions by using a smaller jet cone radius or by using another jet finder, such as the
$k_t$ algorithm~\cite{Butterworth:2002xg,Ellis:1993tq}.  The impact of detector-level jet finding and lepton identification must also be assessed.

%\section*{ACKNOWLEDGEMENTS}

%We would like to thank R.~Godbole, D.P.~Roy and particularly D.~Ghosh for useful discussions and for the calculations and code they have provided.  We would also like to thank 
%our CMS colleagues, A.~Nayak and A.~Nikitenko, for discussions about detector-level reconstruction issues.  Thanks also to M.~Schumacher and M.~Lehmacher from the ATLAS
%experiment for useful discussions and documentation.  Finally, we would like to thank the organizers of the workshop for their help and support.%

%\bibliography{cpvHiggs}
%\end{document}

\section[Search for the $t \to bH^{+}$, $H^{+} \to H_{1}W$, $H_{1}\to b\bar{b}$
       channel in CPX MSSM scenario in CMS]
{SEARCH FOR THE  $t \to bH^{+}$, $H^{+} \to H_{1}W$, $H_{1}\to b\bar{b}$
       CHANNEL IN  CPX MSSM SCENARIO IN CMS
       ~\protect\footnote{Contributed by: A. K. Nayak, T. Aziz, and A. 
Nikitenko}
       ~\protect\footnote{Results are preliminary and must not be shown 
         at conferences}}
\label{sec:cpx}
%\documentclass[11pt]{cernrep}
%\usepackage{graphicx,epsfig}
%\usepackage{cite}
%\bibliographystyle{lesHouches}
%\begin{document}
%\title{Search for the $t \to bH^{+}$, $H^{+} \to H_{1}W$, $H_{1}\to b\bar{b}$
%       channel in CPX MSSM scenario in CMS
%       ~\footnote{results are preliminary and must not be shown 
%         on the conferences}}
%\author{A. K. Nayak$^1$, T. Aziz$^1$, A. Nikitenko$^2$}
%\institute{$^1$Tata Institute of Fundamental Research, Mumbai,India \\
%$^2$ Imperial College, London, on leave from ITEP, Moscow, Russia}
%\maketitle
%%\begin{abstract}
%\end{abstract}

\subsection{Introduction}

CP violation (CPX) in the Higgs sector of the Minimal Supersymmetric Standard Model (MSSM), 
when the Higgsino mass parameter $\mu$, the gaugino mass parameters $M_{i}$ and 
the trilinear couplings $A_{f}$ are complex, allows the existence of the light neutral Higgs boson
($m_{H_{1}} \leq$ 50 GeV) and relatively light charged Higgs boson 
($m_{H^{+}} \le m_{t}$) in low tan$\beta$ region not excluded by the LEP 
data because of the reduction of $H_{1}ZZ$ coupling \cite{Schael:2006cr}. 
In CPX scenario the usual search channels may not be useful, because 
of the simultaneous reduction in the couplings of the Higgs boson to the vector 
boson pair and to the top quark pair, as it affects the Higgs boson 
production and decays rates. The one of the promising search channels in 
the CPX scenario proposed in \cite{Ghosh:2004cc} is the $t \bar{t}$ production 
when one of the top quarks decays as 
$t \to bH^{+}$, $H^{+} \to H_{1}W$, $H_{1}\to b\bar{b}$. It is due to
the suppression of the $H_{1}ZZ$ coupling leads to the enhancement of the 
$H^{+}W^{-}H_{1}$ coupling in order to satisfy the coupling sum-rule. 
We investigated a feasibility for the discovery of the Higgs bosons in 
this channel using the full CMS detector simulated data. The results
shown are preliminary.

\subsection{Event generation}

The signal events were generated using PYTHIA \cite{Sjostrand:2006za} 
with $m_{t}$=175 GeV, $m_{H_{1}}$=51 GeV and $m_{H^{+}}$=133 GeV, 
corresponding to tan$\beta$=5 and CP mixing angle $\Phi$(CP)=$90^0$ in 
the CPX MSSM. The following decays were forced in PYTHIA:
$t_{1} \to bW$, $t_{2} \to bH^{+}$, $H^{+} \to WH_{1}$, $H_{1} \to b\bar{b}$ 
and both $W$ bosons from the top decays were allowed to decay into all possible 
modes. The decay branching fractions were calculated using CPsuperH 
program~\cite{Lee:2003nta}. The total cross section was calculated taking 
the next-to-leading order cross section for an inclusive $t\bar{t}$ 
production 840 pb ~\cite{Beneke:2000hk} and multiplying by the branching 
ratios, Br($t \to bH^{+}$)=0.01, Br($H^{+} \to H_{1}W$)=0.567, 
Br($t \to bW$)=0.99, Br($H_{1} \to b \bar{b}$)=0.92 which gives the 
cross section 8.68 pb.

The major background processes for this channel are  $t\bar{t} + \rm jets$ 
and $t\bar{t}b\bar{b}$. The $t\bar{t} + \rm jets$ background  
was generated using ALPGEN~\cite{Mangano:2002ea} with the MLM prescription 
for jet-parton matching~\cite{Mangano:2006rw,Hoche:2006ph} at the PYTHIA 
shower simulation. The $t \bar{t}+\rm 2~jets$ (exclusive), 
$t \bar{t}+\rm 3~jets$ and $t \bar{t}+\rm 4~jets$(inclusive) with 
jet $p_{T}>$20 GeV were generated. The cross sections for these processes 
are shown in Table~\ref{table1}.  The $t\bar{t}b\bar{b}$ background was not 
considered yet in this study.

\subsection{Simulation and Reconstruction}

The CMS detector was simulated using full GEANT4 \cite{Agostinelli:2002hh} 
simulation and the reconstruction was done using the CMS simulation and 
reconstruction software CMSSW. No pileup events were included. We summarize
briefly the object reconstruction methods~\cite{ptdr1:2006cms} used in this 
analysis. Muons are reconstructed from the muon chambers and the silicon tracker and 
electrons are reconstructed from the tracks in the silicon tracker and the
clusters in the electromagnetic calorimeter. The loose electron identification 
criteria were applied. The lepton isolation was done using the tracker isolation such 
that leptons are selected if sum $p_{T}$ of the tracks in a cone around the lepton 
(inner radius 0.015 and outer radius 0.25) is less than 3 GeV.  The jets were 
reconstructed from the calorimeter towers using an iterative cone algorithm with the
cone size 0.5. The jet energy was corrected using the Monte Carlo jet energy corrections. 
The missing $E_{T}$ was reconstructed from the calorimeter towers and corrected for the 
jet energy scale. The missing $E_{T}$ was also corrected for the muons by adding the muon 
momenta to the calorimeter missing $E_{T}$.

\subsection{Event selection}

\subsubsection{Primary selections}

The final state considered in this analysis consists of two light quarks, four b quarks,
one lepton and neutrino: $\ell\nu+qq'+b\bar{b}b\bar{b}$. Since the neutral Higgs boson $H_{1}$ is very light 
(51 GeV), the b quarks from the $H_{1} \to b \bar{b}$ decay are very soft as seen in 
Fig.\ref{cpvh1} (a,b). Only $\simeq 36\%$ of events have both b quarks from the $H_{1}$ decay 
with $p_{T}^{b}>$20 GeV. The final state quarks in the event fall very close to each other 
in ($\eta, \phi$) space.  Fig.~\ref{cpvh1} (c) shows the separation in ($\eta, \phi$) space 
between two closest quarks. Because of these reasons it is difficult to reconstruct six jets 
in the event corresponding to the six final states quarks.
\begin{figure}[htbp]
\begin{center}
\includegraphics[width = 5cm, height = 5cm]{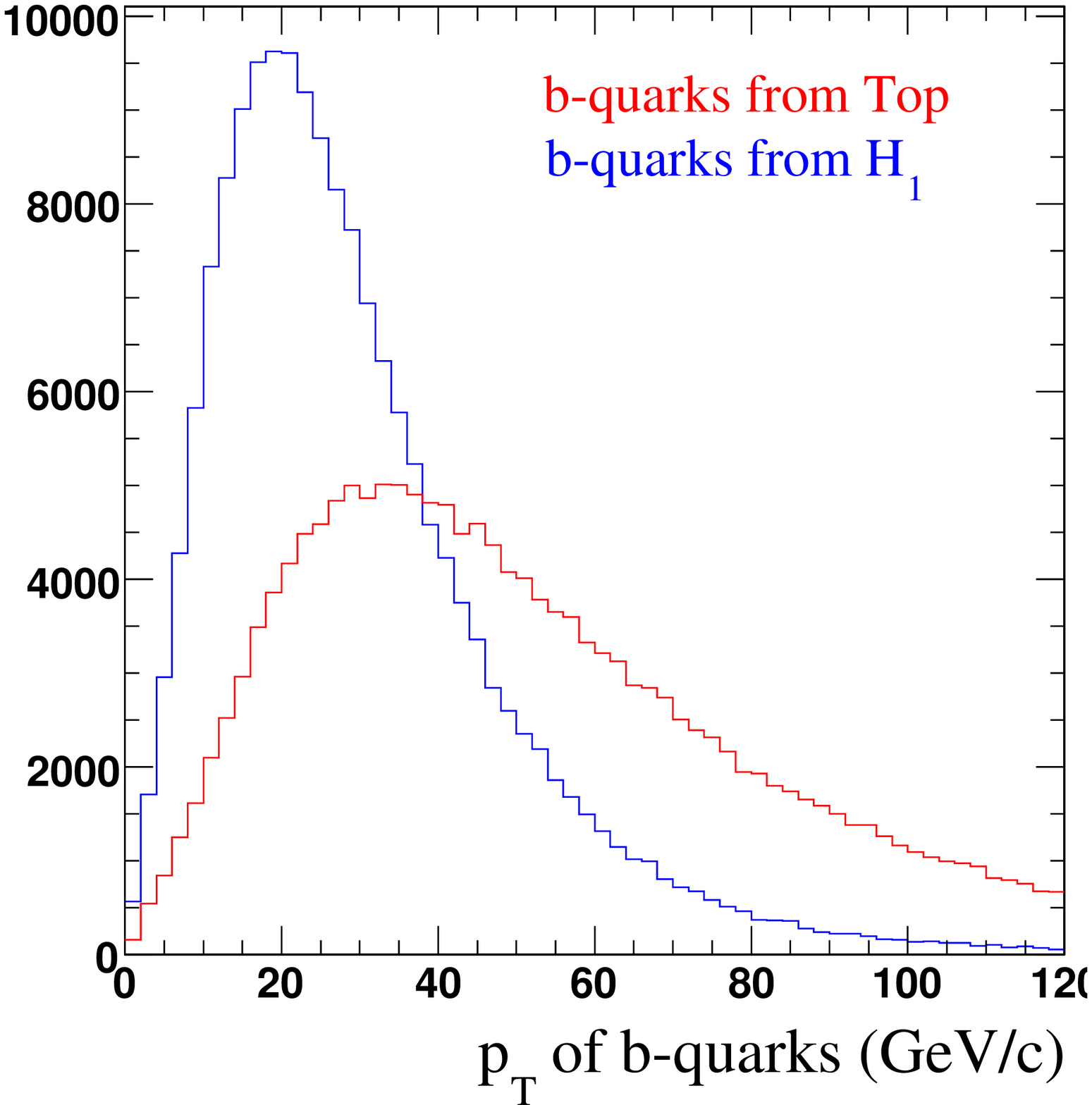}\includegraphics[width = 5cm, height = 5cm]{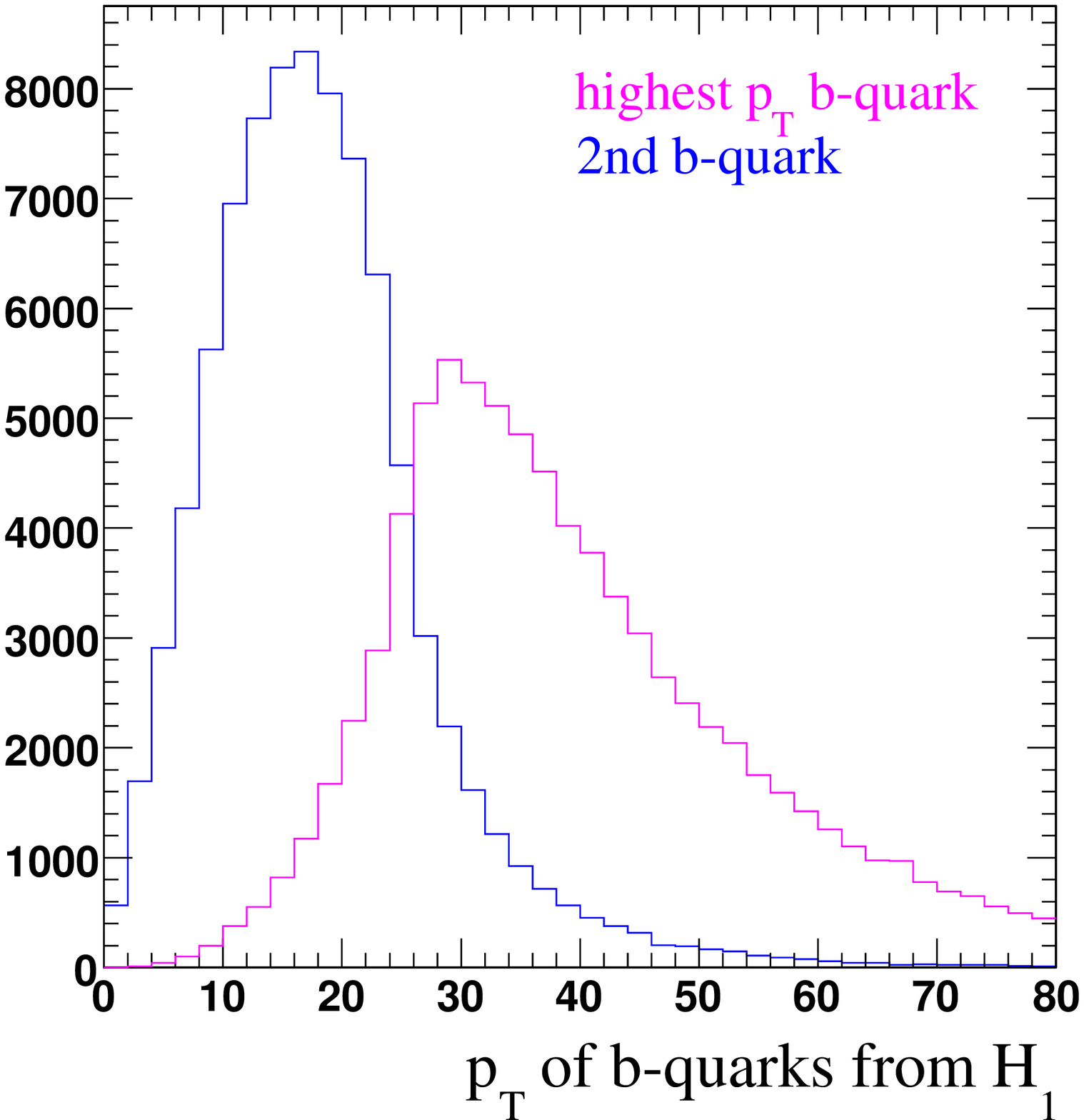}\includegraphics[width = 5cm, height = 5cm]{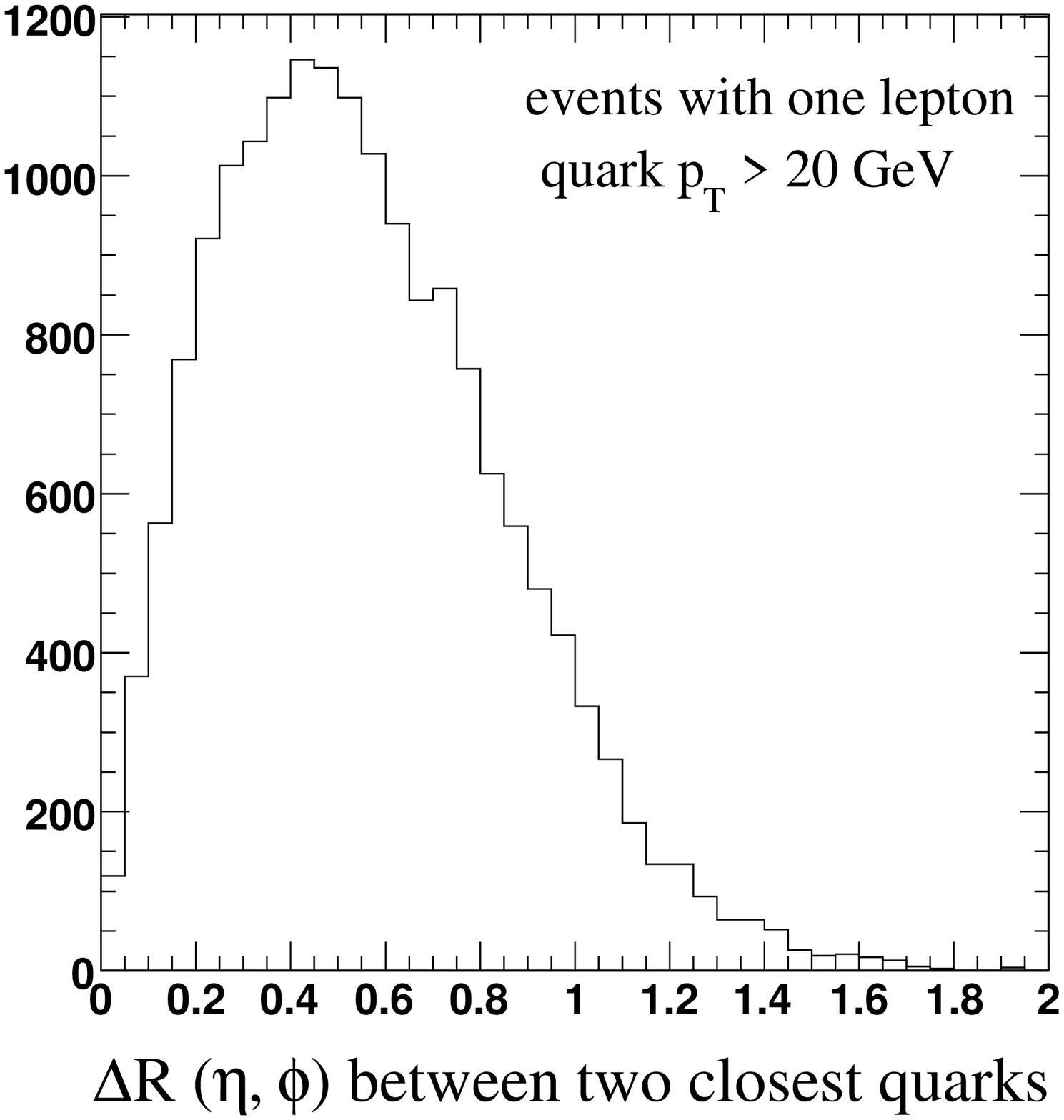}
$\mbox{(a)} \hspace{4.5cm}  \mbox{(b)} \hspace{4.5cm} \mbox{(c)}$
\caption{(a) $p_{T}$ distribution of b quarks from top quark and from $H_{1}$ decays, 
         (b) $p_{T}$ distribution of b quarks from $H_{1}$ decay, 
         (c) the $\Delta R$ separation in ($\eta, \phi$) space between two closest final state 
             quarks in the event.}
\label{cpvh1}
\end{center}
\end{figure}
The events with one isolated lepton with $p_{T}>$20 GeV and six or more jets with $E_{T}>$20 GeV
were selected.  The number of leptons in the event (electrons with $p_{T}>$ 10 GeV and
muons with $p_{T}>$5 GeV) passing the identification and the isolation were counted and
the events with more than one lepton were rejected. The jets were b tagged using the track counting 
b-tagging algorithm. The three dimensional impact parameter significance of the second highest 
significance track in the jet was used as the b-discriminator parameter. The four highest 
discriminator jets with the discriminator value greater than 2.95 were tagged as b jets. 

\subsubsection{Top mass reconstruction}

One $W$ boson in the event was reconstructed from the lepton and the missing $E_{T}$. 
The z-component of the missing energy was calculated using the $W$ mass constraint. 
This yields the real solutions in nearly 66\% events. The events with the imaginary solutions 
were rejected. There are two possible solutions for the z-component of the missing energy which gives 
two possible candidates for the leptonically decaying $W$ boson. The $W$ boson decaying hadronically 
was reconstructed from the jets not tagged as b jets. All jet pairs with the invariant mass within 
the $m_{W}\pm$ 20 GeV mass window were considered as possible candidates. 
The di-jets invariant mass for the jets matching to quarks from the $W$ boson decay is shown in 
Fig.\ref{cpvh2} (a). 
\begin{figure}[h!]
\begin{center}
\includegraphics[width = 5cm, height = 5cm]{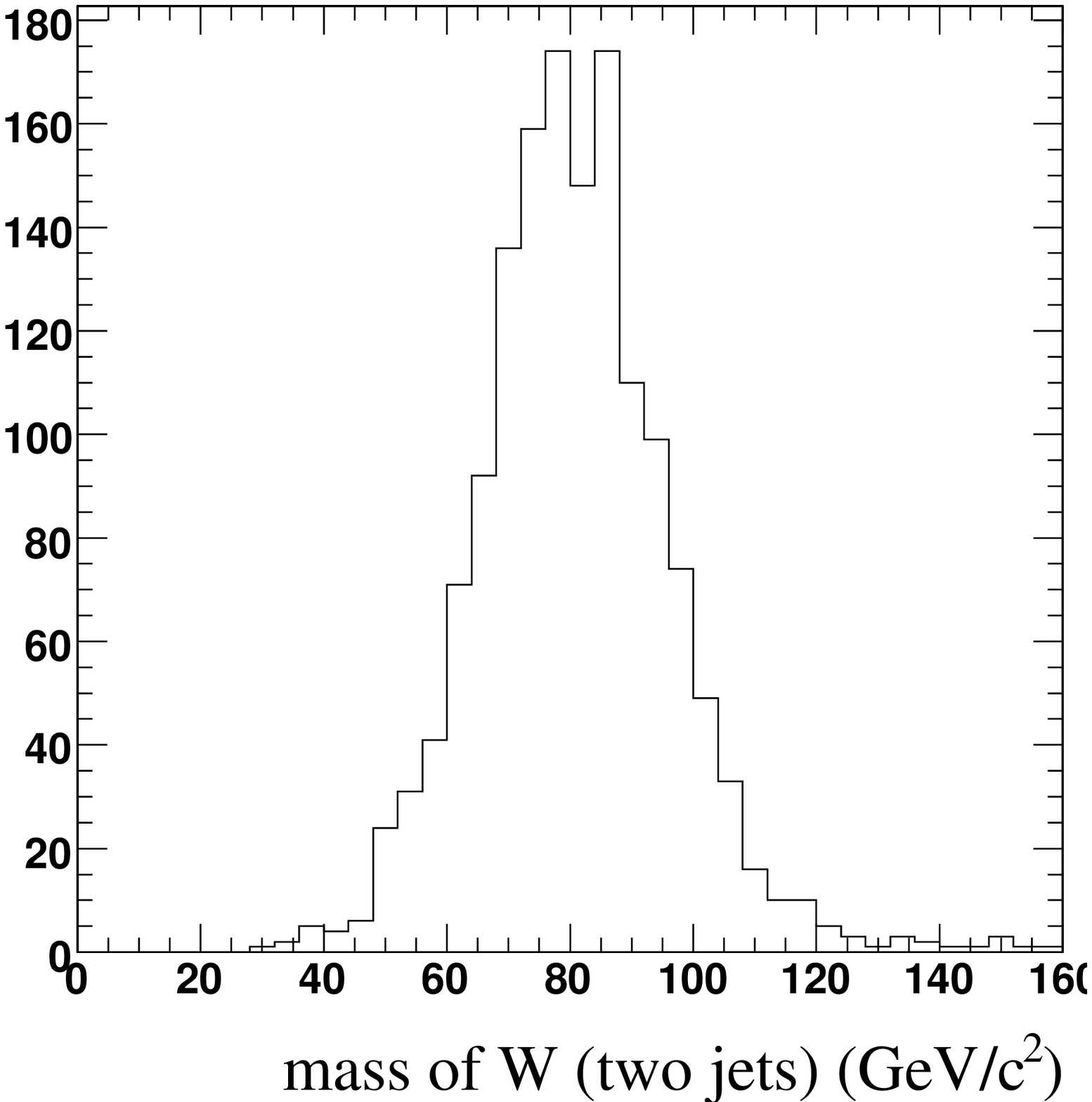}
\includegraphics[width = 5cm, height = 5cm]{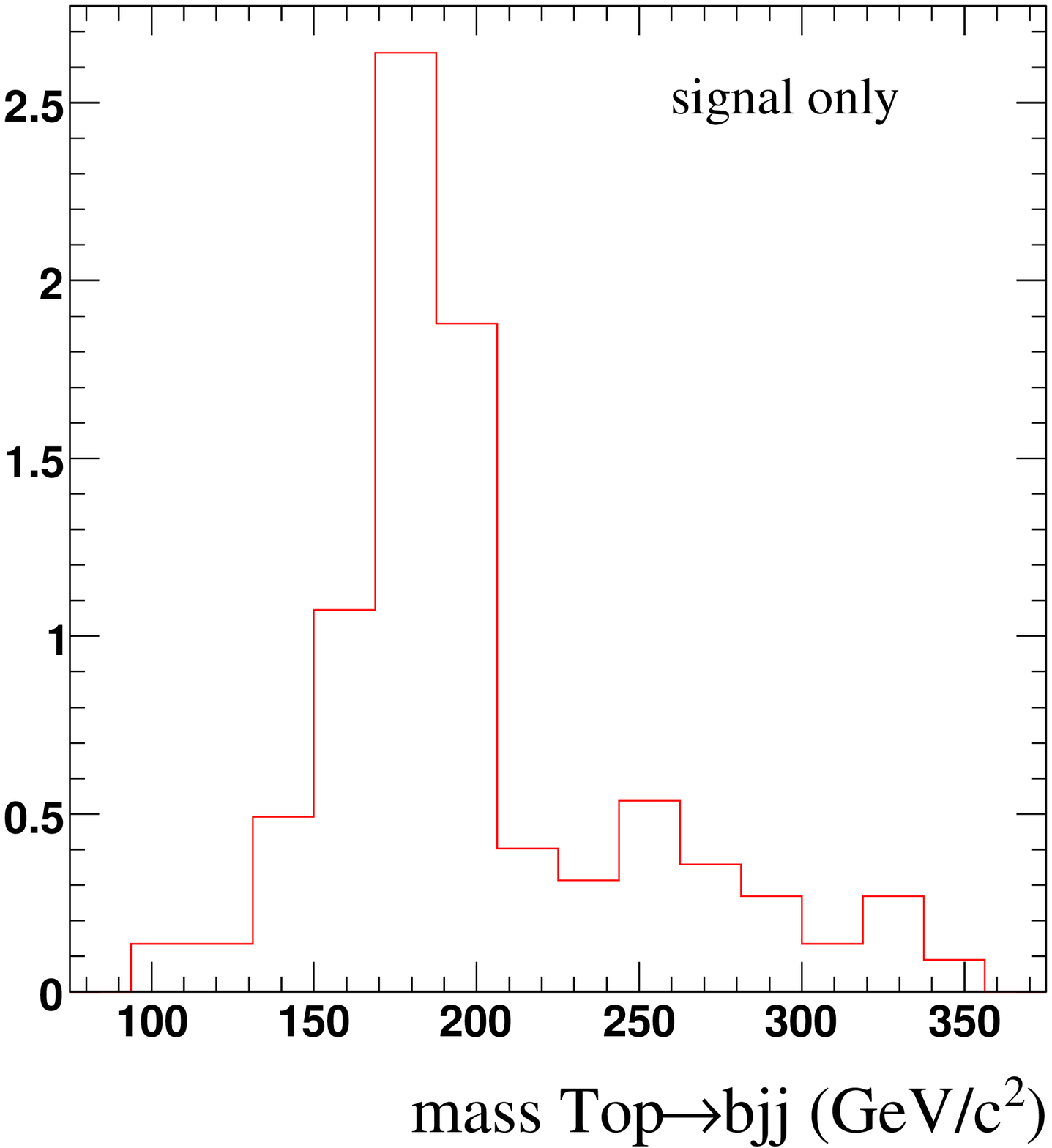}
\includegraphics[width = 5cm, height = 5cm]{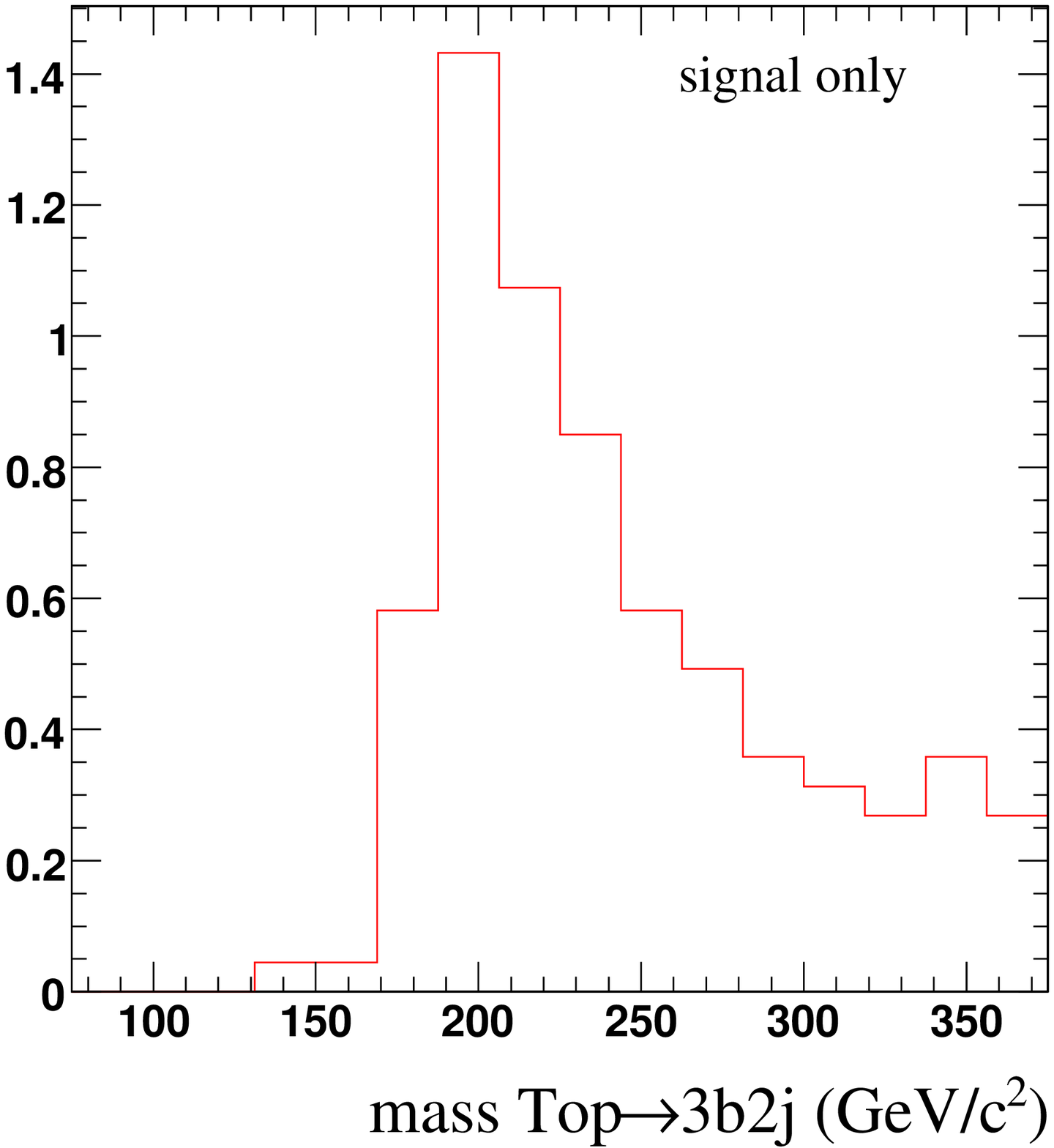} \\
$\mbox{(a)} \hspace{4.5cm}  \mbox{(b)}\hspace{4.5cm}  \mbox{(c)}$
\caption{(a) the di-jet invariant mass of the jets matching to quarks from $W$ decay. 
         (b) the top-quark mass reconstructed from the bjj final state after the minimization of $\delta M$.
         (c) the top-quark mass reconstructed from bbbjj final state after the minimization of $\delta M$.}
\label{cpvh2}
\end{center}
\end{figure}
The momenta of the two top quarks were reconstructed simultaneously from four b-tagged jets, 
two $W \to \ell \nu$ candidates and candidates for the hadronically decaying $W$ boson. 
The jets and the $W$ boson candidates were assigned to the two top quarks by minimizing the $\delta$M, 
where $\delta$M is defined as
\begin{eqnarray}
\delta \rm M = \sqrt{(m_{top1} - m_{top})^{2} + (m_{top2} - m_{top})^{2} +(m_{W(hadronic)} - m_{W})^{2}},   
\end{eqnarray}
there $m_{top1}$ is reconstructed from one b-tagged jet and one $W$ boson candidate, 
$m_{top2}$ is reconstructed from three b-tagged jets and one $W$ boson candidate, $m_{top}$ is the
generated top-quark mass (175 GeV) and the $m_{W}$ is the $W$ boson mass (PDG value). 

The top-quark mass distributions reconstructed from three jets ($bjj$)
and five jets ($bbbjj$) after the minimization of $\delta M$ are shown in Fig.\ref{cpvh2} (b,c). 
One can see that the top-quark mass distribution from the $bbbjj$ final state is very wide and has 
a big tail. It is because of wrong assignment of the jets or $W$ candidates to the top while 
minimizing $\delta M$. The events with the two top-quark reconstructed masses within the $m_{top} \pm$ 30 GeV 
mass window were selected. Table~\ref{table1} shows the initial cross sections for the signal and 
background processes, the number of Monte-Carlo events remaining after each selection step and 
the cross sections after all selections. 
~\footnote{The $W \to \ell \nu$ and $W \to \rm jj$ reconstruction step selects events with the positive solution 
           for z-component of $E_{T}^{miss}$ and with at least one jet pair having the di-jet mass within 
           the $m_{W}\pm$ 20 GeV mass window; the top-quark mass reconstruction step requires that the two top-quark 
           reconstructed masses are within the $m_{top} \pm$ 30 GeV mass window.}
\begin{table}
\caption{The initial cross sections for the signal and background processes, 
         the number of Monte Carlo events remaining after each selection steps and 
         the cross section after all selections.}
\label{table1}
\begin{center}
\begin{tabular*}{0.95\textwidth}{@{\extracolsep{\fill}}|c|c|c|c|c|}\hline
   & signal & $t \bar{t}+\rm 2~jets$ & $t \bar{t}+\rm 3~jets$ & $t \bar{t}+\rm 4~jets$ \\
   &        &(exclusive)&(exclusive)&(inclusive)\\
\hline
\hline
cross section, pb                        & 8.68    &    100 &     40  &     61  \\
\hline
\hline
number of MC events analyzed             & 193884  & 241000 &  71000  &  94000  \\
(corresponding luminosity, fb$^{-1}$)    & (22.35) & (2.41) & (1.775) & (1.54)  \\
\hline
isolated lepton $p_{T}>$ 20 GeV          & 41035   & 57920  &   16915 &  22214  \\
\hline
$\geq$ 6 jets $E_{T}>$ 20 GeV              & 21389   & 36315  &   14479 &  21866  \\
\hline
4 b-tagged jets with discr. $>$2.95      &   881   &   371  &     248 &   1069  \\
\hline
$W \to \ell \nu$ and $W \to \rm jj$ reco &   379   &   158  &     132 &    602  \\
\hline
top-quark mass reconstruction            &    83   &     4  &       1 &      7  \\
\hline
\hline
cross section after all selections, fb   &   3.71  &   1.66 &    0.56 &   4.54 \\
\hline
\end{tabular*}
\end{center}
\end{table}

\subsubsection{Reconstruction of the neutral $H_{1}$ and charged $H^{+}$ Higgs bosons}

Since it is not known what pair of the b-tagged jets from the reconstructed top quark decay 
chain $t \to bbbW$ is coming from the $H_{1} \to b \bar{b}$ decay, all three b-tagged jet 
pairs were considered as the possible candidates. The invariant mass of b-tagged jet pairs, 
$m_{bb}$ is shown in Fig.~\ref{cpvh3} (left plot) for the background and the signal plus background.
The right plot in Fig.~\ref{cpvh3} shows, fitted by the Gaussian the $m_{bb}$ distribution of the 
signal plus background. The mean value of the fitted distribution is close to the generated mass of 
$H_{1}$. The charged Higgs boson was reconstructed from the two b-tagged jets and $W$ boson, where the 
b-tagged jet pair was chosen as the jet pair with the invariant mass closest to the peak of the $m_{bb}$ 
mass distribution and within the window $\pm$20 GeV around the fitted mean value. 
The invariant mass distribution of the charged Higgs boson reconstructed in this way, $m_{bbW}$ is 
shown in Fig.\ref{cpvh4}.
\begin{figure}[h!]
\begin{center}
\includegraphics[width = 6cm, height = 6cm]{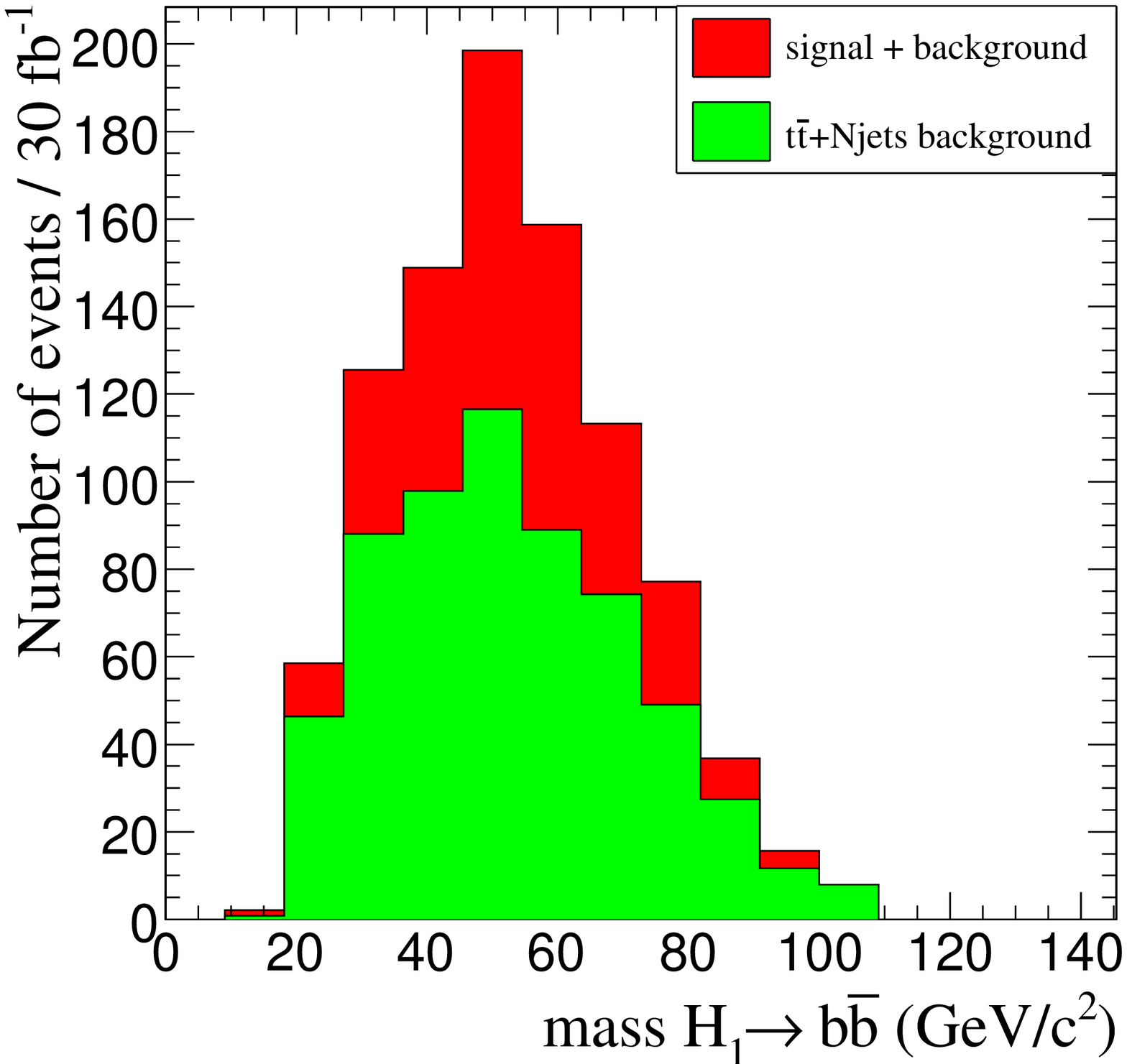}
\includegraphics[width = 6cm, height = 6cm]{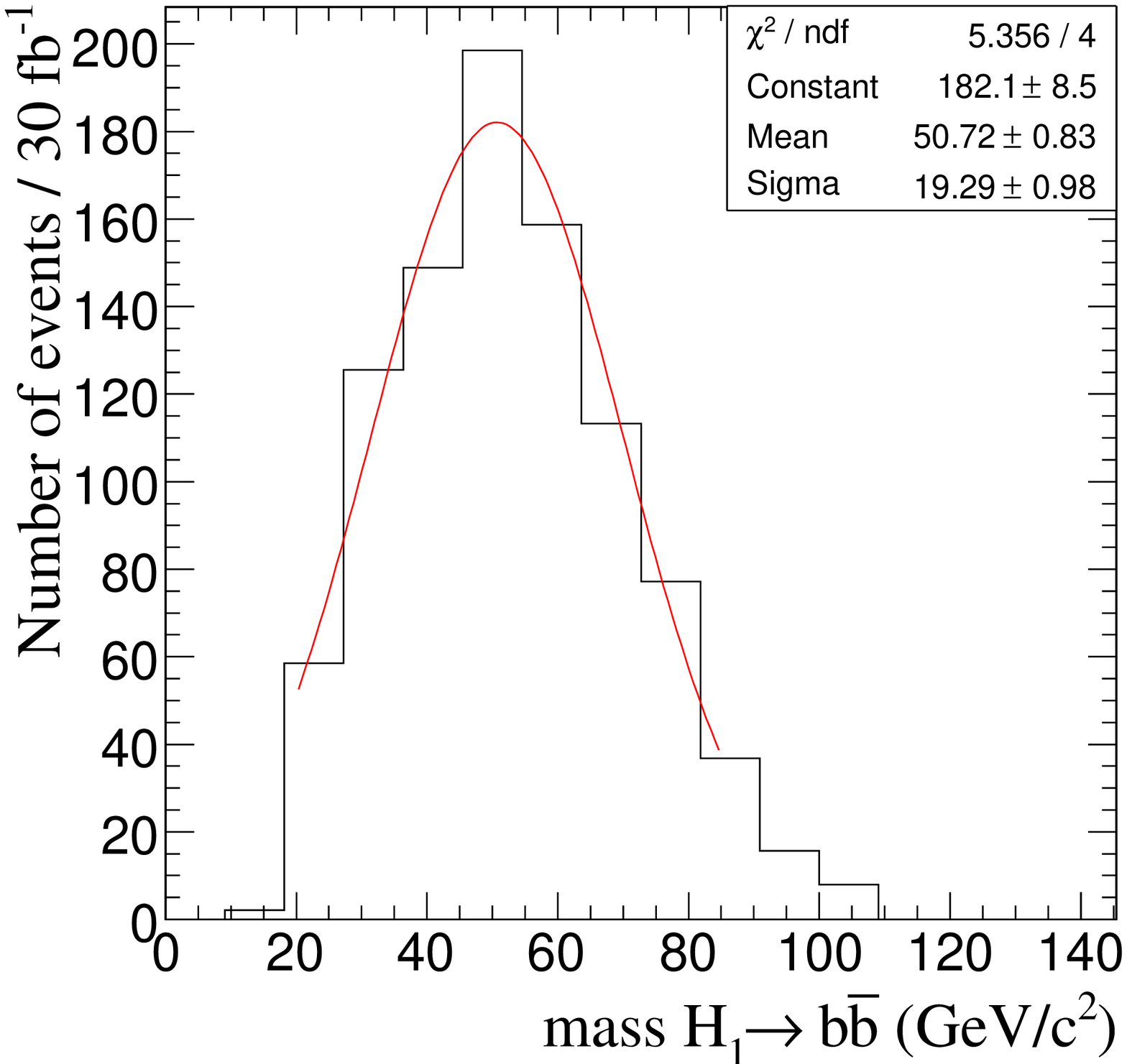} 
\caption{The invariant mass of the b-tagged jet pairs from the reconstructed top quark decay chain $t \to bbbW$}
\label{cpvh3}
\end{center}
\end{figure}
\begin{figure}[h!]
\begin{center}
\includegraphics[width = 6cm, height = 6cm]{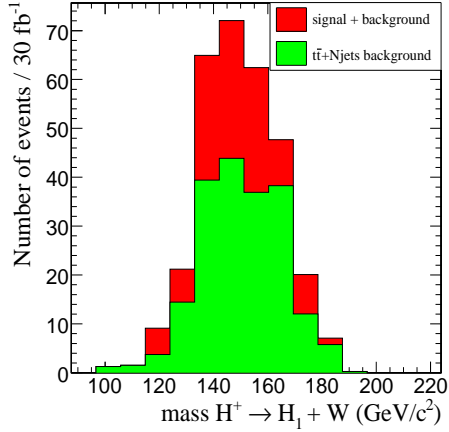}
\caption{The invariant mass of two b-tagged jets and $W$ boson, where two b-tagged jets were chosen
         with the mass closest to the peak of the $m_{bb}$ mass distribution and within the window 
         $\pm$20 GeV around the fitted mean value }
\label{cpvh4}
\end{center}
\end{figure}

The available Monte Carlo statistics of $t \bar{t}+ \rm jets$ background events for this study 
was only order of $\simeq$ 2 fb$^{-1}$, thus it can not be simply rescaled in order to produce the
smooth shape of $m_{bb}$ and $m_{bbW}$ distributions expected for 30 fb$^{-1}$ after all selections. 
We have obtained, however that the shape of $m_{bb}$ and $m_{bbW}$ distributions is almost the same 
after relaxing the cut on the b-discriminator value. Fig.~\ref{cpvh5} shows the $m_{bb}$ (left plot) and 
$m_{bbW}$ (right plot) distributions for four different b-discriminator cuts: 0, 1.0, 1.5 and 2.0. 
\begin{figure}[h!]
\begin{center}
\includegraphics[width = 6cm, height = 6cm]{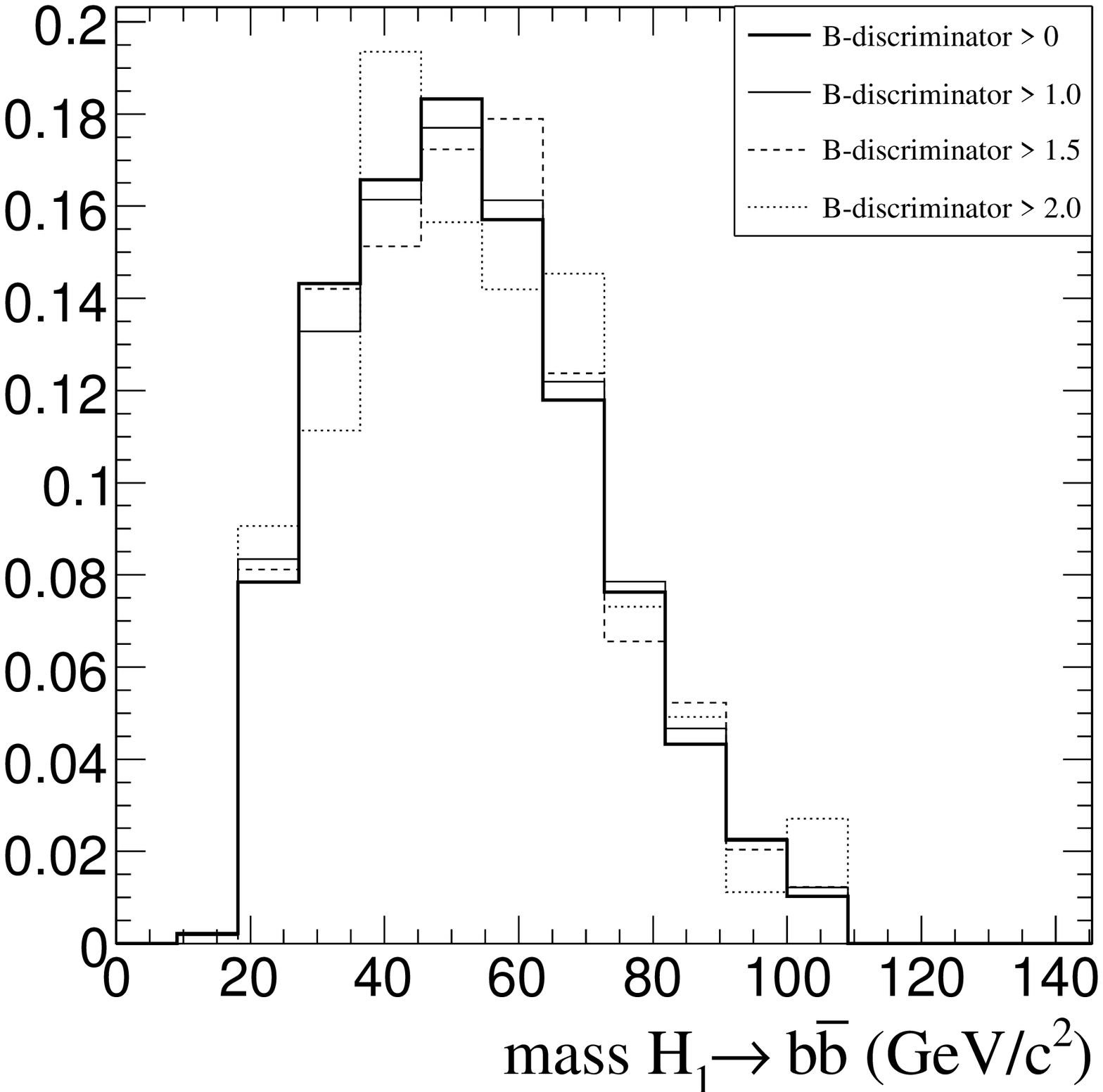}
\includegraphics[width = 6cm, height = 6cm]{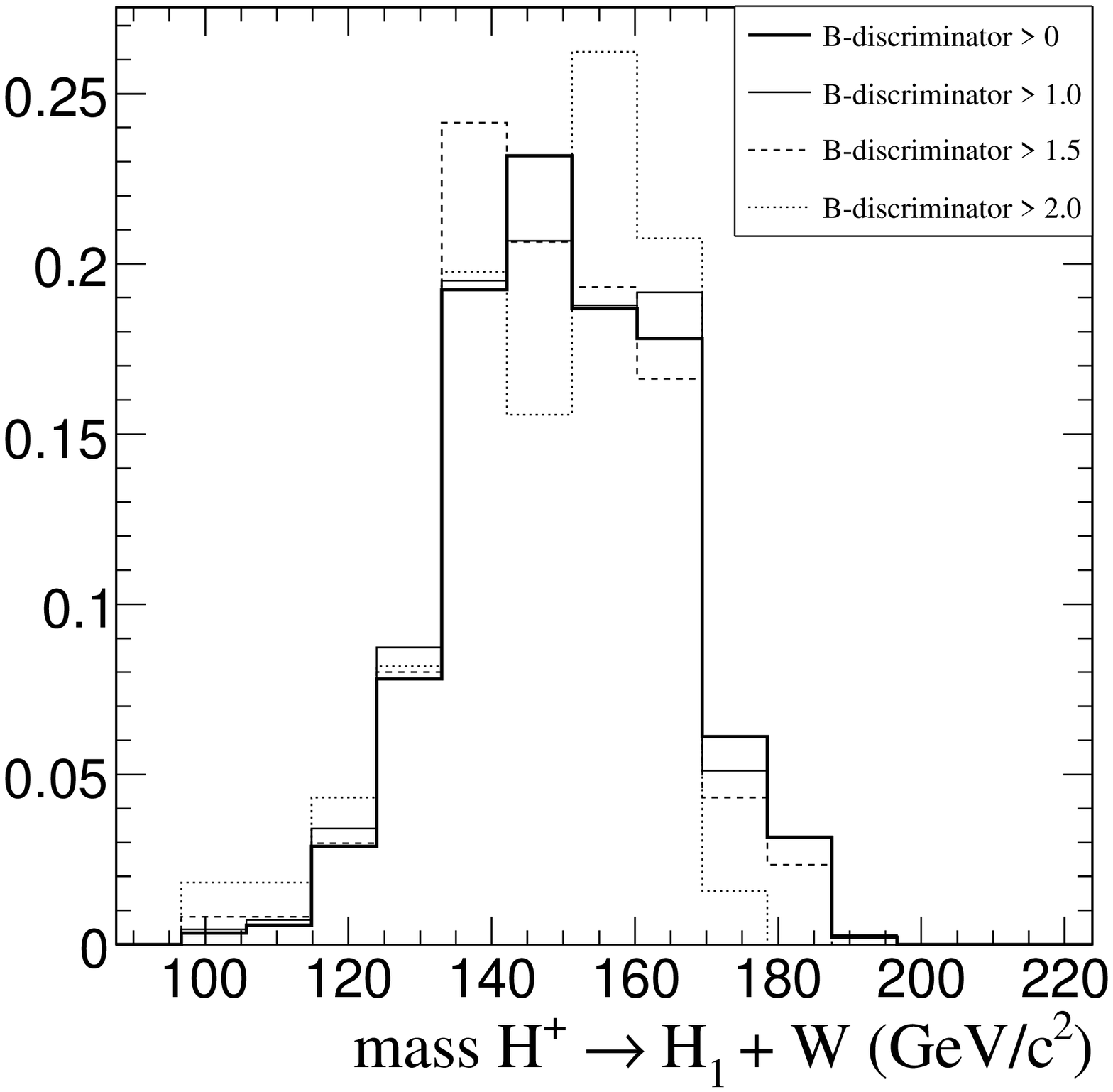}
\caption{The $m_{bb}$ (left) and $m_{bbW}$ (right) distributions after all selections for four different
         cuts on the b-discriminator value: 0, 1.0, 1.5 and 2.0.}
\label{cpvh5}
\end{center}
\end{figure}

\subsection{Results}

The simple selection strategy described in the 
previous sections yields S=110 signal events and B=203 $t\bar{t}+ \rm jets$ 
background events expected with 30 fb$^{-1}$.
The $t\bar{t}b\bar{b}$ background still need to be taken into account. 
The uncertainty due to the Monte Carlo statistics on the $t \bar{t}+ \rm jets$ 
backgrounds is $\simeq$ 30$\%$. The experimental systematic uncertainty was 
estimated by taking into account the systematic uncertainties on the lepton 
identification (2$\%$), the b-jet tagging (5$\%$ per jet), the jet energy 
scale (5$\%$ per jet), the missing transeverse energy scale (10$\%$ on the
raw calorimeter energy scale and 5$\%$ on the jet energy scale) and the 
luminosity uncertainty (5$\%$). It leads to the total systematic uncertainty 
22.5$\%$ (the uncertainty due to the jet and the missing $E_{T}$ scale only 
is 8.8\%).  The significance is calculated as $S/\sqrt{B+ \Delta{B}^{2}}$, 
where $\Delta{B}$ is the experimental systematic uncertainty on the background. 
In order to get the pessimistic value for the significance, the Monte Carlo statistical 
uncertainty was added to the total background: B=203+60=263 events. The signal significance 
is then 110/$\sqrt{263+59^{2}}$=1.8. The uncertainty on the theoretical leading-order cross 
section of the $t \bar{t} + \rm n~jets$, (n$\geq$2) processes is $\geq$50\%. 

One can see that the discovery potential is restricted by both the experimental and the
theoretical uncertainties. The uncertainties can be partially
reduced if the number of the background events and the $m_{bb}$ and $m_{bbW}$ mass shapes
can be extracted from the data. The shapes can be evaluated from the data with the ratio 
S/B$<<$1 when the relaxed cut on the b-discriminator value is used (see Fig.~\ref{cpvh5}). The background 
normalization on the number of events with the relaxed b-jet tagging will eliminate the jet and the  
missing $E_{T}$ scale uncertainties, the luminosity uncertainty and partially reduce the b-tagging 
uncertainty which dominates the experimental uncertainty. It will also reduce the
absolute background prediction uncertainty from the theory, since only the ratio of
$t \bar{t} + \rm jets$ and $t \bar{t} b \bar{b}$ cross sections need to be used. The further, 
more detailed investigations of this channel is foreseen in CMS. 

%expected no. of events for 30 fb^-1
%                        tt~         signal
%B-discriminator > 0.    5716         746  shape of bkg from data
%B-discriminator > 1.0   4477         665
%B-discriminator > 1.5   2455         456
%B-discriminator > 2.0   1110         280

\section*{Acknowledgements}

We would like to thank R.~Godbole for very useful discussions and providing the
code calculating Br($t \to bH^{+}$). We would like to thank our CMS colleagues
Joanne Cole and Claire Shepherd-Themistocleous for the discussions about the optimal
strategy for the jet reconstruction and assignment. Finally A.K.N. and A.N. thank 
organizers of Les Houches Workshop 2007 for the warm hospitality and the friendly 
and stimulating atmosphere.%

%\bibliography{cpvhiggs}

%\end{document}

\part[NMSSM HIGGS BOSONS]{NMSSM HIGGS BOSONS}

\section[Les Houches Benchmark Scenarios for the NMSSM]
{LES HOUCHES BENCHMARK SCENARIOS FOR THE NMSSM
~\protect\footnote{Contributed by: A. Djouadi, M. Drees, U. Ellwanger, 
R.Godbole, C. Hugonie,  
S.F. King,  S. Lehti, S. Moretti, A. Nikitenko, 
I. Rottl\"ander,  M. Schumacher, A. M. Teixeira} 
}
\label{sec:bench}

\subsection{Introduction}

The next-to-minimal supersymmetric extension of the Standard Model (NMSSM) 
\cite{reviewNMSSM,analyses}, in which the spectrum of the minimal supersymmetric
extension (MSSM) is extended by one singlet superfield, is interesting in many
respects. Compared to the MSSM, it solves in an elegant way the so--called $\mu$
problem, has less fine tuning and can induce a rather different phenomenology in
the Higgs and neutralino sectors. Given the possibility of a quite different
phenomenology, it is important to  address the question whether such NMSSM
specific scenarios will be probed at the LHC. In particular, it would be crucial
to make sure that at least one Higgs particle should be observed at the LHC for
the planned integrated luminosity or try to  define regions of the NMSSM
parameter space in which more Higgs states than those available within the MSSM
are visible. However, a potential drawback of the NMSSM, at least in its
non-constrained versions, is that it leads to a larger number of input
parameters  to deal with. In particular, it is clearly unfeasible to make 
multi-dimensional scans over the free inputs of the NMSSM when performing 
complete/realistic simulations to address the two points mentioned above. 

An alternative approach is to resort to a few benchmark scenarios which
embodying the most peculiar/representative  phenomenological features of the
model's parameter space, which can be subject to full experimental
investigation, without loss of substantial theoretical information. Building on
the experience of Ref.~\cite{egh1}, we define in this note benchmark points
which fulfill the present collider and cosmological constraints using the
most--up to date tools to calculate the particle spectra. We  work in the
framework of a semi--constrained NMSSM (cNMSSM) where the soft Supersymmetry
(SUSY) breaking parameters are defined at some high scale, typically that of
grand unification theories (GUTs). This approach leads to a much more plausible
sparticle spectrum, allows to relate features of the Higgs sector to properties
of the neutralino sector and, at the same time, still contains the distinctive
phenomenological features of the NMSSM that are suitable for intensive
phenomenological/experimental  investigation. The emphasis is primarily on the
different possible scenarios within the Higgs sector and the implication for
Higgs searches at the LHC. In particular, we  propose five benchmark points
which lead to Higgs-to-Higgs decays or a light Higgs spectrum but with reduced
Higgs--gauge boson couplings, which are known to be rather difficult to probe at
the LHC.

\subsection{The Model and Its Spectrum}

We confine ourselves to the NMSSM with a scale invariant superpotential given,
in terms of (hatted) superfields with only the third generation (s)fermions
included, by
\begin{equation}
{\cal W} = \lambda \widehat{S} \widehat{H}_u \widehat{H}_d +
\frac{\kappa}{3} \, \widehat{S}^3 + h_t
\widehat{Q}\widehat{H}_u\widehat{t}_R^c - h_b \widehat{Q}
\widehat{H}_d\widehat{b}_R^c  - h_\tau \widehat{L} \widehat{H}_d
\widehat{\tau}_R^c. 
\label{supot}
\end{equation}
The first two terms substitute the $\mu \widehat H_u  \widehat H_d$ term in the
MSSM superpotential, while the three last terms  are the usual generalization of
the Yukawa interactions. The soft SUSY breaking terms consist of  the scalar
mass terms for the Higgs,  sfermion  and gaugino fields and the trilinear 
interactions between the sfermion and Higgs fields.  In an unconstrained NMSSM 
with non--universal soft terms at the GUT scale, the three SUSY breaking masses
squared for $H_u$, $H_d$ and $S$ are  determined through the minimization
conditions of the scalar potential.  Thus,  the Higgs sector of the NMSSM is
described by the six parameters
\begin{equation}
\lambda\ , \ \kappa\ , \ A_{\lambda} \ , \ A_{\kappa}, \ 
\tan \beta =\ \langle H_u \rangle / \langle H_d \rangle \ \mathrm{and}
\ \mu_\mathrm{eff} = \lambda \langle S \rangle\; .
\end{equation}
As the number of input parameters is rather large, one can attempt to define a
constrained (cNMSSM) model, similar to the minimal supergravity model or cMSSM, in
which the soft SUSY breaking  parameters are fixed at the GUT scale, leading to
only a handful of inputs. One can thus impose  unification  of the 
gaugino, sfermion and Higgs mass parameters and the trilinear couplings at
$M_{\rm GUT}$: $M_{1,2,3} \equiv M_{1/2}, m_{\tilde{F}_i} = m_{H_i} \equiv  m_0
, A_{i} \equiv  A_0$. The fully constrained cNMSSM has two additional
parameters, $\lambda$ and $\kappa$, beyond the above and the correct $M_Z$ 
value imposes one constraint. Hence, a  priori, the number of inputs in 
the cMSSM and the fully constrained cNMSSM is exactly the same.

In practice, it is convenient to use the analytic form of the three minimization
conditions of the NMSSM effective potential and, for given $M_Z$, $\tan\beta$,
$\lambda$ and all soft terms at the weak scale except for $m_S^2$, these can be
solved for $|\mu_{\rm eff}|$ (or $|\left< S\right> |$), $\kappa$ and $m_S^2$;
sign($\mu_{\rm eff}$) can still be chosen at will. Here, we will relax the
hypothesis of complete unification of the soft terms in the singlet sector,
$m_S^2 \neq m_0^2$ and $A_\kappa \neq A_0$ at $M_{\rm GUT}$,  since the singlet
could play a special r\^ole. In addition, for some of the benchmark points, we
will also relax the unification hypothesis for $m_{H_u}^2$ and $m_{H_d}^2$ and  
for one scenario, the hypothesis $A_{\lambda} = A_0$. Such points in parameter
space can have additional unconventional properties, whose phenomenology should
also be investigated.

Following the procedure employed by the routine NMSPEC within NMSSMTools
\cite{nmssmtools}, which calculates the spectra of the Higgs and  SUSY particles
in the NMSSM, a point in the parameter space of the cNMSSM is defined by the
soft SUSY breaking terms at $M_{\rm GUT}$ (except for the parameter $m_S^2$),
$\tan\beta$ at the weak scale, $\lambda$ at the SUSY scale (defined as an
average of the first generation squark masses) and the sign of the parameter
$\mu_{\rm eff}$. The parameters $\kappa$, $m_S^2$ and $|\mu_{\rm eff}|$ are
determined at the SUSY scale in terms of the other parameters through the
minimization equations of the scalar potential. The renormalisation group
equations (RGEs) for the gauge and Yukawa couplings and those for the soft 
terms are integrated between $M_Z$ and $M_{\rm GUT}$ defined by gauge couplings
unification. For the most relevant Standard Model parameters, we chose
$\alpha_s(M_Z) = 0.1172$, $m_b(m_b)^{ \overline{\rm MS}} = 4.214$ GeV and
$m_{\mathrm{top}}^{\mathrm{pole}} = 171.4$ GeV. 

After RGE running is completed, the Higgs, gluino, chargino, neutralino  and
sfermion masses are computed including dominant one-loop corrections to their
pole masses. All the Higgs decay branching ratios (BRs) into SM and SUSY
particles are determined including dominant radiative  corrections.
Subsequently, the following Tevatron and LEP constraints are applied: $i$)
Direct searches for the LSP neutralino and invisible $Z$ decay  width, $ii)$ 
direct bounds on the masses of the charged particles $h^\pm$, $ \chi^\pm$,
$\tilde q$,~$\tilde l$ and the gluino;  $iii)$ constraints on the Higgs
production rates from all channels studied at LEP. 

Light $h_i$ ($i=1,2$) scalar states (with $M_{h_i} \lsim 114$ GeV) can still be
allowed by LEP constraints, if  the $Z$--$Z$-$h_i$ coupling is heavily
suppressed or the lightest pseudoscalar $a_1$ state has $M_{a_1} \lsim 10\, {\rm
GeV}$ such that $h_i$ decays dominantly into $a_1 a_1$ states but the $b\bar b$
decay of  the $a_1$  is  impossible. Constraints from the decays $h_i \to a_1
a_1 \to 4\tau$ allow for $M_{h_i}$ down to $\sim 86$~GeV. Note that LEP
constraints are implemented only for individual processes and that combinations
of different processes could potentially rule out seemingly viable scenarios.
Finally, experimental constraints from B physics  are taken into account, and we
require that the relic abundance of the NMSSM dark matter (DM) candidate, the
lightest neutralino $\chi_1^0$ which can be singlino-like, matches the  WMAP
constraint  $0.094 \lsim \Omega_{\rm CDM} h^2 \lsim 0.136$ at the $2\sigma$
level.

\subsection{The Benchmark Points}

In the Higgs sector of the NMSSM, two different types of difficult scenarios
have been pointed out, depending on whether  Higgs-to-Higgs decays are
kinematically allowed or forbidden; see e.g. Ref.~\cite{egh1}. 

Within the first category, there are two possibilities, each associated with
light scalar/pseudo\-scalar Higgs states: (i) The lightest CP--odd $a_1$ state
is rather light, $M_{a_1} \lsim$ 40--50 GeV, and  the lightest CP--even $h_1$
particle has enough phase space for the decay into two $a_1$ particles,  $h_1
\to a_1 a_1$, to be allowed and dominant. The $a_1$ state will mainly decay into
$\tau^+\tau^-$ if $M_{a_1} \lsim 10$~GeV or to  $\tau^+\tau^-$ ($\sim 10\%$) and
$b\bar b$ (90\%) states if $M_{a_1} \gsim 10$~GeV. One would have then the
possibilities $h_1 \to a_1 a_1 \to 4\tau$ and $h_1 \to a_1 a_1 \to 4\tau, 4b$
and $2\tau 2b$ for the $h_1$ state  which can   have a mass that is either close
to its theoretical upper limit of 130 GeV or to the lower limit of 90 GeV.  (ii)
The lightest CP--even $h_1$ boson is relatively light, $M_{h_1} \lsim 50$ GeV,
and decays into $b\bar b$ pairs (the situation where $M_{h_1} \lsim 10$ GeV is
very constrained by LEP data). In this case, the next-to-lightest CP-even $h_2$
state is SM--like with a mass below $\sim 140$ GeV and can decay into two $h_1$
bosons leading to the final topologies $h_2 \to h_1 h_1 \to 4\tau$, $2\tau 2b$
and $4b$. 

The second category of scenarios, where Higgs-to-Higgs decays are suppressed,
includes regions of the parameter space where the five neutral Higgs particles
are relatively light,  with masses in the range 90--180 GeV, which opens the
possibility of producing all of the them at the LHC, but with  couplings to
gauge bosons that are reduced compared to the SM Higgs case.  This scenario is
similar to the so--called  ``intense coupling regime" of the MSSM \cite{ICR} but
with two more neutral Higgs particles. 

We propose five benchmark points of the NMSSM  parameter space, P1 to P5, in
which the above mentioned scenarios are realized (see Ref.~\cite{Djouadi:2008uw} 
for more details). Each point is representative of distinctive NMSSM features. 
Points P1 to P3 exemplify scenarios where $h_1$ decays into light pseudoscalar
states decaying, in turn, into  $b \bar b$ or $\tau^+ \tau^-$ final states; 
these points can be realized within the cNMSSM with nearly universal soft terms
at $M_{\rm GUT}$, the exception being the parameters $m_S^2$ and $A_\kappa$.  P4
illustrates the NMSSM possibility of a very light $h_1$ and can be obtained 
once one relaxes the universality conditions on the soft SUSY  breaking Higgs
mass terms, $M_{H_d} \neq M_{H_u} \neq m_0$. Point P5 corresponds to the case
where all Higgs bosons are rather light and can be obtained if one allows
additionally for the inequality $A_\lambda \neq A_0$. In all cases, the input
parameters as well as the resulting Higgs masses and some decay information are
given in Table~\ref{table:NMP}; the main characteristics of the $\chi_1^0$   DM
candidate are also given.  Next, we summarize the most relevant phenomenological
properties of the benchmark points.

\begin{table}[!ht]
\caption{Input and output parameters for the five benchmark NMSSM points.}
\vspace*{-5mm}
\label{table:NMP}
\vspace{3mm}
\footnotesize
\begin{center}
\begin{tabular}{|l|r|r|r|r|r|}
\hline
{\bf Point} & P1 & P2 &  P3 & P4 & P5
\\\hline
{\bf GUT/input parameters }
\\\hline
sign($\mu_\mathrm{eff}$)  & + &+ &+ &-- &+
\\\hline
$\tan \beta$  & 10 & 10 & 10 & 2.6& 6
\\\hline
$m_0$ (GeV)  & 174& 174& 174& 775&1500
\\\hline
$M_{1/2}$ (GeV) &500 & 500& 500& 760& 175
\\\hline
$A_0$ & -1500&-1500 & -1500& -2300& -2468
\\\hline
$A_\lambda$ & -1500&-1500 & -1500& -2300& -800
\\\hline
$A_\kappa$ & -33.9& -33.4& -628.56& -1170& 60
\\\hline
NUHM: $M_{H_d}$ (GeV) &-&-&-& 880&-311
\\\hline 
NUHM: $M_{H_u}$ (GeV) &-&-&-& 2195&1910
\\\hline\hline
{\bf Parameters at the SUSY scale } 
\\\hline
$\lambda$ (input parameter) & 0.1& 0.1& 0.4& 0.53&0.016
\\\hline 
$\kappa$ & 0.11 & 0.11& 0.31& 0.12&-0.0029
\\\hline
$A_\lambda$ (GeV) & -982 & -982& -629& -510& 45.8
\\\hline
$A_\kappa$ (GeV) & -1.63& -1.14& -11.4& 220& 60.2
\\\hline
$M_2$ (GeV)& 392 & 392 & 393 & 603 & 140
\\\hline
$\mu_{\rm eff}$ (GeV) & 968 &968 & 936& -193& 303
\\\hline\hline
{\bf CP-even Higgs bosons}
\\\hline
$m_{h_1}$ (GeV) & 120.2& 120.2& 89.9&32.3 &90.7
\\\hline
BR($h_1 \to b \bar b$) & 0.072& 0.056& $7 \times 10^{-4}$& 0.918&0.895
\\\hline
BR($h_1 \to \tau^+ \tau^-$) & 0.008& 0.006& $7 \times 10^{-5}$&
0.073&0.088 \\\hline
BR($h_1 \to a_1 a_1$) & 0.897& 0.921& 0.999& 0.0 & 0.0 
\\\hline\hline
$m_{h_2}$ (GeV) & 998 & 998& 964& 123&118
\\\hline \hline
$m_{h_3}$ (GeV) & 2142& 2142& 1434&547 &174
\\\hline\hline
{\bf CP-odd Higgs bosons}
\\\hline\hline
$m_{a_1}$ (GeV) & 40.5& 9.09& 9.13 &185&99.6
\\\hline
BR($a_1 \to b \bar b$) & 0.91& 0.& 0.& 0.62&0.91
\\\hline
BR($a_1 \to \tau^+ \tau^-$) & 0.085& 0.88& 0.88& 0.070&0.090
%\\\hline
%BR($a_1 \to jj$) & & & & &
\\\hline\hline
$m_{a_2}$ (GeV) & 1003& 1003 & 996& 546&170
\\\hline\hline
{\bf Charged Higgs boson} 
\\\hline\hline
$m_{h^\pm}$ (GeV) & 1005& 1005& 987& 541&188
\\\hline
{\bf LSP} 
\\\hline\hline
$m_{\tilde \chi^0_1}$ (GeV)  & 208 & 208 & 208& 101&70.4 \\ \hline
$\Omega_{CDM} h^2$  & 0.099& 0.099& 0.130& 0.099& 0.105
\\\hline
\end{tabular}\end{center}
\end{table}

In the first two points P1 and P2, the lightest $h_1$ CP--even state has a mass
of $M_{h_1}\!\simeq\!120$ GeV and is SM--like with couplings (relative to that
of the SM Higgs) to gauge bosons $R_1$, top quarks $t_1$ and bottom quarks
$b_1$, which are almost equal to unity. The lightest CP--odd $a_1$ boson has a
mass of, respectively, 40.5~GeV and 9.09~GeV. In both cases P1 and P2, the decay
channel $h_1 \to a_1 a_1$ is largely dominating with a BR very close to 90\%,
while the decays $h_1 \to  b\bar b$ and $\tau^+ \tau^-$ are suppressed by an
order of magnitude when compared to the SM case. The most relevant difference
between the two scenarios concerns the mass and decays of the lightest
pseudoscalar state. In P1 the $a_1$ boson decays into $b$ quarks and $\tau$
leptons with  rates  of $\sim 90\%$ and $\sim 10\%$, respectively. In contrast,
in P2 the pseudoscalar $a_1$ state with its mass $M_{a_1}\!\simeq\!9.09$ GeV
decays dominantly into $\tau^+\tau^-$ pairs, with a rate above 80\%.

For point P3, the same inputs of points P1 and P2 are chosen except for the
$A_\kappa$ and $\lambda$ parameters, which are now varied as to have a lighter
$h_1$ state. This again leads to a pseudoscalar $a_1$ boson  which has
approximately the same mass as in scenario P2, $M_{a_1}\!\simeq\!9.96$~GeV, and
which decays almost exclusively into $\tau^+ \tau^-$ final states. The
difference between P3 and P2 is the lightest CP--even Higgs boson $h_1$, which
has a mass  $M_{h_1}\simeq 90$~GeV, lower than in scenarios P1 and P2. In this
case, and although $h_1$ is still SM--like, i.e. exhibiting couplings to gauge
bosons, top and bottom quarks that are very close to those of the SM Higgs
boson, it decays nevertheless almost exclusively into $a_1$ pairs, with a rate
close to 100\%.  Another difference between P2 and P3 is that in the former
case, the interesting decay mode $h_1 \to a_1 Z$ is kinematically possible but
the rate is rather small, BR($h_1 \to a_1 Z) \sim 3\%$. 

Note that in all these first three points, the heaviest neutral Higgs particles
$h_2, h_3$ and $a_2$, as well as the charged Higgs states $h^\pm$, all have
masses close to, or above, 1 TeV. The main decay modes are into $b\bar b$ and 
$t\bar t$ for the neutral and $tb$ for the charged states, as $\tan\beta$ is not
too large and the $t\bar t$--Higgs couplings are not very strongly suppressed,
while the BRs for the neutral Higgs-to-Higgs decays, in particular the channels
$h_2 \to h_1 h_1$ and $h_2 \to  a_1 a_1$, are very tiny, not exceeding the
permille level. Regarding the properties of the DM candidate, P1, P2 and P3
exhibit a  lightest neutralino which is bino--like, with mass is
$m_{\chi_1^0}\!\simeq\!208$ GeV. In all three cases, the correct cosmological
density, $\Omega_{\rm CDM} h^2\simeq 0.1$, is achieved through the
co--annihilation  with the $\tilde \tau_1$ slepton, which has a mass comparable
to that of the LSP.

Point P4 corresponds to a scenario in which the CP--even boson $h_1$ is very
light, $M_{h_1}=32.3$ GeV and singlet--like and  predominantly decays into
$b\bar b$ pairs, with  BR$(h_1 \to b\bar b)=92\%$, and to a smaller extent into 
$\tau$ pairs with BR$(h_1 \to \tau^+\tau^-)\simeq 7\%$.  The CP--even $h_2$
boson has a mass of $M_{h_2}\!\simeq\!123$ GeV and is SM--like, with normalized
couplings to $W/Z$ and $t/b$ states close to unity.  However, it mostly decays
into two $h_1$  bosons, BR$(h_2 \to h_1 h_1)\simeq 88\%$ and the dominant
SM--like $b\bar b$ decay mode occurs only at a rate  less than 10\%. The
lightest CP--odd particle is not very heavy, $M_{a_1}=185$ GeV, and decays
mostly into fermion pairs, with  BR$(a_1 \to b\bar b) \sim 61\%$ and  BR$(a_1
\to \tau^+\tau^-) \sim 7\%$; the other dominant decay is the interesting channel
$a_1 \to h_1 Z$ which has a rate of the order of 30\%. Finally, the heaviest
CP-even $h_3$, CP--odd $a_2$ and the charged $h^\pm$ particles have masses in
the 500 GeV range and will mostly decay, as $\tan\beta$ is small, into $t\bar
t/tb$ final states for the neutral/charged states. All these features make the 
phenomenology of point P4 rather different from that of points  P1 to P3
discussed above.  To achieve a correct cosmological relic density, the common
sfermion  and gaugino mass parameters at $M_{\rm GUT}$ are close to 1 TeV. At the
SUSY scale, one thus finds a higgsino-singlino-like neutralino LSP, whose mass
is $m_{\chi_1^0} \sim 100$ GeV and  LSP annihilation essentially leads to $WW$
and $Zh_1$ final states. 

Finally, point P5 is characterized by having all Higgs particles relatively
light with masses in the range 90 to 190 GeV.  Here, both $\lambda$ and $\kappa$
are relatively small. The three CP--even Higgs bosons with masses of 91, 118 and
174~GeV, respectively, share the couplings of the SM  Higgs boson to the gauge
bosons with the dominant component being taken by the $h_2$ state. The
pseudoscalar Higgs bosons have masses $M_{a_1}\simeq 100$ GeV and $M_{a_2}\simeq
170$ GeV, while the charged Higgs particle is the heaviest one with 
$M_{h^\pm}\simeq 188$ GeV. Here, all the neutral  Higgs-to--Higgs decays are
kinematically disfavored; this is also the case of neutral Higgs decays into
into lighter Higgs states with opposite parity and gauge bosons. The only
non--fermionic two--body Higgs decays are thus $h^\pm \to W h_1$ and $h_3 \to
WW$, but as the involved Higgs--gauge boson couplings are small, the BRs are
tiny.  Here, the LSP with a mass $m_{\chi_1^0} \sim 70$~GeV, is a bino--like
neutralino but it has a small non--negligible higgsino component. The value 
$\Omega_{\rm CDM} h^2\simeq 0.1$  is achieved through the annihilation
processes  $\chi_1^0 \,\chi_1^0 \to b \bar b ,\tau^+ \tau^-$, with $s$--channel
exchange of Higgs bosons.

\subsection{Expectations at the LHC}

In the cases discussed here, at least one CP--even Higgs particle $h_i$ has
strong enough couplings to massive gauge bosons and top quarks, $R_i, t_i \sim
1$, to  allow for the production at the LHC in one of the main channels which
are advocated  for the search of the SM Higgs particle  \cite{Hreviews}: $i)$
gluon--gluon fusion, $gg \to h_i$, $ii)$ vector boson fusion (VBF), $qq\to qq
W^*W^*,qqZ^*Z^*\to qq h_i$ with two forward jets and a centrally decaying Higgs
boson,  $iii)$  Higgs--strahlung (HS), $q\bar q' \to W h_i$ and  $q\bar q \to
Zh_i$, with the gauge boson decaying leptonically, $iv)$  associated production
with heavy top quark pairs $q\bar q/gg \to t\bar t h_i$.

In scenarios P1 to P3, this CP--even $h_i$ particle is the $h_1$ boson which has
$R_1 \simeq t_1 \simeq b_1 \simeq 1$,  but which decays  most of the time into a
pair of light pseudoscalar Higgs particles, $h_1 \to a_1 a_1$, which
subsequently decay into light fermion pairs, $a_1 \to b\bar  b$ and
$\tau^+\tau^-$. In scenario P4, this particle is the $h_2$ boson which decays 
most of the time into a pair of $h_1$ particles, $h_2 \to h_1 h_1$, which
again   decay into  light fermion pairs. In these four cases, the backgrounds
in  both $gg  \to h_i \to 4f$ and $qq/gg \to t\bar t h_i \to t\bar t +4 f$, 
with $f=b,\tau$, processes will be extremely large and only the VBF (owing to
the forward jet tagging) and eventually HS (due to the leptons coming from  the
decays of the gauge bosons) can be viable at the LHC. In P5, the particle that
has couplings to gauge bosons and top quarks  close to those of the SM Higgs
boson is the $h_2$ boson which decays into $b\bar b$ and $\tau^+\tau^-$ final
states with BRs close  to 90\% and 10\%, respectively. Here again, the $gg$
fusion and presumably  associated production with top quarks cannot be used
since the interesting decays  such as $h_2 \to W W^*, ZZ^*$ and $\gamma \gamma$
are  suppressed compared  to the SM case.  Thus, in this case, only the channels
$qq\to qqh_2 \to qq \tau^+ \tau^-$ and eventually $q\bar q' \to Wh_2 \to \ell
\nu b \bar b$ seem feasible. The state $h_1$ has still  non--negligible
couplings to gauge bosons and top quarks which lead to cross  sections that are
``only" one order of magnitude smaller than in the SM. Since here again, only
the decays $h_1 \to b\bar b$ (90\%) and $\tau^+\tau^-$ (10\%)  are relevant, the
only channels which can be used are the VBF  and HS processes discussed above,
but one needs a luminosity 10 times larger to have the same event samples as in
the SM.

Several theoretical studies have been performed in the past to assess the
potential of the LHC to observe NMSSM Higgs particles in some scenarios close to
those presented here; see Ref.~\cite{Djouadi:2008uw} for an account. Recently, the
ATLAS and CMS collaborations started  investigating some channels, the main
focus being on the VBF production process $pp \to qq h_1$ and to a lesser
extent  HS via $q\bar q \to W h_1 \to \ell \nu h_1$, with the SM--like $h_1$
state decaying into  $h_1 \to a_1 a_1  \to 4\tau$, corresponding to scenario P2
and P3. The ATLAS collaboration is analyzing the  $4\mu + 4\nu_\tau + 4\nu_\mu$
channel from VBF, requiring three leptons to be observed and, for triggering,
one or two high--$p_T$ leptons ($p_T >20$ or 10 GeV) \cite{nmssm-ATLAS}.  CMS is
currently considering the  $\mu^{\pm} \mu^{\pm} \tau _{\rm jet}^{\mp} \tau _{\rm
jet}^{\mp}$  final state containing two same sign muons and two $\tau$ jets
\cite{nmssm-CMS}. Despite of the missing energy and the possibility of missing
one lepton, the mass of the $h_2$ state could be reconstructed  with the help of
the collinear approximation. The performance of the algorithms to observe the
signals and the effects of the various backgrounds are under study.

\subsection{Conclusions}

The NMSSM is a very interesting supersymmetric extension  of the SM as it solves
the notorious $\mu$ problem of the MSSM and it has less fine tuning. It also
leads to an interesting collider phenomenology in some cases, in particular in
the Higgs sector, which is extended to contain an additional CP--even and a
CP--odd state. Compared to the SM and MSSM, the  searches for the NMSSM Higgs
bosons will be rather challenging at the LHC in scenarios in which some neutral
Higgs particles are very light, opening the possibility of dominant
Higgs-to-Higgs decays, or when all Higgs bosons have reduced couplings to the
electroweak gauge bosons and to the top quarks. These scenarios, for which we
have provided benchmark points in a semi--unified NMSSM which involves a rather
limited number of input parameters at the grand unification scale and which
fulfills all  present collider and cosmological constraints,  require  much more
detailed phenomenological  studies and experimental simulations  to make sure
that at least one Higgs  particle of the NMSSM will be observed at the LHC. 
\smallskip

\subsection*{Acknowledgments}
 We acknowledge support from The Leverhulme
Trust,  the Royal Society (London, UK), the A. von-Humboldt Foundation, the FP7 
RTN  MRTN-CT-2006-035505 HEPTOOLS, the  Indo--French Center IFCPAR for the
project  3004-2 and the French ANR for the project PHYS@COL\&COS.

%\bibliography{LH-benchmark-nmssm}

%\end{document}

\section[Parameter Scans in Two
Interesting NMSSM Scenarios]
{PARAMETER SCANS IN TWO INTERESTING NMSSM SCENARIOS
~\protect\footnote{Contributed by: I. Rottl\"{a}nder
and  M.~ Schumacher}
}
%\documentclass[11pt]{cernrep}
%\usepackage{graphicx,epsfig}
%\bibliographystyle{lesHouches}
%\begin{document}

%\title{PARAMETER SCANS IN TWO INTERESTING NMSSM SCENARIOS}

%\author{I. Rottl\"{a}nder$^1$, M. Schumacher$^2$}
%\institute{$^1$Physikalisches Institut, University of Bonn, Nussallee 12, 53115% Bonn, Germany
%\\$^2$Fachbereich Physik, University of Siegen, Walter-Flex-Stra\ss{}e 3, 57068% Siegen, Germany
%}

%\maketitle

%%\begin{abstract}
%%Hope this will work out - otherwise what to do with it??
%%Do I need now an abstract or dont I???
%%\end{abstract}

\subsection{Introduction}
In the past, proposals for interesting points in the parameter space of  the {\it Next-to-Minimal Super\-sym\-me\-tric Standard Model} (NMSSM) \cite{Nilles:1982dy, Frere:1983ag, Derendinger:1983bz, Ellis:1988er, Drees:1988fc, Franke:1995tc}
have been made (see e.g. Refs. \cite{Ellwanger:2004gz, Miller:2004uh, Ellwanger:2005uu}).
A new study proposes benchmark points for the constrained NMSSM \cite{Djouadi:BMpoints}.  
To evaluate the discovery potential of NMSSM particles at collider experiments like 
the {\it Large Hadron Collider} (LHC)\footnote{A proton-proton collider with a design center-of-mass energy of 14 TeV. First physics runs are expected for 2008.},  
it is furthermore desirable to define two-dimensional benchmark scans
which include regions of typical
and experimentally challenging NMSSM phenomenology. 
%%Unlike points, scans provide a smooth 
%%variation of model parameters and can give a better feeling of region 
%%sizes and the development of the discovery potential for new 
%%particles, for example the Higgs boson. 
In the following, two such
parameter scans over the Higgs sector of the NMSSM are 
proposed for this model. Both scans include
a benchmark point from Ref. \cite{Ellwanger:2005uu}.

\begin{figure}[tbh]
\begin{minipage}[t]{.5\linewidth} % [b] => Ausrichtung an \caption
\centerline{\includegraphics[width=8cm, height=4.3cm]{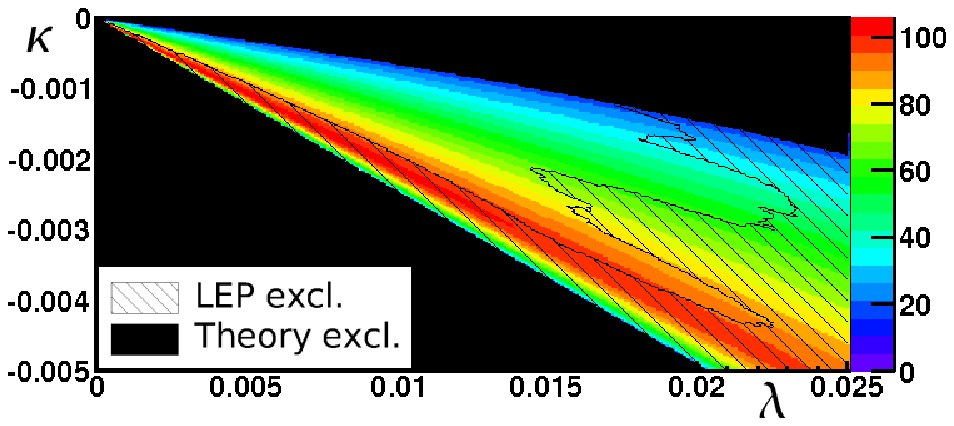}}
\caption{$H_1$ mass [GeV] in the {\it Reduced Couplings Scenario} \label{benchmark_EW1H1mass}}
\end{minipage}
\hspace{.05\linewidth}% Abstand zwischen Bilder
\begin{minipage}[t]{.5\linewidth} % [b] => Ausrichtung an \caption
\centerline{\includegraphics[width=8cm, height=4.3cm]{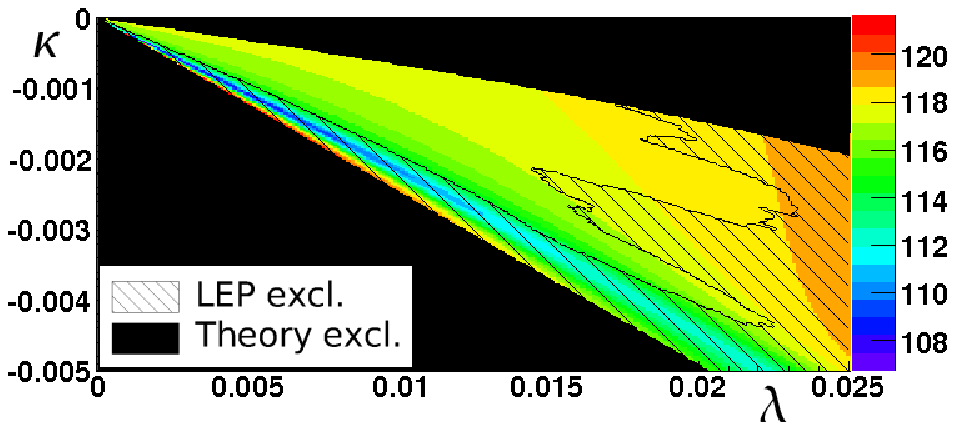}}
\caption{$H_2$ mass [GeV] in the {\it Reduced Couplings Scenario}\label{benchmark_EW1H2mass}}
\end{minipage}
\end{figure}
\begin{figure}[h]
\begin{minipage}[t]{.5\linewidth} % [b] => Ausrichtung an \caption
\centerline{\includegraphics[width=8cm, height=4.3cm]{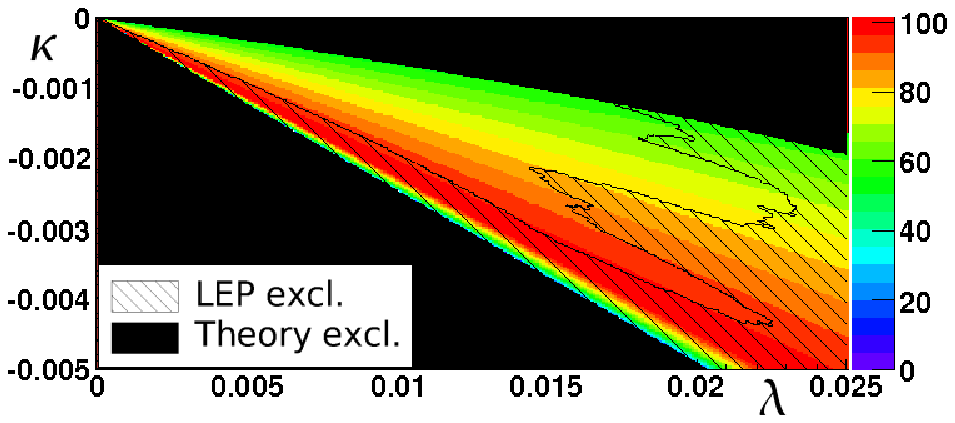}}
\caption{$A_1$ mass [GeV] in the {\it Reduced Couplings Scenario} \label{benchmark_EW1A1mass}}
\end{minipage}
\hspace{.05\linewidth}% Abstand zwischen Bilder
\begin{minipage}[t]{.5\linewidth} % [b] => Ausrichtung an \caption
\centerline{\includegraphics[width=8cm, height=4.3cm]{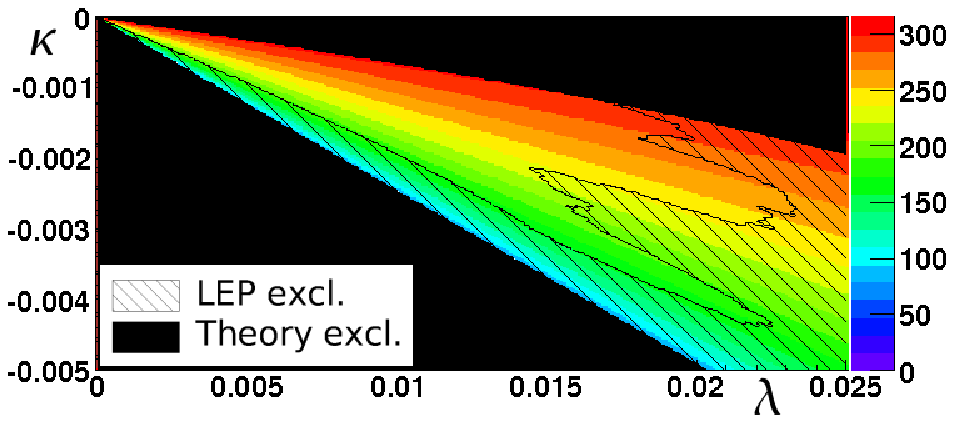}}
\caption{$H^\pm$ mass [GeV] in the {\it Reduced Couplings Scenario}\label{benchmark_EW1CHmass}}
\end{minipage}
\end{figure}
\begin{figure}[h]
\begin{minipage}[t]{.5\linewidth} % [b] => Ausrichtung an \caption
\centerline{\includegraphics[width=8cm, height=4.3cm]{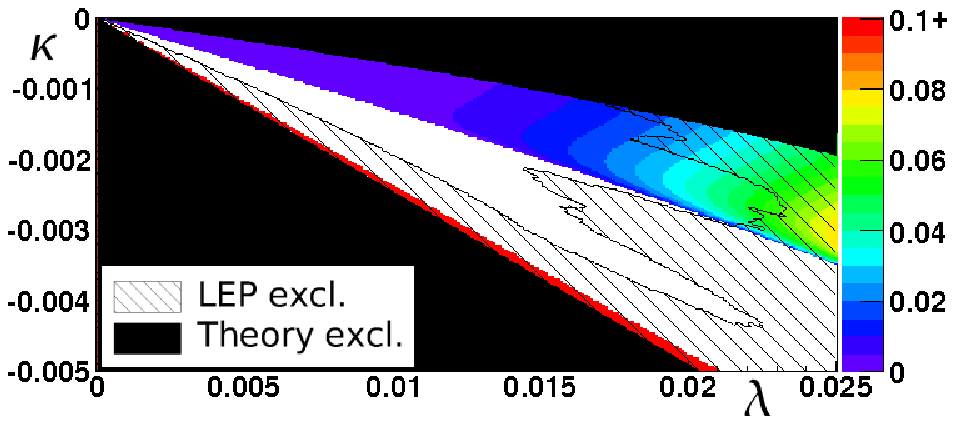}}
\caption{$H_2$$\rightarrow$$H_1H_1$ branching ratio in the {\it Reduced Couplings Scenario}\label{benchmark_EW1BR}}
\end{minipage}
\hspace{.05\linewidth}% Abstand zwischen Bilder
\begin{minipage}[t]{.5\linewidth} % [b] => Ausrichtung an \caption
\centerline{\includegraphics[width=8cm, height=4.3cm]{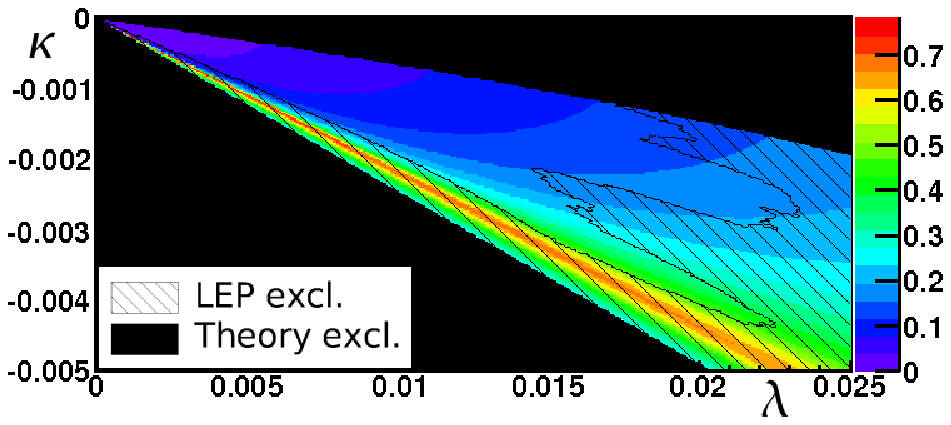}}
\caption{$H_1$ vector boson coupling relative to its SM-value in the {\it Reduced Couplings Scenario} \label{benchmark_EW1H1CV}}
\end{minipage}
\end{figure}
\begin{figure}[h]
\begin{minipage}[t]{.5\linewidth} % [b] => Ausrichtung an \caption
\centerline{\includegraphics[width=8cm, height=4.3cm]{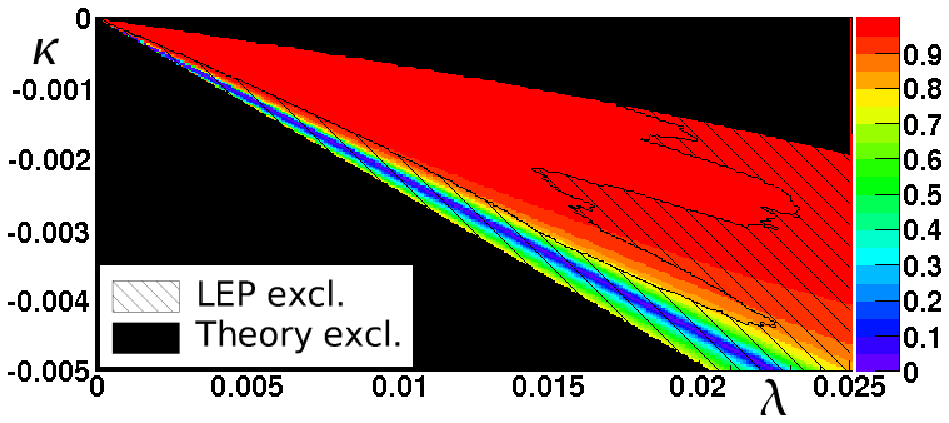}}
\caption{$H_2$ vector boson coupling relative to its SM-value in the {\it Reduced Couplings Scenario}\label{benchmark_EW1H2CV}}
\end{minipage}
\hspace{.05\linewidth}% Abstand zwischen Bilder
\begin{minipage}[t]{.5\linewidth} % [b] => Ausrichtung an \caption
\centerline{\includegraphics[width=8cm, height=4.3cm]{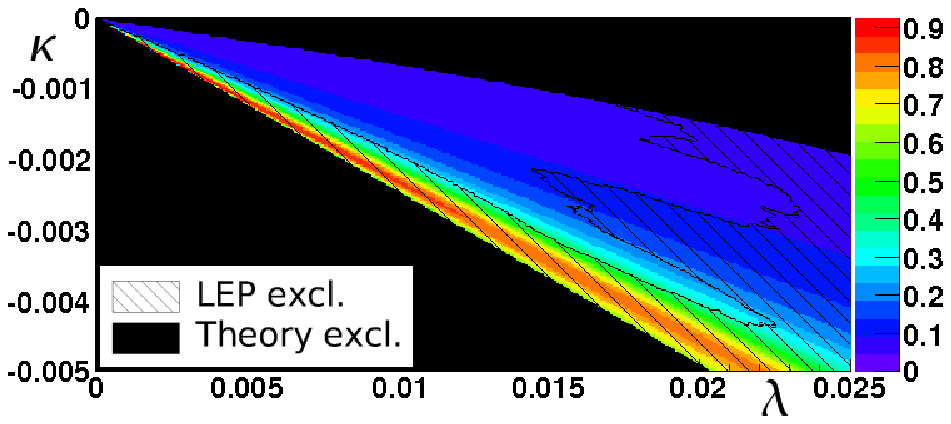}}
\caption{$H_3$ vector boson coupling relative to its SM-value in the  {\it Reduced Couplings Scenario} \label{benchmark_EW1H3CV}}
\end{minipage}
\end{figure}
\begin{figure}[h]
\begin{minipage}[t]{.5\linewidth} % [b] => Ausrichtung an \caption
\centerline{\includegraphics[width=8cm, height=4.3cm]{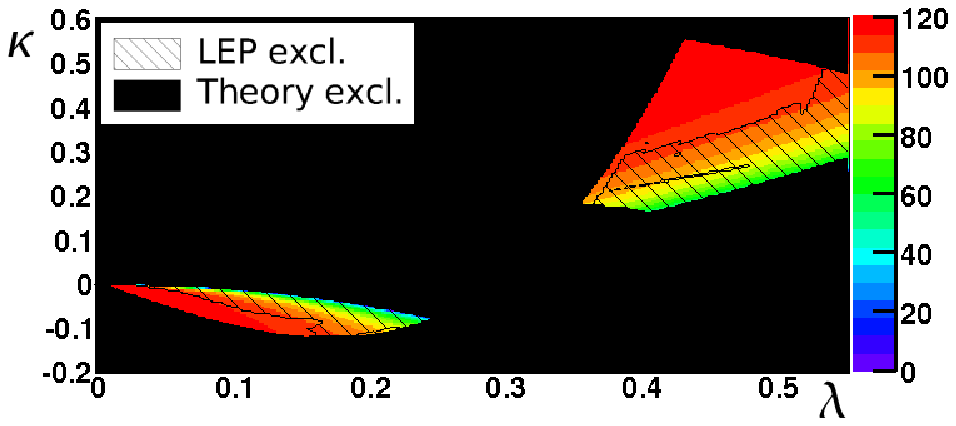}}
%\centerline{\epsfig{file=Effs2.eps, width=6.24.3cm}}
\caption{$H_1$ mass [GeV] in the {\it Light $A_1$ Scenario}\label{benchmark_EW3H1mass}}
\end{minipage}
\hspace{.05\linewidth}% Abstand zwischen Bilder
\begin{minipage}[t]{.5\linewidth} % [b] => Ausrichtung an \caption
\centerline{\includegraphics[width=8cm, height=4.3cm]{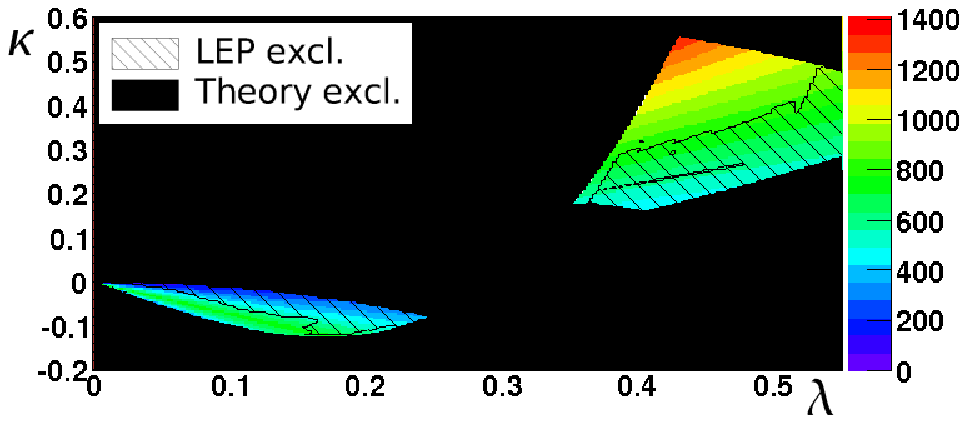}}
\caption{$H_2$ mass [GeV] in the {\it Light $A_1$ Scenario}\label{benchmark_EW3H2mass}}
\end{minipage}
\end {figure}
\subsection{The NMSSM}
In the {\it Minimal Supersymmetric Standard Model} (MSSM), the value of the Higgs-Higgsino mass 
parameter $\mu$ is not confined by theory, but it is experimentally constrained
to lie at the weak scale or else large fine-tuning is required 
(the so called {\it $\mu$-problem}).
%%One of the main unsolved question in the theoretical framework of the MSSM is 
%%the so-called {\it $\mu$-problem}. 
%%The Higgs-Higgsino mass parameter $\mu$ of the MSSM is experimentally contrained
%%to lie approximately at the weak scale. Although this value is in principle 
%%not in contradiction with the theory, $\mu$ is not constrained by the model and 
%%its natural value is thus on the planck-scale.
In the NMSSM, an 
additional neutral singlet superfield $S$ is added to the MSSM.  
After symmetry breaking, $\mu$ is then
given by the product of the vacuum expectation value of the bosonic component 
of $S$ ($<$$s$$>$) and a new coupling constant $\lambda$.
%%\begin{equation}
%%\mu_{eff}=\lambda\cdot <s>
%%\end{equation} 
Constraints from the Higgs potential minimization 
strongly prefer $<$$s$$>$ to lie at the weak scale. The right value of $\mu$ is thus obtained
naturally.\\ 
The resulting model contains the whole particle spectrum of the MSSM with an 
additional neutral scalar boson, a pseudoscalar boson and a neutral fermion 
("singlino"). 
The two additional neutral scalar bosons contained in $S$ mix with the MSSM Higgs bosons to
form the five neutral Higgs bosons of the NMSSM: three CP-even bosons $H_1$, $H_2$, $H_3$ and 
two CP-odd Higgs bosons $A_1$, $A_2$.
The neutral fermion mixes with the four neutralinos of the MSSM, thus, the model contains in total
five neutral fermion states.
%These additional particles mix with the neutral MSSM Higgs 
%bosons and Neutralinos to form the five neutral Higgs bosons and the five
%Neutralinos of the NMSSM:
%\begin{itemize}
%\item Scalar bosons: $h, H \rightarrow H_1, H_2, H_3$
%\item Pseudoscalar bosons: $A \rightarrow  A_1, A_2$
%\item Fermions: $\chi_1^0, \chi_2^0, \chi_3^0, \chi_4^0 \rightarrow \chi_1^0, \chi_2^0, \chi_3^0, \chi_4^0, \chi_5^0$
%%\end{itemize}
Since no charged particles are added, the features of the other MSSM 
particles, including the charged Higgs boson $H^\pm$, are only modified marginally. 
The maximally allowed mass of the lightest NMSSM scalar $H_1$ is about 10 GeV higher
than the bound for $h$ in the MSSM \cite{Ellwanger:2006rm}.\\
In the NMSSM, the Higgs sector can at tree level be described by six parameters. 
Usually, these are chosen to be the coupling parameters of $S$   
($\lambda$, $\kappa$, $A_\lambda$, $A_\kappa$), 
$\mu$ and the ratio of the vacuum expectation values of the Higgs fields, $\tan\beta$.
In the here defined two-dimensional parameter scans, $\lambda$ and $\kappa$ are varied. 
Variation of the other parameters also changes the features of the Higgs sector, 
however, a $\lambda$-$\kappa$ variation was found to be sufficient 
to cover the most important phenomenology types in the two scans described here.\\
To calculate the NMSSM particle spectra and exclusion constraints from theory and LEP\footnote{The {\it Large Electron Positron Collider}, which ran until 2000 at center-of mass energies up to 209 GeV.},
the program NMHDECAY \cite{Ellwanger:2004xm,Ellwanger:2005dv,Domingo:2007dx} was used. The mass parameters were chosen as $M_1 = 500$ GeV, \mbox{$M_2=1$ TeV}, $M_3=3$ TeV and $M_{susy}=1$ TeV. The tri\-li\-ne\-ar soft su\-per\-sym\-me\-try-brea\-king pa\-ra\-me\-ters were set to $A_t=A_b=A_\tau=1.5$ TeV, the top quark mass to 172 GeV.

\subsection{The Reduced Couplings Scenario}
Due to the mixing with the gauge singlet states, the NMSSM Higgs bosons 
can have reduced gauge couplings and thus reduced production cross sections 
compared to the {\it Standard Model} (SM) or the MSSM case. 
A light scalar with reduced gauge couplings and a mass below 114 GeV is still unexcluded by LEP.\\
%%This scenario is a $\lambda$-$\kappa$-scan / includes a point from / around a benchmark point from [EW]. The 
%%point is described in this paper as having the lowest statistical significance
%%found in a region without higgs-to-higgs decays. For exact model parameters, 
%%see Table \ref{benchmark_parameters}.\\
The here proposed scenario is a $\lambda$-$\kappa$ scan with parameters given in Table \ref{benchmark_parameters}.
The point with $\lambda=0.0163$ and $\kappa=-0.0034$ is described as having the lowest statistical significance found in a region without Higgs-to-Higgs decays in Ref. \cite{Ellwanger:2005uu}.\\
%In that region close to the LEP exclusion bounds, the gauge couplings are reduced down to 80\% of their SM-value Fig.XX.
The masses of all six Higgs bosons in this scenario are smaller than about 300 GeV. 
The $H_1$ is very light, down to values of about 20 GeV in an unexcluded region with 
small negative $\kappa$ (Fig.\ref{benchmark_EW1H1mass}). Since the
$H_2$ has a SM-like mass around 120 GeV in the entire plane (Fig.\ref{benchmark_EW1H2mass}), 
there is a region where the decay
$H_2$$\rightarrow$$H_1$$H_1$ is allowed with a small branching ratio of at maximum 6\% in the unexcluded region (Fig.\ref{benchmark_EW1BR}).
The $A_1$ mass ranges from about 55-100 GeV (Fig.\ref{benchmark_EW1A1mass}) in the allowed parameter region, 
whereas the $H_3$, $A_2$, and $H^\pm$ are approximately degenerate in the entire plane, but
with small differences in mass for large negative $\kappa$.
The mass of the $H_3$ ranges from about 150 to 300 GeV, the mass of the $A_2$ from about 
140 to 300 GeV and the 
charged Higgs boson mass from about 165 to 300 GeV in the unexcluded region (Fig.\ref{benchmark_EW1CHmass}).\\
%Then, for example an otherwise SM like 
%light $H_2$ with a mass around e.g. 120 GeV can decay to a fraction via 
%$H_2$$\rightarrow$$H_1$$H_1$. 
In Figures \ref{benchmark_EW1H1CV}, \ref{benchmark_EW1H2CV} \& \ref{benchmark_EW1H3CV}, the vector boson couplings of the scalar bosons are given as an example gauge 
coupling. Higgs boson couplings to gluons and up-type fermions vary similarly. The $H_1$ and 
$H_3$ gauge couplings\footnote{The term 'gauge couplings' here and in the following always excludes the Higgs boson coupling to down-type fermions which may be enhanced with respect to the SM-value, but are still too small to have an impact on the Higgs boson discovery potential with the here used $\tan\beta$ values around 5.} 
are highly suppressed in most of the parameter plane, reaching sizeable values only in the 
LEP excluded region at large negative $\kappa$. The $H_2$ has SM-like gauge couplings in large 
parts of the parameter plane. In the unexcluded region close to the benchmark point from Ref. 
\cite{Ellwanger:2005uu}, the vector boson couplings are reduced down to about 80\% of their 
SM-value. Gauge couplings of the $A_1$ and $A_2$ are highly suppressed for all considered parameter
values.\\
%%The same scan scenario has a otherwise SM/MSSM like light scalar $H_2$, 
%%but which has slightly reduced couplings and might thus be difficult to 
%%discover at future experiments like the LHC - down to 80\%.\\
%%light H1 with reduced couplings - invisible\\
%%H2 with SM-like couplings \& 120 GeV much SM like BR(H2-H1H1) possible but small\\
%%middle region very "SM-like" H2!\\
%%Other Higgs bosons - rather small masses and couplings\\
%%"intense coupling region"\\
%%region with reduced H2 couplings, H1 couplings variing inversly, but nevertheless problematic!\\
%%behaviour with alambda, akappa, mu tanbeta
To summarize, this scenario is characterized by a region with a very light $H_1$ close to the upper
exclusion bound, where  $H_2$$\rightarrow$$H_1$$H_1$ decays are possible, a region with a SM-like $H_2$ 
in the middle of the allowed parameter space, and a region with reduced couplings of the $H_2$ at
large negative $\kappa$ close to the lower exclusion bound.

\begin{table}[tbh]
\caption{Higgs sector parameters of the proposed scenarios\label{benchmark_parameters}}
\begin{center}
\begin{tabular}{|l|c|c|c|c|c|c|} \hline
Scenario & $\lambda$-range & $\kappa$-range& $A_\lambda$ [GeV] &$A_\kappa$ [GeV]&$\mu$ [GeV] & $\tan\beta$\\ \hline 
Reduced couplings & 0 - 0.025 & -0.005 - 0 & -70 & -54 & -284 & 5.7 \\
Light $A_1$ & 0 - 0.55 & -0.2 - 0.6 & -580 & -2.8 & -520 & 5.0 \\ \hline
\end{tabular}
\end{center}
\end{table}
%%\begin{table}[bt]
%%\caption{Used Mass Parameters\label{benchmark_parameters}}
%%\begin{center}
%%\begin{tabular}{|c|c|c|c|c|} \hline
%%$M_1$ [GeV] & $M_2$ [GeV] & $M_3$ [GeV] & $M_{Susy}$ [GeV] & $A_t=A_b=A_\tau$ [GeV]\\ \hline
%%500&1000&3000&1000&1500\\\hline
%%\end{tabular}
%%\end{center}
%%\end{table}

%%by XXX
\begin{figure}[h]
\begin{minipage}[t]{.5\linewidth} % [b] => Ausrichtung an \caption
\centerline{\includegraphics[width=8cm, height=4.3cm]{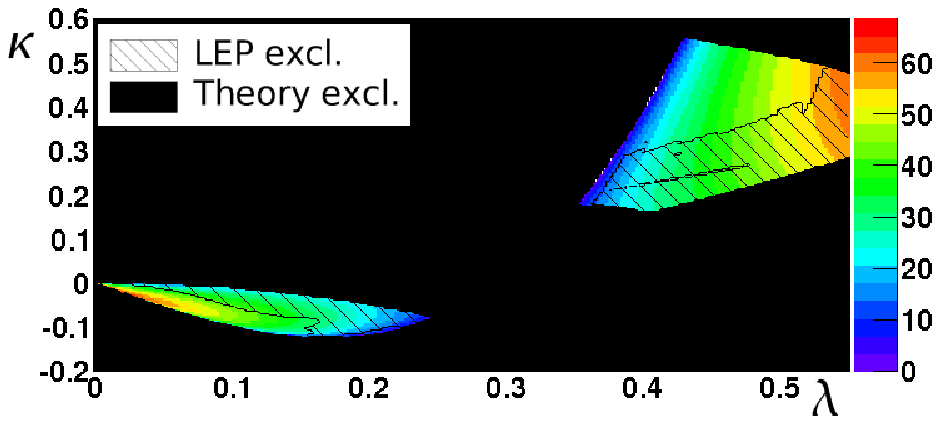}}
%\centerline{\epsfig{file=Effs2.eps, width=6.24.3cm}}
\caption{$A_1$ mass [GeV] in the {\it Light $A_1$ Scenario}\label{benchmark_EW3A1mass}}
\end{minipage}
\hspace{.05\linewidth}% Abstand zwischen Bilder
\begin{minipage}[t]{.5\linewidth} % [b] => Ausrichtung an \caption
\centerline{\includegraphics[width=8cm, height=4.3cm]{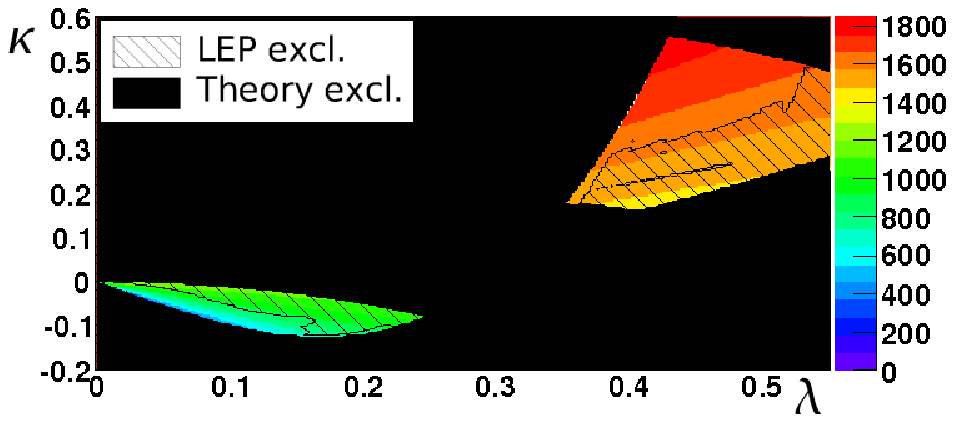}}
\caption{$H^\pm$ mass [GeV] in the {\it Light $A_1$ Scenario}\label{benchmark_EW3CHmass}}
\end{minipage}
\end {figure}
\begin{figure}[h]
\begin{minipage}[t]{.5\linewidth} % [b] => Ausrichtung an \caption
\centerline{\includegraphics[width=8cm, height=4.3cm]{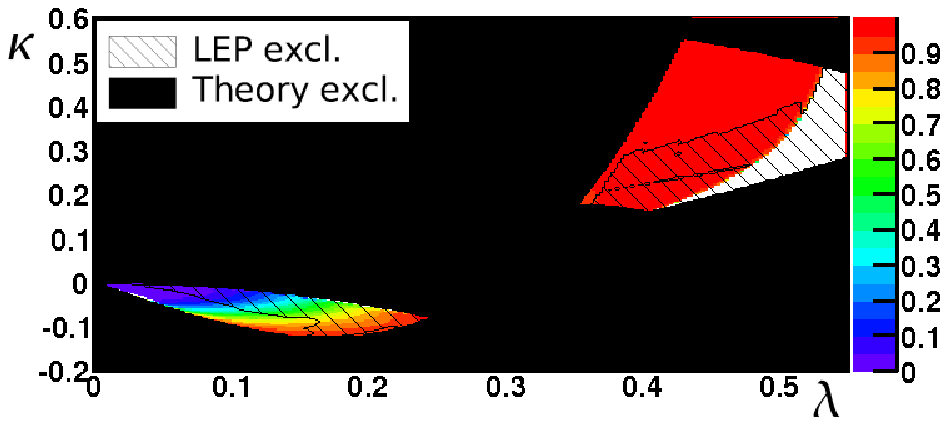}}
%\centerline{\epsfig{file=Effs2.eps, width=6.24.3cm}}
\caption{$H_1$$\rightarrow$$A_1A_1$ branching ratio in the {\it Light $A_1$ Scenario}\label{benchmark_EW3BR1}}
\end{minipage}
\hspace{.05\linewidth}% Abstand zwischen Bilder
\begin{minipage}[t]{.5\linewidth} % [b] => Ausrichtung an \caption
\centerline{\includegraphics[width=8cm, height=4.3cm]{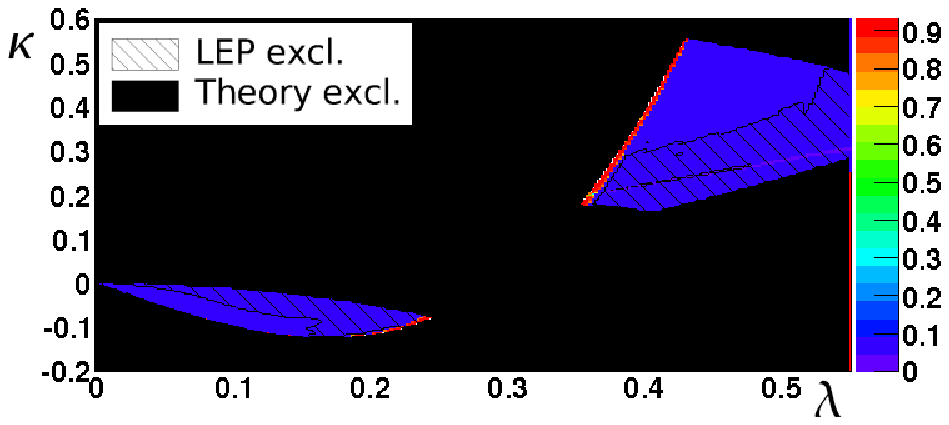}}
\caption{$A_1$$\rightarrow$$\tau\tau$ branching ratio in the {\it Light $A_1$ Scenario}\label{benchmark_EW3BR2}}
\end{minipage}
\end {figure}
\subsection{The Light $A_1$ Scenario}
Unlike in the MSSM, the mass of the lightest pseudoscalar $A_1$ is in the NMSSM 
not closely coupled to the masses of the scalar Higgs bosons and might thus 
lie well below the $H_1$/$H_2$ masses. In such a case, 
the decay chain $H_{1/2}$$\rightarrow$$A_1$$A_1$ can be the dominant decay mode of the lightest 
scalars.\\
%%$A_1$ decays to bottom quarks, and if these are energetically forbidden, to 
%%$\tau$-leptons, are its dominant decay channel. The Higgs boson 
%%phenomenology is thus much dependent on the $A_1$ mass.\\
The here described scenario is also  a $\lambda$-$\kappa$ scan with parameters given in Table \ref{benchmark_parameters}.
The point with $\lambda=0.22$ and $\kappa=-0.1$ has been described 
in Ref. \cite{Ellwanger:2005uu}.\\
Here, the lightest scalar $H_1$ has a mass around 120 GeV in the unexcluded region (Fig.\ref{benchmark_EW3H1mass}). The $A_1$ is very light with masses up to about 60 GeV (Fig.\ref{benchmark_EW3A1mass}), so that the decay $H_1$$\rightarrow$$A_1$$A_1$ is possible in the entire parameter plane with exception of a small region at very small $\lambda$ and $\kappa$ (Fig.\ref{benchmark_EW3BR1}). In the unexcluded region with large $\lambda$ and $\kappa$, this decay reaches branching ratios above 90\%. 
Areas with a smaller branching ratio exists for smaller $\lambda$ and $\kappa$.
The other Higgs bosons are rather heavy (Figs.\ref{benchmark_EW3H2mass},\ref{benchmark_EW3CHmass}), with the $H_3$, $A_2$ and $H^\pm$ being approximately degenerate in large parts of the parameter plane.\\
For $A_1$ masses larger than $2m_b$, about 90\% of the lightest pseudoscalar bosons decay to bottom quarks. In these regions, the decay chains $H_1$$\rightarrow$$A_1$$A_1$$\rightarrow$$b\bar{b}b\bar{b}$ and  $H_1$$\rightarrow$$A_1$$A_1$$\rightarrow$$b\bar{b}\tau\tau$ are important.
In small regions 
at the borders of the unexcluded region, the $A_1$ is so light that the decay chain  $H_1$$\rightarrow$$A_1$$A_1$$\rightarrow$$\tau\tau\tau\tau$ prevails (Fig.\ref{benchmark_EW3BR2}).
In the narrow unexcluded band around $\lambda\approx 0.25$, the couplings of the $A_1$ to fermions are heavily suppressed. Here, the decay chain $H_1$$\rightarrow$$A_1$$A_1\rightarrow\gamma\gamma\gamma\gamma$ is dominant.\\
The gauge couplings of the $H_1$ are SM-like in the entire allowed parameter region (Fig.\ref{benchmark_EW3H1CV}). 
The gauge couplings of the $H_2$
are sizeable only in a small excluded region with $\kappa$-values close to zero (Fig.\ref{benchmark_EW3H2CV}). All other Higgs bosons
have highly suppressed gauge couplings in the entire parameter plane.
%%In short, this scenario is characterized by a the scalar $H_1$ having SM-like couplings and a 
%%'SM-like' mass of 120 GeV, but possible $H_1$$\rightarrow$$A_1$$A_1$ decays. Both regions with 
%%dominant decays of the $A_1$ to bottom-quarks and to $\tau$-leptons were covered. Please note the existence of these types of phenomenology are very sensitive to the value of $A_\kappa$.
%%\section{OTHER PHENOMENOLOGY TYPES}
%%Additionally, it should be mentioned that the upper limit on the lightest 
%%scalar mass in the NMSSM is less severe than in the MSSM, allowing Higgs 
%%masses up to around 140-150 GeV XXX. Although is is surely also an interesting 
%feature of the NMSSM, it is not looked into detail in this proposal.
%%A5 more detailed description of the model can be found in the according 
%%literature, exemplementary ?? gibt es das wort?? This description was inspired 

\begin{figure}[htb]
\begin{minipage}[t]{.5\linewidth} % [b] => Ausrichtung an \caption
\centerline{\includegraphics[width=8cm, height=4.3cm]{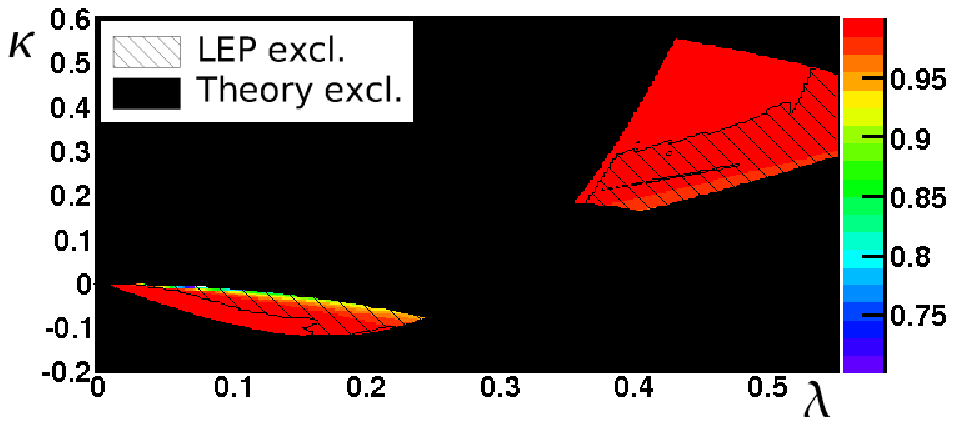}}
%\centerline{\epsfig{file=Effs2.eps, width=6.24.3cm}}
\caption{$H_1$ vector boson coupling relative to its SM-value in the {\it Light
 $A_1$ Scenario}\label{benchmark_EW3H1CV}}
\end{minipage}
\hspace{.05\linewidth}% Abstand zwischen Bilder
\begin{minipage}[t]{.5\linewidth} % [b] => Ausrichtung an \caption
\centerline{\includegraphics[width=8cm, height=4.3cm]{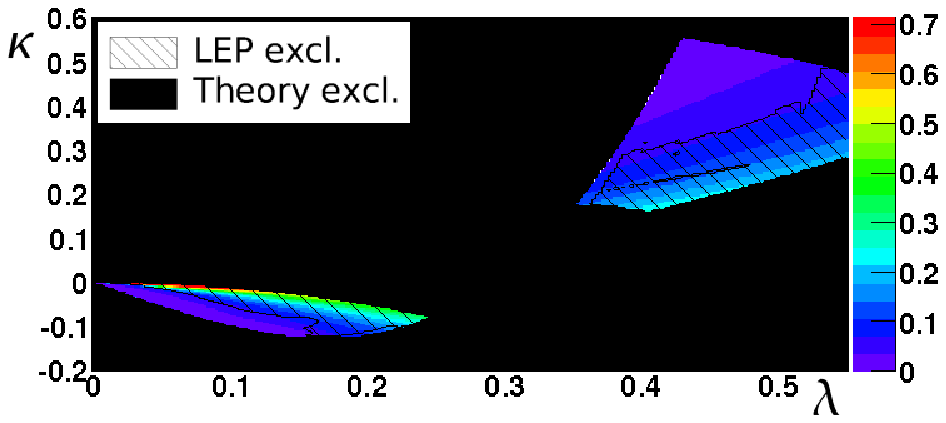}}
\caption{$H_2$ vector boson coupling relative to its SM-value in the {\it Light $A_1$ Scenario}\label{benchmark_EW3H2CV}}
\end{minipage}
\end {figure}

\subsection{Conclusions}

Two interesting two-dimensional NMSSM scans were described and proposed as possible benchmarks 
for NMSSM Higgs boson searches. These two scans cover the four main, for the NMSSM
typical phenomenology types, for which a discovery of Higgs bosons at future experiments 
like the LHC might be difficult:
\begin{itemize}
\item A region with very light scalar $H_1$.
\item A region with reduced gauge couplings of an otherwise SM-like scalar $H_2$.
\item Regions with dominant $H_1$$\rightarrow$$A_1$$A_1$$\rightarrow$$b\bar{b}b\bar{b}$/$b\bar{b}\tau\tau$ decays of an otherwise SM-like scalar
$H_1$.
\item Regions with dominant $H_1$$\rightarrow$$A_1$$A_1$$\rightarrow$$\tau\tau\tau\tau$ decay of an otherwise SM-like scalar $H_1$.
%\item Regions with dominant $H_1$$\rightarrow$$A_1$$A_1$$\rightarrow$$\gamma\gamma\gamma\gamma$ decay of an otherwise SM-like scalar $H_1$.
\end{itemize}
Another example of an experimentally challenging phenomenology type not covered here 
is a dominant $H_1$$\rightarrow$$c\bar{c}$ decay \cite{Miller:2004uh}. 
Also the region where the mass of the lightest
scalar is maximal \cite{Ellwanger:2006rm} could prove interesting for Higgs boson discovery .

%\section*{ACKNOWLEDGEMENTS}
%We would like to thank A. Djouadi and R. Godbole for their valuable help and numerous discussions during this workshop and later on.%

%%For both scenarios, the quartic couplings $\lambda$ and $\kappa$ were varied and 
%%the trilinear couplings $A_\lambda$, $A_\kappa$, $\mu$ and $\tan\beta$ are kept fixed.
%%In general, it was found that these typical regions with unique 
%%phenomenologies are rather marginal, closely surrounded by experimentally 
%%excluded regions or regions with no special NMSSM-character, but rather MSSM- 
%%and SM-like behaviour.
%%other NMSSM-like things like H1-cc, maximal mass

%%\section*{ACKNOWLEDGEMENTS}
%%Whom would we like to thank?

%\bibliography{benchmark}

%\end{document}

\section[The NMSSM No-Lose Theorem at the LHC: 
The Scope of the 4$\tau$ Channel
In Higgs-strahlung and Vector Boson Fusion]
{THE NMSSM NO-LOSE THEOREM AT THE LHC: THE SCOPE OF THE 4$\tau$
CHANNEL IN HIGGS-STRAHLUNG AND VECTOR BOSON FUSION
~\protect\footnote{Contributed by: 
Alexander Belyaev, Stefan Hesselbach, Sami Lehti,  
Stefano Moretti, Alexander Nikitenko, 
Claire H. Shepherd-Themistocleous}
}
\label{sec:4tau}
\subsection{Introduction}
\noindent
As emphasised in Sect. \ref{sec:bench}
(see also \cite{Djouadi:2008uw}), the NMSSM
has obvious advantages with respect to the MSSM. In constrast, 
it is not certain that at least one Higgs boson can be found at the LHC
in such a scenario. In this respect,
of particular relevance are $h_1\to a_1a_1$ decays,
as they have been claimed to be the only means to establish
a no-lose theorem at the CERN machine for the NMSSM
\cite{Gunion:1996fb,Ellwanger:2001iw,Azuelos:2002qw,Ellwanger:2003jt,Ellwanger:2004gz,Miller:2004uh,Assamagan:2004mu,Weiglein:2004hn,Moretti:2006hq,Carena:2007jk}, 
at least over a
region of parameter space where Supersymmetry (SUSY) partners of ordinary 
Standard 
Model (SM) objects are made suitable heavy. Here, $a_1$ states are rather
light (of 10 GeV or less) while $h_1$ ones could well be below the LEP limit
on the SM Higgs mass, of 114 GeV (albeit with weakened couplings to ordinary
matter). The scope of $h_1\to a_1a_1$ decays into $jj\tau^+\tau^-$ pairs
(where j represents a jet of either heavy or light flavour and where
the $\tau$'s decay leptonically) has been found to be rather questionable
\cite{Baffioni:2004gdr}. Hence, in this contribution we 
investigate the scope of the $4\tau$ channel, wherein two $\tau$'s are searched
for in their muonic decays while the other two are selected via their
hadronic ones. We will consider both HS and VBF as production channels.
Finally, to enhance the yield of $a_1\to \tau^+\tau^-$ decays, we
limit ourselves to regions of NMSSM parameter space where 
$M_{a_1}< 2 m_b$ (light $a_1$ scenario). 

\subsection{The Low-Energy NMSSM Parameter Space
For The Light $a_1$ Scenario}
\noindent
In this section we investigate the NMSSM parameter space setups which yield
the $M_{a_1}< 2 m_b$ mass configuration, with particular interest
to the cases where the aforementioned $h_1\to a_1a_1\to 4\tau$ decays
may be visible at the LHC if happening in conjunction with HS and/or VBF 
production  processes of Higgs bosons. Notice that 
there are altogether fourteen parameters that uniquely define at the Electro-Weak (EW) scale
the NMSSM Higgs
sector for the purposes of our analysis. With reference to notation already
defined elsewhere in this report, these are: 
$\tan\beta,  \lambda,  \kappa,  A_\lambda, 
A_\kappa, M_1, M_2, M_3,  A_{t},  A_{b},  A_\tau, 
 M_{f_L}$ and $M_{f_R}$, where $M_{f_L}$ and $M_{f_R}$ denote the soft SUSY breaking slepton
and squark mass parameters.
We will start by establishing the portion of NMSSM parameter space, defined
in terms of the above inputs,
 that survives present theoretical and 
experimental constraints.

\subsubsection{Full NMSSM Parameter Scan}
\noindent
The numerical values over which the parameters introduced above have been
scanned are:
$$
-1000~{\rm{GeV}}<A_\kappa<100~{\rm{GeV}}, ~~
-10~{\rm{TeV}}<A_\lambda<10~{\rm{TeV}}, ~~
100~{\rm{GeV}}<\mu<1000~{\rm{GeV}},
$$
\begin{equation}
10^{-5}<\lambda,\kappa<0.7, ~~
  1.5<\tan\beta<50
\label{par-space}
\end{equation}
while the remaining parameters were fixed at
\begin{equation}
M_1/M_2/M_3=150/300/1000~{\rm{GeV}},~~
A_{t}=A_{b}=A_\tau=2.5~{\rm{TeV}},~~
M_{f_L}=M_{f_R} =1~{\rm{TeV}}.
\label{par-fixed}
\end{equation}
We will call the scan performed over such intervals a `wide' scan. This (as well
as all those in the remainder of this note) has been performed by using the
NMSSMTools 
package \cite{Ellwanger:2004xm,Ellwanger:2005dv,Domingo:2007dx},
which calculates NMSSM spectra (masses, couplings and decay rates) and takes into account
experimental inputs including LEP limits, $B$-physics bounds as well as (cold) DM constraints. In Fig.~\ref{fig:scan-wide} we present the 
results of  this  scan.
 Though only a few  $M_{a_1}<10$~GeV points survive, one can see from  Fig.~\ref{fig:scan-wide}(a) 
the preference for a  large positive $A_\lambda$
while 
Fig.~\ref{fig:scan-wide}(b)  indicates
 that small $|A_\kappa|$'s are preferred.

\begin{figure}[!t]
  \includegraphics[width=0.33\textwidth]{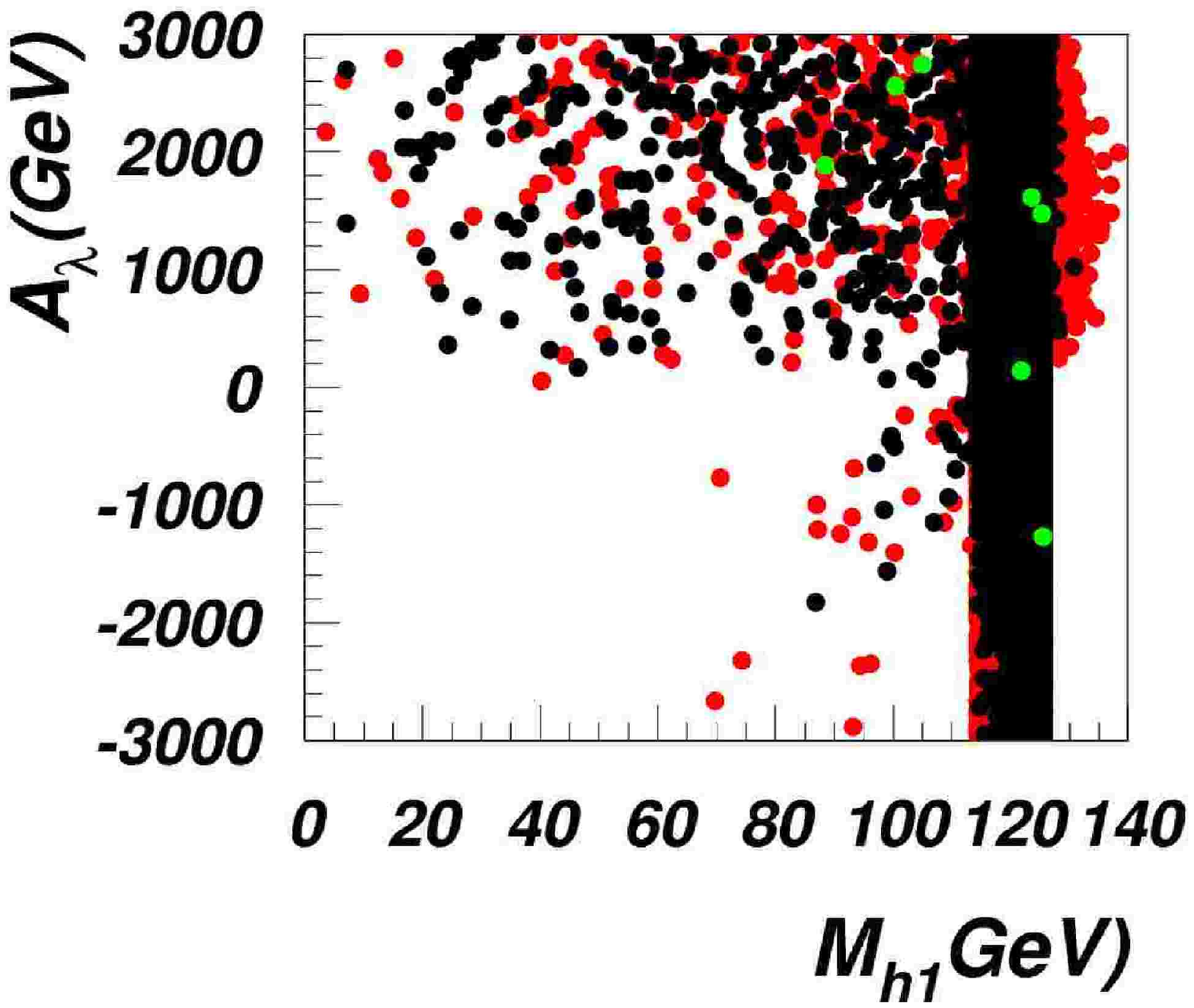}%
  \includegraphics[width=0.33\textwidth]{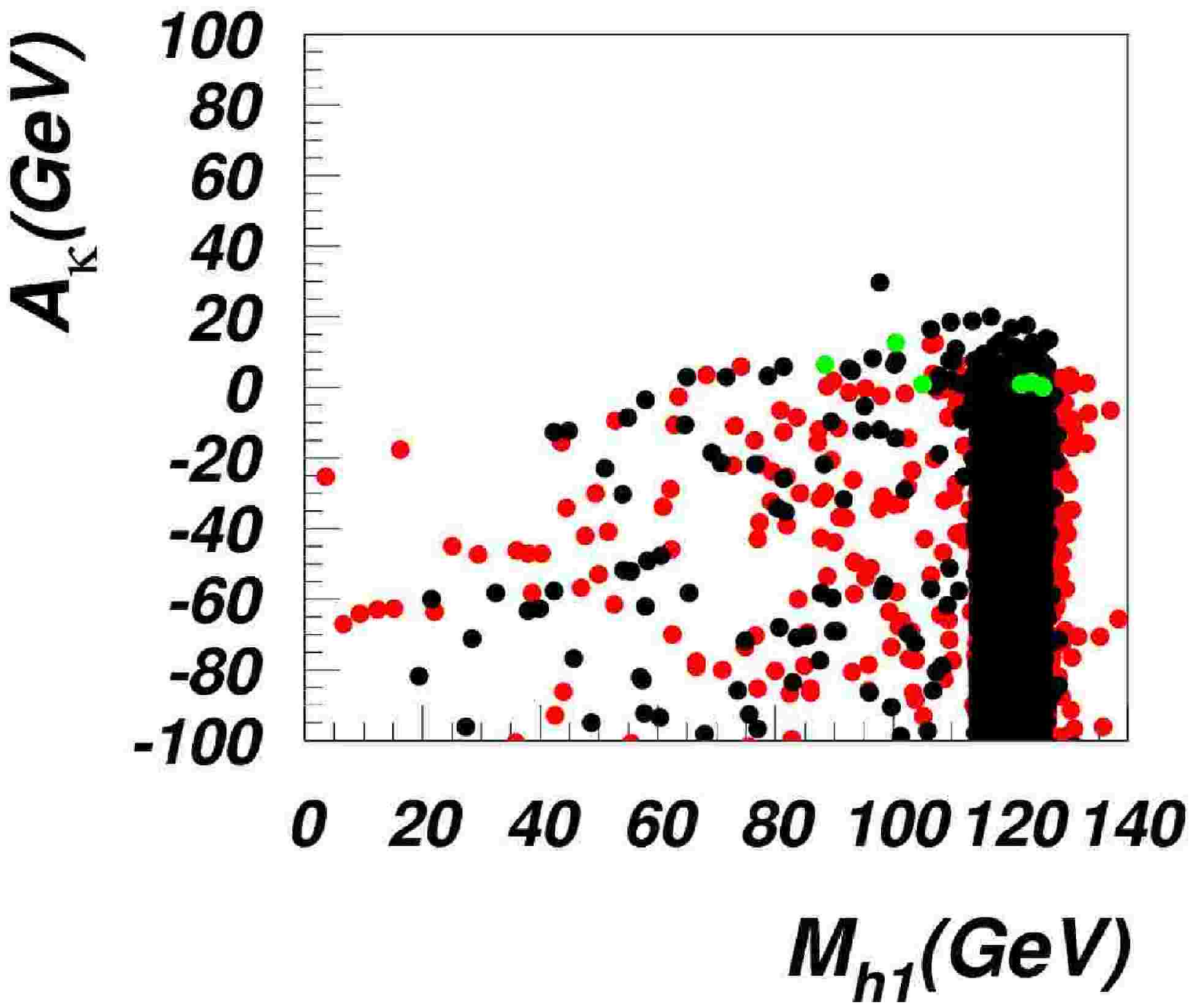}%
  \includegraphics[width=0.33\textwidth]{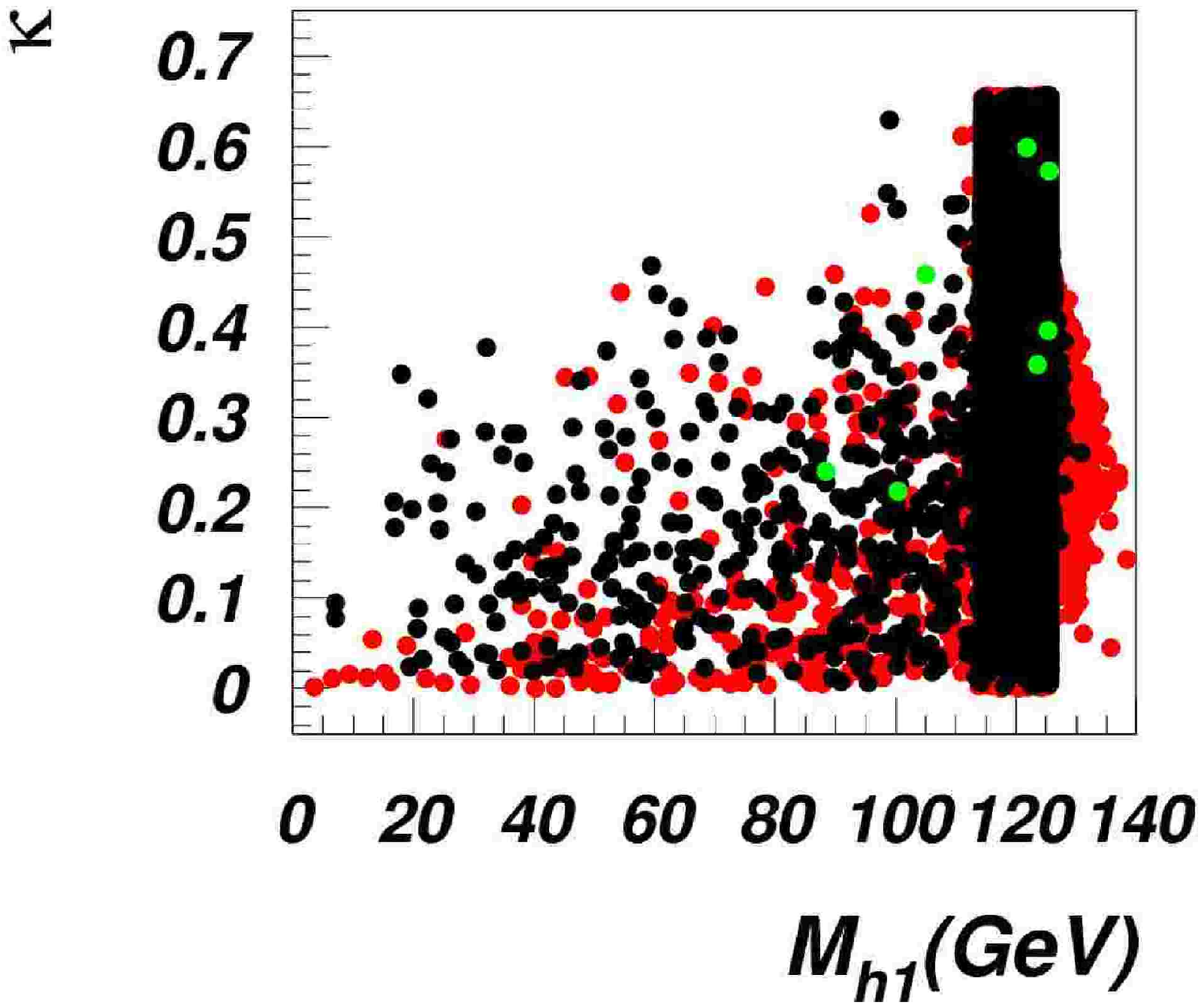}\\
\vskip -3.cm
\hspace*{0.33\textwidth}\hspace*{-5.4cm}{\bf (a)}
\hspace*{0.33\textwidth}\hspace*{-0.5cm}{\bf (b)}
\hspace*{0.33\textwidth}\hspace*{-0.5cm}{\bf (c)}
\vskip  2cm
  \includegraphics[width=0.33\textwidth]{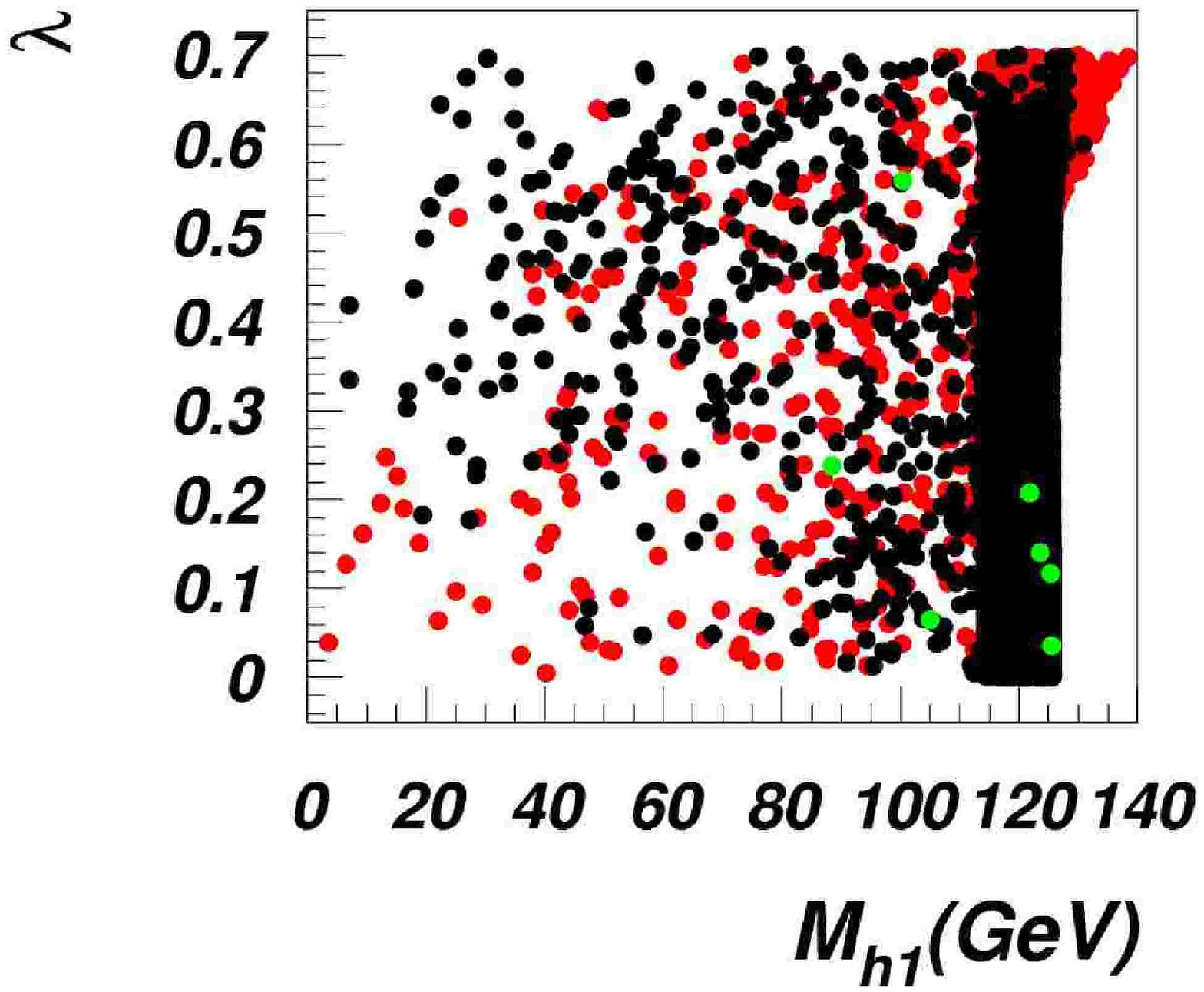}%
  \includegraphics[width=0.33\textwidth]{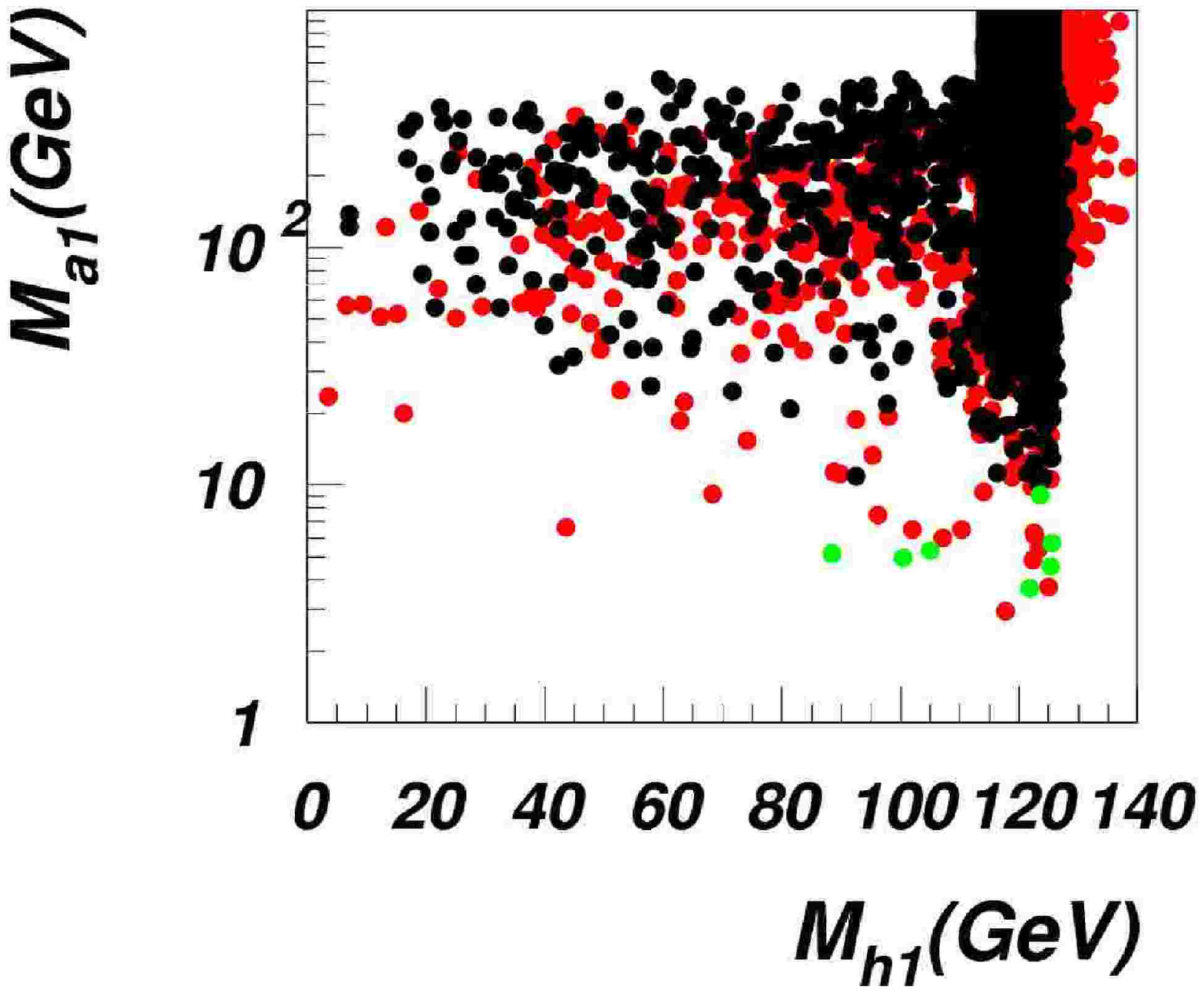}%
  \includegraphics[width=0.33\textwidth]{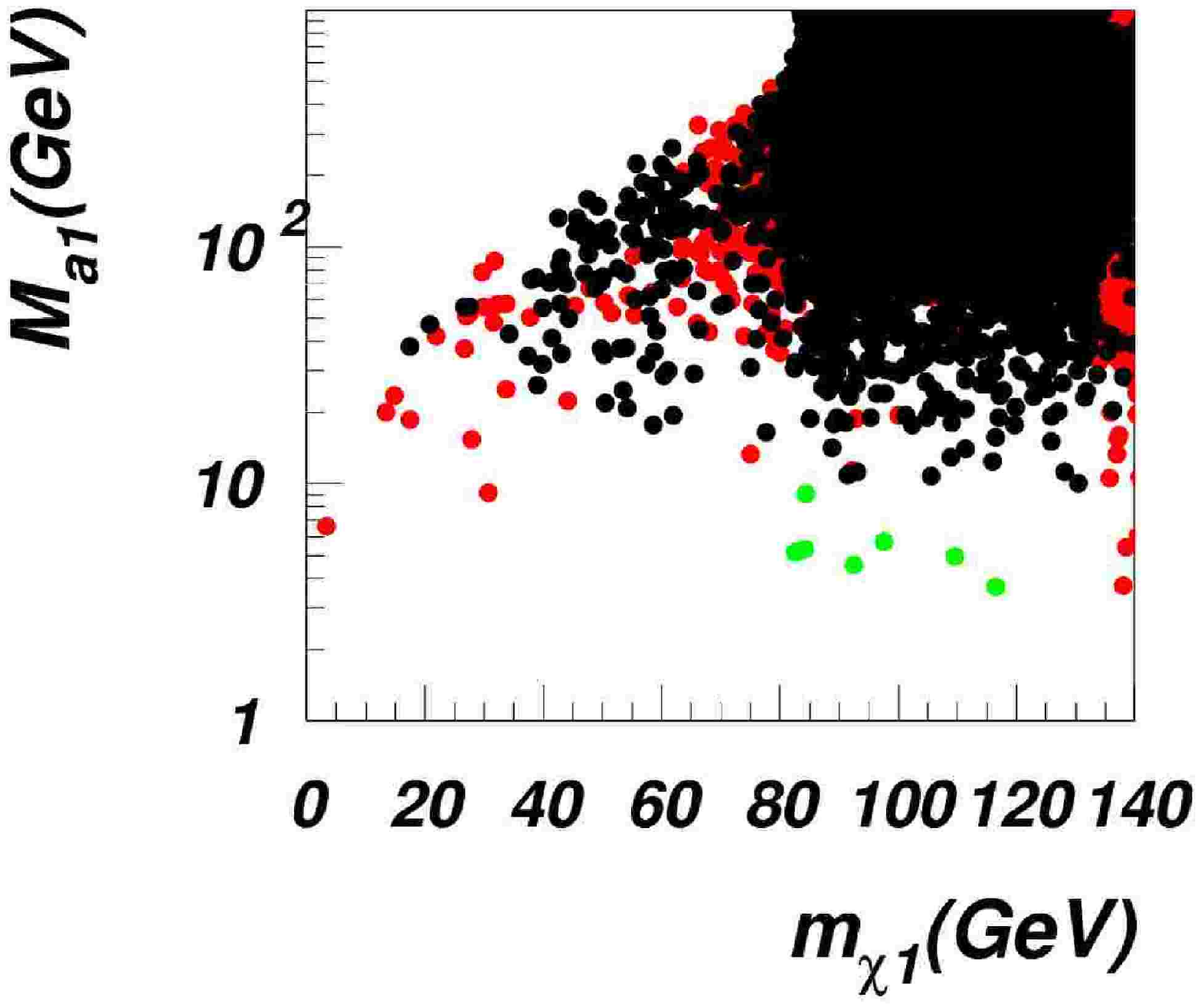}%
\vskip -2.5cm
\hspace*{0.33\textwidth}\hspace*{-5.4cm}{\bf (d)}
\hspace*{0.33\textwidth}\hspace*{-0.5cm}{\bf (e)}
\hspace*{0.33\textwidth}\hspace*{-0.5cm}{\bf (f)}
\vskip  2cm
\caption{Result of the NMSSM `wide' scan mapped onto the planes:
 	 (a) [$A_\lambda, M_{h_1}$],
 	 (b) [$A_\kappa, M_{h_1}$],
 	 (c) [$\kappa, M_{h_1}$],
 	 (d) [$\lambda, M_{h_1}$],
 	 (e) [$M_{a_1}, M_{h_1}$],
         (f) [$M_{a_1}, m_{\chi^0_1}$].
 Colour code:
 red    -- all constraints are satisfied but relic density (above WMAP constraint: $\Omega h^2>0.11$);
 black  -- all constraints are satisfied, $M_A>10$~GeV;
 green -- all constraints are satisfied, $M_A<10$~GeV. 
\label{fig:scan-wide}}
\end{figure}

\subsubsection{Scan for Narrowed $A_\kappa$}
\noindent
The results of Fig.~\ref{fig:scan-wide}
(specifically, the preference for small $A_\kappa$'s) motivated us
to `narrow' the range of the parameters, by scanning it over the intervals
\begin{equation}
-15~{\rm{GeV}}<A_\kappa<20~{\rm{GeV}},~~
-2~{\rm{TeV}}<A_\lambda<4~{\rm{TeV}}, ~~
100~{\rm{GeV}}<\mu<300~{\rm{GeV}},
  \label{cut:ak_narrow}
\end{equation}
and the rest of the parameters as in Eq.~(\ref{par-space}).
Fig.~\ref{fig:scan-narrow} makes the point that this is precisely the region where a
large portion of NMSSM  points  with $M_{a_1}<10$~GeV are found, consistent with all known constraints.
Now we can clearly see certain correlations onsetting in the $M_{a_1}<10$~GeV region:
1. values of $A_\lambda>0$ are preferred,
    see Fig.~\ref{fig:scan-narrow}(a);
2. points with low $M_{h_1}$,  $A_k\sim 0$
   (Fig.~\ref{fig:scan-narrow}(b)) and 
   small values of $\kappa$ 
   (Fig.~\ref{fig:scan-narrow}(c)) are preferred;
3. we can see interesting  $M_{a_1}<10$~GeV points 
   with also low, down to 20 GeV, $M_{h_1}$ values 
(Fig.~\ref{fig:scan-narrow}(e)).

\begin{figure}[!t]
  \includegraphics[width=0.33\textwidth]{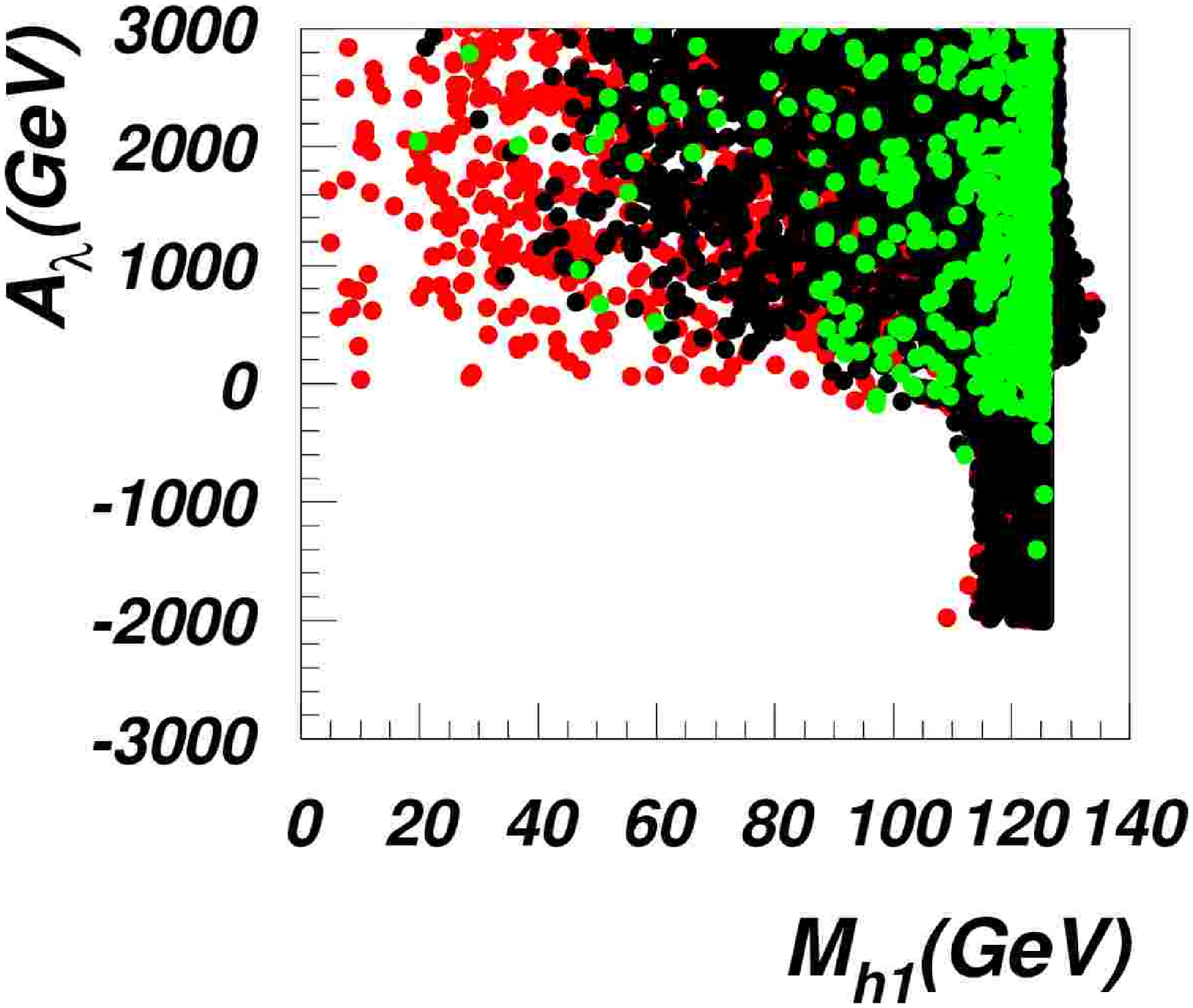}%
  \includegraphics[width=0.33\textwidth]{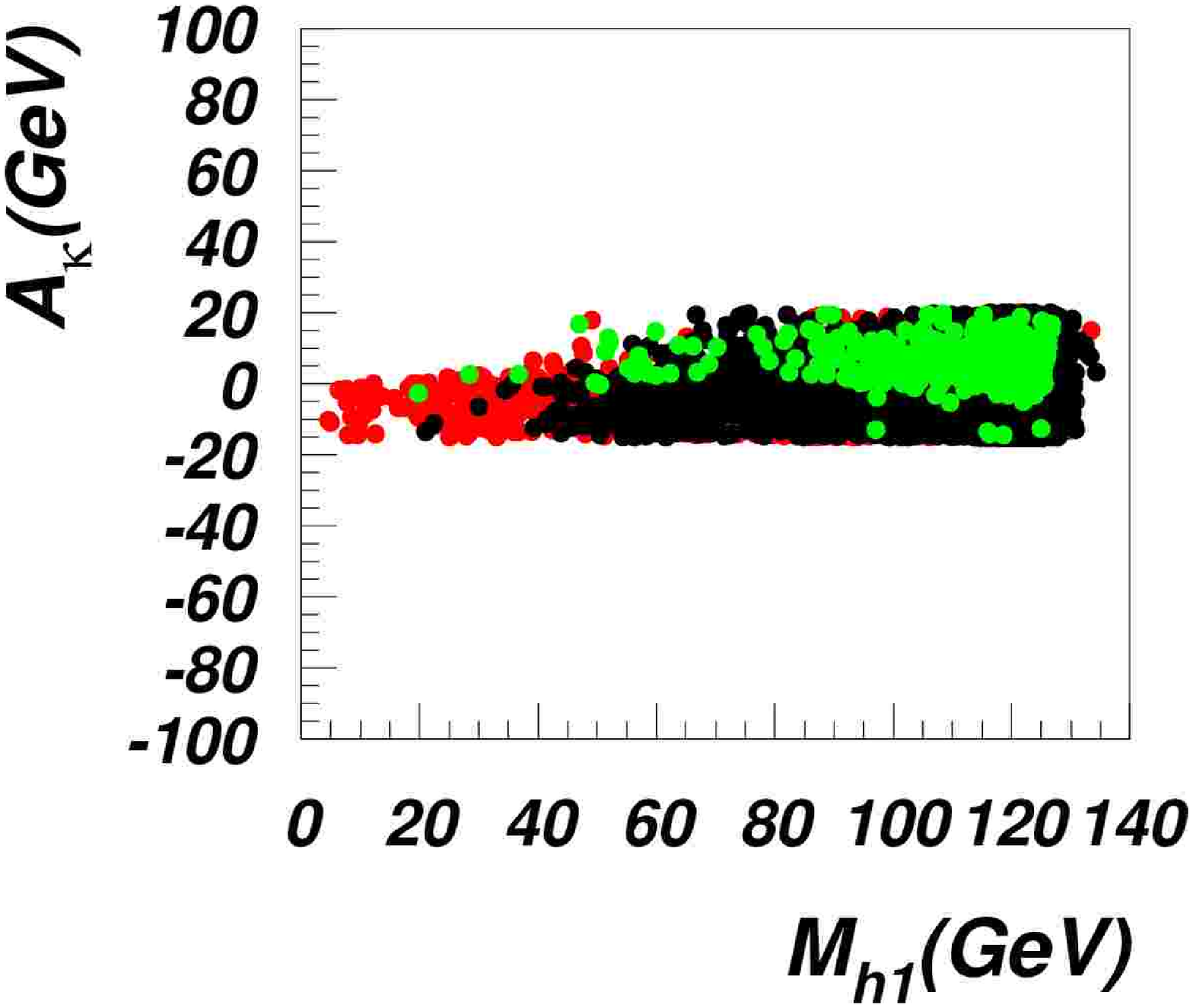}%
  \includegraphics[width=0.33\textwidth]{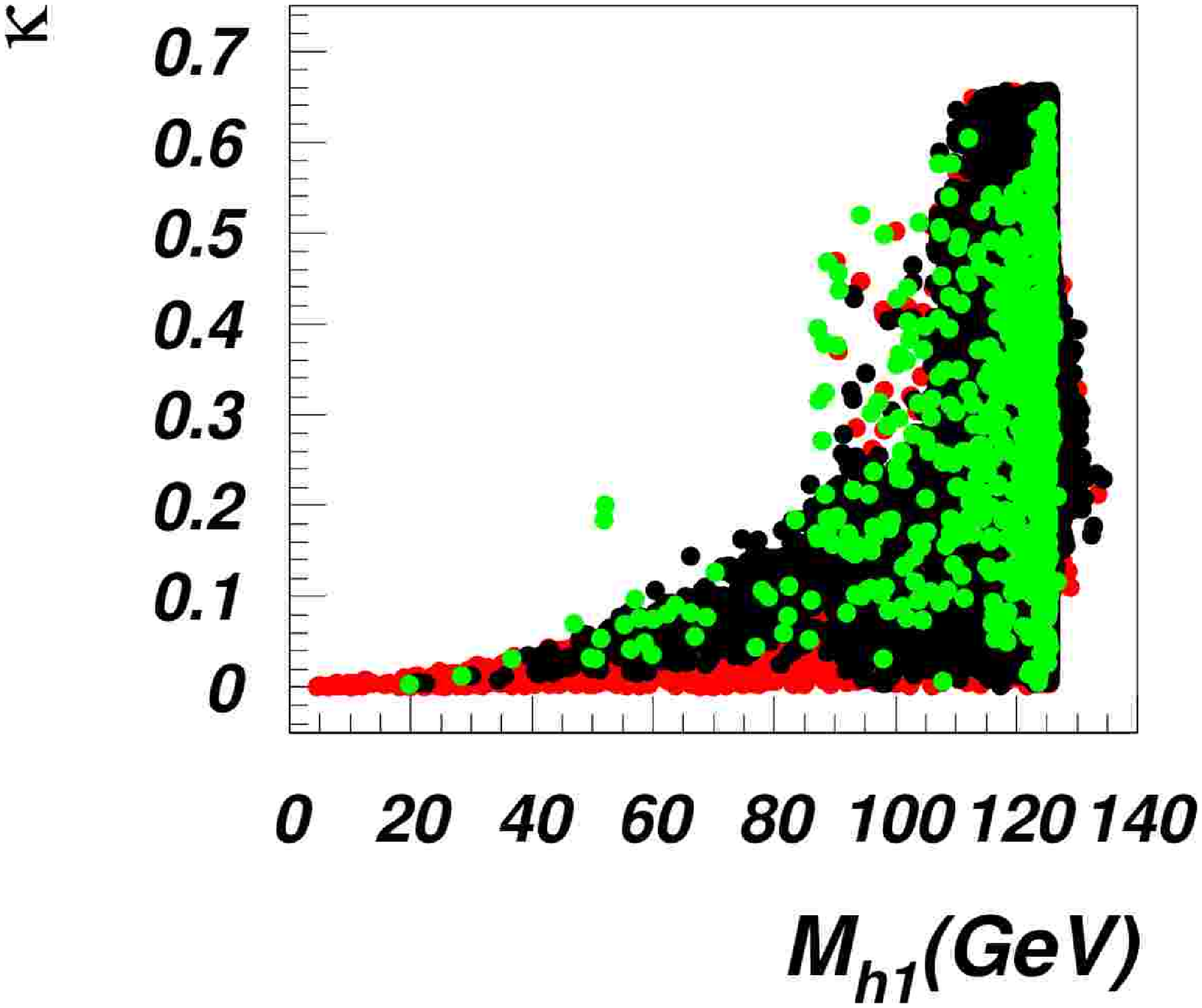}\\
\vskip -3.cm
\hspace*{0.33\textwidth}\hspace*{-5.4cm}{\bf (a)}
\hspace*{0.33\textwidth}\hspace*{-0.5cm}{\bf (b)}
\hspace*{0.33\textwidth}\hspace*{-0.5cm}{\bf (c)}
\vskip  2cm
  \includegraphics[width=0.33\textwidth]{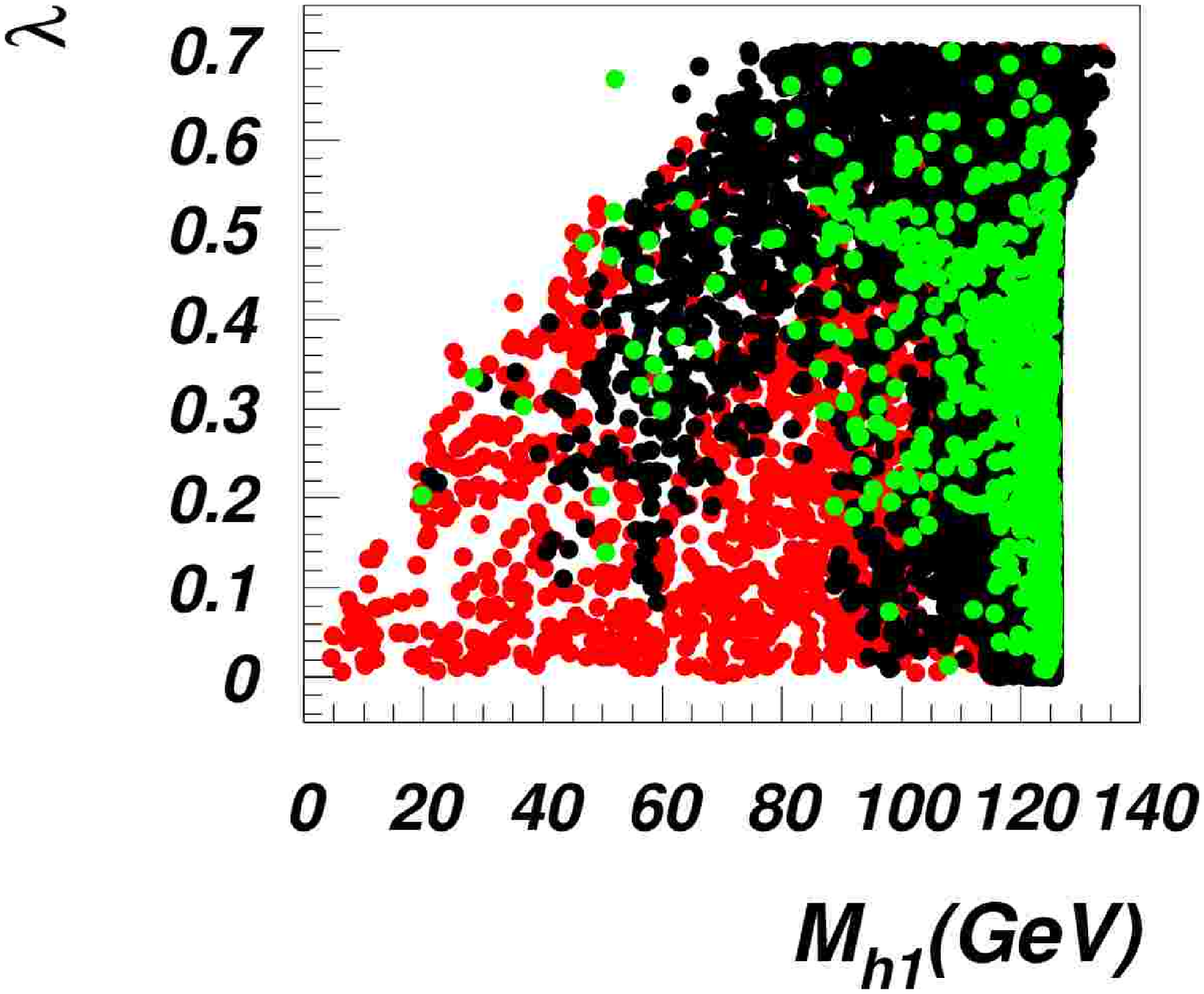}%
  \includegraphics[width=0.33\textwidth]{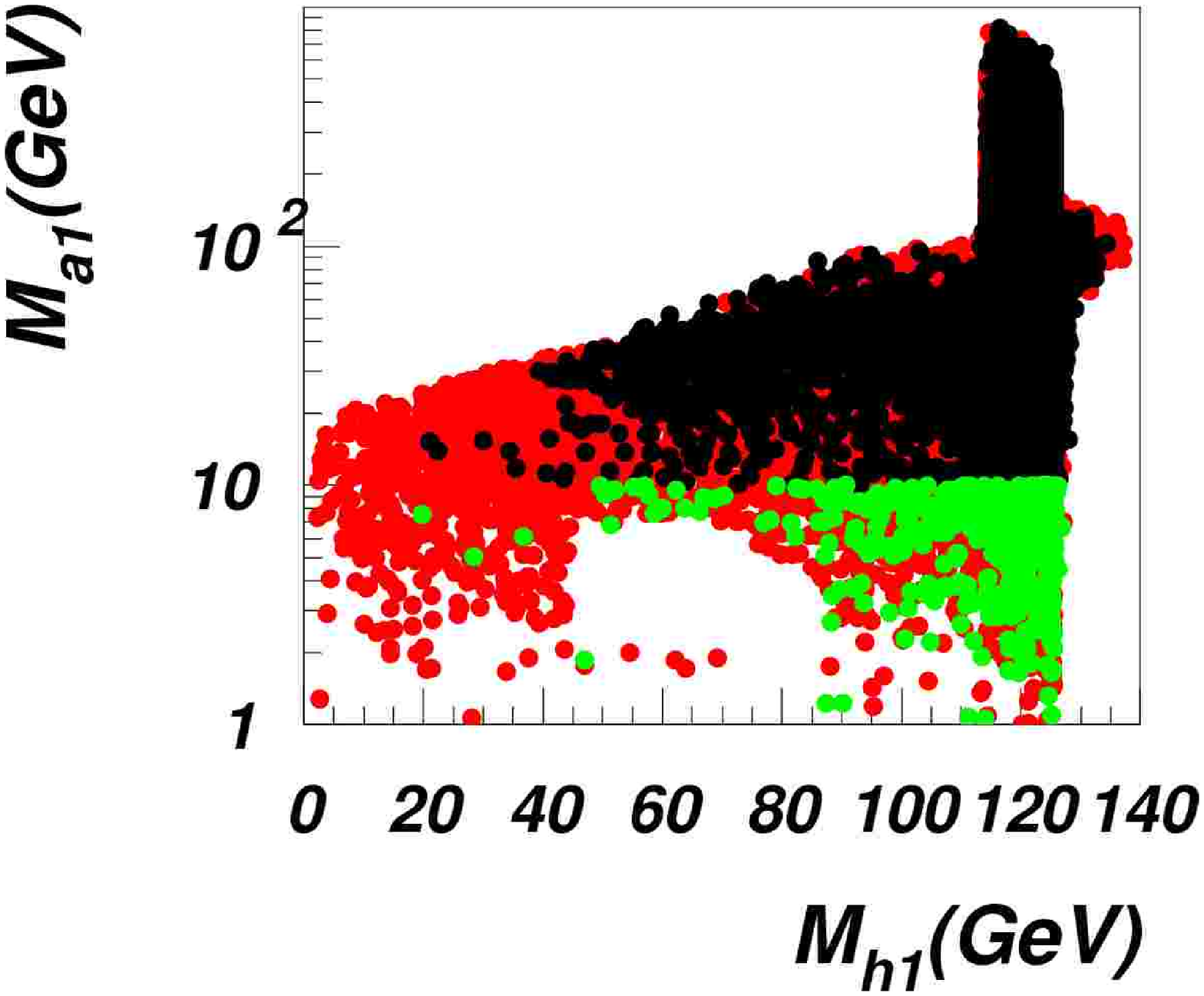}%
  \includegraphics[width=0.33\textwidth]{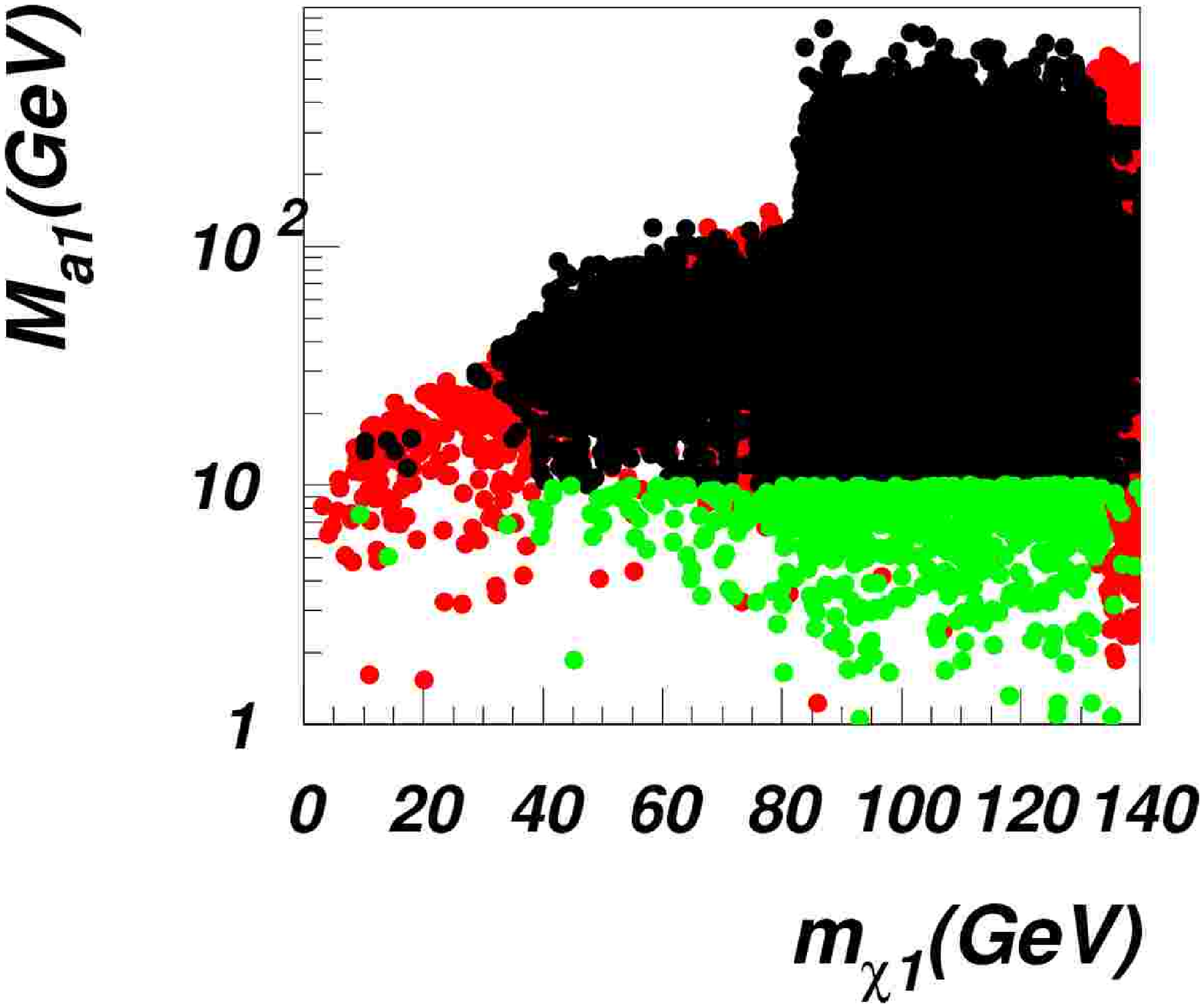}%
\vskip -2.5cm
\hspace*{0.33\textwidth}\hspace*{-5.4cm}{\bf (d)}
\hspace*{0.33\textwidth}\hspace*{-0.5cm}{\bf (e)}
\hspace*{0.33\textwidth}\hspace*{-0.5cm}{\bf (f)}
\vskip  2cm
\caption{Results of the NMSSM `narrow' scan, i.e., analogous to  Fig.~\ref{fig:scan-wide}
but for $-15~{\rm{GeV}}<A_\kappa<20~{\rm{GeV}}$,
$-2~{\rm{TeV}}<A_\lambda<4~{\rm{TeV}}$,
$100~{\rm{GeV}}<\mu<300~{\rm{GeV}}$.
The individual plots and the colour code are the same as in Fig.~\ref{fig:scan-wide}.
\label{fig:scan-narrow}}
\end{figure}

\subsubsection{Final Scan For the Light $a_1$ Scenario}
\noindent
We have then performed one `final' scan over the
NMSSM parameter space by requiring at the same time $M_{a_1}< 10$~GeV and Eq.~(\ref{cut:ak_narrow}). 
Having already learnt the size of such portion of the entire NMSSM
parameter space after experimental constraints,
we now want to characterise it  in terms
of the quantities which enter the event rates for $h_1\to a_1a_1\to 4\tau$ decays
produced via HS and VBF.
The results of this scan are  shown in Fig.~\ref{fig:scan-final}. 
Note that, here, the colours were chosen
 to indicate the measure of decoupling of the lightest CP-even Higgs boson, $h_1$,
 from the SM limit (denoted simply by $H$). To this aim, we have defined the measure $R_{ZZh}=\left(g_{ZZh_1}^{\rm{NMSSM}}/g_{{ZZH}}^{\rm{SM}}\right)^2$,
i.e., the ratio of the coupling strength (squared) of the $ZZh_1$ vertex in the NMSSM relative to the SM case (in fact,
this is the same for couplings to $W^\pm$ gauge bosons).
 One should notice that both HS,
 $pp\to V h_1$, and VBF,
 $pp\to jjV^*V^*\to jjh_1$, rates ($V=Z,W^\pm$) are directly proportional to  $R_{ZZh}$
 and are suppressed in the non-decoupling regime
 whenever $R_{ZZh}$ is essentially smaller then unity.
From Fig.~\ref{fig:scan-final} one can see the following important features of the
$M_{a_1}< 2m_b\approx10$ GeV  scenario:
1. the lighter the Higgs the more significant  
   should be the NMSSM deviations from the SM case, e.g.,
   for any $M_{h_1}<50$~GeV any $R_{ZZh}$ is limited  to be
   $<0.5$, as dictated by LEP 
   constraints~\cite{Barate:2003sz,Schael:2006cr}
   (this correlation is illustrated in a more clear way  in
   Fig.~\ref{final-scan1}(a), presenting the $R_{ZZh}$ versus $M_{h_1}$
   plane,
   which exhibits the typical pattern of the LEP Higgs exclusion 
   curve~\cite{Schael:2006cr});
2. in the $M_{h_1}<40$ GeV region $A_\lambda$ is always positive
   (Fig.~\ref{fig:scan-final}(a)), $\kappa <0.1$ (Fig.~\ref{fig:scan-final}(c))
   while  $\lambda<0.45$ (Fig.~\ref{fig:scan-final}(d)).
\begin{figure}[!t]
  \includegraphics[width=0.33\textwidth]{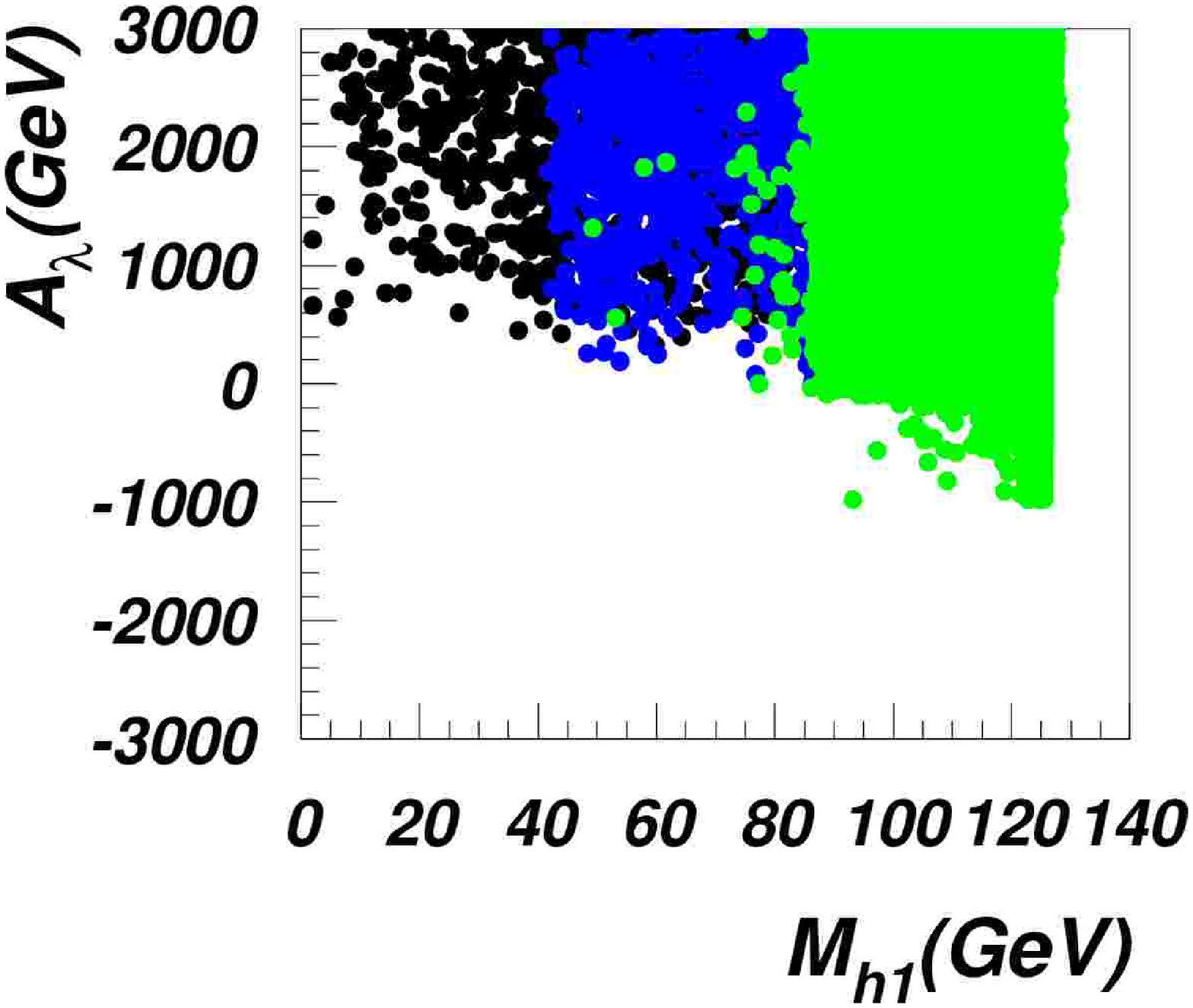}%
  \includegraphics[width=0.33\textwidth]{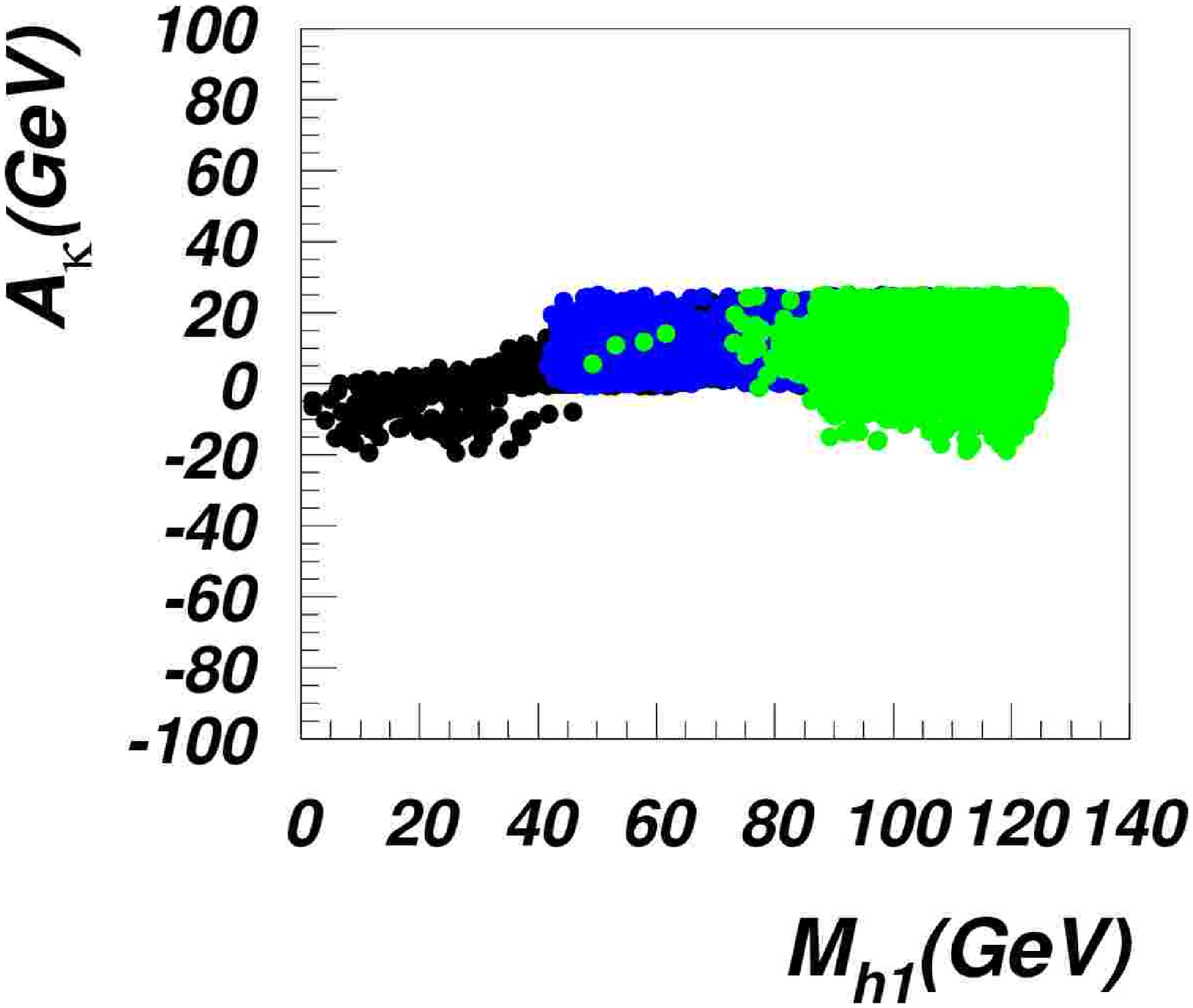}%
  \includegraphics[width=0.33\textwidth]{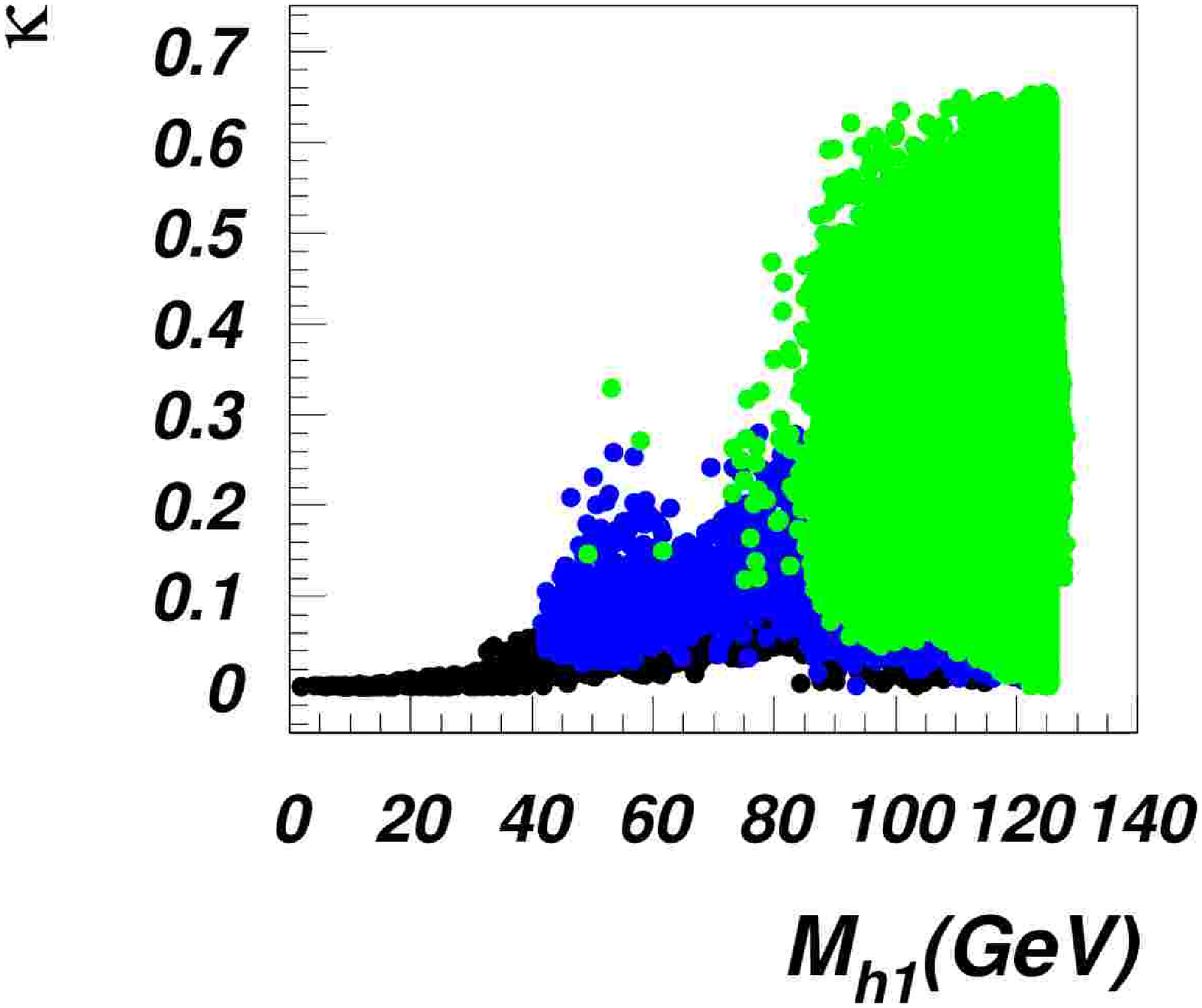}\\
\vskip -3.cm
\hspace*{0.33\textwidth}\hspace*{-5.4cm}{\bf (a)}
\hspace*{0.33\textwidth}\hspace*{-0.5cm}{\bf (b)}
\hspace*{0.33\textwidth}\hspace*{-0.5cm}{\bf (c)}
\vskip  2cm
  \includegraphics[width=0.33\textwidth]{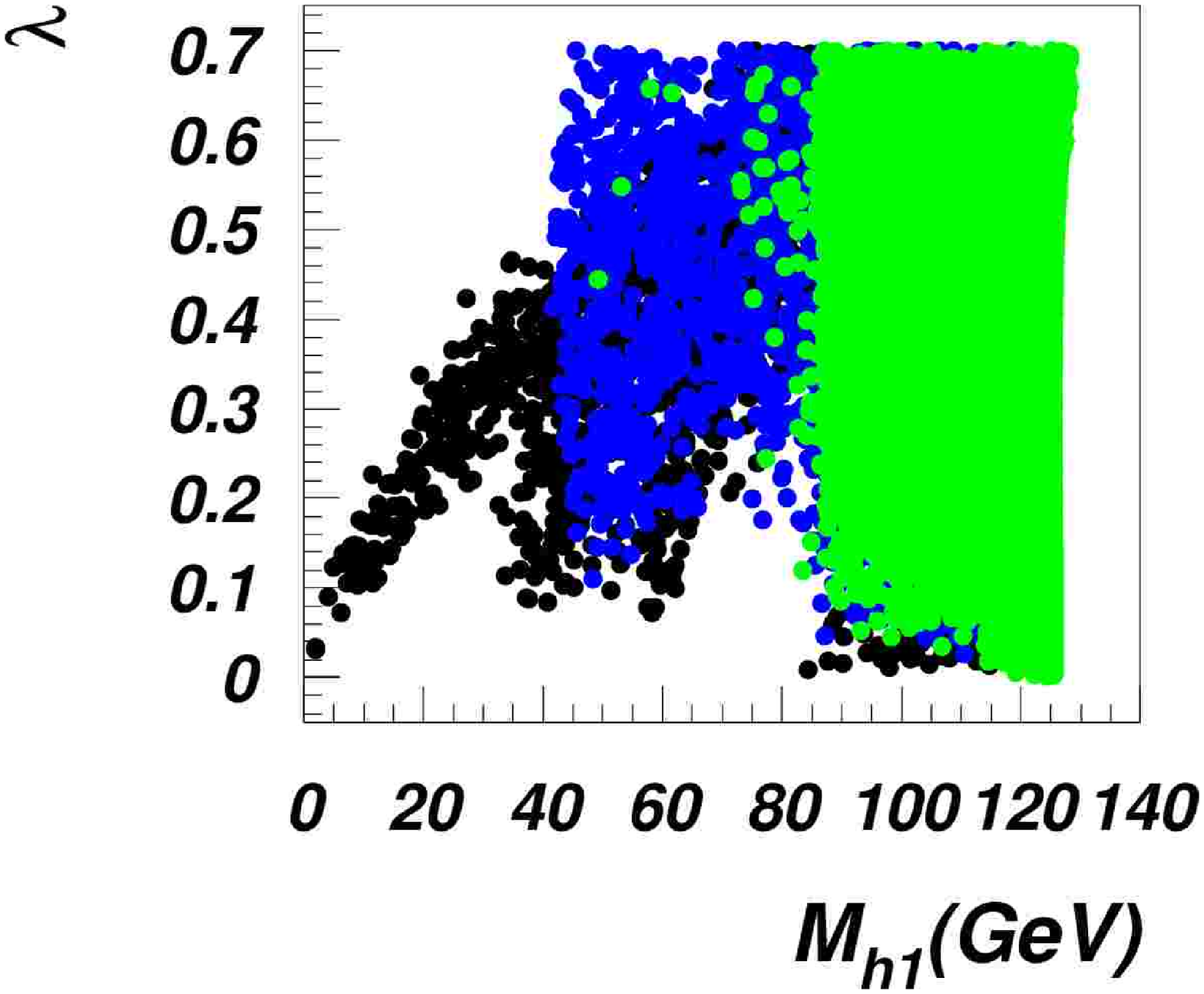}%
  \includegraphics[width=0.33\textwidth]{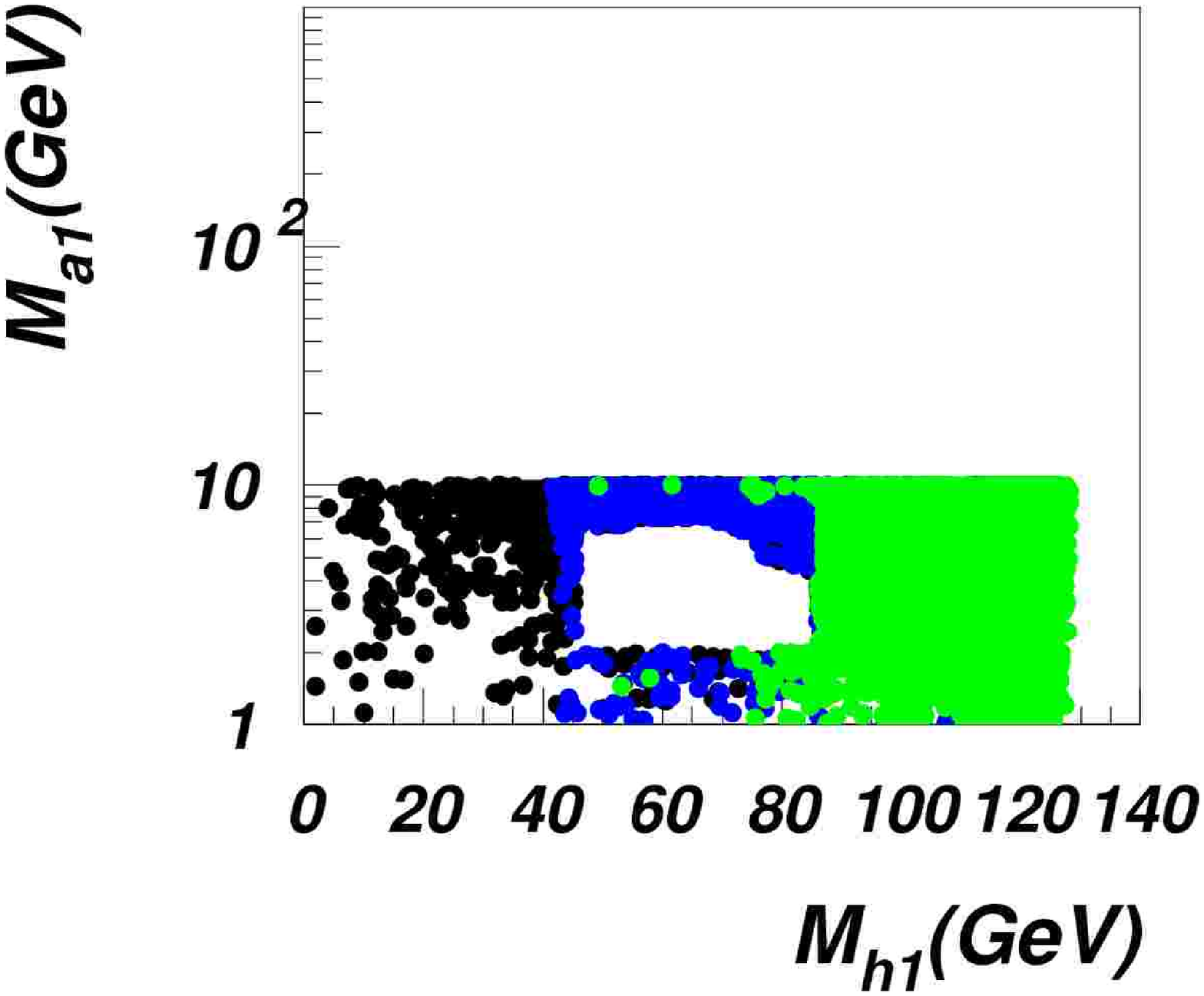}%
  \includegraphics[width=0.33\textwidth]{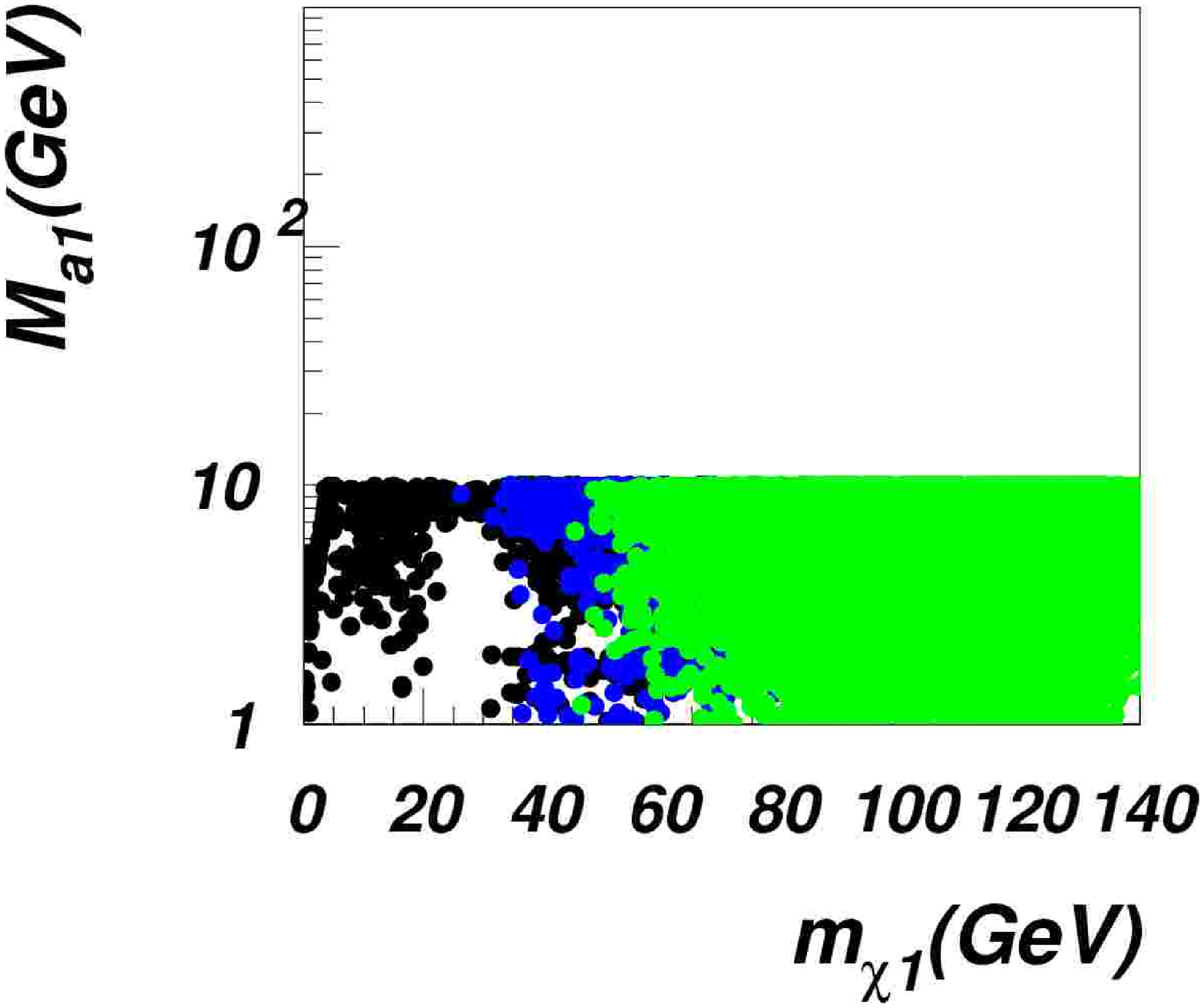}%
\vskip -2.5cm
\hspace*{0.33\textwidth}\hspace*{-5.4cm}{\bf (d)}
\hspace*{0.33\textwidth}\hspace*{-0.5cm}{\bf (e)}
\hspace*{0.33\textwidth}\hspace*{-0.5cm}{\bf (f)}
\vskip  2cm
\caption{\label{fig:scan-final}
Results of the NMSSM `final' scan, i.e., with $M_{a_1}< 10$ GeV and with Eq.~(\ref{cut:ak_narrow})
enforced. The black, blue and green colours  indicate  the cases
 $R_{ZZh}<0.1$,
 $0.1<R_{ZZh}<0.5$ and
 $R_{ZZh}>0.5$, respectively (where $R_{ZZh}$ is defined in the text).
The individual plots are the same as in Fig.~\ref{fig:scan-wide}.}
\end{figure}
\begin{figure}[!t]
  \includegraphics[width=0.33\textwidth]{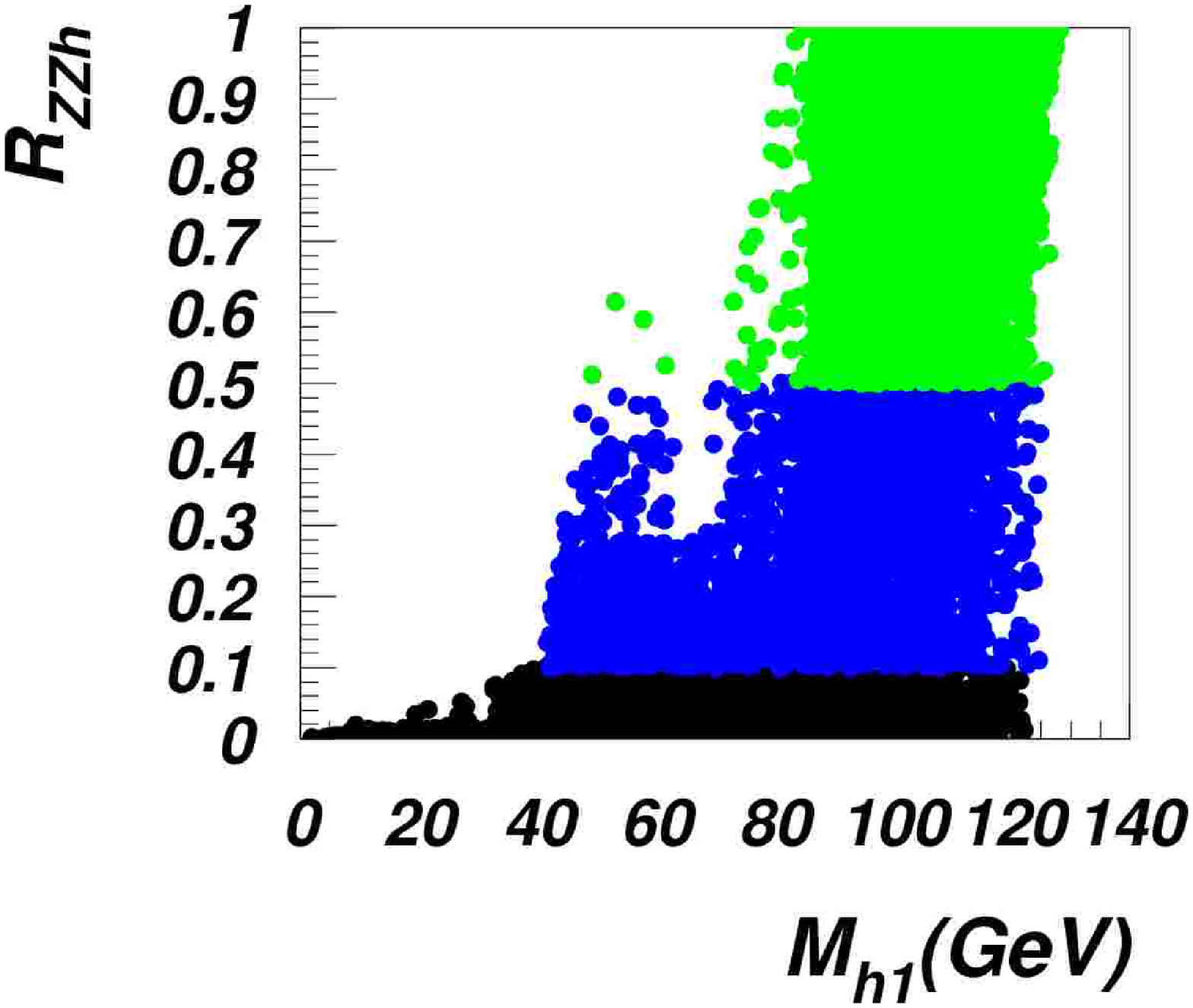}%
  \includegraphics[width=0.33\textwidth]{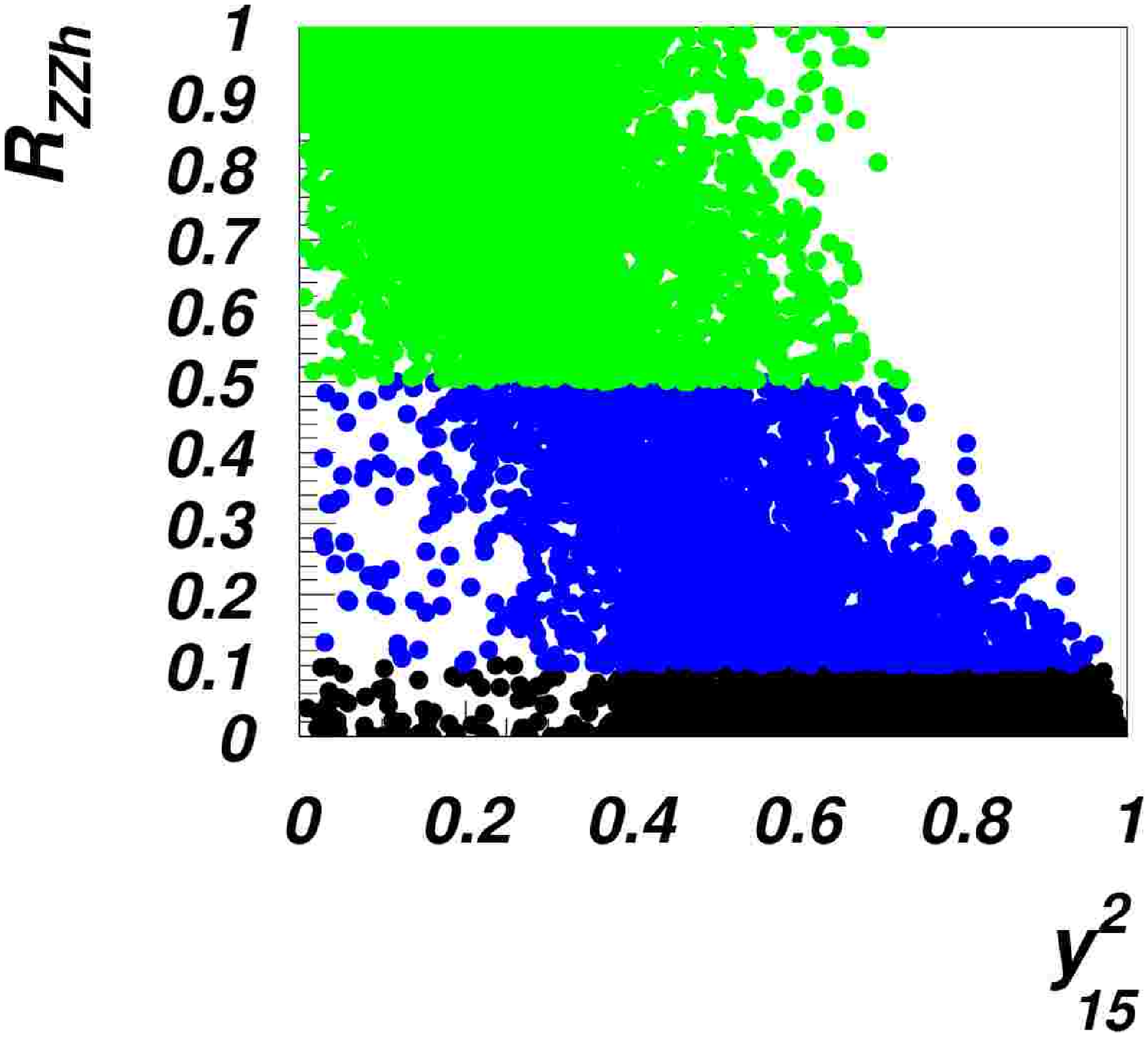}%
  \includegraphics[width=0.33\textwidth]{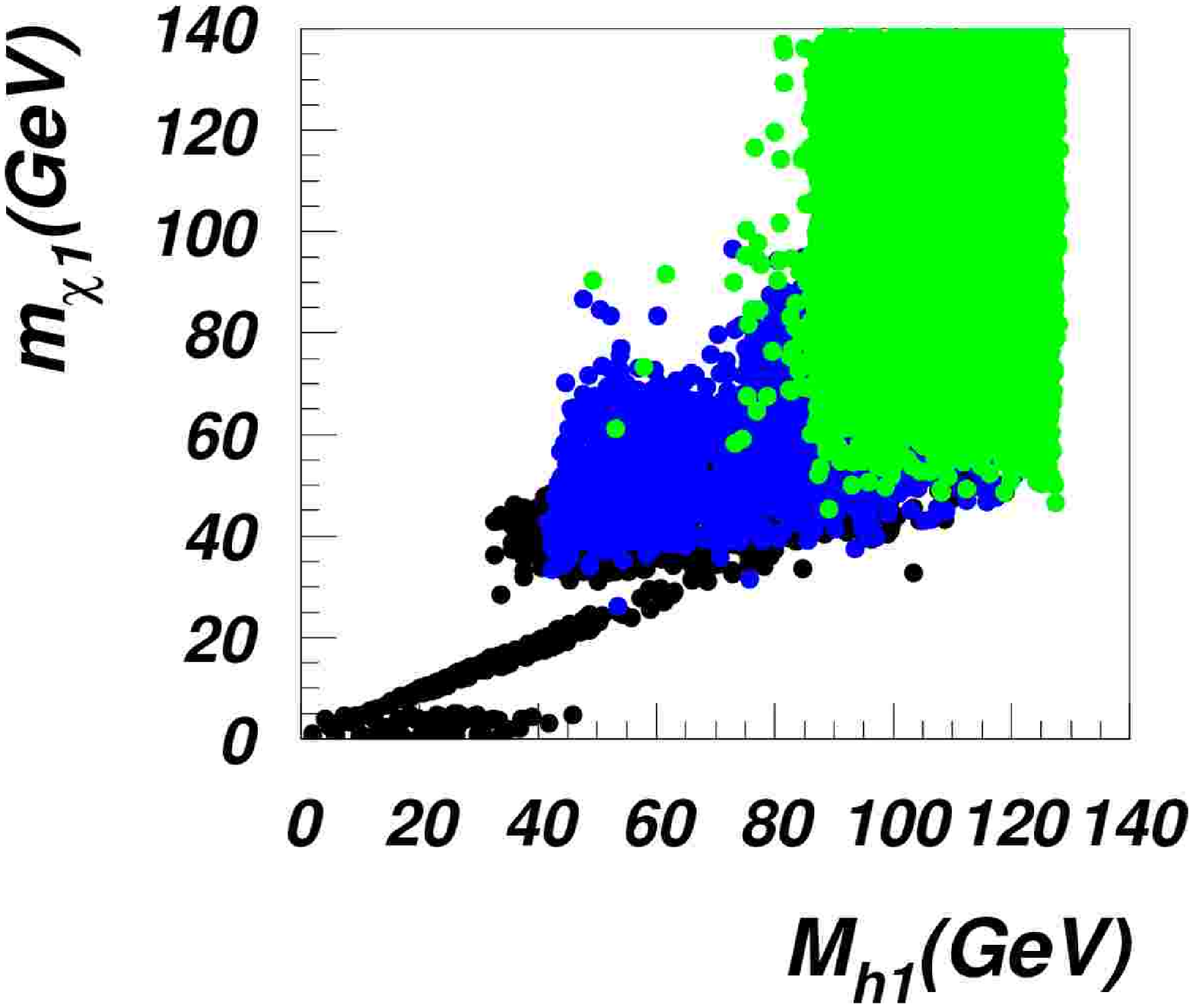}%
\vskip -2.5cm
\hspace*{0.33\textwidth}\hspace*{-5.4cm}{\bf (a)}
\hspace*{0.33\textwidth}\hspace*{-0.5cm}{\bf (b)}
\hspace*{0.33\textwidth}\hspace*{-0.5cm}{\bf (c)}
\vskip  2cm
\caption{\label{final-scan1}
Results of the NMSSM `final' scan,
i.e., with $M_{a_1}< 10$~GeV and with Eq.~(\ref{cut:ak_narrow}) enforced:
(a) $R_{ZZh}$ (see text) versus $M_{h_1}$,
(b) $R_{ZZh}$ versus the singlino component $y_{15}^2$ of the lightest
neutralino $\chi^0_1$,
(c) $m_{\chi^0_1}$ versus $M_{h_1}$.
The colour coding is the same as in Fig.~\ref{fig:scan-final}.}
\end{figure}
In this case, one should  notice the correlation between the singlet
 nature of the $h_1$ and the singlino component of the
 lightest neutralino, which is visually depicted in
 Fig.~\ref{final-scan1}(b).
Finally,
it is also worth to point out the correlation
between their masses in Fig.~\ref{final-scan1}(c). From these
plots, one can see a striking correlation between the 
lightest neutralino and  Higgs boson whenever one has that $M_{h_1}<50$~GeV.
In this connection, one should stress that the NMSSM model structure
requires $h_1$ to be a singlet
and $\chi^0_1$ to be a singlino (for $M_{a_1}<10$~GeV and  $M_{h_1}<10$~GeV)
in order to have a relic density consistent with current experimental constraints.
In fact, over the NMSSM parameter space restricted to having $M_{a_1}<10$~GeV and  $M_{h_1}<10$~GeV,
the $\chi^0_1$-pair annihilation in the early Universe proceeds through the 
$h_1$-funnel region. So, in this region, $ 2m_{\chi^0_1} \simeq M_{h_1}$
as we observe from the lower-left part of  Fig.~\ref{final-scan1}(c).

\subsection{Phenomenology of the Light $a_1$ Scenario}
\noindent

\begin{figure}[!t]
  \includegraphics[width=0.50\textwidth]{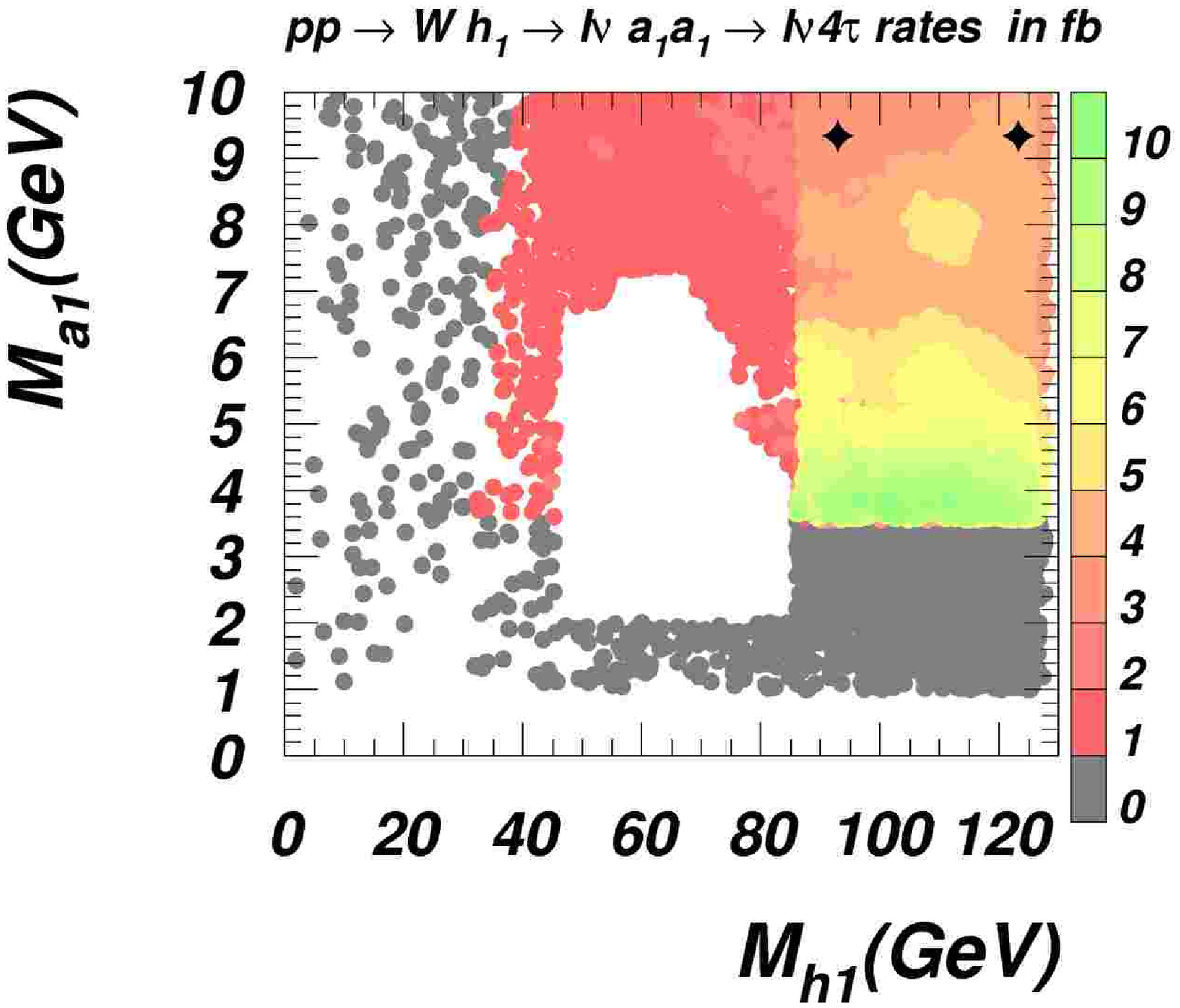}%
  \includegraphics[width=0.50\textwidth]{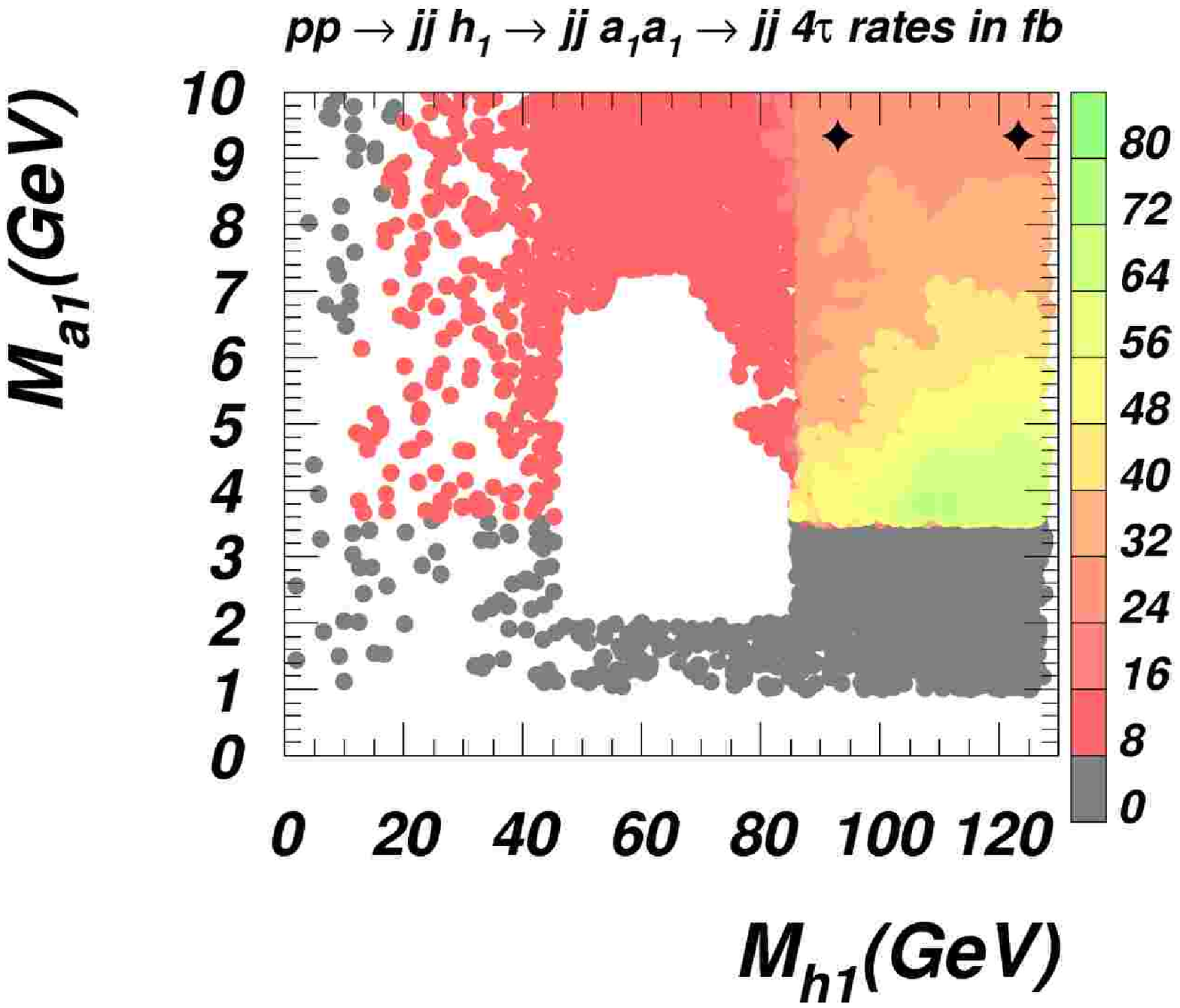}%
\vskip -3.cm
\hspace*{0.50\textwidth}\hspace*{-7.2cm}{\bf (a)}
\hspace*{0.50\textwidth}\hspace*{-0.6cm}{\bf (b)}
\vskip  2cm
  \caption{\label{final-rates}
Cross sections (including all relevant branching ratios) for HS
(a) and VBF (b) after the selection cuts described in the text. 
The population of points used correspond
to that of the `final scan' described previously.
Black diamonds correspond to the benchmark P2 (right) and P3 (left) from Ref.~\cite{Djouadi:2008uw}.}
\end{figure}
\noindent
As final step of our analysis, we combined the production rates of HS
and VBF with selection efficiencies determined by generating these
processes within the PYTHIA Monte Carlo (including smearing effects). The latter have been estimated
in presence of cuts, after parton shower, hadronisation plus
heavy hadrons decays (and with underlying event turned on). 
For HS we enforced (assuming $e,\mu$ decays of the $W^{\pm}$)
$\tau^{+}\rightarrow\mu\nu\nu$ and $\tau^{-}\rightarrow$ hadrons and the
selection cuts were

\noindent
$\bullet$ Trigger selection: isolated single muon or single electron found
with thresholds 19 and 26 GeV, respectively, $|\eta| < $ 2.5.
$\bullet$ Muon $\rm p_{\rm T} > $ 7 GeV, $|\eta| < $ 2.1.
$\bullet$ Tau jet $\rm E_{\rm T} > $ 10 GeV, $|\eta| < $ 2.1.
$\bullet$ Isolated 1-prong $\tau$'s within $\Delta R <$ 0.6 from the muon using
tracker isolation for tracks $p_{\rm T} > $ 2 GeV.
$\bullet$ Tau and muon oppositely charged.
$\bullet$ Two tau + muon pairs found.

\noindent
For VBF the selection cuts were

\noindent
$\bullet$ Two same sign muons with $\rm p_{\rm T} > $ 7 GeV, $|\eta| < $ 2.1
      and with one track of $\rm p_{\rm T} > $ 2 GeV in cone 0.6 around
      each muon.
$\bullet$ Two $\tau$ jets with $\rm E_{\rm T} > $ 10 GeV, $|\eta| < $ 2.1.
$\bullet$ Two jets with $\rm E_{\rm T} > $ 30 GeV, $|\eta| < $ 4.5.

\noindent
The results in Fig.~\ref{final-rates} show that, after our final scan, the 
population of parameter points is such that in both channels the highest
cross sections are found for $M_{h_1}\gsim80$ GeV, although
in the case of VBF also lower $h_1$ masses can yield sizable rates, but never
for values less than 40 GeV. Independently
of $M_{h_1}$, the $a_1$ mass  enables sizable event rates anywhere above
$2m_\tau$, but particularly just above the threshold. 
At high luminosity, 100 fb$^{-1}$, the highest rates would correspond to
1000 events per year for HS and 8000 for VBF. 

\subsection{Conclusions}
\noindent
We have shown that there is significant potential in establishing a no-lose theorem for the NMSSM
at the LHC via (marginally) HS and (primarily) VBF production of the lightest CP-even Higgs boson $h_1$ decaying
into $a_1a_1$ pairs in turn yielding four $\tau$ leptons, searched for through their
semi-leptonic/hadronic decays into muons and jets. To enhance the decay fraction into $\tau$'s of the
lightest CP-odd Higgs boson $a_1$ we have restricted ourselves to the case $M_{a_1}<2m_b$ (otherwise
$a_1\to b\bar b$ decays are dominant). We have also found that the $h_1$ state can be very light, indeed at
 times lighter than the 
${a_1}$. However, this last configuration  can only be achieved in a low-energy NMSSM setup, 
with no unification assumptions at the high scale. In fact, we are currently investigating whether such light $h_1$ masses
can be found in a less constrained version of the cNMSSM discussed in \cite{Djouadi:2008uw}. Finally, with 
reference to the NMSSM benchmarks discussed in  \cite{Djouadi:2008uw}, we should like to point out here that those
relevant to our $4\tau$ channels are P2 and P3. We have reported the cross section times efficiency rates for these
two points in Fig.~\ref{final-rates} ({black diamond} symbols, P2 to the right and P3 to the left). As it can be
appreciated, they correspond to event rates that are mid range amongst all those explored, hence not particularly biased
towards a far too favourable NMSSM setup, yet susceptible to experimental discovery. 
Our summary is preliminary, as only signal processes have been considered and only in presence of MC simulations,
with no backgrounds and detector performance enabled. The latter clearly ought to be investigated before drawing
any firm conclusions and this is currently being done. 

\section*{Acknowledgements}
\noindent
 SM thanks the Royal Society (London, UK) 
for financial support in the form of a Conference Grant to
attend the workshop.

\section[Investigation of the LHC Discovery Potential for Higgs Bosons in the
NMSSM]{INVESTIGATION 
OF THE LHC DISCOVERY POTENTIAL FOR HIGGS BOSONS IN THE 
NMSSM
~\protect\footnote{Contributed by:I. Rottl\"{a}nder
and  M.~ Schumacher}
}
%\documentclass[11pt]{cernrep}
%\usepackage{graphicx,epsfig}
%\bibliographystyle{lesHouches}
%\begin{document}

%\title{INVESTIGATION OF THE LHC DISCOVERY POTENTIAL FOR HIGGS BOSONS IN THE NMSSM}

%\author{I. Rottl\"{a}nder$^1$, M. Schumacher$^2$}
%\institute{$^1$Physikalisches Institut, University of Bonn, Nussallee 12, 53115 Bonn, Germany
%\\$^2$Fachbereich Physik, University of Siegen, Walter-Flex-Stra\ss{}e 3, 57068 Siegen, Germany}

%\maketitle

%\begin{abstract}
%%%The NMSSM is a powerful extension to the MSSM, solving the so-called {\it $\mu$-problem}. 
%The discovery potential of the six NMSSM Higgs bosons with the 
%ATLAS detector is evaluated within two benchmark scenarios. A detailed description
%of the applied scanning procedure and its results is given.
%\end{abstract}

\subsection{Introduction}
The {\it Large Hadron Collider} (LHC) will deliver proton-proton collisions at a
center-of-mass energy of 14 TeV. First physics runs are expected for 2008. 
First, the LHC will operate at low luminosity \mbox{($2\cdot10^{33}$cm$^{-2}$s$^{-1}$).} 
Later, the luminosity will be increased to its design value of $10^{34}$cm$^{-2}$s$^{-1}$.
One of the main aims of the ATLAS \cite{Atlas:TDR} and CMS \cite{CMStdr} experiments 
at the LHC is the search for the Higgs boson.
In the Standard Model (SM) electroweak symmetry breaking is achieved via the introduction of 
one Higgs doublet. Only one neutral Higgs boson is predicted. Extended Higgs sectors, 
with additonal Higgs doublets and Higgs singlets give rise to several neutral and charged Higgs bosons,
e.g. the two Higgs doublets of the Minimal Supersymmetric Extension of the SM (MSSM) yield three neutral 
and two charged Higgs bosons. Detailed studies have shown that the SM Higgs  boson will be observable at 
ATLAS and CMS \cite{Atlas:TDR, CMStdr, Asai:2004ws}. The discovery of one or more Higgs bosons of the 
CP-conserving MSSM will be possible \cite{Schumacher:2004da}.
Previous studies claim that at least one Higgs boson of the Next-to-Minimal Supersymmetric 
Standard Model (NMSSM) will most likely 
be observable at the LHC \cite{Ellwanger:2004gz, Ellwanger:2001iw}. 
Here, we present an evaluation of the discovery potential for NMSSM Higgs bosons based on current ATLAS studies \cite{Atlas:TDR, Asai:2004ws, Cranmer:2004uz, 
Cammin:685523, Trefzger:683987, Thomas:685421, Cavalli:685488, González:685407, Cavalli:683878, Mohn:2007fd,
Assamagan:2002ne, Biscarat:681548}. 

\subsection{The NMSSM Higgs Sector}
In the framework of the NMSSM, the {\it $\mu$-problem} of the MSSM is solved by the 
introduction of an additional neutral singlet superfield $S$ \cite{Ellis:1988er}. 
The two additional neutral scalar bosons contained in $S$ mix with the MSSM Higgs bosons to
form the five neutral Higgs bosons of the NMSSM: three CP-even bosons $H_1$, $H_2$, $H_3$ and 
two CP-odd Higgs bosons $A_1$, $A_2$. The phenomenology of the charged Higgs boson $H^\pm$ 
is only modified marginally with respect to the MSSM. The Higgs sector of the NMSSM at Born level is determined
by the four coupling parameters of the singlet superfield, $\lambda$, $\kappa$, $A_\lambda$, $A_\kappa$,
and the two parameters $\mu$ and $\tan\beta$.
For a more detailed description of the NMSSM Higgs sector see e.g. Refs. \cite{Ellis:1988er, Miller:2003ay}.

\subsection{Evaluation of the Discovery Potential}
Two two-dimensional benchmark scenarios are investigated in this study: the {\it Reduced Couplings Scenario} and the 
{\it Light $A_1$ Scenario} which were proposed during this workshop 
(for details see these proceedings). The parameters $\lambda$ and $\kappa$ are varied in meaningful ranges 
whereas the other parameters are fixed as described previously in this report. 
The method of evaluation of the discovery potential is similar to the study performed for the MSSM 
in Ref. \cite{Schumacher:2004da}. 

\subsubsection{Calculation of masses and events rates in the NMSSM}
NMHDECAY \cite{Ellwanger:2004xm, Ellwanger:2005dv} was used to calculate the masses, branching ratios 
and decay widths of the NMSSM Higgs bosons and the couplings of the neutral Higgs bosons to fermions and gauge bosons, relative to the respective SM couplings. 
Couplings to gluons relative to the SM couplings were calculated from the ratio of partial widths of $H$$\rightarrow$$gg$ in the NMSSM and the SM \cite{Djouadi:1997yw} as 
in Eq.\ref{nmssmscan_Hgg}.
%to include effects of higher order corrections. 
\begin{equation} \label{nmssmscan_Hgg}
\frac{g_{Hgg,NMSSM}^2}{g_{Hgg,SM}^2}=\frac{\Gamma(H\rightarrow gg)_{NMSSM}}{\Gamma(H\rightarrow gg)_{SM}}
\end{equation}
%%The branching ratios for the SM were provided by HDECAY \cite{Djouadi:1997yw}. 
%The $H^\pm$$tb$-coupling was obtained from the ratio of the
%partial widths of $H^\pm$$\rightarrow$$tb$ in the NMSSM and a defined MSSM case
%calculated with NMHDECAY in the {\it Decoupling Limit} (see e.g. Ref. \cite{Miller:2003ay}).
For the neutral Higgs bosons, leading order SM cross sections \cite{Spira:1997dg} were scaled according to
Eq.\ref{nmssmscan_scale}.
\begin{equation} \label{nmssmscan_scale}
\sigma_{NMSSM}= \sigma_{SM} \cdot \frac{g_{NMSSM}^2}{g_{SM}^2}
\end{equation}
The charged Higgs boson $gb$$\rightarrow$$tH^\pm$ cross sections in leading order were taken from Ref.\cite{Plehn:2002vy} and were modified according to the $H^\pm$$tb$-couplings obtained with NMHDECAY.
The branching ratio $t$$\rightarrow$$H^\pm$$b$ was calculated with Feynhiggs \cite{Heinemeyer:1998yj}.
For $t\bar{t}$-production, a leading order  cross section of 482 pb was assumed.
%%For the charged Higgs boson, 
%MSSM $gb$$\rightarrow$$tH^\pm$ cross sections \cite{Plehn:2002vy} in leading order were scaled similarly.
%For $tt$$\rightarrow$$H^\pm$$bWb$, a leading order $tt$-production cross section of 482 pb and $t$$\rightarrow$$H^\pm$$b$ branching ratios \cite{Heinemeyer:1998yj} were used.
%If required, other NMSSM branching ratios were applied to the cross sections.
The top quark mass was set to 172 GeV.
Theoretical and LEP\footnote{The {\it Large Electron Positron Collider}, which ran until 2000 at center-of mass energies up to 209 GeV.} exclusion criteria (bounds from hZ and hA searches) were calculated by NMHDECAY.

\subsubsection{Significance Calculation}
The expected number of signal events is derived from the above discussed NMSSM cross sections.
Signal efficiencies are taken from published ATLAS Monte-Carlo studies (Table \ref{nmssmscan_searches}).
The expected numbers of background events are also taken from published ATLAS MC studies. 
If MC studies at design luminosity exist, a data volume of 300 fb$^{-1}$ is assumed;
if only low luminosity studies are available, 30 fb$^{-1}$ are used, and 
if both scenarios have been investigated, 30 fb$^{-1}$ taken at low luminosity and 
270 fb$^{-1}$ taken at design luminosity are assumed. 
The current results only include SM background processes. Systematic uncertainties are neglected.
For the significance calculation, the profile likelihood method \cite{Rolke:2006ve} with asymptotic 
approximation \cite{Gross:nmssmscan} is used. To claim a discovery, a significance of 
at least 5$\sigma$ is required. The number of expected signal events is corrected for the 
effects of increased Higgs boson decay widths and the possibility of degenerate Higgs boson masses as described in the following.
\begin{table}[bt]
\caption{Included search topologies with allowed mass ranges.\label{nmssmscan_searches}}
\begin{center}
\begin{tabular}{|l|c|c|} \hline
\textbf{Search Channel} & \textbf{Mass Range [GeV]} & \textbf{Refs.}\\ \hline 
VBF, $H$$\rightarrow$$\tau\tau$ & 110-150 & \cite{Asai:2004ws}\\
VBF, $H$$\rightarrow$$WW$$\rightarrow$$ll$$\nu\nu$&110-200&\cite{Asai:2004ws}\\
VBF, $H$$\rightarrow$$WW$$\rightarrow$$l$$\nu$$h$&130-200&\cite{Asai:2004ws}\\
VBF, $H$$\rightarrow$$\gamma\gamma$&110-160&\cite{Cranmer:2004uz}\\ \hline
ttH, $H$$\rightarrow$$b\bar{b}$&70-150&\cite{Cammin:685523} \\ \hline
GGF, $H$$\rightarrow$$ZZ$$\rightarrow$$4l$&120-420&\cite{Atlas:TDR}\\
GGF, $H$$\rightarrow$$WW$$\rightarrow$$ll$$\nu\nu$&140-200&\cite{Trefzger:683987}\\
WH, $H$$\rightarrow$$WW$$\rightarrow$$ll$$\nu\nu$, W$\rightarrow$l$\nu$&150-190&\cite{Atlas:TDR}\\ \hline
Inclusive $H$$\rightarrow$$\gamma\gamma$&70-160&\cite{Atlas:TDR}\\
Inclusive $A$$\rightarrow$$\gamma\gamma$&200-450&\cite{Atlas:TDR}\\
WH, ZH, ttH, $H$$\rightarrow$$\gamma\gamma$&70-150&\cite{Atlas:TDR}\\ \hline
bbH, $H/A$$\rightarrow$$\tau\tau$$\rightarrow$hh&450-800&\cite{Thomas:685421}\\
GGF, bbH, $H/A$$\rightarrow$$\tau\tau$$\rightarrow$$l$$\nu$h&150-800&\cite{Cavalli:685488}\\
GGF, bbH, $H/A$$\rightarrow$$\mu\mu$&70-450&\cite{González:685407,Cavalli:683878}\\ \hline
GGF, $H$$\rightarrow$$hh$$\rightarrow$$\gamma\gamma$bb&230-270 / 70-100&\cite{Atlas:TDR}\\
GGF, $H$$\rightarrow$$ZA$$\rightarrow$$llb\bar{b}$&200-250 / 70-100&\cite{Atlas:TDR}\\ \hline
$gb$$\rightarrow$$H$$^\pm$$t$, $H$$^\pm$$\rightarrow$$\tau\nu$&175-600&\cite{Mohn:2007fd}\\
$gb$$\rightarrow$$H$$^\pm$$t$, $H$$^\pm$$\rightarrow$$tb$&190-400&\cite{Assamagan:2002ne}\\
$t\bar{t}$$\rightarrow$$H$$^\pm$$bWb$$\rightarrow$$\tau\nu$$l$$\nu$$b\bar{b}$&90-165&\cite{Atlas:TDR}\\
$t\bar{t}$$\rightarrow$$H$$^\pm$$bWb$$\rightarrow$$\tau\nu$$q\bar{q}b\bar{b}$&80-165&\cite{Biscarat:681548}\\ \hline
\end{tabular}
\end{center}
\end{table}
\newline\newline{\textit{Corrections for large Higgs bosons widths} \newline}
In the NMSSM, the natural line width of the Higgs boson may be enhanced relative to the SM case. 
Thus, a larger fraction of signal events may lie outside a mass window cut than in the SM. To correct
for this, the Higgs boson peak was described by a Voigt-function whose Breit-Wigner 
part is given by the natural line width, the Gaussian part by the detector resolution. 
The ratio of the integral values over the mass window for the SM and the NMSSM case was 
used as a correction factor.\newline\newline
{\textit{ Corrections for degenerate Higgs boson masses} \newline}
Higgs boson peaks were described by a Voigt function as previously. 
The peaks were assumed to be indistinguishable 
if their separation was smaller than $2.355\cdot$FWHM. 
In case of negligible Higgs boson width, this corresponds to a $2\sigma$ separation of two Gaussians. 
Higgs bosons with overlapping mass windows were also 
considered indistinguishable to avoid double counting of events.
In case of inseparable peaks, contributions from all Higgs bosons were added up for each 
boson's mass window. Only the highest observed significance was 
kept and assigned to the Higgs boson with the largest 
fraction of signal events in that mass window. 
\subsection{Search Topologies}
The combinations of production mechanisms and decay modes considered in the evaluation 
of the disco\-very potential and the considered mass ranges 
are summarised in Table \ref{nmssmscan_searches}.\footnote{
Production modes are abbreviated GGF for gluonfusion, VBF for vector boson fusion and ttH, bbH, 
WH and ZH for associated production with top quarks, bottom quarks and vector bosons.}
Within the scenarios examined here, only the VBF, $H$$\rightarrow$$\tau\tau$; 
ttH, $H$$\rightarrow$$b\bar{b}$ and $H$$\rightarrow$$\gamma\gamma$ channels  
show significances greater 5$\sigma$ at the given integrated luminosities 
in the theoretically allowed and yet unexcluded regions (see section \ref{nmssmscan_results}).
\begin{figure}[b]
\begin{minipage}[t]{.5\linewidth} % [b] => Ausrichtung an \caption
\centerline{\includegraphics[width=8cm, height=5.4cm]{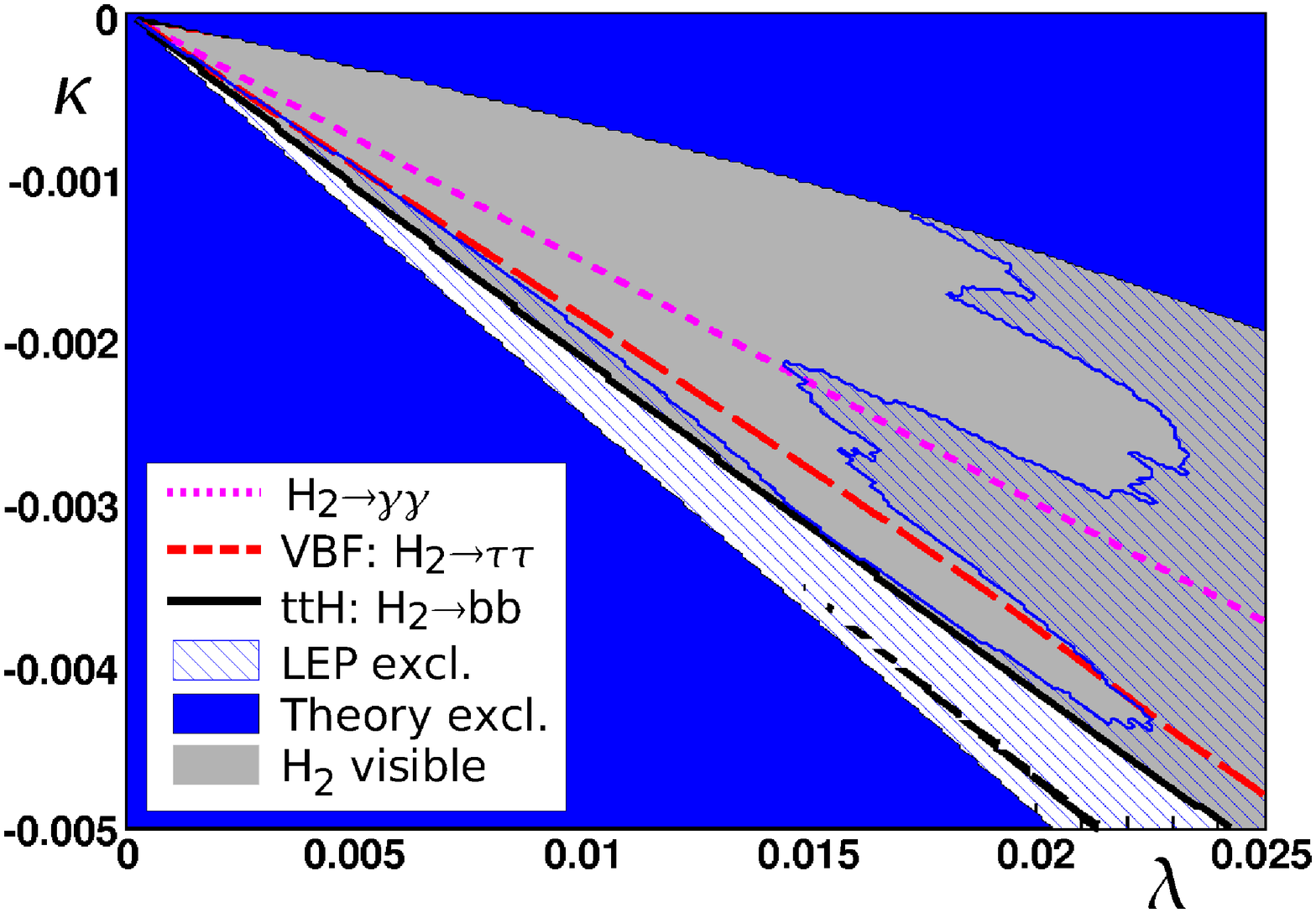}}
\caption{5$\sigma$ discovery contours of the $H_2$ in the $\lambda$-$\kappa$ plane 
for the {\it Reduced Couplings Scenario} \label{nmssmscan_EW1H2}}
\end{minipage}
\hspace{.05\linewidth}% Abstand zwischen Bilder
\begin{minipage}[t]{.5\linewidth} % [b] => Ausrichtung an \caption
\centerline{\includegraphics[width=8cm, height=5.4cm]{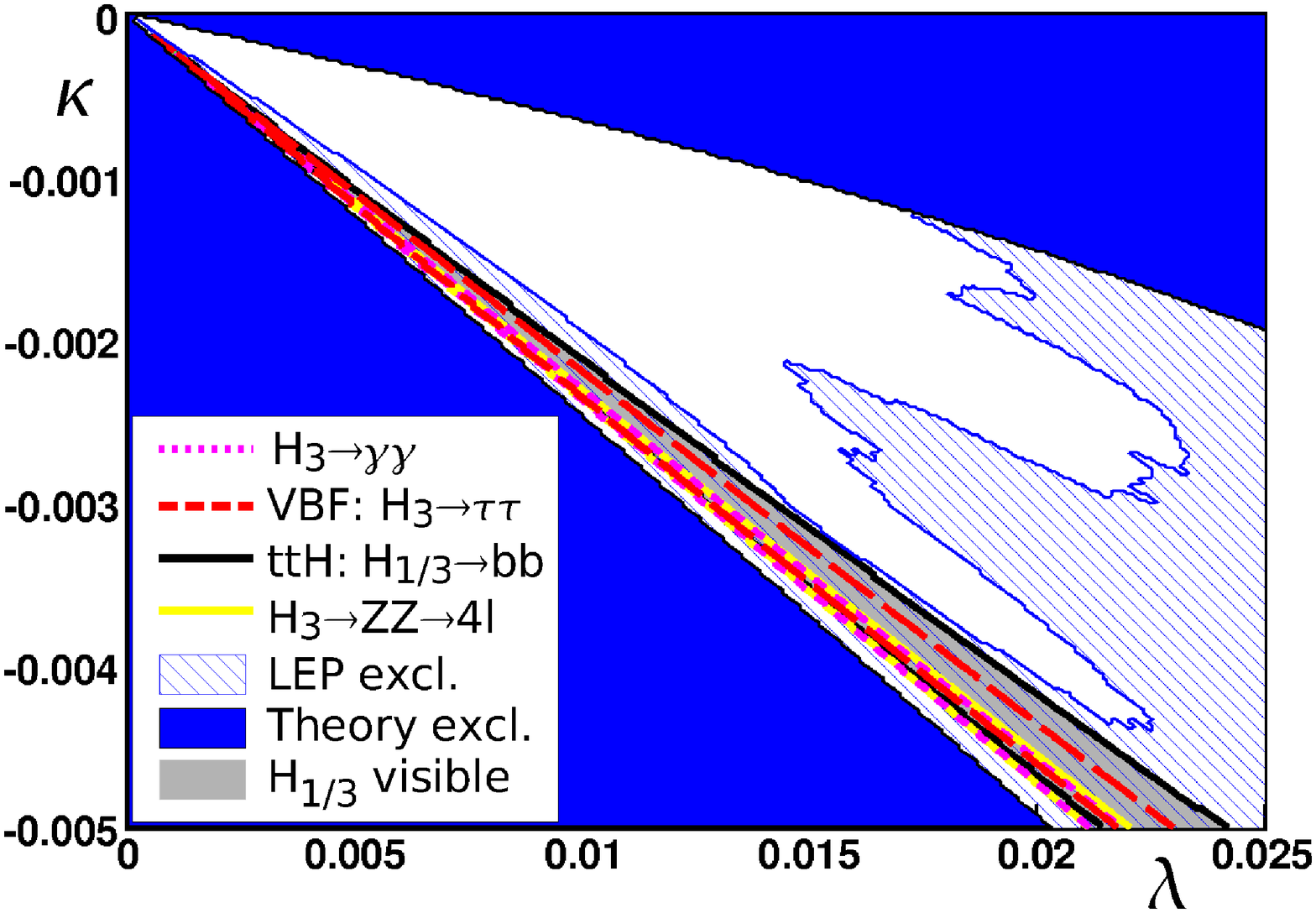}}
\caption{5$\sigma$ discovery contours of the $H_1$ and $H_3$ in the $\lambda$-$\kappa$ 
plane for the {\it Reduced Couplings Scenario}\label{nmssmscan_EW1H3}}
\end{minipage}
\end{figure}
\begin{figure}[tb]
\begin{minipage}[t]{.5\linewidth} % [b] => Ausrichtung an \caption
\centerline{\includegraphics[width=8cm, height=5.4cm]{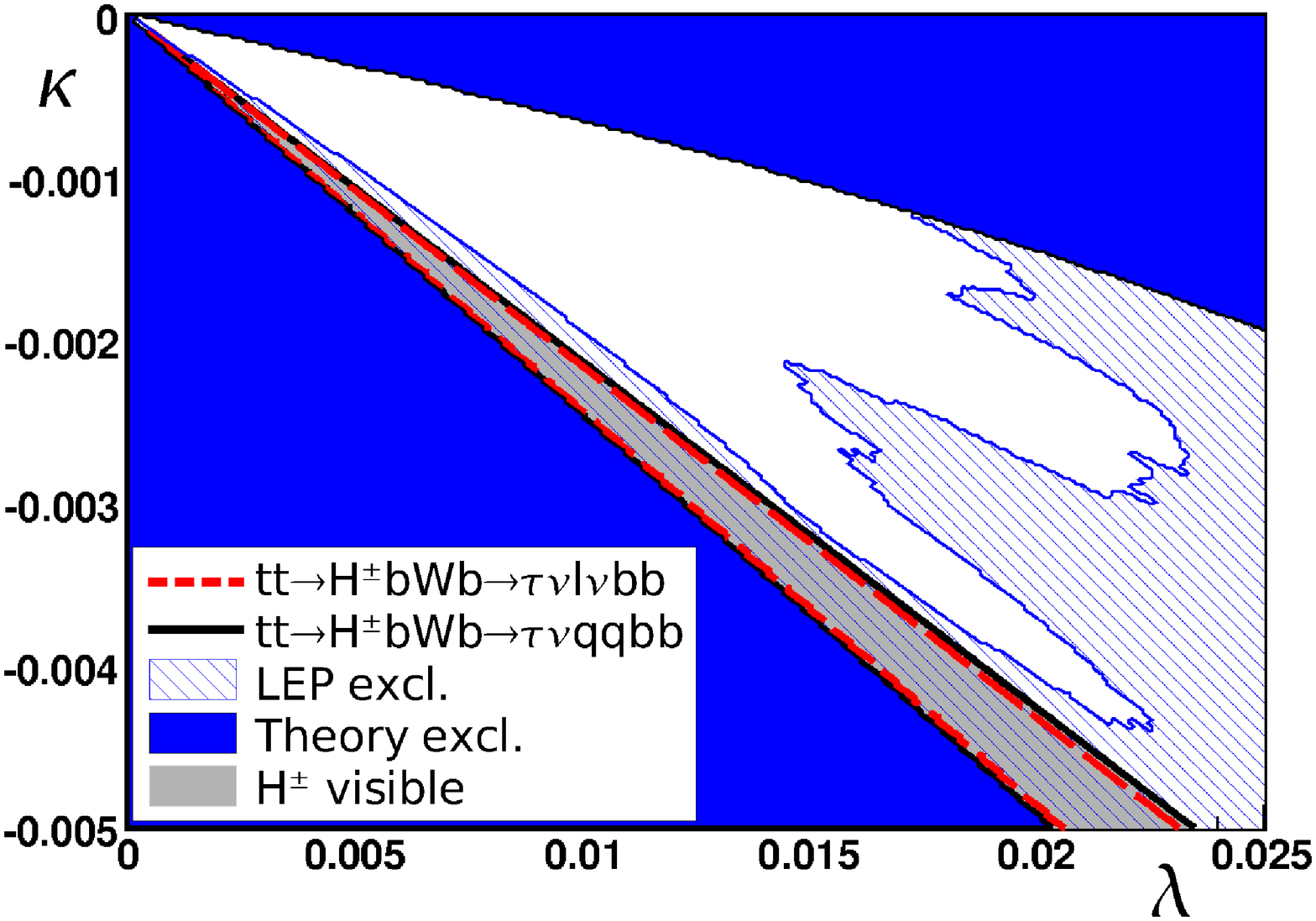}}
%\centerline{\epsfig{file=Effs2.eps, width=6.25cm}}
\caption{5$\sigma$ discovery contours of the $H^\pm$ in the $\lambda$-$\kappa$ plane
for the {\it Reduced Couplings Scenario}\label{nmssmscan_EW1CH}}
\end{minipage}
\hspace{.05\linewidth}% Abstand zwischen Bilder
\begin{minipage}[t]{.5\linewidth} % [b] => Ausrichtung an \caption
\centerline{\includegraphics[width=8cm, height=5.4cm]{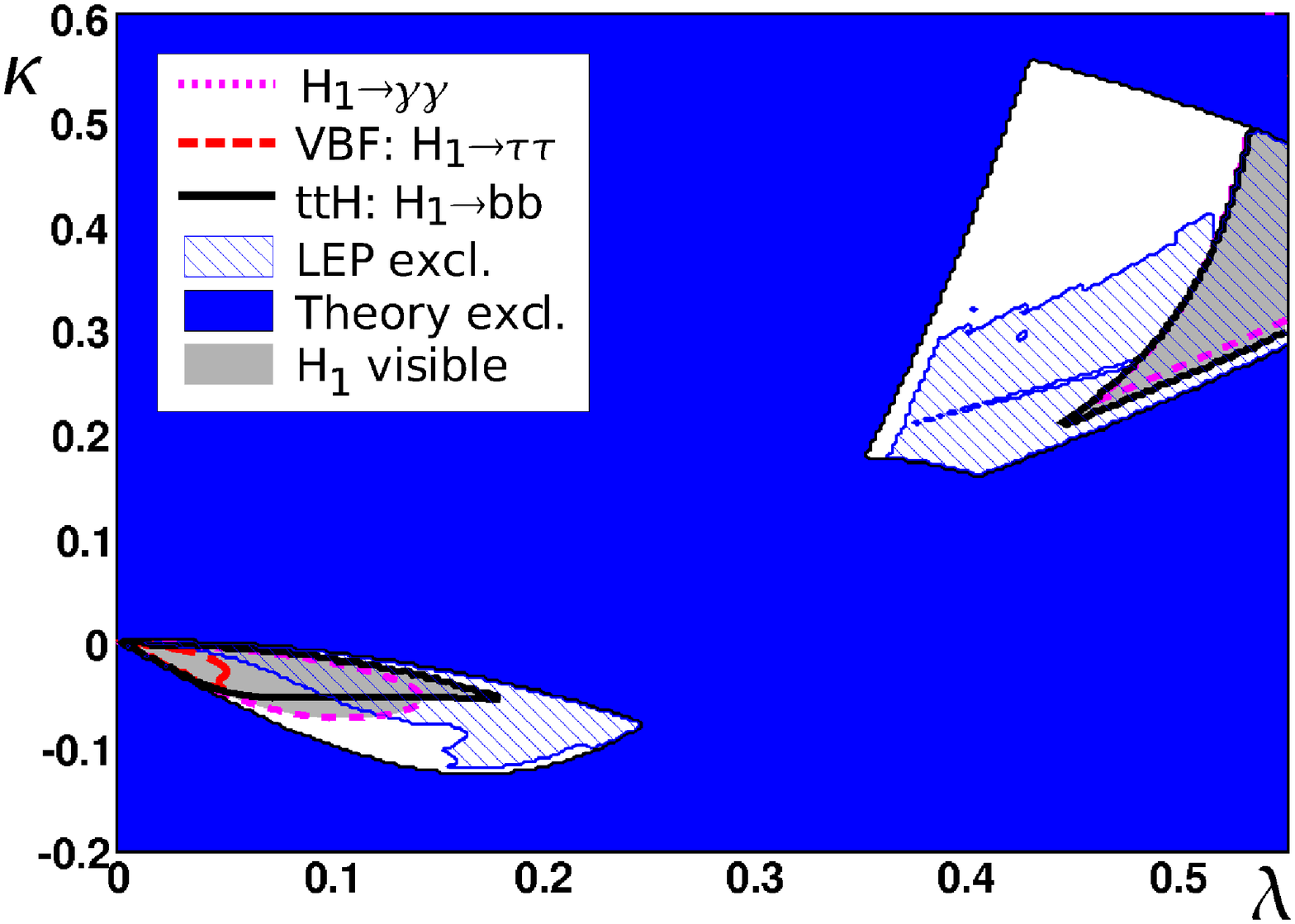}}
\caption{5$\sigma$ discovery contours of the $H_1$ in the $\lambda$-$\kappa$ plane
for the {\it Light $A_1$ Scenario}\label{nmssmscan_EW3H1}}
\end{minipage}
\end{figure}
\begin{figure}[b]
\begin{minipage}[t]{.5\linewidth} % [b] => Ausrichtung an \caption
\centerline{\includegraphics[width=8cm, height=5.4cm]{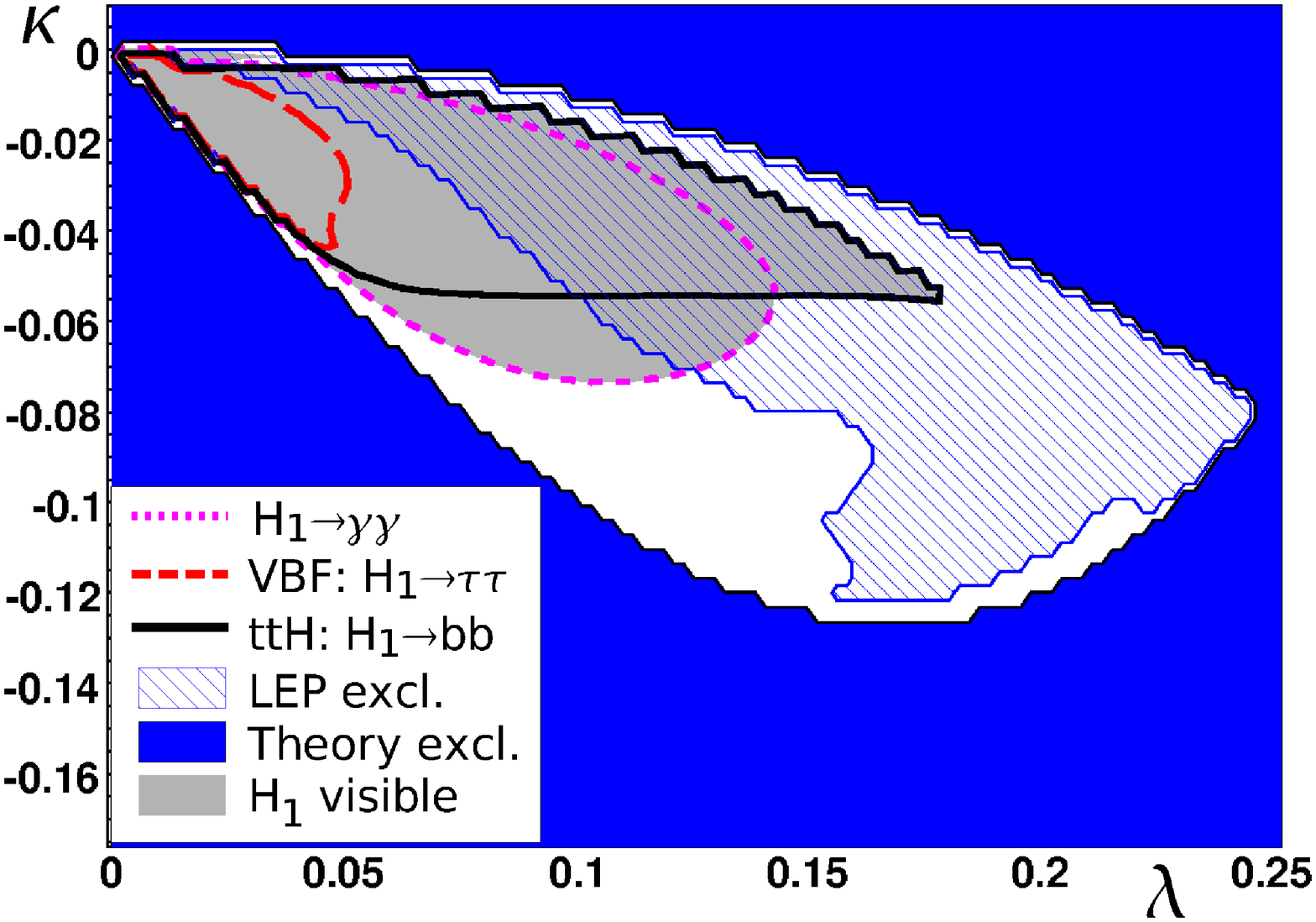}}
%\centerline{\epsfig{file=Effs2.eps, width=6.25cm}}
\caption{5$\sigma$ discovery contours of the $H_1$ in the $\lambda$-$\kappa$ plane
for the {\it Light $A_1$ Scenario}, restricted to low $\lambda$ and $\kappa$ values.\label{nmssmscan_EW3H1b}}
\end{minipage}
\hspace{.05\linewidth}% Abstand zwischen Bilder
\begin{minipage}[t]{.5\linewidth} % [b] => Ausrichtung an \caption
\centerline{\includegraphics[width=8cm, height=5.4cm]{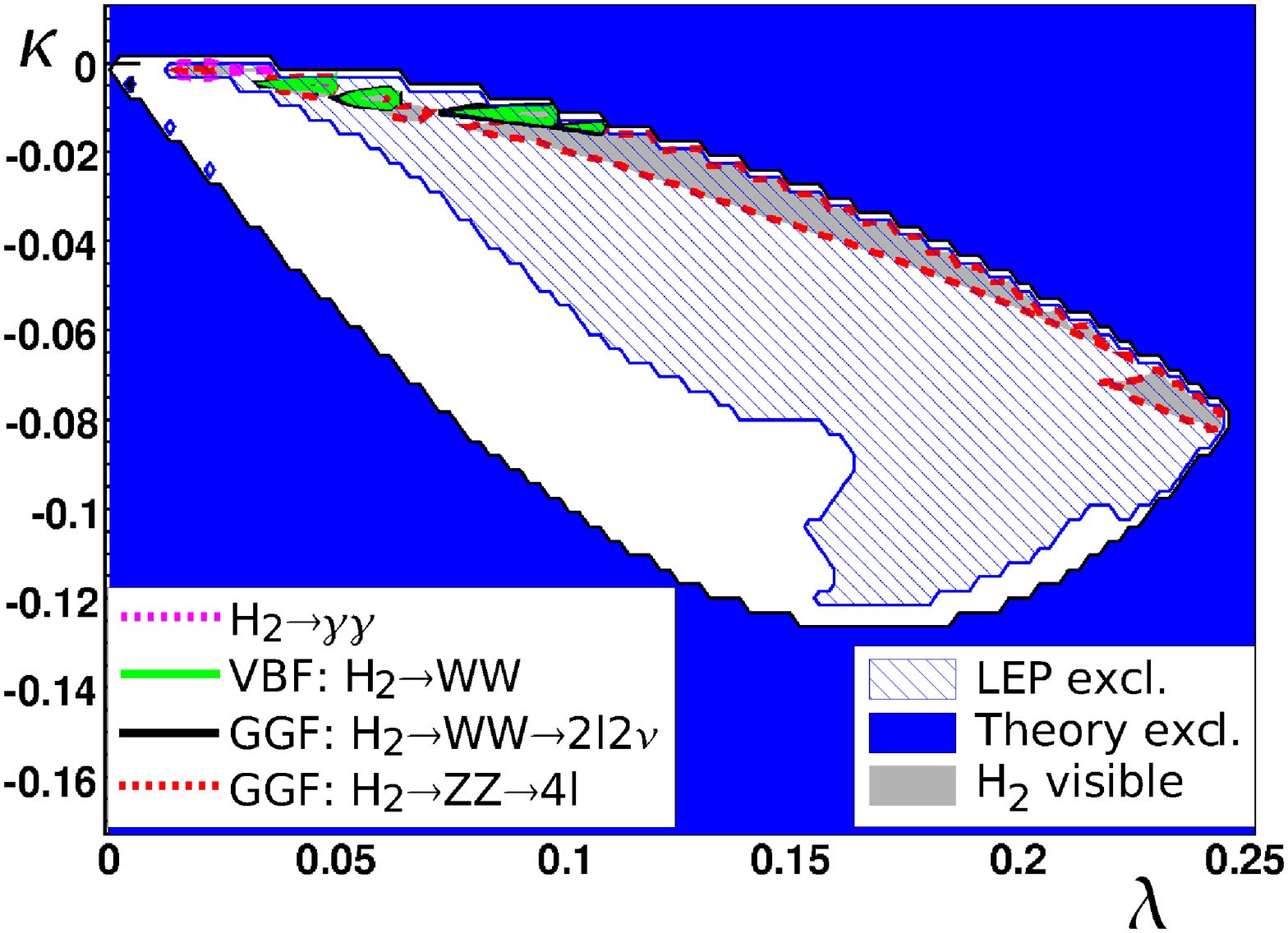}}
\caption{5$\sigma$ discovery contours of $H_2$ in the $\lambda$-$\kappa$ plane
for the {\it Light $A_1$ Scenario}, restricted to low $\lambda$ and $\kappa$ values.\label{nmssmscan_EW3H2}}
\end{minipage}
\end {figure}

\subsection{Results\label{nmssmscan_results}}
\subsubsection{The Reduced Couplings Scenario}
In the {\it Reduced Couplings Scenario}, the $H_2$ with a mass of about 120 GeV is SM-like 
in large parts of the parameter space. 
In an unexcluded region with large negative $\kappa$ 
close to the lower exclusion bound, the gauge couplings of $H_2$ are reduced  to about 80\% 
of their SM-value. The $H_1$ gets very light at the region close to the upper exclusion bound, 
so that the decay $H_2$$\rightarrow$$H_1H_1$ is kinematically allowed. However, due to the small 
branching ratio for this decay mode of at maximum 6\%, its effect on the discovery potential is negligible.
The discovery potential for the $H_2$ is shown in Fig.\ref{nmssmscan_EW1H2}. 
The entire unexcluded region is covered by the ttH, $H_2$$\rightarrow$$b\bar{b}$ channel 
despite the coupling reduction. The inclusive $H_2$$\rightarrow$$\gamma\gamma$ and the VBF, 
$H_2$$\rightarrow$$\tau\tau$ channels also contribute. With 30 fb$^{-1}$, the search for
$H_2$$\rightarrow$$\tau\tau$ will be the only sensitive channel. The region with reduced couplings 
will not be covered in that case.
The gauge couplings of the $H_1$ and $H_3$ are sizable only at large negative $\kappa$.
Here, the channels $H_3$$\rightarrow$$\gamma\gamma$; VBF, $H_3$$\rightarrow$$\tau\tau$; 
ttH, $H_{1/3}$$\rightarrow$$b\bar{b}$ and GGF, $H_3$$\rightarrow$$ZZ$$\rightarrow$$4l$ contribute in a 
region ruled out by LEP (Fig.\ref{nmssmscan_EW1H3}). 
Since the charged Higgs boson is lighter than the top quark in the same region, 
it can be detected via the $t\bar{t}$$\rightarrow$$H^\pm$$bW^\pm$$b$$\rightarrow$$\tau\nu$$l\nu$$b\bar{b}$ and
$t\bar{t}$$\rightarrow$$H^\pm$$b$$W^\pm$$b$$\rightarrow$$\tau\nu$$q\bar{q}b\bar{b}$ 
searches only in the LEP excluded region also (Fig.\ref{nmssmscan_EW1CH}).
All other Higgs bosons
have highly reduced gauge couplings and are therefore unobservable.

%%In short, in the  {\it Reduced Couplings Scenario}, only the $H_2$ will be discoverable. 
%%The whole parameter plane can be 
%%covered with 30+270fb$^{-1}$ of integrated luminosity. 
%%It can be shown that at higher values at tanbeta,
%%up to two neutral and the charged higgs boson can be discoverable.

\subsubsection{The Light $A_1$ Scenario}
In this scenario, the $H_1$ has a mass of about 120 GeV and SM-like gauge couplings.
Since the $A_1$ is light, $H_1$$\rightarrow$$A_1A_1$ decays are kinematically possible and 
often dominant. In the upper right unexcluded region, the branching ratio of 
$H_1$$\rightarrow$$A_1A_1$ is larger than 90\%. Here, the $H_1$ cannot be observed 
(see Fig.\ref{nmssmscan_EW3H1}). The branching ratio of $H_1$$\rightarrow$$A_1A_1$ drops 
for small $\lambda$ and $\kappa$. Therefore, a discovery via the inclusive and associated 
$H_1$$\rightarrow$$\gamma\gamma$; VBF, $H_1$$\rightarrow$$\tau\tau$ and ttH, $H_1$$\rightarrow$$b\bar{b}$ 
modes is possible in that region (Fig.\ref{nmssmscan_EW3H1b}). The outermost discovery contour of 
$H_1$$\rightarrow$$\gamma\gamma$ follows approximately the 60\% line of the branching ratio of 
$H_1$$\rightarrow$$A_1A_1$. The $H_2$ has contributions from the channels $H_2$$\rightarrow$$\gamma\gamma$; 
VBF, $H_2$$\rightarrow$$WW$; GGF, $H_2$$\rightarrow$$ZZ$$\rightarrow$$4l$ and GGF, 
$H_2$$\rightarrow$$WW$$\rightarrow$$2l2\nu$ in the excluded region where it is light enough to be accessible 
(Fig.\ref{nmssmscan_EW3H2}). All other Higgs bosons have either highly reduced couplings or are too heavy 
to be observed in this scenario.

\subsection{Conclusions}
An evaluation of the ATLAS discovery potential for NMSSM Higgs bosons within two benchmark scenarios 
was performed. At least one Higgs boson was found to be observable in regions without a light $A_1$ 
or  where the branching ratio of $H_{1/2}$$\rightarrow$$A_1A_1$ is smaller than about 60\%. 
In the other cases, searches for the decay chains $H_{1/2}$$\rightarrow$$A_1A_1$$\rightarrow$$\tau\tau$$b\bar{b}$ 
or $H_{1/2}$$\rightarrow$$A_1A_1$$\rightarrow$$4\tau$ could be considered. 

\section*{Acknowledements}
We would like to thank U. Ellwanger, E. Gross, S. Heinemeyer, T. Plehn and M. Spira for helpful discussions. 
The authors thank U. Ellwanger, E. Gross, S. Heinemeyer, T. Plehn and M. Spira for helpful discussions.
%\bibliography{nmssmscan}

%\end{document}

\section[The $h^0\rightarrow A^0 A^0\rightarrow b\overline{b}\tau^+\tau^-$ 
Signal in Vector Boson Fusion Production at the LHC]{THE 
$h^0\rightarrow A^0 A^0\rightarrow b\overline{b}\tau^+\tau^-$ 
SIGNAL IN VECTOR BOSON FUSION PRODUCTION AT THE LHC
~\protect\footnote{Contributed by: N. E. Adam, V. Halyo, M. Herquet, 
and S. Gleyzer }}
%\documentclass[11pt]{cernrep}
%\usepackage{graphicx,epsfig}
%\usepackage{amsmath}
%\usepackage{bm}
%\usepackage{color,colordvi}  
%\bibliographystyle{lesHouches}
%\begin{document}

%\title{The $h^0\rightarrow A^0 A^0\rightarrow b\overline{b}\tau^+\tau^-$ signal in Vector Boson Fusion production at the LHC}

%\author{N. E. Adam$^1$, V. Halyo$^1$, M. Herquet$^2$, S. Gleyzer$^3$}
%\institute{$^1$Department of Physics, Princeton University, Princeton, NJ 08544, USA
%\\$^2$Centre for Particle Physics an Phenomenology, Universit\'e catholique de Louvain, 2 Chemin du Cyclotron, 1348 Louvain-la-Neuve, Belgique
%\\$^3$Department of Physics, Florida State University, Tallahassee, Florida 32306, USA}

%\maketitle

%\begin{abstract}

%We examine the $h^0\rightarrow A^0 A^0\rightarrow
%b\overline{b}\tau^+\tau^-$ signal after production of a Standard Model
%(SM) like Higgs particle $h^0$ in Vector Boson Fusion (VBF) at the
%LHC. This first analysis is restricted to a particular mass spectrum,
%i.e. $m_{h^0}$=120 GeV and $m_{A^0}$=50 GeV, and assumes a reasonable
%hypothesis for the involved branching ratios. We demonstrate that a simple 
%cut based analysis with an integrated luminosity of about 100 fb$^{-1}$
%results in significance ($S/\sqrt{B}$) of 2. This parton-level result
%shows the potential feasibility of this channel and motivates improvement
%using a multivariate approach, which is currently under investigation.

%\end{abstract}

\subsection{Motivation}

In the Minimal Supersymmetric extension of the Standard Model (MSSM) at least
one additional $SU(2)_L$ Higgs doublet is required compared to the SM in order to cancel
gauge anomalies of the superpartners and to allow Yukawa couplings for all fermions.
In order to address the fine-tuning ``$\mu$-problem'' that appears in the MSSM, one can also add 
an extra complex singlet to these doublets. This last possibility, known in the literature as the Next-to-Minimal Supersymmetric Standard Model (NMSSM), has an interesting phenomenology (e.g. see \cite{Accomando:2006ga}).

In the NMSSM, one of the pseudoscalar states ($A^0$) is the Goldstone boson of either a global $U(1)$ R-symmetry or a $U(1)$ Peccei-Quinn symmetry in some limit of the model parameters. Since low-fine tuning scenarios predict a moderate breaking of these symmetries, the mass of $A^0$ is expected to be relatively small compared to the mass of the lightest scalar ($h^0$) such that the $h^0\rightarrow A^0 A^0$ decay is kinematically allowed. In \cite{Dermisek:2005ar}, two different types of scenarios are considered, depending on if $m_{A^0}>2 m_b$ or $m_{A^0}<2 m_b$ . Scenarios with $m_{A^0}>2 m_b$ are disfavored when LEP data for $Z2b$ and $Z4b$ final states are taken into account. Indeed, the simultaneous analysis of both these channels excludes at better than 99\% the possibility for $h^0$ to be lighter than $\sim 108$ GeV, and a heavier $h^0$ in turn requires a higher fine-tuning of model parameters. On the contrary, scenarios with $m_{A^0}<2 m_b$ are favored by the same data and can even account for the $2\sigma$ excess observed in the $Z2b$ final state in the $m_{h^0}\sim 100$ GeV vicinity. As a consequence, many NMSSM related analyses focus on the $h^0\rightarrow A^0A^0\rightarrow \tau^+\tau^-\tau^+\tau^-$ decay which has the most favorable branching ratio if $m_{A^0}<2 m_b$.

Nevertheless, besides the particular context of the NMSSM, many other
possibilities remain open. If the MSSM scalar sector violates the $CP$ symmetry, standard mass relations do not hold anymore and the decay of $h^0$ into two lighter Higgs bosons may be allowed \cite{Carena:2002bb}. In \cite{Dobrescu:2000jt}, a
light $A^0$ (i.e., between 0.1 and a few tens of GeV)  decays predominantly into pairs of photons (or gluons) thanks to a vector-like quark loop.  Another possibility arises in the context of the generic two-Higgs-doublet model (2HDM). As shown in \cite{Gerard:2007kn}, a moderately light $A^0$ (i.e., between 10 and 100 GeV) can \textit{naturally} satisfy the $\rho$ parameter constraints thanks to a twisted realization of the custodial (or equivalently $CP$) symmetry. As emphasized in \cite{Krawczyk:2001pe}, a pseudoscalar in this mass range together with a moderate value of $\tan\beta$ can also account (in type II 2HDMs) for the observed discrepancy between the experimental measurement of the muon anomalous magnetic moment and the SM predictions.

Assuming $m_{A^0}>2 m_b$, and that the coupling of $A^0$ to fermions is
proportional to the mass for down-type quarks and leptons (the up-type
quark couplings being negligible for $\tan\beta\gg 1$), the main decay
modes are $A^0\rightarrow b\overline{b}$ (BR $\sim$ 0.92) and
$A^0\rightarrow \tau^+\tau^-$ (BR $\sim$ 0.08).   Under the hypothesis
that BR($h^0\rightarrow A^0A^0$)$\sim1$ (which is a reasonable
approximation in many models), this gives a total branching ratio of
$\sim 0.85$ for $h^0\rightarrow A^0A^0\rightarrow 4b$, $\sim 0.15$ for
$h^0\rightarrow A^0A^0\rightarrow 2b2\tau$ and less than one percent for
$h^0\rightarrow A^0A^0\rightarrow 4\tau$. Since the four $\tau$ final state
signal is suppressed at least by a factor of a hundred compared to the
$m_{A^0}<2 m_b$ scenario studied in Section\ref{sec:4tau}, the LHC discovery
of $h^0$ and $A^0$ in this channel is probably difficult. On the other
hand, the four $b$ final state has a large BR, but suffers from important QCD backgrounds. This final state has been investigated in direct production mode at the Tevatron (where it is overwhelmed by the backgrounds  \cite{Stelzer:2006sp})  and in $W/Z$ associated production \cite{Carena:2007jk}. At the LHC, a discovery significance may still be reached in this last mode \cite{Carena:2007jk,Cheung:2007sva}.

In the current work, we focus on the intermediate $2b2\tau$ final state,
which has a smaller but still sizable BR than the $4b$ final state, together with a much lower background. This final state has been considered in the framework of the associated production of $h^0$ with a $W/Z$ boson at the Tevatron in \cite{Carena:2007jk,Aglietti:2006ne}. However, in this case, only a few events could be observed  after a few fb$^{-1}$ due to the cuts and $b/\tau$ tagging necessary to remove the large reducible background. Similar difficulties with the reducible background are also expected at LHC \cite{Carena:2007jk}. In the present study, we concentrate on the Vector Boson Fusion (VBF) production mode for $h^0$, which has been shown to be a promising channel at the LHC for the SM decay $h^0\rightarrow \tau^+\tau^-$ both in parton-level analysis \cite{Rainwater:1998kj,Plehn:1999nw} and after full detector simulation \cite{Asai:2004ws,Cavalli:2002vs,Klute2002}.
After the end of the redaction of this work, it has been brought to our knowledge that 
a study on similar lines in the context of the NMSSM, using parton shower
based simulations, can be found in \cite{Ellwanger:2003jt,Ellwanger:2004gz}.

\subsection{Signal and Background}

The signal and background Monte-Carlo simulation has been carried out at
tree level using \textsc{MadGraph/MadEvent v4} \cite{Alwall:2007st} for
the parton-level event generation.

In the framework of this preliminary analysis, some simplifying
assumptions are made. The SM-like Higgs, $h^0$, shares all SM Higgs boson
couplings plus an additional coupling to the pseudoscalar $A^0$ large
enough to ensure BR($h^0\rightarrow A^0A^0$) $\sim 1$. Its mass is fixed at
120 GeV, i.e. this is above the best LEP limit to avoid \textit{de facto} all
possible direct constraints, but is still light enough to ensure a sizable
production cross-section. The light pseudoscalar mass is fixed at 50 GeV
in order to lie below the $m_{h^0}/2$ threshold, while still being large
enough to guarantee a good angular separation of decay products.

As mentioned in the previous section, the coupling of $A^0$ to fermions
is assumed to be proportional to their mass for down-type quarks and
charged leptons, giving a total branching ratio for $h^0\rightarrow
A^0A^0\rightarrow 2b2\tau$ of about 0.15, which may be compared with the
SM expectation BR($h^0\rightarrow\tau^+\tau^-$) of $\sim 0.08$. This is
only true if the coupling to up-type quarks is strongly suppressed, for
example, due to an additional $\tan\beta$ factor in a type II 2HDM. If
this is not the case, the considered total branching ratio can be reduced
by up to a factor two.

In order to improve signal to background separation, few kinematical cuts such as 
minimum $p_T$ of 10 GeV for all $b$-jets, 20 GeV for all non-$b$ jets, and 10 GeV for all
 leptons have been applied. In addition, acceptance cuts such as the maximum pseudorapidity 
of 5 for jets, and of 2.5 for  $b$-jets and leptons, and  a minimal separation cut,
i.e. $\Delta R>0.3$, on all objects pairs have been applied at the parton level. 
Furthermore, to narrow ourselves to the particular kinematic configuration of signal events,
standard VBF cuts are applied, i.e. $|\Delta\eta|>4$ and $m_{jj}>700$ GeV
for the two forward jets. Finally, a maximum invariant mass cut,
$m_{\tau\tau}<80$ GeV, is imposed on all leptons pairs to avoid the $Z$
peak in some backgrounds.

The signal is characterized by a final state populated with two central
$b$ jets, two central $\tau$'s and two forward jets. To avoid triggering
issues, we focus on final states in which both $\tau$'s decay leptonically. 
The associated tree level cross-section (after $\tau$ decays and cuts) is 9.5 fb. The
irreducible background, where the $\tau$ pair is coming from an off-shell
photon or $Z$, and the $b$ pair from a gluon splitting, is rather low,
with a 1 fb cross-section. The same process with an $e$ or $\mu$ pair
replacing the $\tau$ pair has a more sizable cross section of around 8.7
fb, due to the absence of the $\tau$ branching ratio suppression. The
most dangerous background is $t\overline{t} + 2$jets with fully leptonic top
decays (through an intermediate $\tau$ or not). However, even though the total
cross-section is almost three orders of magnitude larger than the signal
(3.2 pb), the associated kinematics, and in particular the invariant mass distribution of
$b$'s and leptons, are very different.

\subsection{Results}
Figure~\ref{fig:invmass} shows the invariant mass combination of any
oppositely charged di-lepton pairs and any bottom quark pairs. Only the
cuts described in the previous section have been applied. 

\begin{figure}
\begin{center}
  \includegraphics[scale=0.37]{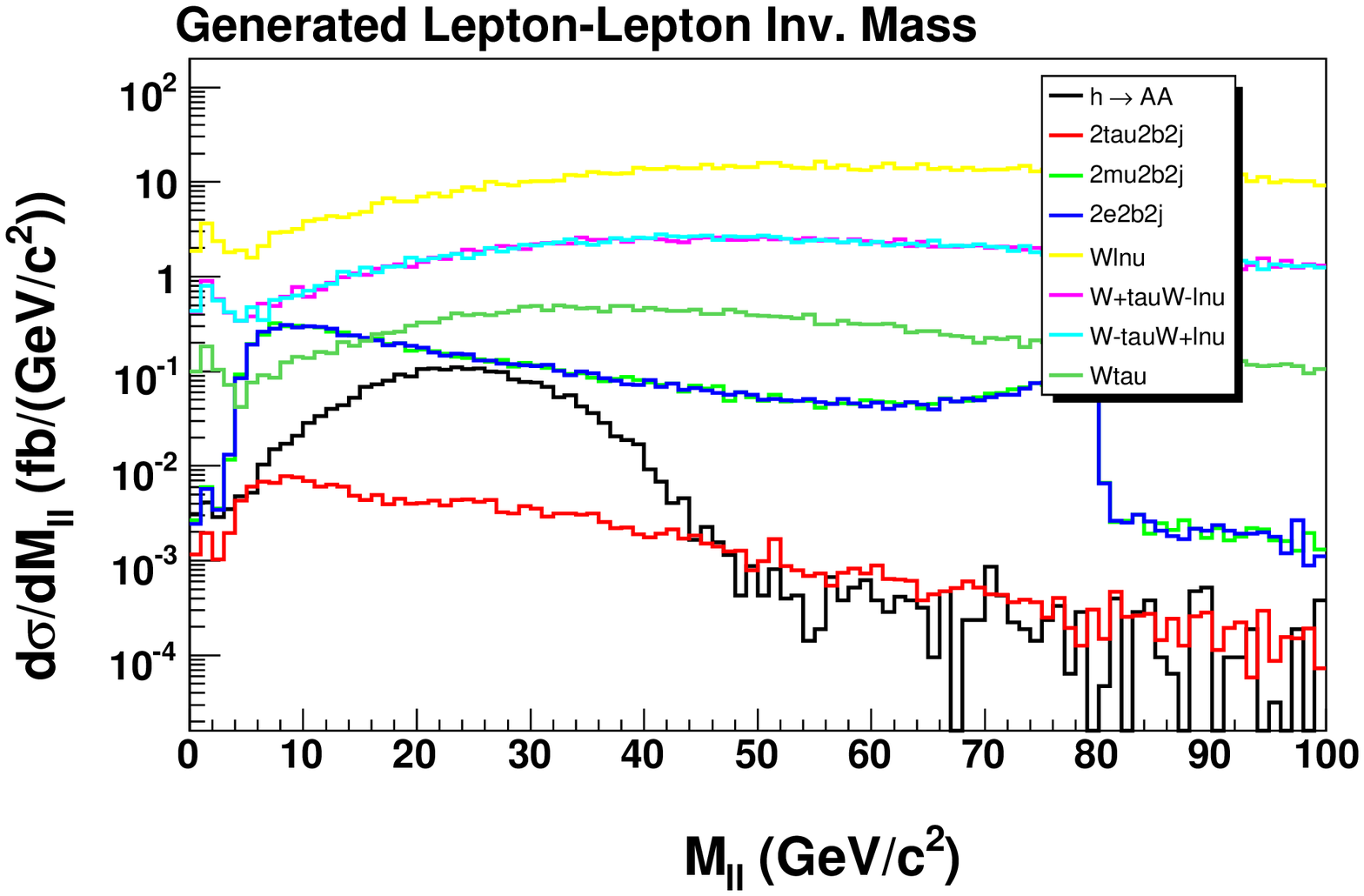}
  \includegraphics[scale=0.37]{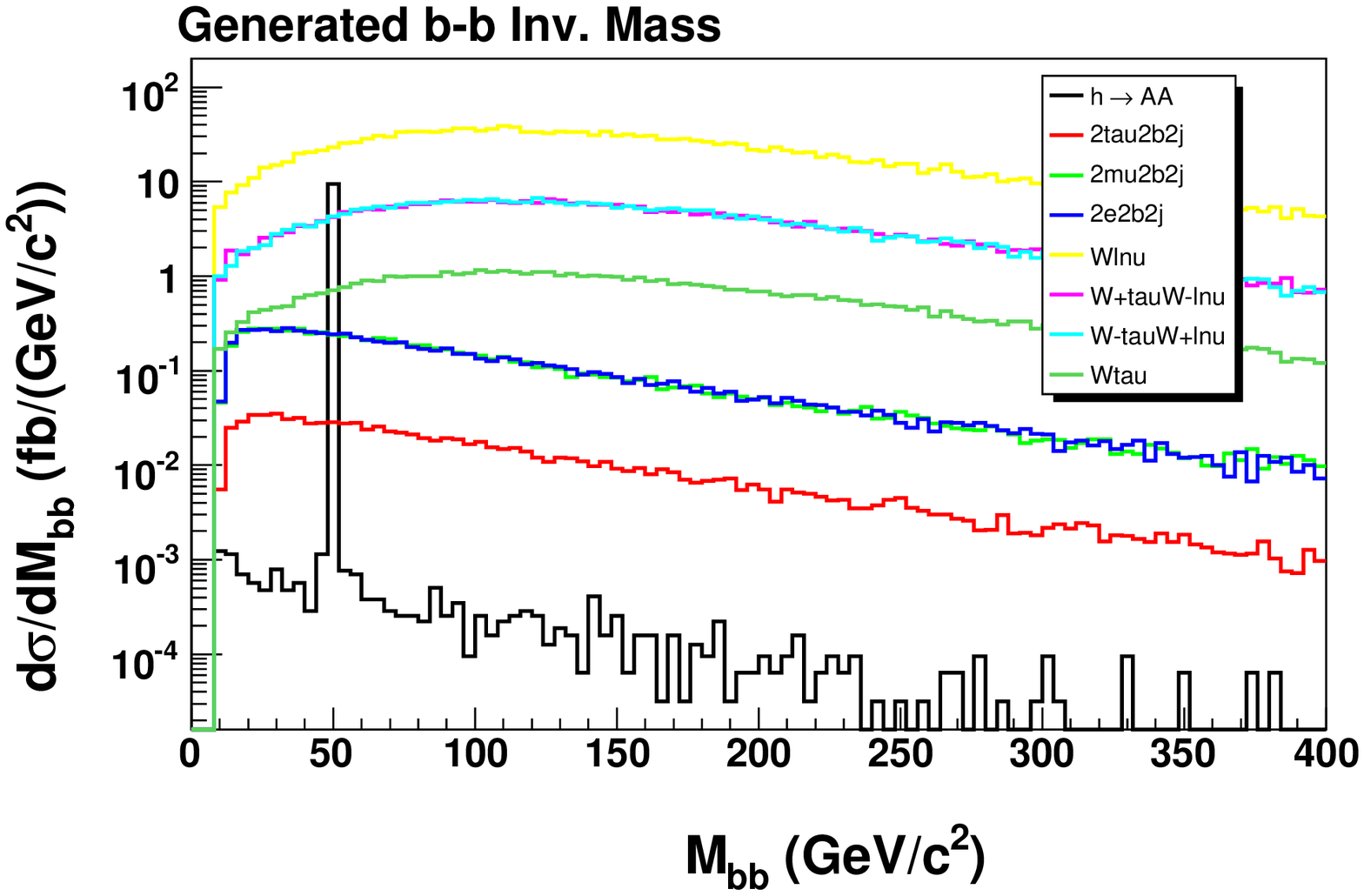}
\caption[]{Invariant mass of the final-state (oppositely charged) di-lepton
  pairs and di-bottom quarks before the final kinematical
  cuts~\ref{eq:cuts}. Each signal or background contribution is
  normalized by cross-section.}
\label{fig:invmass}
\end{center}
\end{figure}

\begin{figure}
\begin{center}
  \includegraphics[scale=0.4]{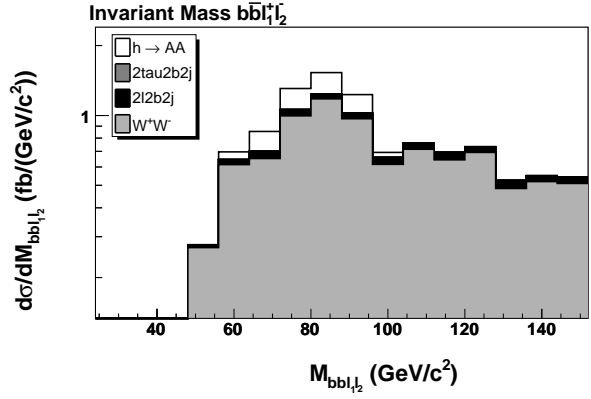}
\caption[]{Invariant mass $M_{bbll}$ of the four-body final state. The
  signal and background histograms are stacked and normalized by their
  corresponding cross-sections.}
\label{fig:VBF_hAA2b2ta_fig1}
\end{center}
\end{figure}

Looking at the kinematic distributions of the signal and background samples
(described in the previous section), it is evident that a cut based
technique can be defined to achieve separation. The chosen selection criteria are:
\begin{equation}
 M_{ll} \le 30 \mathrm{,} \;\;\;\;\;\;  40 \le M_{bb}  \le 60\mathrm{,} \;\;\;\;\;   \Delta R_{ll} \le 2\mathrm{,} \;\;\;\;\mathrm{and} \;\;\;\;   \Delta R_{bb} \le 2.
\label{eq:cuts}
\end{equation}

Figure~\ref{fig:VBF_hAA2b2ta_fig1} shows the invariant mass $M_{bbll}$, of
the four body final state after these simple cuts. The signal and the
background considered are stacked and normalized by cross-section. A
simple estimate of the significance around $M_{bbll}$, in the region $50
\le M_{bbll} \le 110$, yields $S/\sqrt{B} = 4$ for an integrated
luminosity of 100 fb$^{-1}$, with approximatively 100 signal
events. B-tagging efficiency will impact the number of both signal and background
events, and reduces this significance by a factor of $\sim$ 2 if a
b-tagging efficiency of $50\%$ is assumed. Of course this simple
generator-level estimate is merely a crude check on the feasibility of
studying $h^0\rightarrow A^0 A^0\rightarrow b\overline{b}\tau^+\tau^-$ in VBF. 

Since after reconstruction we expect the significance to decrease even
further, this parton-level result motivates the use of a statistics-based
multivariate approach in order to further discriminate between signal and
background. Preliminary results demonstrating the discriminating power of the technique
between the signal, the irreducible background and part of the $t \bar{t}
+2j$ background are shown in Figure~\ref{fig:VBF_hAA2b2ta_fig2}.

\begin{figure}
\begin{center}
\begin{tabular}{cc}
  \includegraphics[scale=0.4]{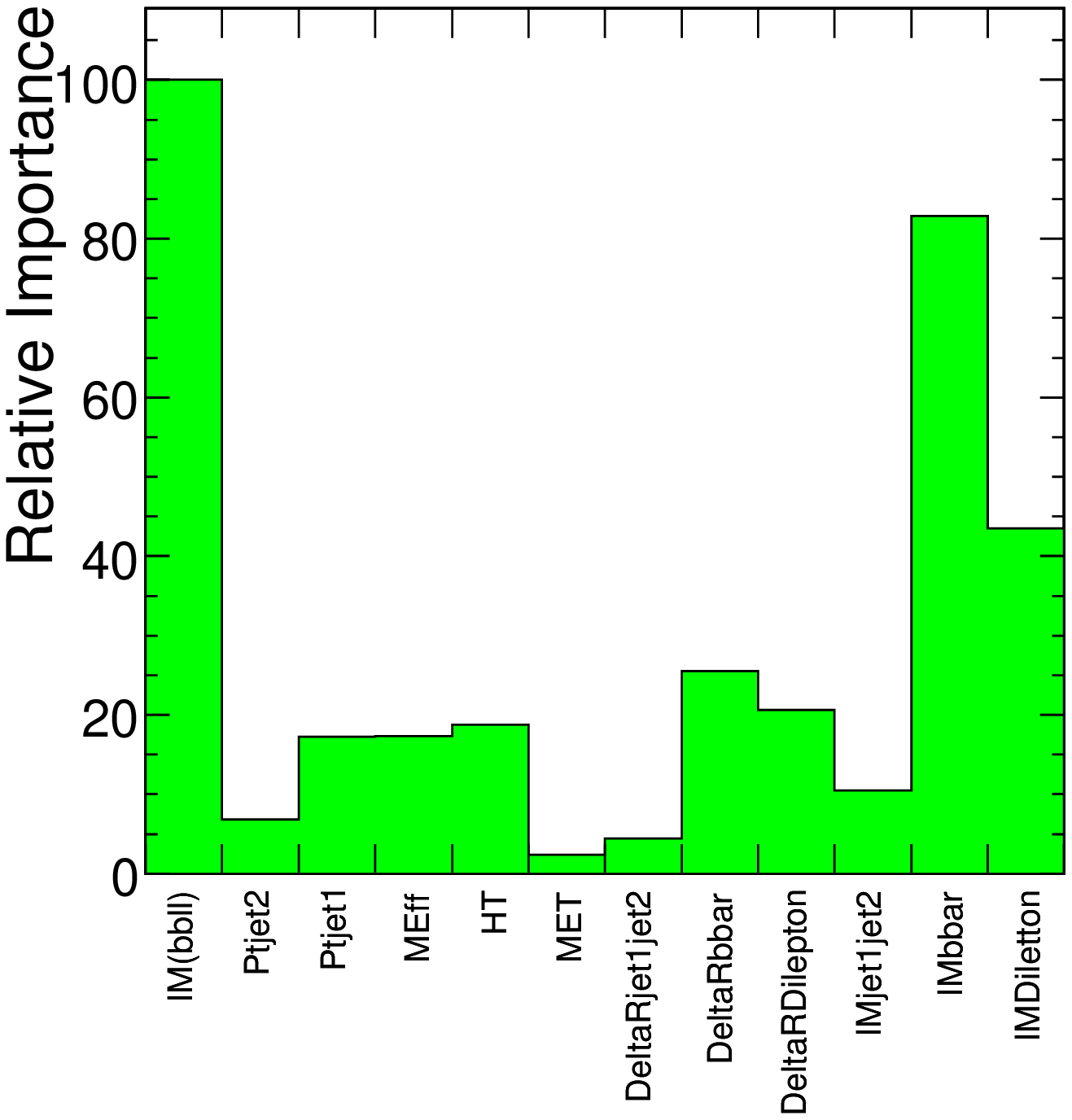} &
  \includegraphics[scale=0.4]{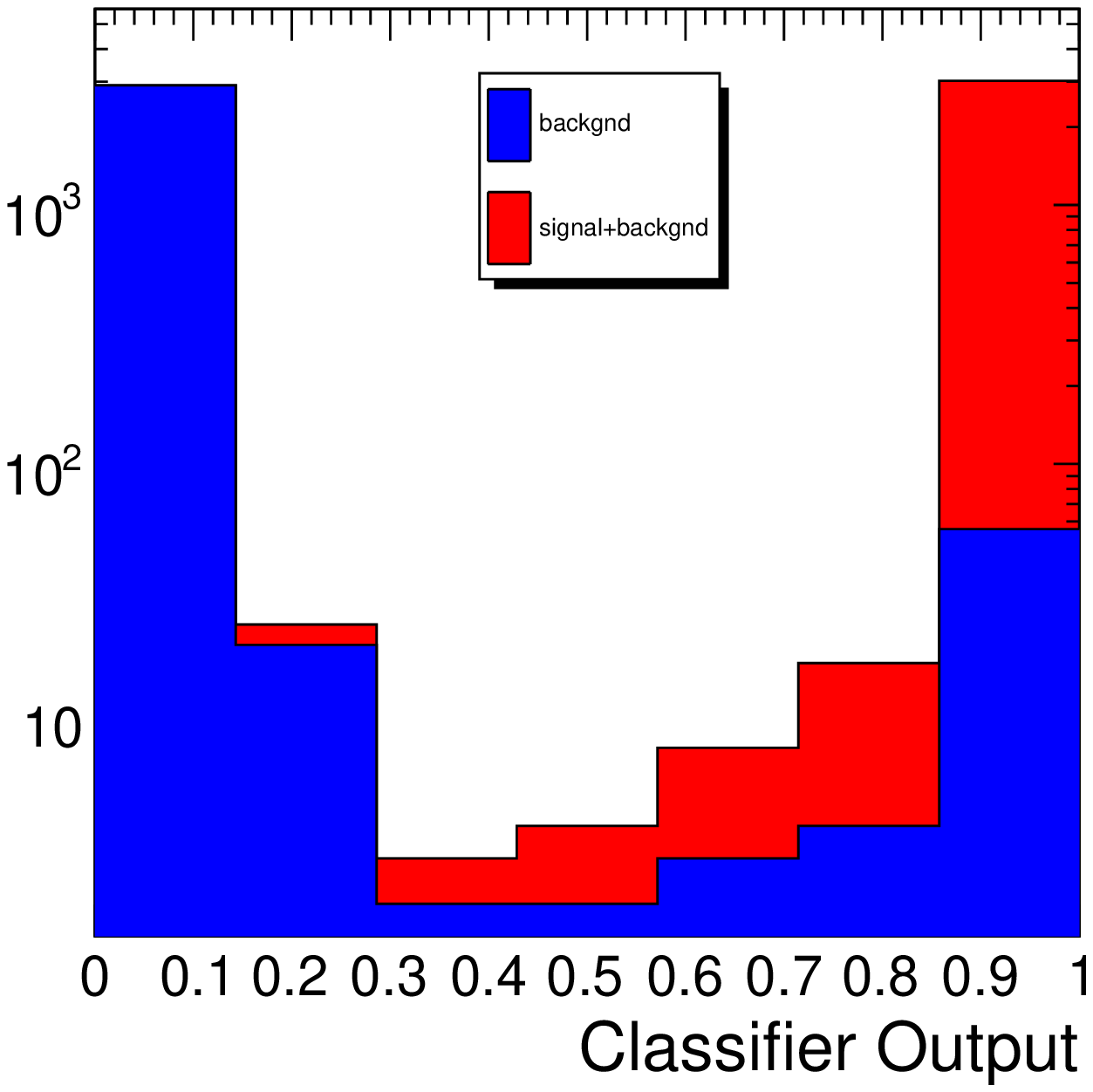} \\
(a) & (b) \\
\end{tabular}
\caption[]{ a) Relative importance of the various kinematic
  variables, as determined using the \emph{ PARADIGM }
  algorithm. b) The decision tree classifier output.}
\label{fig:VBF_hAA2b2ta_fig2}
\end{center}
\end{figure}

Figure~\ref{fig:VBF_hAA2b2ta_fig2} (a) shows the relative contribution of
the various input variables to signal and background separation. 
A framework for parameter space optimization,
\textsc{paradigm}, is utilized for the above task~\cite{Gleyzer:2007}. The
two most effective variables for signal/background separation in this
decay mode are the invariant mass of the $b$-jets and leptons
$M_{bbll}$ and the invariant mass of the $b$-jets
$M_{bb}$, as was also observed in the cut based study. 
Although \textsc{paradigm} allows the reduction of parameter
space, we do not eliminate any of the variables since the
dimensionality of the initial feature space considered is lower than the
degrees of freedom of the model. Therefore, it is likely that the
classifier performance can be further enhanced by the addition of more variables.

The decision tree classifier output is shown in
Figure~\ref{fig:VBF_hAA2b2ta_fig2} (b). The measure of discrepancy between
the background-only hypothesis and the background plus signal hypothesis
(assuming a normal error distribution and using the classifier itself as
the test statistic in a two-tailed test\cite{Freund:1984}) is found to be
0.0086 $\pm$ 0.0058 at 95\% CL. This is a statistically significant
result.

\subsection{Conclusions}
We showed that the $h^0\rightarrow A^0 A^0\rightarrow
b\overline{b}\tau^+\tau^-$ signal in VBF production at the LHC is
potentially feasible with an integrated luminosity of 100
fb$^{-1}$. Using a simple cut based technique, we
found approximately 25 signal events and a significance of $\sim$2 for
this luminosity (taking into account a 50\% $b$-tagging efficiency)
. This result motivates the use other techniques, such as a multivariate
analysis, to further enhance the feasibility of this search at the LHC. A
more robust multivariate analysis that includes different mass hypotheses, a full set of
reducible backgrounds as well as fast detector simulation and evaluation of
systematic uncertainties is envisaged by the authors.

\section*{Acknowledgments}
We thank the organizers and conveners of the Les Houches workshop where this work originated. We also thank Fabio Maltoni and Simon de Visscher for useful discussion. The work of M.H. was supported by the Institute Interuniversitaire des Sciences Nucl\'eaires and by the Belgian Federal Office for Scientific, Technical and Cultural Affairs through the Interuniversity Attraction Pole P6/11.
This work was also supported in part by USDOE grant DE-FG02-91ER40671

%\bibliography{VBF_hAA2b2ta}

%\end{document}

\bibliography{higgs_summary}

\providecommand{\href}[2]{#2}\begingroup\raggedright\begin{thebibliography}{10%
0}

\bibitem{Higgs:1964pj}
P.~W. Higgs, {\em Phys. Rev. Lett.} {\bf 13} (1964) 508--509.

\bibitem{Djouadi:2005gi}
A.~Djouadi, {\em Phys. Rept.} {\bf 457} (2008) 1--216,
  [\href{http://xxx.lanl.gov/abs/hep-ph/0503172}{{\tt hep-ph/0503172}}].

\bibitem{Carena:2002es}
M.~S. Carena and H.~E. Haber, {\em Prog. Part. Nucl. Phys.} {\bf 50} (2003)
  63--152, [\href{http://xxx.lanl.gov/abs/hep-ph/0208209}{{\tt
  hep-ph/0208209}}].

\bibitem{Barate:2003sz}
R.~Barate {\em et.~al.},, {\bf LEP} Collaboration {\em Phys. Lett.} {\bf B565}
  (2003) 61--75, [\href{http://xxx.lanl.gov/abs/hep-ex/0306033}{{\tt
  hep-ex/0306033}}].

\bibitem{Alcaraz:2007ri}
J.~Alcaraz {\em et.~al.},, {\bf LEP} Collaboration
  \href{http://xxx.lanl.gov/abs/arXiv:0712.0929 [hep-ex]}{{\tt arXiv:0712.0929
  [hep-ex]}}.

\bibitem{lepewwg}
L.~E.~W. group, {\em /lepewwg.web.cern.ch/LEPEWWG}.

\bibitem{:2007gz}
{\bf TEVNPH Working Group} Collaboration
  \href{http://xxx.lanl.gov/abs/arXiv:0712.2383 [hep-ex]}{{\tt arXiv:0712.2383
  [hep-ex]}}.

\bibitem{Duhrssen:2004cv}
M.~Duhrssen {\em et.~al.}, {\em Phys. Rev.} {\bf D70} (2004) 113009,
  [\href{http://xxx.lanl.gov/abs/hep-ph/0406323}{{\tt hep-ph/0406323}}].

\bibitem{Dawson:1991zj}
S.~Dawson, {\em Nucl. Phys.} {\bf B359} (1991) 283--300.

\bibitem{Djouadi:1991tka}
A.~Djouadi, M.~Spira, and P.~M. Zerwas, {\em Phys. Lett.} {\bf B264} (1991)
  440--446.

\bibitem{Spira:1995rr}
M.~Spira, A.~Djouadi, D.~Graudenz, and P.~M. Zerwas, {\em Nucl. Phys.} {\bf
  B453} (1995) 17--82, [\href{http://xxx.lanl.gov/abs/hep-ph/9504378}{{\tt
  hep-ph/9504378}}].

\bibitem{Harlander:2000mg}
R.~V. Harlander, {\em Phys. Lett.} {\bf B492} (2000) 74--80,
  [\href{http://xxx.lanl.gov/abs/hep-ph/0007289}{{\tt hep-ph/0007289}}].

\bibitem{Catani:2001ic}
S.~Catani, D.~de~Florian, and M.~Grazzini, {\em JHEP} {\bf 05} (2001) 025,
  [\href{http://xxx.lanl.gov/abs/hep-ph/0102227}{{\tt hep-ph/0102227}}].

\bibitem{Harlander:2001is}
R.~V. Harlander and W.~B. Kilgore, {\em Phys. Rev.} {\bf D64} (2001) 013015,
  [\href{http://xxx.lanl.gov/abs/hep-ph/0102241}{{\tt hep-ph/0102241}}].

\bibitem{Harlander:2002wh}
R.~V. Harlander and W.~B. Kilgore, {\em Phys. Rev. Lett.} {\bf 88} (2002)
  201801, [\href{http://xxx.lanl.gov/abs/hep-ph/0201206}{{\tt
  hep-ph/0201206}}].

\bibitem{Anastasiou:2002yz}
C.~Anastasiou and K.~Melnikov, {\em Nucl. Phys.} {\bf B646} (2002) 220--256,
  [\href{http://xxx.lanl.gov/abs/hep-ph/0207004}{{\tt hep-ph/0207004}}].

\bibitem{Ravindran:2003um}
V.~Ravindran, J.~Smith, and W.~L. van Neerven, {\em Nucl. Phys.} {\bf B665}
  (2003) 325--366, [\href{http://xxx.lanl.gov/abs/hep-ph/0302135}{{\tt
  hep-ph/0302135}}].

\bibitem{Catani:2003zt}
S.~Catani, D.~de~Florian, M.~Grazzini, and P.~Nason, {\em JHEP} {\bf 07} (2003)
  028, [\href{http://xxx.lanl.gov/abs/hep-ph/0306211}{{\tt hep-ph/0306211}}].

\bibitem{Moch:2005ky}
S.~Moch and A.~Vogt, {\em Phys. Lett.} {\bf B631} (2005) 48--57,
  [\href{http://xxx.lanl.gov/abs/hep-ph/0508265}{{\tt hep-ph/0508265}}].

\bibitem{Laenen:2005uz}
E.~Laenen and L.~Magnea, {\em Phys. Lett.} {\bf B632} (2006) 270--276,
  [\href{http://xxx.lanl.gov/abs/hep-ph/0508284}{{\tt hep-ph/0508284}}].

\bibitem{Idilbi:2005ni}
A.~Idilbi, X.-d. Ji, J.-P. Ma, and F.~Yuan, {\em Phys. Rev.} {\bf D73} (2006)
  077501, [\href{http://xxx.lanl.gov/abs/hep-ph/0509294}{{\tt
  hep-ph/0509294}}].

\bibitem{Djouadi:1994ge}
A.~Djouadi and P.~Gambino, {\em Phys. Rev. Lett.} {\bf 73} (1994) 2528--2531,
  [\href{http://xxx.lanl.gov/abs/hep-ph/9406432}{{\tt hep-ph/9406432}}].

\bibitem{Aglietti:2004nj}
U.~Aglietti, R.~Bonciani, G.~Degrassi, and A.~Vicini, {\em Phys. Lett.} {\bf
  B595} (2004) 432--441, [\href{http://xxx.lanl.gov/abs/hep-ph/0404071}{{\tt
  hep-ph/0404071}}].

\bibitem{Degrassi:2004mx}
G.~Degrassi and F.~Maltoni, {\em Phys. Lett.} {\bf B600} (2004) 255--260,
  [\href{http://xxx.lanl.gov/abs/hep-ph/0407249}{{\tt hep-ph/0407249}}].

\bibitem{Harlander:2004tp}
R.~V. Harlander and M.~Steinhauser, {\em JHEP} {\bf 09} (2004) 066,
  [\href{http://xxx.lanl.gov/abs/hep-ph/0409010}{{\tt hep-ph/0409010}}].

\bibitem{Harlander:2003bb}
R.~V. Harlander and M.~Steinhauser, {\em Phys. Lett.} {\bf B574} (2003).

\bibitem{Harlander:2003kf}
R.~Harlander and M.~Steinhauser, {\em Phys. Rev.} {\bf D68} (2003) 111701,
  [\href{http://xxx.lanl.gov/abs/hep-ph/0308210}{{\tt hep-ph/0308210}}].

\bibitem{Muhlleitner:2006wx}
M.~Muhlleitner and M.~Spira, {\em Nucl. Phys.} {\bf B790} (2008) 1--27,
  [\href{http://xxx.lanl.gov/abs/hep-ph/0612254}{{\tt hep-ph/0612254}}].

\bibitem{Bonciani:2007ex}
R.~Bonciani, G.~Degrassi, and A.~Vicini, {\em JHEP} {\bf 11} (2007) 095,
  [\href{http://xxx.lanl.gov/abs/arXiv:0709.4227 [hep-ph]}{{\tt arXiv:0709.4227
  [hep-ph]}}].

\bibitem{Anastasiou:2006hc}
C.~Anastasiou, S.~Beerli, S.~Bucherer, A.~Daleo, and Z.~Kunszt, {\em JHEP} {\bf
  01} (2007) 082, [\href{http://xxx.lanl.gov/abs/hep-ph/0611236}{{\tt
  hep-ph/0611236}}].

\bibitem{deFlorian:1999zd}
D.~de~Florian, M.~Grazzini, and Z.~Kunszt, {\em Phys. Rev. Lett.} {\bf 82}
  (1999) 5209--5212, [\href{http://xxx.lanl.gov/abs/hep-ph/9902483}{{\tt
  hep-ph/9902483}}].

\bibitem{Campbell:2006xx}
J.~M. Campbell, R.~K. Ellis, and G.~Zanderighi, {\em JHEP} {\bf 10} (2006) 028,
  [\href{http://xxx.lanl.gov/abs/hep-ph/0608194}{{\tt hep-ph/0608194}}].

\bibitem{Brein:2003df}
O.~Brein and W.~Hollik, {\em Phys. Rev.} {\bf D68} (2003) 095006,
  [\href{http://xxx.lanl.gov/abs/hep-ph/0305321}{{\tt hep-ph/0305321}}].

\bibitem{Field:2003yy}
B.~Field, S.~Dawson, and J.~Smith, {\em Phys. Rev.} {\bf D69} (2004) 074013,
  [\href{http://xxx.lanl.gov/abs/hep-ph/0311199}{{\tt hep-ph/0311199}}].

\bibitem{Catani:2001cr}
S.~Catani, D.~de~Florian, and M.~Grazzini, {\em JHEP} {\bf 01} (2002) 015,
  [\href{http://xxx.lanl.gov/abs/hep-ph/0111164}{{\tt hep-ph/0111164}}].

\bibitem{Anastasiou:2005qj}
C.~Anastasiou, K.~Melnikov, and F.~Petriello, {\em Nucl. Phys.} {\bf B724}
  (2005) 197--246, [\href{http://xxx.lanl.gov/abs/hep-ph/0501130}{{\tt
  hep-ph/0501130}}].

\bibitem{Anastasiou:2007mz}
C.~Anastasiou, G.~Dissertori, and F.~Stockli, {\em JHEP} {\bf 09} (2007) 018,
  [\href{http://xxx.lanl.gov/abs/arXiv:0707.2373 [hep-ph]}{{\tt arXiv:0707.2373
  [hep-ph]}}].

\bibitem{Catani:2007vq}
S.~Catani and M.~Grazzini, {\em Phys. Rev. Lett.} {\bf 98} (2007) 222002,
  [\href{http://xxx.lanl.gov/abs/hep-ph/0703012}{{\tt hep-ph/0703012}}].

\bibitem{Grazzini:2008tf}
M.~Grazzini, \href{http://xxx.lanl.gov/abs/arXiv:0801.3232 [hep-ph]}{{\tt
  arXiv:0801.3232 [hep-ph]}}.

\bibitem{Langenegger:2006wu}
U.~Langenegger, M.~Spira, A.~Starodumov, and P.~Trueb, {\em JHEP} {\bf 06}
  (2006) 035, [\href{http://xxx.lanl.gov/abs/hep-ph/0604156}{{\tt
  hep-ph/0604156}}].

\bibitem{Balazs:2000wv}
C.~Balazs and C.~P. Yuan, {\em Phys. Lett.} {\bf B478} (2000) 192--198,
  [\href{http://xxx.lanl.gov/abs/hep-ph/0001103}{{\tt hep-ph/0001103}}].

\bibitem{Berger:2002ut}
E.~L. Berger and J.-w. Qiu, {\em Phys. Rev.} {\bf D67} (2003) 034026,
  [\href{http://xxx.lanl.gov/abs/hep-ph/0210135}{{\tt hep-ph/0210135}}].

\bibitem{Kulesza:2003wn}
A.~Kulesza, G.~Sterman, and W.~Vogelsang, {\em Phys. Rev.} {\bf D69} (2004)
  014012, [\href{http://xxx.lanl.gov/abs/hep-ph/0309264}{{\tt
  hep-ph/0309264}}].

\bibitem{Bozzi:2003jy}
G.~Bozzi, S.~Catani, D.~de~Florian, and M.~Grazzini, {\em Phys. Lett.} {\bf
  B564} (2003) 65--72, [\href{http://xxx.lanl.gov/abs/hep-ph/0302104}{{\tt
  hep-ph/0302104}}].

\bibitem{Bozzi:2005wk}
G.~Bozzi, S.~Catani, D.~de~Florian, and M.~Grazzini, {\em Nucl. Phys.} {\bf
  B737} (2006) 73--120, [\href{http://xxx.lanl.gov/abs/hep-ph/0508068}{{\tt
  hep-ph/0508068}}].

\bibitem{Bozzi:2007pn}
G.~Bozzi, S.~Catani, D.~de~Florian, and M.~Grazzini, {\em Nucl. Phys.} {\bf
  B791} (2008) 1--19, [\href{http://xxx.lanl.gov/abs/arXiv:0705.3887
  [hep-ph]}{{\tt arXiv:0705.3887 [hep-ph]}}].

\bibitem{Ravindran:2002dc}
V.~Ravindran, J.~Smith, and W.~L. Van~Neerven, {\em Nucl. Phys.} {\bf B634}
  (2002) 247--290, [\href{http://xxx.lanl.gov/abs/hep-ph/0201114}{{\tt
  hep-ph/0201114}}].

\bibitem{Glosser:2002gm}
C.~J. Glosser and C.~R. Schmidt, {\em JHEP} {\bf 12} (2002) 016,
  [\href{http://xxx.lanl.gov/abs/hep-ph/0209248}{{\tt hep-ph/0209248}}].

\bibitem{Balazs:2004rd}
C.~Balazs, M.~Grazzini, J.~Huston, A.~Kulesza, and I.~Puljak,
  \href{http://xxx.lanl.gov/abs/hep-ph/0403052}{{\tt hep-ph/0403052}}.

\bibitem{Djouadi:1990aj}
A.~Djouadi, M.~Spira, J.~J. van~der Bij, and P.~M. Zerwas, {\em Phys. Lett.}
  {\bf B257} (1991) 187--190.

\bibitem{Degrassi:2005mc}
G.~Degrassi and F.~Maltoni, {\em Nucl. Phys.} {\bf B724} (2005) 183--196,
  [\href{http://xxx.lanl.gov/abs/hep-ph/0504137}{{\tt hep-ph/0504137}}].

\bibitem{Passarino:2007fp}
G.~Passarino, C.~Sturm, and S.~Uccirati, {\em Phys. Lett.} {\bf B655} (2007)
  298--306, [\href{http://xxx.lanl.gov/abs/arXiv:0707.1401 [hep-ph]}{{\tt
  arXiv:0707.1401 [hep-ph]}}].

\bibitem{Steinhauser:1996wy}
M.~Steinhauser, \href{http://xxx.lanl.gov/abs/hep-ph/9612395}{{\tt
  hep-ph/9612395}}.

\bibitem{Binoth:1999qq}
T.~Binoth, J.~P. Guillet, E.~Pilon, and M.~Werlen, {\em Eur. Phys. J.} {\bf
  C16} (2000) 311--330, [\href{http://xxx.lanl.gov/abs/hep-ph/9911340}{{\tt
  hep-ph/9911340}}].

\bibitem{Bern:2002jx}
Z.~Bern, L.~J. Dixon, and C.~Schmidt, {\em Phys. Rev.} {\bf D66} (2002) 074018,
  [\href{http://xxx.lanl.gov/abs/hep-ph/0206194}{{\tt hep-ph/0206194}}].

\bibitem{Dittmar:1996ss}
M.~Dittmar and H.~K. Dreiner, {\em Phys. Rev.} {\bf D55} (1997) 167--172,
  [\href{http://xxx.lanl.gov/abs/hep-ph/9608317}{{\tt hep-ph/9608317}}].

\bibitem{Anastasiou:2008ik}
C.~Anastasiou, G.~Dissertori, F.~Stoeckli, and B.~R. Webber,
  \href{http://xxx.lanl.gov/abs/arXiv:0801.2682 [hep-ph]}{{\tt arXiv:0801.2682
  [hep-ph]}}.

\bibitem{Davatz:2004zg}
G.~Davatz, G.~Dissertori, M.~Dittmar, M.~Grazzini, and F.~Pauss, {\em JHEP}
  {\bf 05} (2004) 009, [\href{http://xxx.lanl.gov/abs/hep-ph/0402218}{{\tt
  hep-ph/0402218}}].

\bibitem{Dixon:1999di}
L.~J. Dixon, Z.~Kunszt, and A.~Signer, {\em Phys. Rev.} {\bf D60} (1999)
  114037, [\href{http://xxx.lanl.gov/abs/hep-ph/9907305}{{\tt
  hep-ph/9907305}}].

\bibitem{Campbell:1999ah}
J.~M. Campbell and R.~K. Ellis, {\em Phys. Rev.} {\bf D60} (1999) 113006,
  [\href{http://xxx.lanl.gov/abs/hep-ph/9905386}{{\tt hep-ph/9905386}}].

\bibitem{Grazzini:2005vw}
M.~Grazzini, {\em JHEP} {\bf 01} (2006) 095,
  [\href{http://xxx.lanl.gov/abs/hep-ph/0510337}{{\tt hep-ph/0510337}}].

\bibitem{Frixione:2002ik}
S.~Frixione and B.~R. Webber, {\em JHEP} {\bf 06} (2002) 029,
  [\href{http://xxx.lanl.gov/abs/hep-ph/0204244}{{\tt hep-ph/0204244}}].

\bibitem{Frixione:2003ei}
S.~Frixione, P.~Nason, and B.~R. Webber, {\em JHEP} {\bf 08} (2003) 007,
  [\href{http://xxx.lanl.gov/abs/hep-ph/0305252}{{\tt hep-ph/0305252}}].

\bibitem{Binoth:2005ua}
T.~Binoth, M.~Ciccolini, N.~Kauer, and M.~Kramer, {\em JHEP} {\bf 03} (2005)
  065, [\href{http://xxx.lanl.gov/abs/hep-ph/0503094}{{\tt hep-ph/0503094}}].

\bibitem{Duhrssen:2005bz}
M.~Duhrssen, K.~Jakobs, J.~J. van~der Bij, and P.~Marquard, {\em JHEP} {\bf 05}
  (2005) 064, [\href{http://xxx.lanl.gov/abs/hep-ph/0504006}{{\tt
  hep-ph/0504006}}].

\bibitem{Bernreuther:2001rq}
W.~Bernreuther, A.~Brandenburg, Z.~G. Si, and P.~Uwer, {\em Phys. Rev. Lett.}
  {\bf 87} (2001) 242002, [\href{http://xxx.lanl.gov/abs/hep-ph/0107086}{{\tt
  hep-ph/0107086}}].

\bibitem{Kauer:2001sp}
N.~Kauer and D.~Zeppenfeld, {\em Phys. Rev.} {\bf D65} (2002) 014021,
  [\href{http://xxx.lanl.gov/abs/hep-ph/0107181}{{\tt hep-ph/0107181}}].

\bibitem{Frederix:2008vb}
R.~Frederix and M.~Grazzini, \href{http://xxx.lanl.gov/abs/arXiv:0801.2229
  [hep-ph]}{{\tt arXiv:0801.2229 [hep-ph]}}.

\bibitem{Bredenstein:2006rh}
A.~Bredenstein, A.~Denner, S.~Dittmaier, and M.~M. Weber, {\em Phys. Rev.} {\bf
  D74} (2006) 013004, [\href{http://xxx.lanl.gov/abs/hep-ph/0604011}{{\tt
  hep-ph/0604011}}].

\bibitem{Bredenstein:2006ha}
A.~Bredenstein, A.~Denner, S.~Dittmaier, and M.~M. Weber, {\em JHEP} {\bf 02}
  (2007) 080, [\href{http://xxx.lanl.gov/abs/hep-ph/0611234}{{\tt
  hep-ph/0611234}}].

\bibitem{Han:1992hr}
T.~Han, G.~Valencia, and S.~Willenbrock, {\em Phys. Rev. Lett.} {\bf 69} (1992)
  3274--3277, [\href{http://xxx.lanl.gov/abs/hep-ph/9206246}{{\tt
  hep-ph/9206246}}].

\bibitem{Figy:2003nv}
T.~Figy, C.~Oleari, and D.~Zeppenfeld, {\em Phys. Rev.} {\bf D68} (2003)
  073005, [\href{http://xxx.lanl.gov/abs/hep-ph/0306109}{{\tt
  hep-ph/0306109}}].

\bibitem{Figy:2004pt}
T.~Figy and D.~Zeppenfeld, {\em Phys. Lett.} {\bf B591} (2004) 297--303,
  [\href{http://xxx.lanl.gov/abs/hep-ph/0403297}{{\tt hep-ph/0403297}}].

\bibitem{Berger:2004pca}
E.~L. Berger and J.~Campbell, {\em Phys. Rev.} {\bf D70} (2004) 073011,
  [\href{http://xxx.lanl.gov/abs/hep-ph/0403194}{{\tt hep-ph/0403194}}].

\bibitem{Ciccolini:2007jr}
M.~Ciccolini, A.~Denner, and S.~Dittmaier, {\em Phys. Rev. Lett.} {\bf 99}
  (2007) 161803, [\href{http://xxx.lanl.gov/abs/arXiv:0707.0381 [hep-ph]}{{\tt
  arXiv:0707.0381 [hep-ph]}}].

\bibitem{Ciccolini:2007ec}
M.~Ciccolini, A.~Denner, and S.~Dittmaier, {\em Phys. Rev.} {\bf D77} (2008)
  013002, [\href{http://xxx.lanl.gov/abs/arXiv:0710.4749 [hep-ph]}{{\tt
  arXiv:0710.4749 [hep-ph]}}].

\bibitem{DelDuca:2001fn}
V.~Del~Duca, W.~Kilgore, C.~Oleari, C.~Schmidt, and D.~Zeppenfeld, {\em Nucl.
  Phys.} {\bf B616} (2001) 367--399,
  [\href{http://xxx.lanl.gov/abs/hep-ph/0108030}{{\tt hep-ph/0108030}}].

\bibitem{DelDuca:2006hk}
V.~Del~Duca {\em et.~al.}, {\em JHEP} {\bf 10} (2006) 016,
  [\href{http://xxx.lanl.gov/abs/hep-ph/0608158}{{\tt hep-ph/0608158}}].

\bibitem{Campbell:2002tg}
J.~Campbell and R.~K. Ellis, {\em Phys. Rev.} {\bf D65} (2002) 113007,
  [\href{http://xxx.lanl.gov/abs/hep-ph/0202176}{{\tt hep-ph/0202176}}].

\bibitem{Oleari:2003tc}
C.~Oleari and D.~Zeppenfeld, {\em Phys. Rev.} {\bf D69} (2004) 093004,
  [\href{http://xxx.lanl.gov/abs/hep-ph/0310156}{{\tt hep-ph/0310156}}].

\bibitem{Jager:2006zc}
B.~Jager, C.~Oleari, and D.~Zeppenfeld, {\em JHEP} {\bf 07} (2006) 015,
  [\href{http://xxx.lanl.gov/abs/hep-ph/0603177}{{\tt hep-ph/0603177}}].

\bibitem{Dittmaier:2007wz}
S.~Dittmaier, P.~Uwer, and S.~Weinzierl, {\em Phys. Rev. Lett.} {\bf 98} (2007)
  262002, [\href{http://xxx.lanl.gov/abs/hep-ph/0703120}{{\tt
  hep-ph/0703120}}].

\bibitem{Dittmaier:2003ej}
S.~Dittmaier, M.~Kramer, and M.~Spira, {\em Phys. Rev.} {\bf D70} (2004)
  074010, [\href{http://xxx.lanl.gov/abs/hep-ph/0309204}{{\tt
  hep-ph/0309204}}].

\bibitem{Dawson:2003kb}
S.~Dawson, C.~B. Jackson, L.~Reina, and D.~Wackeroth, {\em Phys. Rev.} {\bf
  D69} (2004) 074027, [\href{http://xxx.lanl.gov/abs/hep-ph/0311067}{{\tt
  hep-ph/0311067}}].

\bibitem{Dawson:2004sh}
S.~Dawson, C.~B. Jackson, L.~Reina, and D.~Wackeroth, {\em Phys. Rev. Lett.}
  {\bf 94} (2005) 031802, [\href{http://xxx.lanl.gov/abs/hep-ph/0408077}{{\tt
  hep-ph/0408077}}].

\bibitem{Dawson:2005vi}
S.~Dawson, C.~B. Jackson, L.~Reina, and D.~Wackeroth, {\em Mod. Phys. Lett.}
  {\bf A21} (2006) 89--110, [\href{http://xxx.lanl.gov/abs/hep-ph/0508293}{{\tt
  hep-ph/0508293}}].

\bibitem{Harlander:2003ai}
R.~V. Harlander and W.~B. Kilgore, {\em Phys. Rev.} {\bf D68} (2003) 013001,
  [\href{http://xxx.lanl.gov/abs/hep-ph/0304035}{{\tt hep-ph/0304035}}].

\bibitem{Campbell:2002zm}
J.~Campbell, R.~K. Ellis, F.~Maltoni, and S.~Willenbrock, {\em Phys. Rev.} {\bf
  D67} (2003) 095002, [\href{http://xxx.lanl.gov/abs/hep-ph/0204093}{{\tt
  hep-ph/0204093}}].

\bibitem{Campbell:2004pu}
J.~Campbell {\em et.~al.}, \href{http://xxx.lanl.gov/abs/hep-ph/0405302}{{\tt
  hep-ph/0405302}}.

\bibitem{Buttar:2006zd}
C.~Buttar {\em et.~al.}, \href{http://xxx.lanl.gov/abs/hep-ph/0604120}{{\tt
  hep-ph/0604120}}.

\bibitem{Dittmaier:2006cz}
S.~Dittmaier, M.~Kramer, A.~Muck, and T.~Schluter, {\em JHEP} {\bf 03} (2007)
  114, [\href{http://xxx.lanl.gov/abs/hep-ph/0611353}{{\tt hep-ph/0611353}}].

\bibitem{Boudjema:2007uh}
F.~Boudjema and L.~D. Ninh, \href{http://xxx.lanl.gov/abs/arXiv:0711.2005
  [hep-ph]}{{\tt arXiv:0711.2005 [hep-ph]}}.

\bibitem{Dawson:2007ur}
S.~Dawson and C.~B. Jackson, {\em Phys. Rev.} {\bf D77} (2008) 015019,
  [\href{http://xxx.lanl.gov/abs/arXiv:0709.4519 [hep-ph]}{{\tt arXiv:0709.4519
  [hep-ph]}}].

\bibitem{Beenakker:2001rj}
W.~Beenakker {\em et.~al.}, {\em Phys. Rev. Lett.} {\bf 87} (2001) 201805,
  [\href{http://xxx.lanl.gov/abs/hep-ph/0107081}{{\tt hep-ph/0107081}}].

\bibitem{Beenakker:2002nc}
W.~Beenakker {\em et.~al.}, {\em Nucl. Phys.} {\bf B653} (2003) 151--203,
  [\href{http://xxx.lanl.gov/abs/hep-ph/0211352}{{\tt hep-ph/0211352}}].

\bibitem{Dawson:2002tg}
S.~Dawson, L.~H. Orr, L.~Reina, and D.~Wackeroth, {\em Phys. Rev.} {\bf D67}
  (2003) 071503, [\href{http://xxx.lanl.gov/abs/hep-ph/0211438}{{\tt
  hep-ph/0211438}}].

\bibitem{Dawson:2003zu}
S.~Dawson, C.~Jackson, L.~H. Orr, L.~Reina, and D.~Wackeroth, {\em Phys. Rev.}
  {\bf D68} (2003) 034022, [\href{http://xxx.lanl.gov/abs/hep-ph/0305087}{{\tt
  hep-ph/0305087}}].

\bibitem{CMStdr}
 {\em CMS Physics, Technical Design Report, Vol.~II Physics Performance}
  [\href{http://xxx.lanl.gov/abs/report CERN/LHCC 2006-021}{{\tt report
  CERN/LHCC 2006-021}}].

\bibitem{Belyaev:2002ua}
A.~Belyaev and L.~Reina, {\em JHEP} {\bf 08} (2002) 041,
  [\href{http://xxx.lanl.gov/abs/hep-ph/0205270}{{\tt hep-ph/0205270}}].

\bibitem{Aglietti:2006ne}
U.~Aglietti {\em et.~al.}, \href{http://xxx.lanl.gov/abs/hep-ph/0612172}{{\tt
  hep-ph/0612172}}.

\bibitem{Han:1991ia}
T.~Han and S.~Willenbrock, {\em Phys. Lett.} {\bf B273} (1991) 167--172.

\bibitem{Hamberg:1990np}
R.~Hamberg, W.~L. van Neerven, and T.~Matsuura, {\em Nucl. Phys.} {\bf B359}
  (1991) 343--405.

\bibitem{Brein:2003wg}
O.~Brein, A.~Djouadi, and R.~Harlander, {\em Phys. Lett.} {\bf B579} (2004)
  149--156, [\href{http://xxx.lanl.gov/abs/hep-ph/0307206}{{\tt
  hep-ph/0307206}}].

\bibitem{Ciccolini:2003jy}
M.~L. Ciccolini, S.~Dittmaier, and M.~Kramer, {\em Phys. Rev.} {\bf D68} (2003)
  073003, [\href{http://xxx.lanl.gov/abs/hep-ph/0306234}{{\tt
  hep-ph/0306234}}].

\bibitem{hnnlo}
 {\em {\tt http://theory.fi.infn.it/grazzini/codes.html}}.

\bibitem{Martin:2004ir}
A.~D. Martin, R.~G. Roberts, W.~J. Stirling, and R.~S. Thorne, {\em Phys.
  Lett.} {\bf B604} (2004) 61--68,
  [\href{http://xxx.lanl.gov/abs/hep-ph/0410230}{{\tt hep-ph/0410230}}].

\bibitem{Catani:1997xc}
S.~Catani and B.~R. Webber, {\em JHEP} {\bf 10} (1997) 005,
  [\href{http://xxx.lanl.gov/abs/hep-ph/9710333}{{\tt hep-ph/9710333}}].

\bibitem{Asai:2004ws}
S.~Asai {\em et.~al.}, {\em Eur. Phys. J.} {\bf C32S2} (2004) 19--54.

\bibitem{Abdullin:2005yn}
S.~Abdullin {\em et.~al.}, {\em Eur. Phys. J.} {\bf C39S2} (2005) 41--61.

\bibitem{Hankele:2006ma}
V.~Hankele, G.~Klamke, D.~Zeppenfeld, and T.~Figy, {\em Phys. Rev.} {\bf D74}
  (2006) 095001, [\href{http://xxx.lanl.gov/abs/hep-ph/0609075}{{\tt
  hep-ph/0609075}}].

\bibitem{Barger:1994zq}
V.~D. Barger, R.~J.~N. Phillips, and D.~Zeppenfeld, {\em Phys. Lett.} {\bf
  B346} (1995) 106--114, [\href{http://xxx.lanl.gov/abs/hep-ph/9412276}{{\tt
  hep-ph/9412276}}].

\bibitem{Rainwater:1997dg}
D.~L. Rainwater and D.~Zeppenfeld, {\em JHEP} {\bf 12} (1997) 005,
  [\href{http://xxx.lanl.gov/abs/hep-ph/9712271}{{\tt hep-ph/9712271}}].

\bibitem{Rainwater:1998kj}
D.~L. Rainwater, D.~Zeppenfeld, and K.~Hagiwara, {\em Phys. Rev.} {\bf D59}
  (1999) 014037, [\href{http://xxx.lanl.gov/abs/hep-ph/9808468}{{\tt
  hep-ph/9808468}}].

\bibitem{Rainwater:1999sd}
D.~L. Rainwater and D.~Zeppenfeld, {\em Phys. Rev.} {\bf D60} (1999) 113004,
  [\href{http://xxx.lanl.gov/abs/hep-ph/9906218}{{\tt hep-ph/9906218}}].

\bibitem{Spira:1997dg}
M.~Spira, {\em Fortsch. Phys.} {\bf 46} (1998) 203--284,
  [\href{http://xxx.lanl.gov/abs/hep-ph/9705337}{{\tt hep-ph/9705337}}].

\bibitem{Zeppenfeld:vbfnlo}
M.~B{\"a}hr {\em et.~al.},
  \href{http://xxx.lanl.gov/abs/http://www-itp.particle.uni-karlsruhe.de/\~{}v%
bfnloweb/}{{\tt http://www-itp.particle.uni-karlsruhe.de/\~{}vbfnloweb/}}.

\bibitem{Spira:vv2h}
M.~Spira,
  \href{http://xxx.lanl.gov/abs/http://people.web.psi.ch/spira/vv2h}{{\tt
  http://people.web.psi.ch/spira/vv2h}}.

\bibitem{Eidelman:2004wy}
S.~Eidelman {\em et.~al.},, {\bf Particle Data Group} Collaboration {\em Phys.
  Lett.} {\bf B592} (2004) 1.

\bibitem{Denner:2005fg}
A.~Denner, S.~Dittmaier, M.~Roth, and L.~H. Wieders, {\em Nucl. Phys.} {\bf
  B724} (2005) 247--294, [\href{http://xxx.lanl.gov/abs/hep-ph/0505042}{{\tt
  hep-ph/0505042}}].

\bibitem{Bardin:1988xt}
D.~Y. Bardin, A.~Leike, T.~Riemann, and M.~Sachwitz, {\em Phys. Lett.} {\bf
  B206} (1988) 539--542.

\bibitem{Jegerlehner:2001ca}
F.~Jegerlehner, \href{http://xxx.lanl.gov/abs/hep-ph/0105283}{{\tt
  hep-ph/0105283}}.

\bibitem{Pumplin:2002vw}
J.~Pumplin {\em et.~al.}, {\em JHEP} {\bf 07} (2002) 012,
  [\href{http://xxx.lanl.gov/abs/hep-ph/0201195}{{\tt hep-ph/0201195}}].

\bibitem{Catani:1992zp}
S.~Catani, Y.~L. Dokshitzer, and B.~R. Webber, {\em Phys. Lett.} {\bf B285}
  (1992) 291--299.

\bibitem{Blazey:2000qt}
G.~C. Blazey {\em et.~al.}, \href{http://xxx.lanl.gov/abs/hep-ex/0005012}{{\tt
  hep-ex/0005012}}.

\bibitem{Cahn:1983ip}
R.~N. Cahn and S.~Dawson, {\em Phys. Lett.} {\bf B136} (1984) 196.

\bibitem{Dicus:1985zg}
D.~A. Dicus and S.~S.~D. Willenbrock, {\em Phys. Rev.} {\bf D32} (1985) 1642.

\bibitem{Altarelli:1987ue}
G.~Altarelli, B.~Mele, and F.~Pitolli, {\em Nucl. Phys.} {\bf B287} (1987)
  205--224.

\bibitem{Plehn:2001nj}
T.~Plehn, D.~L. Rainwater, and D.~Zeppenfeld, {\em Phys. Rev. Lett.} {\bf 88}
  (2002) 051801, [\href{http://xxx.lanl.gov/abs/hep-ph/0105325}{{\tt
  hep-ph/0105325}}].

\bibitem{Berger:2004pc}
E.~L. Berger and J.~Campbell, {\em Phys. Rev.} {\bf D70} (2004) 073011,
  [\href{http://xxx.lanl.gov/abs/hep-ph/0403194}{{\tt hep-ph/0403194}}].

\bibitem{DelDuca:2004wt}
V.~Del~Duca, A.~Frizzo, and F.~Maltoni, {\em JHEP} {\bf 05} (2004) 064,
  [\href{http://xxx.lanl.gov/abs/hep-ph/0404013}{{\tt hep-ph/0404013}}].

\bibitem{Wilczek:1977zn}
F.~Wilczek, {\em Phys. Rev. Lett.} {\bf 39} (1977) 1304.

\bibitem{Andersen:2006ag}
J.~R. Andersen and J.~M. Smillie, {\em Phys. Rev.} {\bf D75} (2007) 037301,
  [\href{http://xxx.lanl.gov/abs/hep-ph/0611281}{{\tt hep-ph/0611281}}].

\bibitem{Andersen:2007mp}
J.~R. Andersen, T.~Binoth, G.~Heinrich, and J.~M. Smillie,
  \href{http://xxx.lanl.gov/abs/arXiv:0709.3513 [hep-ph]}{{\tt arXiv:0709.3513
  [hep-ph]}}.

\bibitem{Binoth:1999sp}
T.~Binoth, J.~P. Guillet, and G.~Heinrich, {\em Nucl. Phys.} {\bf B572} (2000)
  361--386, [\href{http://xxx.lanl.gov/abs/hep-ph/9911342}{{\tt
  hep-ph/9911342}}].

\bibitem{Binoth:2005ff}
T.~Binoth, J.~P. Guillet, G.~Heinrich, E.~Pilon, and C.~Schubert, {\em JHEP}
  {\bf 10} (2005) 015, [\href{http://xxx.lanl.gov/abs/hep-ph/0504267}{{\tt
  hep-ph/0504267}}].

\bibitem{Binoth:2006mf}
T.~Binoth, M.~Ciccolini, N.~Kauer, and M.~Kramer, {\em JHEP} {\bf 12} (2006)
  046, [\href{http://xxx.lanl.gov/abs/hep-ph/0611170}{{\tt hep-ph/0611170}}].

\bibitem{Binoth:2006ym}
T.~Binoth, S.~Karg, N.~Kauer, and R.~Ruckl, {\em Phys. Rev.} {\bf D74} (2006)
  113008, [\href{http://xxx.lanl.gov/abs/hep-ph/0608057}{{\tt
  hep-ph/0608057}}].

\bibitem{Binoth:2003xk}
T.~Binoth, J.~P. Guillet, and F.~Mahmoudi, {\em JHEP} {\bf 02} (2004) 057,
  [\href{http://xxx.lanl.gov/abs/hep-ph/0312334}{{\tt hep-ph/0312334}}].

\bibitem{Dixon:priv}
L.~J. Dixon and Y.~Sofianatos,.
\newblock in preparation.

\bibitem{Giele:1991vf}
W.~T. Giele and E.~W.~N. Glover, {\em Phys. Rev.} {\bf D46} (1992) 1980--2010.

\bibitem{Giele:1993dj}
W.~T. Giele, E.~W.~N. Glover, and D.~A. Kosower, {\em Nucl. Phys.} {\bf B403}
  (1993) 633--670, [\href{http://xxx.lanl.gov/abs/hep-ph/9302225}{{\tt
  hep-ph/9302225}}].

\bibitem{PDBook}
W.-M. {Yao} {\em et.~al.}, {\em {Journal of Physics G}} {\bf 33} (2006).

\bibitem{Forshaw:2007vb}
J.~R. Forshaw and M.~Sjodahl, {\em JHEP} {\bf 09} (2007) 119,
  [\href{http://xxx.lanl.gov/abs/arXiv:0705.1504 [hep-ph]}{{\tt arXiv:0705.1504
  [hep-ph]}}].

\bibitem{Andersen:2008ue}
J.~R. Andersen and C.~D. White, \href{http://xxx.lanl.gov/abs/arXiv:0802.2858
  [hep-ph]}{{\tt arXiv:0802.2858 [hep-ph]}}.

\bibitem{Klamke:2007cu}
G.~Klamke and D.~Zeppenfeld, {\em JHEP} {\bf 04} (2007) 052,
  [\href{http://xxx.lanl.gov/abs/hep-ph/0703202}{{\tt hep-ph/0703202}}].

\bibitem{Barger:1995zq}
V.~D. Barger, R.~J.~N. Phillips, and D.~Zeppenfeld,.

\bibitem{Fadin:1975cb}
V.~S. Fadin, E.~A. Kuraev, and L.~N. Lipatov, {\em Phys. Lett.} {\bf B60}
  (1975) 50--52.

\bibitem{DelDuca:2003ba}
V.~Del~Duca, W.~Kilgore, C.~Oleari, C.~R. Schmidt, and D.~Zeppenfeld, {\em
  Phys. Rev.} {\bf D67} (2003) 073003,
  [\href{http://xxx.lanl.gov/abs/hep-ph/0301013}{{\tt hep-ph/0301013}}].

\bibitem{DelDuca:1995hf}
V.~Del~Duca,
  \href{http://xxx.lanl.gov/abs/http://arXiv.org/abs/hep-ph/9503226}{{\tt
  http://arXiv.org/abs/hep-ph/9503226}}.

\bibitem{Ciafaloni:1987ur}
M.~Ciafaloni, {\em Nucl. Phys.} {\bf B296} (1988) 49.

\bibitem{Catani:1989yc}
S.~Catani, F.~Fiorani, and G.~Marchesini, {\em Phys. Lett.} {\bf B234} (1990)
  339.

\bibitem{Kwiecinski:1996td}
J.~Kwiecinski, A.~D. Martin, and P.~J. Sutton, {\em Z. Phys.} {\bf C71} (1996)
  585--594, [\href{http://xxx.lanl.gov/abs/hep-ph/9602320}{{\tt
  hep-ph/9602320}}].

\bibitem{Alwall:2007st}
J.~Alwall {\em et.~al.}, \href{http://xxx.lanl.gov/abs/arXiv:0706.2334
  [hep-ph]}{{\tt arXiv:0706.2334 [hep-ph]}}.

\bibitem{Andersen:2006sp}
J.~R. Andersen, {\em Phys. Lett.} {\bf B639} (2006) 290--293,
  [\href{http://xxx.lanl.gov/abs/hep-ph/0602182}{{\tt hep-ph/0602182}}].

\bibitem{Butterworth:2002xg}
J.~M. Butterworth, J.~P. Couchman, B.~E. Cox, and B.~M. Waugh, {\em Comput.
  Phys. Commun.} {\bf 153} (2003) 85--96,
  [\href{http://xxx.lanl.gov/abs/hep-ph/0210022}{{\tt hep-ph/0210022}}].

\bibitem{Hankele:2006ja}
V.~Hankele, G.~Klamke, and D.~Zeppenfeld,
  \href{http://xxx.lanl.gov/abs/hep-ph/0605117}{{\tt hep-ph/0605117}}.

\bibitem{Mele:1990bq}
B.~Mele, P.~Nason, and G.~Ridolfi, {\em Nucl. Phys.} {\bf B357} (1991)
  409--438.

\bibitem{Ohnemus:1994ff}
J.~Ohnemus, {\em Phys. Rev.} {\bf D50} (1994) 1931--1945,
  [\href{http://xxx.lanl.gov/abs/hep-ph/9403331}{{\tt hep-ph/9403331}}].

\bibitem{Dicus:1987dj}
D.~A. Dicus, C.~Kao, and W.~W. Repko, {\em Phys. Rev.} {\bf D36} (1987) 1570.

\bibitem{Glover:1988rg}
E.~W.~N. Glover and J.~J. van~der Bij, {\em Nucl. Phys.} {\bf B321} (1989) 561.

\bibitem{Matsuura:1991pj}
T.~Matsuura and J.~J. van~der Bij, {\em Z. Phys.} {\bf C51} (1991) 259--266.

\bibitem{Zecher:1994kb}
C.~Zecher, T.~Matsuura, and J.~J. van~der Bij, {\em Z. Phys.} {\bf C64} (1994)
  219--226, [\href{http://xxx.lanl.gov/abs/hep-ph/9404295}{{\tt
  hep-ph/9404295}}].

\bibitem{Hahn:1998yk}
T.~Hahn and M.~Perez-Victoria, {\em Comput. Phys. Commun.} {\bf 118} (1999)
  153--165, [\href{http://xxx.lanl.gov/abs/hep-ph/9807565}{{\tt
  hep-ph/9807565}}].

\bibitem{Hahn:2000kx}
T.~Hahn, {\em Comput. Phys. Commun.} {\bf 140} (2001) 418--431,
  [\href{http://xxx.lanl.gov/abs/hep-ph/0012260}{{\tt hep-ph/0012260}}].

\bibitem{Berends:1994pv}
F.~A. Berends, R.~Pittau, and R.~Kleiss, {\em Nucl. Phys.} {\bf B424} (1994)
  308--342, [\href{http://xxx.lanl.gov/abs/hep-ph/9404313}{{\tt
  hep-ph/9404313}}].

\bibitem{Kauer:2002sn}
N.~Kauer, {\em Phys. Rev.} {\bf D67} (2003) 054013,
  [\href{http://xxx.lanl.gov/abs/hep-ph/0212091}{{\tt hep-ph/0212091}}].

\bibitem{Dixon:1998py}
L.~J. Dixon, Z.~Kunszt, and A.~Signer, {\em Nucl. Phys.} {\bf B531} (1998)
  3--23, [\href{http://xxx.lanl.gov/abs/hep-ph/9803250}{{\tt hep-ph/9803250}}].

\bibitem{GG2ZZ}
\texttt{http://hepsource.sf.net/programs/GG2ZZ/}.

\bibitem{Boos:2001cv}
E.~Boos {\em et.~al.}, \href{http://xxx.lanl.gov/abs/hep-ph/0109068}{{\tt
  hep-ph/0109068}}.

\bibitem{Alwall:2006yp}
J.~Alwall {\em et.~al.}, {\em Comput. Phys. Commun.} {\bf 176} (2007) 300--304,
  [\href{http://xxx.lanl.gov/abs/hep-ph/0609017}{{\tt hep-ph/0609017}}].

\bibitem{Futyan:2007zz}
D.~Futyan, D.~Fortin, and D.~Giordano, {\em J. Phys.} {\bf G34} (2007)
  N315--N342.

\bibitem{sjostrand-2001-135}
T.~Sjostrand, P.~Eden, C.~Friberg, L.~Lonnblad, G.~Miu, S.~Mrenna, and
  E.~Norrbin, {\em Computer Physics Communications} {\bf 135} (2001) 238.

\bibitem{Maltoni:2002qb}
F.~Maltoni and T.~Stelzer, {\em JHEP} {\bf 02} (2003) 027,
  [\href{http://xxx.lanl.gov/abs/hep-ph/0208156}{{\tt hep-ph/0208156}}].

\bibitem{Bartalini:2006}
P.~Bartalini {\em et.~al.},. CERN-CMS-NOTE-2006-130.

\bibitem{Plehn:1999xi}
T.~Plehn, D.~L. Rainwater, and D.~Zeppenfeld, {\em Phys. Rev.} {\bf D61} (2000)
  093005, [\href{http://xxx.lanl.gov/abs/hep-ph/9911385}{{\tt
  hep-ph/9911385}}].

\bibitem{Takahashi:qqh}
C.~Foudas, A.~Nikitenko, and M.~Takahashi,. CERN-CMS-NOTE-2006-088.

\bibitem{Natasha:CJV}
N.~Ilina, V.~Gavrilov, and A.~Krokhotin,. CERN-CMS-INTERNAL NOTE-2004-040.

\bibitem{ETH:HWW}
G.~Davatz, M.~Dittmar, and A.-S. Giolo-Nicollerat,. CERN-CMS-NOTE-2006-047.

\bibitem{Dokshitzer:1991he}
Y.~L. Dokshitzer, V.~A. Khoze, and T.~Sjostrand, {\em Phys. Lett.} {\bf B274}
  (1992) 116--121.

\bibitem{Mangano:2002ea}
M.~L. Mangano, M.~Moretti, F.~Piccinini, R.~Pittau, and A.~D. Polosa, {\em
  JHEP} {\bf 07} (2003) 001,
  [\href{http://xxx.lanl.gov/abs/hep-ph/0206293}{{\tt hep-ph/0206293}}].

\bibitem{Mangano:2006rw}
M.~L. Mangano, M.~Moretti, F.~Piccinini, and M.~Treccani, {\em JHEP} {\bf 01}
  (2007) 013, [\href{http://xxx.lanl.gov/abs/hep-ph/0611129}{{\tt
  hep-ph/0611129}}].

\bibitem{Hoche:2006ph}
S.~Hoche {\em et.~al.}, \href{http://xxx.lanl.gov/abs/hep-ph/0602031}{{\tt
  hep-ph/0602031}}.

\bibitem{Sjostrand:2006za}
T.~Sjostrand, S.~Mrenna, and P.~Skands, {\em JHEP} {\bf 05} (2006) 026,
  [\href{http://xxx.lanl.gov/abs/hep-ph/0603175}{{\tt hep-ph/0603175}}].

\bibitem{Acosta:2006bp}
D.~Acosta {\em et.~al.},. CERN-CMS-NOTE-2006-067.

\bibitem{Assamagan:2004mu}
K.~A. Assamagan {\em et.~al.},, {\bf Higgs Working Group} Collaboration
  \href{http://xxx.lanl.gov/abs/hep-ph/0406152}{{\tt hep-ph/0406152}}.

\bibitem{Mangano:2002wn}
M.~L. Mangano, M.~Moretti, F.~Piccinini, R.~Pittau, and A.~D. Polosa, {\em
  Phys. Lett.} {\bf B556} (2003) 50--60,
  [\href{http://xxx.lanl.gov/abs/hep-ph/0210261}{{\tt hep-ph/0210261}}].

\bibitem{Gabrielli:2007wf}
E.~Gabrielli {\em et.~al.}, {\em Nucl. Phys.} {\bf B781} (2007) 64--84,
  [\href{http://xxx.lanl.gov/abs/hep-ph/0702119}{{\tt hep-ph/0702119}}].

\bibitem{Djouadi:1997yw}
A.~Djouadi, J.~Kalinowski, and M.~Spira, {\em Comput. Phys. Commun.} {\bf 108}
  (1998) 56--74, [\href{http://xxx.lanl.gov/abs/hep-ph/9704448}{{\tt
  hep-ph/9704448}}].

\bibitem{Lai:1999wy}
H.~L. Lai {\em et.~al.},, {\bf CTEQ} Collaboration {\em Eur. Phys. J.} {\bf
  C12} (2000) 375--392, [\href{http://xxx.lanl.gov/abs/hep-ph/9903282}{{\tt
  hep-ph/9903282}}].

\bibitem{Abe:1997xk}
F.~Abe {\em et.~al.},, {\bf CDF} Collaboration {\em Phys. Rev.} {\bf D56}
  (1997) 3811--3832.

\bibitem{Korotkikh:2004bz}
V.~L. Korotkikh and A.~M. Snigirev, {\em Phys. Lett.} {\bf B594} (2004)
  171--176, [\href{http://xxx.lanl.gov/abs/hep-ph/0404155}{{\tt
  hep-ph/0404155}}].

\bibitem{Cattaruzza:2005nu}
E.~Cattaruzza, A.~Del~Fabbro, and D.~Treleani, {\em Phys. Rev.} {\bf D72}
  (2005) 034022, [\href{http://xxx.lanl.gov/abs/hep-ph/0507052}{{\tt
  hep-ph/0507052}}].

\bibitem{Spira:1993bb}
M.~Spira, A.~Djouadi, D.~Graudenz, and P.~M. Zerwas, {\em Phys. Lett.} {\bf
  B318} (1993) 347--353.

\bibitem{Harlander:2002vv}
R.~V. Harlander and W.~B. Kilgore, {\em JHEP} {\bf 10} (2002) 017,
  [\href{http://xxx.lanl.gov/abs/hep-ph/0208096}{{\tt hep-ph/0208096}}].

\bibitem{Anastasiou:2002wq}
C.~Anastasiou and K.~Melnikov, {\em Phys. Rev.} {\bf D67} (2003) 037501,
  [\href{http://xxx.lanl.gov/abs/hep-ph/0208115}{{\tt hep-ph/0208115}}].

\bibitem{Ravindran:2005vv}
V.~Ravindran, {\em Nucl. Phys.} {\bf B746} (2006) 58--76,
  [\href{http://xxx.lanl.gov/abs/hep-ph/0512249}{{\tt hep-ph/0512249}}].

\bibitem{Ravindran:2006cg}
V.~Ravindran, {\em Nucl. Phys.} {\bf B752} (2006) 173--196,
  [\href{http://xxx.lanl.gov/abs/hep-ph/0603041}{{\tt hep-ph/0603041}}].

\bibitem{Dawson:1996xz}
S.~Dawson, A.~Djouadi, and M.~Spira, {\em Phys. Rev. Lett.} {\bf 77} (1996)
  16--19, [\href{http://xxx.lanl.gov/abs/hep-ph/9603423}{{\tt
  hep-ph/9603423}}].

\bibitem{Harlander:2005if}
R.~V. Harlander and F.~Hofmann, {\em JHEP} {\bf 03} (2006) 050,
  [\href{http://xxx.lanl.gov/abs/hep-ph/0507041}{{\tt hep-ph/0507041}}].

\bibitem{Carena:2002qg}
M.~S. Carena, S.~Heinemeyer, C.~E.~M. Wagner, and G.~Weiglein, {\em Eur. Phys.
  J.} {\bf C26} (2003) 601--607,
  [\href{http://xxx.lanl.gov/abs/hep-ph/0202167}{{\tt hep-ph/0202167}}].

\bibitem{Djouadi:2005gj}
A.~Djouadi, \href{http://xxx.lanl.gov/abs/hep-ph/0503173}{{\tt
  hep-ph/0503173}}.

\bibitem{Ellis:1975ap}
J.~R. Ellis, M.~K. Gaillard, and D.~V. Nanopoulos, {\em Nucl. Phys.} {\bf B106}
  (1976) 292.

\bibitem{Shifman:1979eb}
M.~A. Shifman, A.~I. Vainshtein, M.~B. Voloshin, and V.~I. Zakharov, {\em Sov.
  J. Nucl. Phys.} {\bf 30} (1979) 711--716.

\bibitem{Badelek:2001xb}
B.~Badelek {\em et.~al.},, {\bf ECFA/DESY Photon Collider Working Group}
  Collaboration {\em Int. J. Mod. Phys.} {\bf A19} (2004) 5097--5186,
  [\href{http://xxx.lanl.gov/abs/hep-ex/0108012}{{\tt hep-ex/0108012}}].

\bibitem{Muhlleitner:2005pr}
M.~M. Muhlleitner and P.~M. Zerwas, {\em Acta Phys. Polon.} {\bf B37} (2006)
  1021--1038, [\href{http://xxx.lanl.gov/abs/hep-ph/0511339}{{\tt
  hep-ph/0511339}}].

\bibitem{Muhlleitner:2001kw}
M.~M. Muhlleitner, M.~Kramer, M.~Spira, and P.~M. Zerwas, {\em Phys. Lett.}
  {\bf B508} (2001) 311--316,
  [\href{http://xxx.lanl.gov/abs/hep-ph/0101083}{{\tt hep-ph/0101083}}].

\bibitem{Muhlleitner:2000jj}
M.~M. Muhlleitner, \href{http://xxx.lanl.gov/abs/hep-ph/0008127}{{\tt
  hep-ph/0008127}}.

\bibitem{Asner:2001ia}
D.~M. Asner, J.~B. Gronberg, and J.~F. Gunion, {\em Phys. Rev.} {\bf D67}
  (2003) 035009, [\href{http://xxx.lanl.gov/abs/hep-ph/0110320}{{\tt
  hep-ph/0110320}}].

\bibitem{Niezurawski:2005cr}
P.~Niezurawski, A.~F. Zarnecki, and M.~Krawczyk,
  \href{http://xxx.lanl.gov/abs/hep-ph/0507006}{{\tt hep-ph/0507006}}.

\bibitem{Spira:2006aa}
M.~Spira, P.~Niezurawski, M.~Krawczyk, and A.~F. Zarnecki, {\em Pramana} {\bf
  69} (2007) 931--936, [\href{http://xxx.lanl.gov/abs/hep-ph/0612369}{{\tt
  hep-ph/0612369}}].

\bibitem{Kramer:1996iq}
M.~Kramer, E.~Laenen, and M.~Spira, {\em Nucl. Phys.} {\bf B511} (1998)
  523--549, [\href{http://xxx.lanl.gov/abs/hep-ph/9611272}{{\tt
  hep-ph/9611272}}].

\bibitem{Carena:2007aq}
M.~S. Carena, A.~Menon, and C.~E.~M. Wagner, {\em Phys. Rev.} {\bf D76} (2007)
  035004, [\href{http://xxx.lanl.gov/abs/arXiv:0704.1143 [hep-ph]}{{\tt
  arXiv:0704.1143 [hep-ph]}}].

\bibitem{Carena:1998gk}
M.~S. Carena, S.~Mrenna, and C.~E.~M. Wagner, {\em Phys. Rev.} {\bf D60} (1999)
  075010, [\href{http://xxx.lanl.gov/abs/hep-ph/9808312}{{\tt
  hep-ph/9808312}}].

\bibitem{Dicus:1998hs}
D.~Dicus, T.~Stelzer, Z.~Sullivan, and S.~Willenbrock, {\em Phys. Rev.} {\bf
  D59} (1999) 094016, [\href{http://xxx.lanl.gov/abs/hep-ph/9811492}{{\tt
  hep-ph/9811492}}].

\bibitem{Maltoni:2003pn}
F.~Maltoni, Z.~Sullivan, and S.~Willenbrock, {\em Phys. Rev.} {\bf D67} (2003)
  093005, [\href{http://xxx.lanl.gov/abs/hep-ph/0301033}{{\tt
  hep-ph/0301033}}].

\bibitem{Brein:2007da}
O.~Brein and W.~Hollik, {\em Phys. Rev.} {\bf D76} (2007) 035002,
  [\href{http://xxx.lanl.gov/abs/arXiv:0705.2744 [hep-ph]}{{\tt arXiv:0705.2744
  [hep-ph]}}].

\bibitem{Carena:1999py}
M.~S. Carena, D.~Garcia, U.~Nierste, and C.~E.~M. Wagner, {\em Nucl. Phys.}
  {\bf B577} (2000) 88--120,
  [\href{http://xxx.lanl.gov/abs/hep-ph/9912516}{{\tt hep-ph/9912516}}].

\bibitem{Hall:1993gn}
L.~J. Hall, R.~Rattazzi, and U.~Sarid, {\em Phys. Rev.} {\bf D50} (1994)
  7048--7065, [\href{http://xxx.lanl.gov/abs/hep-ph/9306309}{{\tt
  hep-ph/9306309}}].

\bibitem{Guasch:2003cv}
J.~Guasch, P.~Hafliger, and M.~Spira, {\em Phys. Rev.} {\bf D68} (2003) 115001,
  [\href{http://xxx.lanl.gov/abs/hep-ph/0305101}{{\tt hep-ph/0305101}}].

\bibitem{Carena:1994bv}
M.~Carena, M.~Olechowski, S.~Pokorski, and C.~E.~M. Wagner, {\em Nucl. Phys.}
  {\bf B426} (1994) 269--300,
  [\href{http://xxx.lanl.gov/abs/hep-ph/9402253}{{\tt hep-ph/9402253}}].

\bibitem{Haber:2000kq}
H.~E. Haber {\em et.~al.}, {\em Phys. Rev.} {\bf D63} (2001) 055004,
  [\href{http://xxx.lanl.gov/abs/hep-ph/0007006}{{\tt hep-ph/0007006}}].

\bibitem{Brein:2007ej}
O.~Brein and W.~Hollik, \href{http://xxx.lanl.gov/abs/arXiv:0710.4781
  [hep-ph]}{{\tt arXiv:0710.4781 [hep-ph]}}.

\bibitem{Nilles:1983ge}
H.~P. Nilles, {\em Phys. Rept.} {\bf 110} (1984) 1.

\bibitem{Haber:1985rc}
H.~E. Haber and G.~L. Kane, {\em Phys. Rept.} {\bf 117} (1985) 75.

\bibitem{Barbieri:1987xf}
R.~Barbieri, {\em Riv. Nuovo Cim.} {\bf 11N4} (1988) 1--45.

\bibitem{:2001xy}
{\bf LEP Higgs Working Group for Higgs boson searches} Collaboration
  \href{http://xxx.lanl.gov/abs/hep-ex/0107031}{{\tt hep-ex/0107031}}.

\bibitem{LEPchargedHiggsProc}
{\bf LEP Higgs Working Group for Higgs boson searches} CollaborationP.~Lutz, ,,
  2007.

\bibitem{LEPchargedHiggs}
{\bf LEP Higgs Working Group for Higgs boson searches} Collaboration.
\newblock in preparation.

\bibitem{Abulencia:2005jd}
A.~Abulencia {\em et.~al.},, {\bf CDF} Collaboration {\em Phys. Rev. Lett.}
  {\bf 96} (2006) 042003, [\href{http://xxx.lanl.gov/abs/hep-ex/0510065}{{\tt
  hep-ex/0510065}}].

\bibitem{ATLASTDR}
{ATLAS Collaboration},, {\bf ATLAS} Collaboration. CERN-LHCC-99-14 and
  CERN-LHCC-99-15.

\bibitem{Carena:2005ek}
M.~S. Carena, S.~Heinemeyer, C.~E.~M. Wagner, and G.~Weiglein, {\em Eur. Phys.
  J.} {\bf C45} (2006) 797--814,
  [\href{http://xxx.lanl.gov/abs/hep-ph/0511023}{{\tt hep-ph/0511023}}].

\bibitem{AguilarSaavedra:2001rg}
J.~A. Aguilar-Saavedra {\em et.~al.},, {\bf ECFA/DESY LC Physics Working Group}
  Collaboration \href{http://xxx.lanl.gov/abs/hep-ph/0106315}{{\tt
  hep-ph/0106315}}.

\bibitem{tesla2}
K.~Ackermann {\em et.~al.},. prepared for 4th ECFA / DESY Workshop on Physics
  and Detectors for a 90-GeV to 800-GeV Linear e+ e- Collider, Amsterdam, The
  Netherlands, 1-4 Apr 2003.
\newblock DESY-PROC-2004-01.

\bibitem{Abe:2001npb}
T.~Abe {\em et.~al.},, {\bf American Linear Collider Working Group}
  Collaboration \href{http://xxx.lanl.gov/abs/hep-ex/0106056}{{\tt
  hep-ex/0106056}}.

\bibitem{Abe:2001gc}
K.~Abe {\em et.~al.},, {\bf ACFA Linear Collider Working Group} Collaboration
  \href{http://xxx.lanl.gov/abs/hep-ph/0109166}{{\tt hep-ph/0109166}}.

\bibitem{Heinemeyer:2005gs}
S.~Heinemeyer {\em et.~al.}, {\em ECONF} {\bf C0508141} (2005) ALCPG0214,
  [\href{http://xxx.lanl.gov/abs/hep-ph/0511332}{{\tt hep-ph/0511332}}].

\bibitem{cmsHiggs2}
M.~Hashemi {\em et.~al.},.
\newblock in preparation.

\bibitem{lightHexp}
M.~Baarmand, M.~Hashemi, and A.~Nikitenko,.
\newblock CMS Note 2006/056.

\bibitem{Stelzer:1994ta}
T.~Stelzer and W.~F. Long, {\em Comput. Phys. Commun.} {\bf 81} (1994)
  357--371, [\href{http://xxx.lanl.gov/abs/hep-ph/9401258}{{\tt
  hep-ph/9401258}}].

\bibitem{heavyHexp}
R.~Kinnunen,.
\newblock CMS Note 2006/100.

\bibitem{Alwall:2004xw}
J.~Alwall and J.~Rathsman, {\em JHEP} {\bf 12} (2004) 050,
  [\href{http://xxx.lanl.gov/abs/hep-ph/0409094}{{\tt hep-ph/0409094}}].

\bibitem{Hempfling:1993kv}
R.~Hempfling, {\em Phys. Rev.} {\bf D49} (1994) 6168--6172.

\bibitem{Gennai:2007ys}
S.~Gennai {\em et.~al.}, {\em Eur. Phys. J.} {\bf C52} (2007) 383--395,
  [\href{http://xxx.lanl.gov/abs/arXiv:0704.0619 [hep-ph]}{{\tt arXiv:0704.0619
  [hep-ph]}}].

\bibitem{Plehn:2002vy}
T.~Plehn, {\em Phys. Rev.} {\bf D67} (2003) 014018,
  [\href{http://xxx.lanl.gov/abs/hep-ph/0206121}{{\tt hep-ph/0206121}}].

\bibitem{Berger:2003sm}
E.~L. Berger, T.~Han, J.~Jiang, and T.~Plehn, {\em Phys. Rev.} {\bf D71} (2005)
  115012, [\href{http://xxx.lanl.gov/abs/hep-ph/0312286}{{\tt
  hep-ph/0312286}}].

\bibitem{Heinemeyer:1998yj}
S.~Heinemeyer, W.~Hollik, and G.~Weiglein, {\em Comput. Phys. Commun.} {\bf
  124} (2000) 76--89, [\href{http://xxx.lanl.gov/abs/hep-ph/9812320}{{\tt
  hep-ph/9812320}}].

\bibitem{Heinemeyer:1998np}
S.~Heinemeyer, W.~Hollik, and G.~Weiglein, {\em Eur. Phys. J.} {\bf C9} (1999)
  343--366, [\href{http://xxx.lanl.gov/abs/hep-ph/9812472}{{\tt
  hep-ph/9812472}}].

\bibitem{Degrassi:2002fi}
G.~Degrassi, S.~Heinemeyer, W.~Hollik, P.~Slavich, and G.~Weiglein, {\em Eur.
  Phys. J.} {\bf C28} (2003) 133--143,
  [\href{http://xxx.lanl.gov/abs/hep-ph/0212020}{{\tt hep-ph/0212020}}].

\bibitem{Frank:2006yh}
M.~Frank {\em et.~al.}, {\em JHEP} {\bf 02} (2007) 047,
  [\href{http://xxx.lanl.gov/abs/hep-ph/0611326}{{\tt hep-ph/0611326}}].

\bibitem{ida}
J.~Malmgren and K.~Johansson, {\em Nucl. Inst. Methods A} {\bf 403} (1998) 481.

\bibitem{Sjostrand:2003wg}
T.~Sjostrand, L.~Lonnblad, S.~Mrenna, and P.~Skands,
  \href{http://xxx.lanl.gov/abs/hep-ph/0308153}{{\tt hep-ph/0308153}}.

\bibitem{Jadach:1990mz}
S.~Jadach, J.~H. Kuhn, and Z.~Was, {\em Comput. Phys. Commun.} {\bf 64} (1990)
  275--299.

\bibitem{Golonka:2003xt}
P.~Golonka {\em et.~al.}, {\em Comput. Phys. Commun.} {\bf 174} (2006) 818.

\bibitem{SHW}
J.~S. Conway {\em et.~al.}, in {\em Proceedings of the Workshop on Physics at
  Run II -- Supersymmetry/Higgs}, p.~39, Fermilab, 1998.
\newblock \href{http://xxx.lanl.gov/abs/hep-ph/0010338}{{\tt hep-ph/0010338}}.

\bibitem{Alwall:2003tc}
J.~Alwall, C.~Biscarat, S.~Moretti, J.~Rathsman, and A.~Sopczak, {\em Eur.
  Phys. J.} {\bf C39S1} (2005) 37--39,
  [\href{http://xxx.lanl.gov/abs/hep-ph/0312301}{{\tt hep-ph/0312301}}].

\bibitem{ttbarxsec}
V.~M. Abazov {\em et.~al.}, {\em D\O\ note 4879-CONF} (2005).

\bibitem{Assamagan:2002in}
K.~A. Assamagan and Y.~Coadou, {\em Acta Phys. Polon.} {\bf B33} (2002)
  707--720.

\bibitem{Hesselbach:2007jj}
S.~Hesselbach, S.~Moretti, J.~Rathsman, and A.~Sopczak,
  \href{http://xxx.lanl.gov/abs/arXiv:0708.4394 [hep-ph]}{{\tt arXiv:0708.4394
  [hep-ph]}}.

\bibitem{Roy:1991sf}
D.~P. Roy, {\em Phys. Lett.} {\bf B277} (1992) 183--189.

\bibitem{Raychaudhuri:1995cc}
S.~Raychaudhuri and D.~P. Roy, {\em Phys. Rev.} {\bf D53} (1996) 4902--4908,
  [\href{http://xxx.lanl.gov/abs/hep-ph/9507388}{{\tt hep-ph/9507388}}].

\bibitem{Roy:1999xw}
D.~P. Roy, {\em Phys. Lett.} {\bf B459} (1999) 607--614,
  [\href{http://xxx.lanl.gov/abs/hep-ph/9905542}{{\tt hep-ph/9905542}}].

\bibitem{beneke00}
A.~Beneke {\em et.~al.}, {\em hep-ph/0003033} (2000).

\bibitem{Mohn:2007fd}
B.~Mohn, M.~Flechl, and J.~Alwall,. ATL-PHYS-PUB-2007-006.

\bibitem{biscarat}
C.~Biscarat and M.~Dosil, {\em ATL-PHYS-2003-038} (2003).

\bibitem{abdullin}
S.~Abdullin {\em et.~al.}, {\em CMS Note 2003/033} (2003).

\bibitem{assamagan}
K.~A. Assamagan and N.~Gollub, {\em SN-ATLAS-2004-042} (2004).

\bibitem{kinnunen}
R.~Kinnunen, {\em CMS Note 2000/039} (2000).

\bibitem{salmi}
P.~Salmi, R.~Kinnunen, and N.~Stepanov, {\em CMS Note 2002/024} (2002).

\bibitem{lowette}
S.~Lowette, J.~Heyninck, and P.~Vanlaer, {\em CMS Note 2004/017} (2004).

\bibitem{Schael:2006cr}
S.~Schael {\em et.~al.},, {\bf ALEPH} Collaboration {\em Eur. Phys. J.} {\bf
  C47} (2006) 547--587, [\href{http://xxx.lanl.gov/abs/hep-ex/0602042}{{\tt
  hep-ex/0602042}}].

\bibitem{Ghosh:2004cc}
D.~K. Ghosh, R.~M. Godbole, and D.~P. Roy, {\em Phys. Lett.} {\bf B628} (2005)
  131--140, [\href{http://xxx.lanl.gov/abs/hep-ph/0412193}{{\tt
  hep-ph/0412193}}].

\bibitem{Lee:2003nta}
J.~S. Lee {\em et.~al.}, {\em Comput. Phys. Commun.} {\bf 156} (2004) 283--317,
  [\href{http://xxx.lanl.gov/abs/hep-ph/0307377}{{\tt hep-ph/0307377}}].

\bibitem{Beneke:2000hk}
M.~Beneke {\em et.~al.}, \href{http://xxx.lanl.gov/abs/hep-ph/0003033}{{\tt
  hep-ph/0003033}}.

\bibitem{DellaNegra:922757}
M.~Della~Negra, L.~Foà, A.~Herve, and A.~Petrilli,, {\em CMS physics Technical
  Design Report}.
\newblock Technical Design Report CMS. CERN, Geneva, 2006.

\bibitem{Ellis:1993tq}
S.~D. Ellis and D.~E. Soper, {\em Phys. Rev.} {\bf D48} (1993) 3160--3166,
  [\href{http://xxx.lanl.gov/abs/hep-ph/9305266}{{\tt hep-ph/9305266}}].

\bibitem{Agostinelli:2002hh}
S.~Agostinelli {\em et.~al.},, {\bf GEANT4} Collaboration {\em Nucl. Instrum.
  Meth.} {\bf A506} (2003) 250--303.

\bibitem{ptdr1:2006cms}
 {\em CMS Physics Technical Design Report, volume-I} {\bf CERN/LHCC/2006-001}.

\bibitem{reviewNMSSM}
H.P. Nilles, M. Srednicki and D. Wyler, Phys. Lett. B \textbf{120} (1983) 346;
  J.M. Frere, D.R. Jones and S. Raby, Nucl. Phys. B \textbf{222} (1983) 11;
  J.R. Ellis, J.F. Gunion, H.E. Haber, L. Roszkowski and F. Zwirner, Phys. Rev.
  D \textbf{39} (1989) 844; M. Drees, Int. J. Mod. Phys. A \textbf{4} (1989)
  3635; U. Ellwanger, M. Rausch de Traubenberg and C.A. Savoy, Phys. Lett. B
  \textbf{315} (1993) 331; S.F. King and P.L. White, Phys. Rev. D \textbf{52}
  (1995) 4183.

\bibitem{analyses}
For a recent review, see E. Accomando et al., arXiv:hep-ph/0608079.

\bibitem{egh1}
J.~F.~Gunion and C.~Hugonie, JHEP {\bf 0507} (2005) 041 [arXiv:hep-ph/0503203].

\bibitem{nmssmtools}
U.~Ellwanger and C.~Hugonie, Comput. Phys. Commun. {\bf 177} (2007) 399; {\sf
  http://www.th.u-psud.fr/NMHDECAY/nmssmtools.html}.

\bibitem{ICR}
E. Boos, A. Djouadi and A. Nikitenko, Phys. Lett. B {\bf 578} (2004) 384
  [arXiv:hep-ph/0307079].

\bibitem{Djouadi:2008uw}
A.~Djouadi {\em et.~al.}, \href{http://xxx.lanl.gov/abs/arXiv:0801.4321
  [hep-ph]}{{\tt arXiv:0801.4321 [hep-ph]}}.

\bibitem{Hreviews}
A. Djouadi, arXiv:hep-ph/0503172 and arXiv:hep-ph/0503173, Phys. Repts. in
  press.

\bibitem{nmssm-ATLAS}
I. Rottlander, M. Schumacher et al, to appear.

\bibitem{nmssm-CMS}
S. Lehti, S. Nikitenko et al., to appear.

\bibitem{Nilles:1982dy}
H.~P. Nilles, M.~Srednicki, and D.~Wyler, {\em Phys. Lett.} {\bf B120} (1983)
  346.

\bibitem{Frere:1983ag}
J.~M. Frere, D.~R.~T. Jones, and S.~Raby, {\em Nucl. Phys.} {\bf B222} (1983)
  11.

\bibitem{Derendinger:1983bz}
J.~P. Derendinger and C.~A. Savoy, {\em Nucl. Phys.} {\bf B237} (1984) 307.

\bibitem{Ellis:1988er}
J.~R. Ellis, J.~F. Gunion, H.~E. Haber, L.~Roszkowski, and F.~Zwirner, {\em
  Phys. Rev.} {\bf D39} (1989) 844.

\bibitem{Drees:1988fc}
M.~Drees, {\em Int. J. Mod. Phys.} {\bf A4} (1989) 3635.

\bibitem{Franke:1995tc}
F.~Franke and H.~Fraas, {\em Int. J. Mod. Phys.} {\bf A12} (1997) 479--534,
  [\href{http://xxx.lanl.gov/abs/hep-ph/9512366}{{\tt hep-ph/9512366}}].

\bibitem{Ellwanger:2004gz}
U.~Ellwanger, J.~F. Gunion, C.~Hugonie, and S.~Moretti,
  \href{http://xxx.lanl.gov/abs/hep-ph/0401228}{{\tt hep-ph/0401228}}.

\bibitem{Miller:2004uh}
D.~J. Miller and S.~Moretti, \href{http://xxx.lanl.gov/abs/hep-ph/0403137}{{\tt
  hep-ph/0403137}}.

\bibitem{Ellwanger:2005uu}
U.~Ellwanger, J.~F. Gunion, and C.~Hugonie, {\em JHEP} {\bf 07} (2005) 041,
  [\href{http://xxx.lanl.gov/abs/hep-ph/0503203}{{\tt hep-ph/0503203}}].

\bibitem{Djouadi:BMpoints}
A.~Djouadi {\em et.~al.},. These proceedings.

\bibitem{Ellwanger:2006rm}
U.~Ellwanger and C.~Hugonie, {\em Mod. Phys. Lett.} {\bf A22} (2007)
  1581--1590, [\href{http://xxx.lanl.gov/abs/hep-ph/0612133}{{\tt
  hep-ph/0612133}}].

\bibitem{Ellwanger:2004xm}
U.~Ellwanger, J.~F. Gunion, and C.~Hugonie, {\em JHEP} {\bf 02} (2005) 066,
  [\href{http://xxx.lanl.gov/abs/hep-ph/0406215}{{\tt hep-ph/0406215}}].

\bibitem{Ellwanger:2005dv}
U.~Ellwanger and C.~Hugonie, {\em Comput. Phys. Commun.} {\bf 175} (2006)
  290--303, [\href{http://xxx.lanl.gov/abs/hep-ph/0508022}{{\tt
  hep-ph/0508022}}].

\bibitem{Domingo:2007dx}
F.~Domingo and U.~Ellwanger, {\em JHEP} {\bf 12} (2007) 090,
  [\href{http://xxx.lanl.gov/abs/arXiv:0710.3714 [hep-ph]}{{\tt arXiv:0710.3714
  [hep-ph]}}].

\bibitem{Gunion:1996fb}
J.~F. Gunion, H.~E. Haber, and T.~Moroi,
  \href{http://xxx.lanl.gov/abs/hep-ph/9610337}{{\tt hep-ph/9610337}}.

\bibitem{Ellwanger:2001iw}
U.~Ellwanger, J.~F. Gunion, and C.~Hugonie,
  \href{http://xxx.lanl.gov/abs/hep-ph/0111179}{{\tt hep-ph/0111179}}.

\bibitem{Azuelos:2002qw}
G.~Azuelos {\em et.~al.}, \href{http://xxx.lanl.gov/abs/hep-ph/0204031}{{\tt
  hep-ph/0204031}}.

\bibitem{Ellwanger:2003jt}
U.~Ellwanger, J.~F. Gunion, C.~Hugonie, and S.~Moretti,
  \href{http://xxx.lanl.gov/abs/hep-ph/0305109}{{\tt hep-ph/0305109}}.

\bibitem{Weiglein:2004hn}
G.~Weiglein {\em et.~al.},, {\bf LHC/LC Study Group} Collaboration {\em Phys.
  Rept.} {\bf 426} (2006) 47--358,
  [\href{http://xxx.lanl.gov/abs/hep-ph/0410364}{{\tt hep-ph/0410364}}].

\bibitem{Moretti:2006hq}
S.~Moretti, S.~Munir, and P.~Poulose, {\em Phys. Lett.} {\bf B644} (2007)
  241--247, [\href{http://xxx.lanl.gov/abs/hep-ph/0608233}{{\tt
  hep-ph/0608233}}].

\bibitem{Carena:2007jk}
M.~Carena, T.~Han, G.-Y. Huang, and C.~E.~M. Wagner,
  \href{http://xxx.lanl.gov/abs/arXiv:0712.2466 [hep-ph]}{{\tt arXiv:0712.2466
  [hep-ph]}}.

\bibitem{Baffioni:2004gdr}
S.~Baffioni, talk presented at ``GdR Supersym\'etrie 2004'', 5--7 July 2004,
  Clermont-Ferrand, France.

\bibitem{Atlas:TDR}
{\em ATLAS detector and physics performance Technical Design Report}.
\newblock CERN, Geneva, 1999.

\bibitem{Schumacher:2004da}
M.~Schumacher, \href{http://xxx.lanl.gov/abs/hep-ph/0410112}{{\tt
  hep-ph/0410112}}.

\bibitem{Cranmer:2004uz}
K.~Cranmer, B.~Mellado, W.~Quayle, and S.~L. Wu,
  \href{http://xxx.lanl.gov/abs/hep-ph/0401088}{{\tt hep-ph/0401088}}.

\bibitem{Cammin:685523}
J.~Cammin and M.~Schumacher, Tech. Rep. ATL-PHYS-2003-024.

\bibitem{Trefzger:683987}
T.~M. Trefzger and K.~Jakobs, Tech. Rep. ATL-PHYS-2000-015.

\bibitem{Thomas:685421}
J.~Thomas, Tech. Rep. ATL-PHYS-2003-003.

\bibitem{Cavalli:685488}
D.~Cavalli and G.~Negri, Tech. Rep. ATL-PHYS-2003-009.

\bibitem{González:685407}
S.~Gonz\'{a}lez, E.~Ros, and M.~A. Vos, Tech. Rep. ATL-PHYS-2002-021.

\bibitem{Cavalli:683878}
D.~Cavalli and P.~Bosatelli, Tech. Rep. ATL-PHYS-2000-001.

\bibitem{Assamagan:2002ne}
K.~A. Assamagan, Y.~Coadou, and A.~Deandrea, {\em Eur. Phys. J. direct} {\bf
  C4} (2002) 9.

\bibitem{Biscarat:681548}
C.~Biscarat and M.~Dosil, Tech. Rep. ATL-PHYS-2003-038.

\bibitem{Miller:2003ay}
D.~J. Miller, R.~Nevzorov, and P.~M. Zerwas, {\em Nucl. Phys.} {\bf B681}
  (2004) 3--30.

\bibitem{Rolke:2006ve}
W.~A. Rolke and A.~M. Lopez,
  \href{http://xxx.lanl.gov/abs/physics/0606006}{{\tt physics/0606006}}.

\bibitem{Gross:nmssmscan}
E.~Gross,. Several talks. Source code can be obtained from M. Schumacher.

\bibitem{Accomando:2006ga}
E.~Accomando {\em et.~al.}, \href{http://xxx.lanl.gov/abs/hep-ph/0608079}{{\tt
  hep-ph/0608079}}.

\bibitem{Dermisek:2005ar}
R.~Dermisek and J.~F. Gunion, {\em Phys. Rev. Lett.} {\bf 95} (2005) 041801,
  [\href{http://xxx.lanl.gov/abs/hep-ph/0502105}{{\tt hep-ph/0502105}}].

\bibitem{Carena:2002bb}
M.~S. Carena, J.~R. Ellis, S.~Mrenna, A.~Pilaftsis, and C.~E.~M. Wagner, {\em
  Nucl. Phys.} {\bf B659} (2003) 145--178,
  [\href{http://xxx.lanl.gov/abs/hep-ph/0211467}{{\tt hep-ph/0211467}}].

\bibitem{Dobrescu:2000jt}
B.~A. Dobrescu, G.~L. Landsberg, and K.~T. Matchev, {\em Phys. Rev.} {\bf D63}
  (2001) 075003, [\href{http://xxx.lanl.gov/abs/hep-ph/0005308}{{\tt
  hep-ph/0005308}}].

\bibitem{Gerard:2007kn}
J.~M. Gerard and M.~Herquet, {\em Phys. Rev. Lett.} {\bf 98} (2007) 251802,
  [\href{http://xxx.lanl.gov/abs/hep-ph/0703051}{{\tt hep-ph/0703051}}].

\bibitem{Krawczyk:2001pe}
M.~Krawczyk, \href{http://xxx.lanl.gov/abs/hep-ph/0103223}{{\tt
  hep-ph/0103223}}.

\bibitem{Stelzer:2006sp}
T.~Stelzer, S.~Wiesenfeldt, and S.~Willenbrock, {\em Phys. Rev.} {\bf D75}
  (2007) 077701, [\href{http://xxx.lanl.gov/abs/hep-ph/0611242}{{\tt
  hep-ph/0611242}}].

\bibitem{Cheung:2007sva}
K.~Cheung, J.~Song, and Q.-S. Yan, {\em Phys. Rev. Lett.} {\bf 99} (2007)
  031801, [\href{http://xxx.lanl.gov/abs/hep-ph/0703149}{{\tt
  hep-ph/0703149}}].

\bibitem{Plehn:1999nw}
T.~Plehn, D.~L. Rainwater, and D.~Zeppenfeld, {\em Phys. Lett.} {\bf B454}
  (1999) 297--303, [\href{http://xxx.lanl.gov/abs/hep-ph/9902434}{{\tt
  hep-ph/9902434}}].

\bibitem{Cavalli:2002vs}
D.~Cavalli {\em et.~al.}, \href{http://xxx.lanl.gov/abs/hep-ph/0203056}{{\tt
  hep-ph/0203056}}.

\bibitem{Klute2002}
M.~Klute, {\em ATLAS Note} {\bf ATL-PHYS-2002-018} (2002).

\bibitem{Gleyzer:2007}
S.~Gleyzer and H.~Prosper, {\em See: http://cern.ch/sergei/paradigm.pdf for
  details} (2007).

\bibitem{Freund:1984}
J.~Freund, {\em Prentice-Hall, Englewood Cliffs, New Jersey} (1984) 289.

\end{thebibliography}\endgroup

\end{document}